%% file: Tesis.tex
\documentclass[b5paper,11pt]{book}

\usepackage{amsmath,amssymb,amscd,amsthm,dsfont,mathabx,amsfonts,amsbsy}
\usepackage{placeins}
\usepackage{amscd}
\usepackage{color,graphicx}
\usepackage{subfigure}
\usepackage{mathrsfs}
\usepackage{array}

\usepackage{hyperref,bookmark}
\hypersetup{
    colorlinks = true,
    linkcolor = {blue},
    citecolor={red},
    breaklinks=true,
}
\usepackage{cite}
\usepackage{varioref}[2001/09/04]
\usepackage{refstyle}
\usepackage{caption}

\def\>#1{{\mathbf{#1}}}
\newcommand{\diagdots}[3][-25]{%
  \rotatebox{#1}{\makebox[3pt]{\makebox[#2]{\xleaders\hbox{$\cdot$\hskip#3}\hfill\kern0pt}}}%
}

\newcommand{\Tr}{\operatorname{Tr}}
\newcommand{\e}{\operatorname{e}}
\newcommand{\diff}{\mathrm{d}}
\newcommand{\MM}{\operatorname{\mathcal{M}}}
\newcommand{\AM}{\operatorname{\mathcal{A}}}
\newcommand{\NN}{\operatorname{\mathcal{N}}}

\newcommand{\RED}[1]{{\color{red}#1}}
\newcommand{\BLUE}[1]{{\color{blue}#1}}
\definecolor{mygreen}{rgb}{0, 0.7, 0.4}
\definecolor{mybrown}{rgb}{0.6, 0.1, 0.2}
\newcommand{\MYGREEN}[1]{{\color{mygreen}#1}}
\newcommand{\MYBROWN}[1]{{\color{mybrown}#1}}
\newcommand{\changeurlcolor}[1]{\hypersetup{linkcolor=#1}}

\newtheorem{theorem}{Theorem}
\newtheorem{proposition}[theorem]{Proposition}
\newtheorem{definition}[theorem]{Definition}
\newtheorem{lemma}[theorem]{Lemma}

\makeatletter
\def\@makecaption#1{%
  \vskip\abovecaptionskip
  \vskip\belowcaptionskip}
\makeatother

\makeatletter
\def\tagform@#1{\maketag@@@{\color{blue}(#1)}}
\makeatother

\makeatletter
\newcommand*{\myfnsymbolsingle}[1]{%
  \ensuremath{%
    \ifcase#1
    \or 
      \MYBROWN{\star}%
    \or 
      \MYBROWN{*}
    \or 
      \MYBROWN{\dagger}
    \or 
      \mathsection
    \or 
      \mathparagraph
    \else 
      \@ctrerr  
    \fi
  }%
}   
\makeatother

\usepackage{alphalph}
\newalphalph{\myfnsymbolmult}[mult]{\myfnsymbolsingle}{}

\DeclareMathAlphabet{\Ma}{U}{msa}{m}{n}
\DeclareMathAlphabet{\Mb}{U}{msb}{m}{n}

\DeclareMathAlphabet{\Meuf}{U}{euf}{m}{n}
\DeclareSymbolFont{ASMa}{U}{msa}{m}{n}
\DeclareSymbolFont{ASMb}{U}{msb}{m}{n}

\numberwithin{equation}{section}
\numberwithin{theorem}{section}
\numberwithin{figure}{section}

\usepackage[indentafter,nobottomtitles]{titlesec}
\usepackage{titletoc}
\usepackage{fancyhdr}
\newcommand{\clearemptydoublepage}{\newpage{\pagestyle{empty}\cleardoublepage}}
\newcommand{\numerogordopart}[1]{\definecolor{gris}{gray}{0.75}\fontfamily{phv}\selectfont{\fontsize{30mm}{30mm}\selectfont\color{gris}{#1}}}

\titleformat{\chapter}
	[display]
	{\rmfamily\bfseries\scshape\Large}
	{\filleft \numerogordopart{\thechapter}}
	{2ex}
	{\titlerule \vspace{2ex}\filright}
	[\vspace{2ex}\titlerule]

\titleformat{\section}{\vspace{.8ex} \normalfont\bfseries\upshape} {\bf\thesection.}{.5em}{}[\titlerule]
\titleformat{\subsection}{\vspace{.8ex} \normalfont\bfseries\upshape} {\bf\thesubsection.}{.5em}{}

\titlecontents{chapter}
	[0pt]
	{\addvspace{0.5pc}\scshape}
	{\contentsmargin{0pt}%
	\bfseries
		\makebox[0pt][r]{
		\enspace
		}\markboth{Contents}{Contents}
		}
	{\contentsmargin{0pt}%
		\bfseries
		\makebox[0pt][r]{
		\thecontentslabel
		\enspace
		}
		}
	{\bf\hspace\fill\thecontentspage\hspace*{-1.4pc}}
	[\addvspace{0.5pc}]

\begin{document}

\include{titulo}
\clearemptydoublepage

\pagenumbering{roman}
\thispagestyle{empty}
\vspace*{140mm}
\begin{flushright}
{\large \textit{De lo que siembras, recoges.}}
\end{flushright}

\clearemptydoublepage

\thispagestyle{empty}
\include{agradecimientos}

\clearemptydoublepage

\thispagestyle{empty}
\include{Resumen}
\clearemptydoublepage


{\hypersetup{linkcolor=black,linktoc=all}

\hypertarget{tocpage}{}
\bookmark[dest=tocpage,level=-1]{Contents}
\chapter*{Contents} 
\input{Tesisr.toc}
}
\markboth{Contents}{Contents}
\clearemptydoublepage

\pagenumbering{arabic}
\clearemptydoublepage
\include{Tesischap1}

\clearemptydoublepage

\clearemptydoublepage
\include{Tesischap2}

\clearemptydoublepage

\clearemptydoublepage
\include{Tesischap3}

\clearemptydoublepage

\clearemptydoublepage
\include{Tesischap4}

\clearemptydoublepage

\clearemptydoublepage

\include{Tesischap5}

\clearemptydoublepage

\clearemptydoublepage

\include{Appendix}

\clearemptydoublepage

\clearemptydoublepage
\include{Conclusions}
\clearemptydoublepage

\refstepcounter{chapter}

\end{document}

%% file: titulo.tex
\thispagestyle{empty}
\begin{center}
~
\mbox{\parbox{3cm} {\centering\scriptsize\mbox{\hspace{-0.28cm}\includegraphics[width=3.2cm]{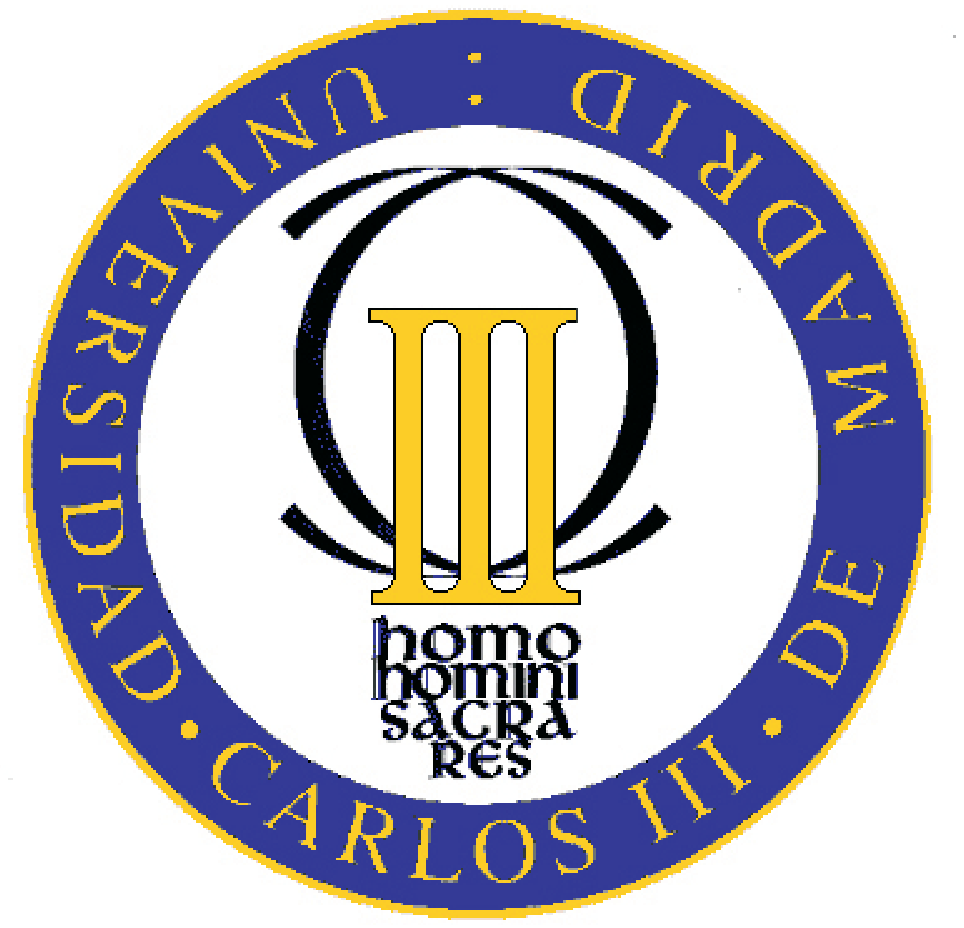}}}}

\vspace{0.5cm}

{\large Departamento de Matem\'aticas}\\
{\large Universidad Carlos III de Madrid}

\vspace{0.75cm}

{\Large Tesis doctoral} \vspace{0.75cm}

\rule{\linewidth}{0.2mm}

\vspace{2mm}


{\Large\sc\textbf{ On the tomographic description of quantum}}
\vspace{1mm}

{\Large\sc\textbf{systems: theory and applications}}
\vspace{1mm}
\rule{\linewidth}{0.2mm}


\vspace{1cm}

{Autor: }

\vspace{0.7mm}

{\textbf{Alberto L\'opez Yela}}

\vspace{3mm}

{Director:}

\vspace{0.7mm}

{\textbf{Alberto Ibort Latre}}\\
{Catedr\'atico de Universidad}\\
Departamento de Matem\'aticas\\
Universidad Carlos III de Madrid

\vspace{1.5cm}

{\large \textrm{Legan\'es, noviembre de 2015}}
\end{center}



%% file: agradecimientos.tex
\chapter*{Agradecimientos}
\markboth{Agradecimientos}{Agradecimientos}


En primer lugar, quiero agradecer el apoyo recibido por todos los compa-\~neros que han pasado por el departamento durante mi periodo en esta universidad, en especial, a todos los que hemos compartido el despacho 2.1D14: Walter, Yadira, Javier, H\'ector, Alejandro y Manuel. Tambi\'en, !`c\'omo no!, no me puedo olvidar de otros amigos con los que he compartido muy buenos momentos: Kenneth, Lino, Javier Gonz\'alez y Mari Francis.

Quisiera tambi\'en agradecer todo lo que me han ense\~nado todos los profesores que me han dado clase y todos con los que he compartido docencia, bueno, a excepci\'on de una que conocemos bien  que supongo que estar\'a en su ``refugio'' evolucionando en el noble arte de incordiar. En concreto, quiero mecionar a Julio Moro, al cual agradezco much\'isimo su inter\'es en el algoritmo num\'erico que se expondr\'a en el \hyperref[section_smily_alg]{{\color{black}\textbf{tercer cap\'itulo}}}, ya que sus aportaciones me ayudaron mucho para conseguir desarrollarlo de manera definitiva, a Fernando Lled\'o del que he aprendido la importante y necesaria aportaci\'on que hace las matem\'aticas a la mec\'anica cu\'antica y por supuesto, a mi jefe Alberto, del que pienso que no podr\'ia haber escogido mejor director de tesis. Espero haber conseguido plasmar en este trabajo todo lo que me has ense\~nado en estos cinco a\~nos. Todav\'ia me sigue sorprendiendo la cantidad de cosas que sabes, espero seguir aprendiendo de ti mucho en el futuro, aunque perm\'iteme decirte que tambi\'en me has ense\~nado lo importante que es la organizaci\'on en el trabajo porque creo, aunque en parte porque siempre tienes trabajo de uno u otro lado, que no te vendr\'ia mal organizarte un poco mejor.

Por supuesto, no puedo olvidarme de Giuseppe Marmo, Franco Ventriglia y Volodya Man'ko de los que he aprendido much\'isimo durante mi estancia en N\'apoles y a partir de ahora espero aportar m\'as a este ``Tomographic Team''. 

Tambi\'en quiero agradecer a mis compa\~neros en el grupo de informaci\'on cu\'antica, Julio de Vicente, Juan Manuel y Leonardo, todo lo aprendido en esas discusiones hablando de temas cient\'ificos y otros m\'as cotidianos como f\'utbol y pol\'itica. Quiero tener un par de palabras m\'as para mis compa\~neros Juan Manuel y Leonardo ya que hemos compartido director y eso creo que genera un v\'inculo especial. Muchas gracias a ti Juanma porque casi has sido un codirector para m\'i, sobretodo ese a\~no en el que Alberto estuvo fuera y trabajamos en ese num\'erico que ahora que termine esta tesis podremos terminar por fin y a Leonardo por todas esas partidas de billar que hemos disfrutado y por haber sido un gran anfitri\'on en mi etapa en N\'apoles.

Tampoco puedo olvidarme de ese gran pianista llamado Fernando y ese clarinetista llamado Diego, mis compa\~neros durante la carrera, espero que os resulte interesante este trabajo y me deis vuestra sincera opini\'on sobre \'el y por supuesto de mi amigo bi\'ologo Alberto, a ti s\'olo decirte que aunque me haya adelantado en terminar la tesis ya sabes que se suele decir que lo bueno siempre se hace esperar.

Aunque est\'en fuera del aspecto cient\'ifico, quiero dedicar unas palabras a mis compa\~neros de karate, aunque supongo que os habr\'eis dado cuenta que la frase escrita como apertura a este texto se refiere a vosotros. Quiero tener estas palabras para reivindicar lo que aporta a la vida diaria ese noble arte marcial, ya que todas las ense\~nanzas que recibimos del maestro cada martes y jueves son un buen camino a seguir porque todo duro esfuerzo termina dando una grata recompensa.

Por \'ultimo, quiero agradecer a mi familia por todo el apoyo recibido. A mi hermano David y mi cu\~nada Ang\'elica de los que me siento muy orgulloso, a mi sobrino Dar\'io que qui\'en sabe, puede que cuando sea mayor se dedique a la f\'isica y trabaje en cosas relacionadas con esta tesis y le pueda servir para aprender, a mi hermana Ana que ser\'a la segunda doctora de la familia aunque todav\'ia le queden unos a\~nitos para conseguirlo, a mi madre porque siempre ha confiado en que alcanzar\'ia este objetivo y a mi padre el culpable de que decidiera aventurarme en este mundo de la f\'isica.

Tambi\'en, me gustar\'ia dedicar unas palabras a mi t\'io Benito, ya que siempre que nos ve\'iamos me preguntaba por mis trabajos. Tu manera de ver la vida es un ejemplo que vale la pena seguir que todos deber\'iamos tratar de hacer, es un orgullo haber sido tu sobrino. Estas palabras son mi homenaje hacia ti.

Disculpadme si hay alguno que no haya mencionado dej\'andome en el tintero, a todos vosotros simplemente os quiero decir que espero que este trabajo os sirva para aprender un poco sobre ese mundo tan extra\~no que parece la mec\'anica cu\'antica.
\vfill
{\small{Este trabajo ha sido posible gracias al proyecto QUITEMAD (Quantum Information Technologies in Madrid) y la beca PIF del Departamento de Matem\'aticas de la Universidad Carlos III de Madrid.}}

%% file: Resumen.tex
\chapter*{Resumen}\label{Resumen_ps}
\markboth{Resumen}{Resumen}
\hyphenation{ima-gen obser-va-bles gene-ra-li-zada repre-sen-ta-ciones ellas ve-re-mos des-cri-ben des-cri-be na-tu-ral re-cons-truir ma-ne-ra}

En este trabajo, se analiza una teor\'{i}a que es en cierto modo una extensi\'{o}n natural del campo de las telecomunicaciones cl\'{a}sicas a la mec\'{a}nica cu\'{a}ntica. Dicha teor\'{i}a se llama \textit{tomograf\'{i}a cu\'{a}ntica} y es de hecho una imagen de la mec\'{a}nica cu\'{a}ntica equivalente a las m\'as habituales que son la imagen de Schr\"{o}dinger \cite{Sc26} o la de Heisenberg \cite{He27}. Esta nueva imagen, difiere de las precedentes en que est\'{a} muy ligada a la capacidad tecn\'{o}logica a la hora de medir observables (momento, energ\'{i}a, etc.) en el campo de la \'{o}ptica cu\'{a}ntica, ya que su objetivo primordial es el de conseguir reconstruir el estado de un sistema cu\'antico a partir de mediciones en el laboratorio, y dado al avance tecnol\'{o}gico del instrumental de laboratorio disponible como l\'{a}seres y fotodetectores, cada vez est\'{a} suscitando mayor inter\'{e}s.

En el \hyperref[chap_birth]{{\color{black}\textbf{primer cap\'itulo}}}, veremos c\'{o}mo nace esta idea de reconstruir estados cu\'{a}nticos discutiendo brevemente la t\'{e}cnica cl\'asica conocida como \hyperref[section_computerized_axial_tom]{{\color{black}\textbf{Tomograf\'{i}a Axial Computerizada}}} (TAC). Esta t\'{e}cnica est\'{a} basada en los trabajos de Johann Karl August Radon~\cite{Ra17} aplicando la transformada que lleva su nombre. Introduciremos la \hyperref[Radon_Transform_ps]{{\color{black}\textbf{transformada de Radon}}} de una funci\'{o}n de probabilidad definida en el espacio de fases para ver c\'{o}mo se aplica en el caso del TAC. Para aplicar esta idea para la reconstrucci\'on de estados cu\'{a}ticos, veremos, en primer lugar, que existe una extensi\'{o}n natural de las t\'{e}cnicas de demodulaci\'{o}n de se\~nales \hyperref[section_telecomm]{{\color{black}\textbf{moduladas en am-}}} \hyperref[section_telecomm]{{\color{black}\textbf{plitud}}} (AM) en el campo de la \'{o}ptica cu\'{a}ntica mostrando que el papel que cumple un \hyperref[ph_section_local]{{\color{black}\textbf{mezclador}}} puede ser reemplazado por una combinaci\'{o}n de \hyperref[section_photodetection]{{\color{black}\textbf{divisores de haz}}} y \hyperref[res_foto]{{\color{black}\textbf{fotodetectores}}} y mostraremos expl\'icitamente c\'omo reconstruir el operador densidad que describe un estado cu\'antico a trav\'es de un proceso reminiscente de la \hyperref[Radon_Transform_ps]{{\color{black}\textbf{transformada de Radon cl\'asica}}}.

El \hyperref[chap_tom_qu]{{\color{black}\textbf{segundo cap\'itulo}}} ser\'{a} el coraz\'{o}n de este trabajo. En \'{e}l, introduciremos de manera formal la descripci\'{o}n tomogr\'{a}fica de la mec\'{a}nica cu\'{a}ntica. Presentaremos una teor\'{i}a general, para ello, trataremos a los observables como elementos de un \hyperref[resumen_C_alg]{{\color{black}\textbf{\'{a}lgebra $C^*$}}} y los estados ser\'{a}n \hyperref[resumen_states]{{\color{black}\textbf{fun-}}} \hyperref[resumen_states]{{\color{black}\textbf{cionales lineales positivos}}} que act\'{u}en en dicha \'{a}lgebra. Veremos que esta descripci\'{o}n de la mec\'anica cu\'antica puede dividirse en dos partes, una primera con el objetivo de obtener una f\'ormula para reconstruir el estado de un sistema cu\'antico a partir de una funci\'on definida sobre un conjunto llamado \hyperref[section_Sampling_C]{{\color{black}\textbf{\textit{conjunto tomogr\'afico}}}}, que ser\'a un conjunto de observables que tendr\'a que cumplir una serie de condiciones que expondremos debidamente. A esta primera parte de la teor\'ia, la bautizaremos como \hyperref[section_Sampling_C]{{\color{black}\textbf{\textit{Teor\'ia de muestreo generalizada}}}} en sistemas cu\'anticos.

La segunda parte estar\'a relacionada con la parte puramente experimental. Lo que hace necesario tener que a\~nadir esta segunda parte es el hecho de que la funci\'on definida anteriormente sobre el \hyperref[section_Sampling_C]{{\color{black}\textbf{conjunto tomogr\'afico}}}, que llamaremos \hyperref[resumen_sampling_fun]{{\color{black}\textbf{\textit{funci\'on de muestreo}}}}, en general, no puede ser medida por medio de los dispositivos con los que contamos en un laboratorio de \'optica cu\'antica, sin embargo, a partir de las mediciones hechas con un fotodetector, podemos obtener distribuciones de probabilidad de cantidades relacionadas con los observables. Entonces, tomando esto \'ultimo como motivaci\'on, esta segunda parte de la teor\'ia consistir\'a en relacionar esa \hyperref[resumen_sampling_fun]{{\color{black}\textbf{funci\'on de muestreo}}} con una distribuci\'on de probabilidad que llamaremos \hyperref[resumen_tomogram]{{\color{black}\textbf{\textit{tomograma}}}}, que ser\'a el resultado directo de un proceso de medida en el laboratorio, y la llamaremos \hyperref[ps_GPT]{{\color{black}\textbf{\textit{Transformada generalizada positiva}}}} por motivos que se expondr\'an convenientemente.

Uno de los problemas m\'as sutiles de esta teor\'ia consiste en hallar un \hyperref[biortho_ps]{{\color{black}\textbf{conjunto tomogr\'afico}}} que cumpla las condiciones necesarias para permitir reconstruir el estado a partir de \'el. Sin embargo, veremos que de manera natural, las representaciones unitarias irreducibles de un grupo finito o de Lie compacto proporcionan \hyperref[biortho_finite_groups_ps]{{\color{black}\textbf{conjuntos tomogr\'aficos}}} que cumplen las condiciones requeridas, por eso, nos centraremos en el estudio de la re\-construcci\'on de estados a partir de grupos relacionados con el sistema f\'isico dado. Aunque tambi\'en destacaremos que existen otras representaciones unitarias que nos permiten reconstruir el estado cu\'antico a partir de ellas, como lo es la representaci\'on unitaria irreducible del grupo de \hyperref[resume_heis]{{\color{black}\textbf{Heisen-}}} \hyperref[resume_heis]{{\color{black}\textbf{berg--Weyl}}} dada por el \'algebra de Lie que forman los operadores momento y posici\'on cu\'anticos que es el ejemplo con el que nace la tomograf\'ia cu\'antica introducido en el \hyperref[section_QM_ps]{{\color{black}\textbf{primer cap\'itulo}}}.

En el \hyperref[chap_smily]{{\color{black}\textbf{tercer cap\'itulo}}} presentaremos un algoritmo num\'erico que se deriva a partir de unos estados que llamaremos \hyperref[adapted_state_a_ps]{{\color{black}\textbf{\textit{estados adaptados}}}} que habremos definido en el \hyperref[chap_tom_qu]{{\color{black}\textbf{cap\'itulo anterior}}}. Este algoritmo nace como un problema inverso, ya que hasta entonces nos habremos centrado en reconstruir estados a partir de un \hyperref[section_Sampling_C]{{\color{black}\textbf{conjunto tomogr\'afico}}}, en especial, cuando el \hyperref[section_particular group]{{\color{black}\textbf{conjunto tomogr\'afico}}} est\'a definido a partir de una \hyperref[resumen_rep]{{\color{black}\textbf{representaci\'on}}} \hyperref[resumen_rep]{{\color{black}\textbf{unitaria de un grupo}}}. Este problema inverso consiste en determinar qu\'e informaci\'on es posible obtener de una \hyperref[resumen_rep]{{\color{black}\textbf{representaci\'on unitaria de}}} \hyperref[resumen_rep]{{\color{black}\textbf{un grupo}}} si se tiene una familia de estados que describe un sistema f\'isico relacionado con un grupo de simetr\'ia. La respuesta a esta pregunta es muy satisfactoria ya que es posible conocer la matriz de \hyperref[resumen_trans]{{\color{black}\textbf{transformaci\'on de}}} \hyperref[resumen_trans]{{\color{black}\textbf{base}}} que nos permita transformar la base, en la que est\'a descrita la representaci\'on unitaria, en una base adaptada a los subespacios invariantes bajo la acci\'on de todos los elementos de la representaci\'on. Este problema se conoce como \hyperref[resumen_CGP]{{\color{black}\textbf{\textit{descomposici\'on de Clebsh--Gordan}}}}, ya que dicha transformaci\'on aplicada a la representaci\'on unitaria la convierte en una matriz \hyperref[resumen_diag]{{\color{black}\textbf{diagonal por bloques}}} en la que cada bloque corresponde a una representaci\'on unitaria irreducible. 

La manera en la que resolveremos este problema es con un algoritmo num\'erico que s\'olo requiere \hyperref[rho_tilde_ps]{{\color{black}\textbf{dos estados adaptados}}} como argumentos de entrada, que pueden ser obtenidos de manera directa si se conoce de forma expl\'icita la representaci\'on unitaria que queremos reducir. Este algoritmo lo hemos bautizado con el nombre de \hyperref[section_smily_alg]{{\color{black}\textbf{SMILY}}}. Hay que destacar que como este algoritmo se ha generado s\'olo aplicando ciertas transformaciones unitarias sobre matrices que representan un estado cu\'antico, tiene una extensi\'on natural que podr\'ia implementarse en un ordenador cu\'antico.

Para acabar esta tesis, generalizaremos la descripci\'on tomogr\'afica a campos cl\'asicos y cu\'anticos. Para el caso \hyperref[chap_clas]{{\color{black}\textbf{cl\'asico}}}, primero realizaremos una descripic\'on tomogr\'afica para sistemas con \hyperref[sec_tom_pic_clas]{{\color{black}\textbf{finitos grados de libertad}}} y obtendremos el equivalente tomogr\'afico de la \hyperref[section_Tom_Liouville]{{\color{black}\textbf{ecuaci\'on de Liouville}}} para una densidad de probabilidad y tras esto, haremos el mismo an\'alisis para sistemas con \hyperref[sec_KG_fields]{{\color{black}\textbf{infinitos grados de libertad}}}.

Para obtener la descripci\'on tomogr\'afica para \hyperref[chap_QFT]{{\color{black}\textbf{campos cu\'anticos}}}, partiremos del concepto de \hyperref[resumen_quantization]{{\color{black}\textbf{\textit{segunda cuantizaci\'on}}}} y mostraremos el equivalente tomogr\'afico de los axiomas de \hyperref[section_WS_axioms]{{\color{black}\textbf{Wightman--Streater}}} para una teor\'ia cu\'antica de campos. Y para terminar, obtendremos un \hyperref[resumen_rec_field]{{\color{black}\textbf{teorema de re-}}} \hyperref[resumen_rec_field]{{\color{black}\textbf{construcci\'on}}} para campos escalares y calcularemos el tomograma de ciertos estados de un \hyperref[resumen_bosonic]{{\color{black}\textbf{campo cu\'antico escalar libre}}}. Comentemos esto \'ultimo diciento que es el inicio de una teor\'ia que permitir\'ia una descripci\'on tomogr\'afica de estados, por ejemplo ligados, para teor\'ias con interacci\'on.

Para concluir este resumen, quisiera resaltar que el lector especializado, si lo considera conveniente, puede comenzar a leer a partir del \hyperref[chap_tom_qu]{{\color{black}\textbf{segundo}}} \hyperref[chap_tom_qu]{{\color{black}\textbf{cap\'itulo}}}, ya que aunque el \hyperref[chap_birth]{{\color{black}\textbf{primer cap\'itulo}}} sirve como motivaci\'on de por qu\'e desarrollar una descripci\'on tomogr\'afica de la mec\'anica cu\'antica, el texto puede comprenderse si previamente no se ha le\'ido.

%% file: Tesischap1.tex
\chapter{The birth of Quantum Tomography}\label{chap_birth}

As it was indicated in the \hyperref[Resumen_ps]{{\color{black}{\textbf{summary}}}}, this chapter will be devoted to provide an informal presentation of \textit{Quantum Tomography} connecting it with the foundations of classical tomography, i.e., the \textit{classical Radon Transform}, and the techniques used in Quantum Optics: homodyne and heterodyne detection. 

Because these ideas have their roots in classical telecommunications, an effort has been made to offer a brief summary of the foundations of classical homodyne and heterodyne detection. So that, by means of the naive canonical \textit{quantization} of the Electromagnetic field, the reader will be able to relate the quantum results with their classical counterparts. Needless to say that the mathematical foundations of Quantum Tomography will be addressed again under much more rigorous grounds in \hyperref[chap_tom_qu]{{\color{black}\textbf{chapter \ref*{chap_tom_qu}}}}, and the extension of the corresponding ideas to classical and quantum \textit{fields} will be the subject of chapters \hyperref[chap_clas]{{\color{black}\textbf{\ref*{chap_clas}}}} and \hyperref[chap_QFT]{{\color{black}\textbf{\ref*{chap_QFT}}}}.\newpage

\section{Radon Transform}\label{section_Radon_Transform}
The process of reconstruction of quantum states, that will be presented in this work, is inspired on the technology for producing tomographic images of sections of scanned bodies for medical purposes, known commonly as CAT (Computerized Axial Tomography).

This technique is based on the mathematical transformation obtained by Radon \cite{Ra17} that allows to recover the value of a regular enough function at any point $(q,p)$ in the plane by averaging the value of that function over all possible lines that pass through it.

\begin{figure}[htbp]
\centering
\includegraphics{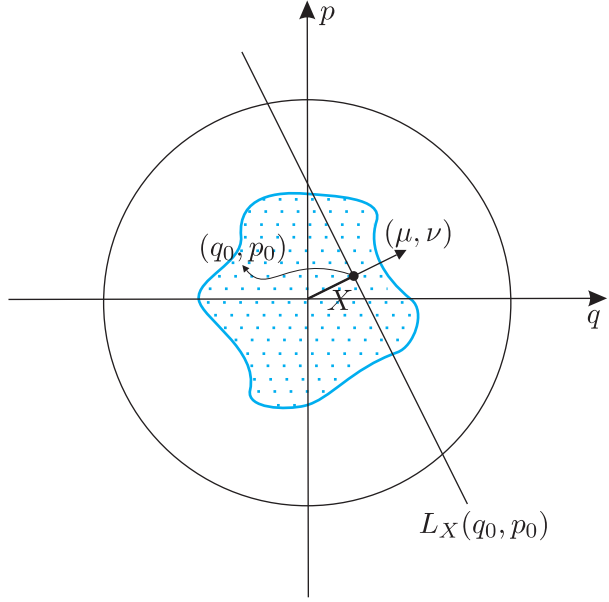}
\caption{\hfil\MYGREEN{Figure 1.1.1}: Geometric scheme of Radon Transform on the plane.\hfil}
\label{Geometric_Radon}
\end{figure}

More formally, let $f(q,p)$ be a Schwarz function on $\mathbb{R}^2$. The Radon Transform of $f$ is defined as:
\phantomsection\label{Radon_Transform_ps}\begin{equation}\label{Radon_Transform}
\mathcal{R}f(X,\mu,\nu)\coloneq\hspace{-.26cm}\int\limits_{\hspace{.5cm}\mathbb{R}^2}\hspace{-.19cm} f(q,p)\delta(X-\mu q-\nu p)\diff q\diff p=\hspace{-1.2cm}\int\limits_{\hspace{1.3cm}L_X(q_0,p_0)}\hspace{-1.07cm}f\big(q(s),p(s)\big)\diff s,
\end{equation}
where $\delta$ is the delta distribution defined on the space of test functions $\mathcal{D}\subset\mathscr{S}$ and $L_X(q_0,p_0)=\left\{\vphantom{a^\dagger}\big(q(s),p(s)\big)|X-\mu q(s)-\nu p(s)=0\right\}$ is the line we integrate over, where $(q_0,p_0)$ is the point of the line closest to the origin, $X$ is a parameter that indicates the distance of the point $(q_0,p_0)$ to the origin, and $s$ is the affine variable that parametrizes such line, {\changeurlcolor{mygreen}\hyperref[Geometric_Radon]{Figure~\ref*{Geometric_Radon}}}.

The Schwartz space $\mathscr{S}$ is the space of smooth rapidly decreasing functions on $\mathbb{R}^n$. The \textit{Fourier Transform} defines a continuous invertible map $\mathscr{F}:\mathscr{S}(\mathbb{R}^n)\rightarrow \mathscr{S}(\mathbb{R}^n)$ by means of:
\phantomsection\label{Fourier_Transform_ps}\begin{equation}\label{Fourier_Transform}
\mathscr{F}(\boldsymbol{k})\coloneq\widehat{f}(\boldsymbol{k})=\frac{1}{(2\pi)^{n/2}}\hspace{-.23cm}\int\limits_{\hspace{.5cm}\mathbb{R}^n}\hspace{-.19cm}f(\boldsymbol{x})\e^{-i\boldsymbol{k}\cdot\boldsymbol{x}}\diff x_1\cdots \diff x_n,
\end{equation}
and the \textit{Inverse Fourier Transform} is the map $\mathscr{F}^{-1}:\mathscr{S}(\mathbb{R}^n)\rightarrow\mathscr{S}({\mathbb{R}^n})$:
\phantomsection\label{Inverse_Fourier_Transform_ps}\begin{equation}\label{Inverse_Fourier_Transform}
\mathscr{F}^{-1}(\boldsymbol{x})\coloneq\widecheck{F}(\boldsymbol{x})=\frac{1}{(2\pi)^{n/2}}\hspace{-.23cm}\int\limits_{\hspace{.5cm}\mathbb{R}^n}\hspace{-.19cm}F(\boldsymbol{k})\e^{i\boldsymbol{k}\cdot\boldsymbol{x}}\diff k_1\cdots \diff k_n.
\end{equation}\newpage

The Fourier Transform can be extended to $\mathscr{S}^\prime$ (the topological dual space of $\mathscr{S}$), which is the space of continuous linear functionals on $\mathbb{R}^n$ ($\mathcal{D}\subset\mathscr{S}\subset\mathscr{S}^\prime\subset\mathcal{D}^\prime$) called the space of \textit{tempered distributions}. Thus, if $T\in\mathscr{S}^\prime$, then $\langle\,\widehat{T},f\,\rangle=\langle\, T,\widehat{f}\,\rangle$ for all $f\in\mathscr{S}$. 

Therefore, we can define the delta function as the distribution whose Fourier Transform is the constant function $
$\,, hence it has the integral representation:
\phantomsection\label{delta_function_ps}\begin{equation}\label{delta_function}
\delta(x)=\frac{1}{2\pi}\hspace{-.2cm}\int\limits_{\hspace{-.04cm}-\infty}^{\hspace{.45cm}\infty} \hspace{-.1cm}\e^{ikx}\diff k.
\end{equation}

Using formula~\hyperref[delta_function_ps]{(\ref*{delta_function})}, we can write:
\phantomsection\label{delta_function_Radon_Transform_ps}\begin{equation*}
\mathcal{R}f(X,\mu,\nu)=\frac{1}{2\pi}\hspace{-.23cm}\int\limits_{\hspace{.5cm}\mathbb{R}^3}\hspace{-.19cm} f(q,p)\e^{ikX}\e^{-ik\mu q}\e^{-ik\nu p}\diff k\diff q\diff p,
\end{equation*}
so that, using the notation $\tilde{*}=k\cdot*$, we get:
\phantomsection\label{delta_function_Radon_Transform_2_ps}\begin{equation*}\label{delta_function_Radon_Transform_2}
\widehat{f}(\tilde{\mu},\tilde{\nu})=\frac{1}{2\pi}\hspace{-.23cm}\int\limits_{\hspace{.5cm}\mathbb{R}^2}\hspace{-.19cm} f(q,p)\e^{-ik\mu q}\e^{-ik\nu p}\diff q\diff p
=\frac{1}{2\pi}\hspace{-.2cm}\int\limits_{\hspace{-.04cm}-\infty}^{\hspace{.45cm}\infty}\hspace{-.1cm}\mathcal{R}f(X,\mu,\nu)\e^{-ikX}\diff X.
\end{equation*}
Hence, the \textit{Inverse Radon Transform} is obtained from the inverse of the Fourier Transform as:
\phantomsection\label{Inverse_Radon_Transform_ps}\begin{align*}
f(q,p)=\widecheck{\widehat{f}}(q,p)&=\frac{1}{2\pi}\hspace{-.23cm}\int\limits_{\hspace{.5cm}\mathbb{R}^2}\hspace{-.19cm}\widehat{f}(\tilde{\mu},\tilde{\nu})\e^{i\tilde{\mu} q}\e^{i\tilde{\nu} p}\diff\tilde{\mu}\diff\tilde{\nu}\nonumber\\
&=\frac{1}{(2\pi)^2}\hspace{-.26cm}\int\limits_{\hspace{.5cm}\mathbb{R}^3}\hspace{-.2cm}\mathcal{R}f(X,\mu,\nu)\e^{-i(kX-\tilde{\mu} q-\tilde{\nu} p)}\diff X\diff\tilde{\mu}\diff\tilde{\nu},
\end{align*}
and evaluating at $k=1$, we get:
\phantomsection\label{Inverse_Radon_Transform_2_ps}\begin{equation}\label{Inverse_Radon_Transform_2}
f(q,p)=\frac{1}{(2\pi)^2}\hspace{-.3cm}\int\limits_{\hspace{.5cm}\mathbb{R}^3}\hspace{-.2cm}\mathcal{R}f(X,\mu,\nu)\e^{-i(X-\mu q-\nu p)}\diff X\diff\mu\diff\nu.
\end{equation}

\section{Computerized Axial Tomography}\label{section_computerized_axial_tom}

The CAT technique consists on obtaining the absorption coefficient of an object by measuring the intensity of a beam before and after crossing through the object at different points \cite{Fa10}. X-rays are the most common radiation used in CAT processes because their wavelength is of the order of atomic size ({\AA}). If radiation with smaller wavelength is used, because the energy of the radiation is proportional to the frequency, the damage to the subjects would be greater.

When a photon interacts with matter, the probability that the photon is absorbed is proportional to the space $\diff s$ travelled by the photon, i.e.,
\phantomsection\label{Photon_absorption_ps}\begin{equation}\label{Photon_absorption}
\diff p(s)=\alpha(s)\diff s,
\end{equation}
where $\alpha(s)$ is the absorption coefficient at point $s$.

If we have a photon beam, the probability that a photon will pass through a portion of matter of length $l$ along the line $L_X(x_0,y_0)$ is equal to the product of the probabilities of not being absorbed along the path:
\phantomsection\label{Photon_absorption_2_ps}\begin{multline}\label{Photon_absorption_2}
P_0(l)=\prod_{s=0}^l \big(1-\diff p(s)\big)\approx\prod_{s=0}^l\exp\big(-\diff p(s)\big)\\
=\exp\bigg(-\hspace{-.2cm}\int\limits_{\hspace{.23cm}0}^{\hspace{.45cm}l}\diff p(s)\bigg)=\exp\bigg(-\hspace{-.2cm}\int\limits_{\hspace{.23cm}0}^{\hspace{.45cm}l}\alpha(s)\diff s\bigg),
\end{multline}
where $s$ is the parameter of the line.
\begin{figure}[h]
\centering
\includegraphics{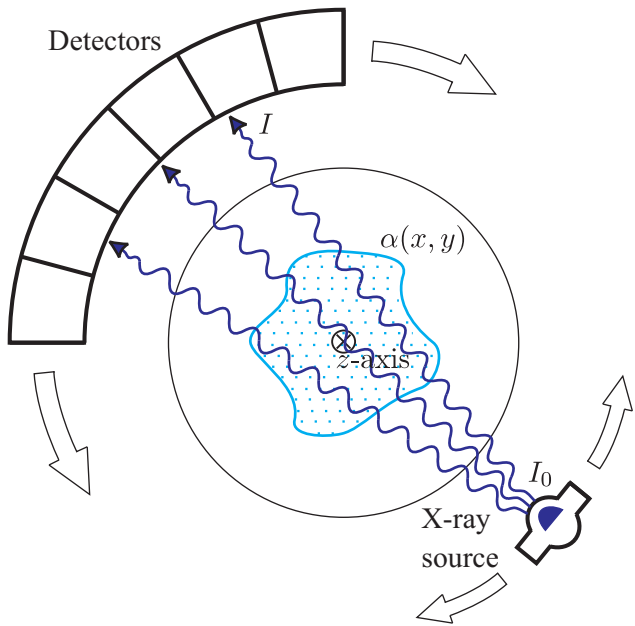}
\caption{\hfil\MYGREEN{Figure 1.2.1}: Section of a scanner of CAT detection.\hfil}
\label{Section_CAT}
\end{figure}

If $N_0$ is the number of photons emitted by the source and $N$ is the number of photons detected at the output, the probability that the photons were not absorbed is
\phantomsection\label{Photon_no_interaction_probability_ps}\begin{equation*}
P_0(l)=\frac{N}{N_0}\,.
\end{equation*}
Hence, because of the intensity $I_0$ of the beam is the number of photons per unit of time, we have that
\phantomsection\label{Photon_no_interaction_probability_2_ps}\begin{equation*}
P_0(l)=\frac{I}{I_0}\,,
\end{equation*}
and finally, we have that the intensity measured at the output is
\phantomsection\label{Photon_no_interaction_probability_3_ps}\begin{equation}\label{Photon_no_interaction_probability_3}
I=I_0\exp\bigg(-\hspace{-.2cm}\int\limits_{\hspace{.23cm}0}^{\hspace{.45cm}l}\alpha(s)\diff s\bigg).
\end{equation}

By definition, $\alpha$ is a probability distribution on $\mathbb{R}^2$, therefore we can use the Inverse Radon Transform to obtain the absorption function $\alpha(x,y)$, where $(x,y)$ are the coordinates of a point in the plane.

Let us insert a body in a cylinder tube, fixing the $z$-axis in the center of the polar section, and let us put a X-ray source and a detector at opposite points with respect to the $z$-axis as represented in {\changeurlcolor{mygreen}\hyperref[Section_CAT]{Figure~\ref*{Section_CAT}}}. Hence, from equation~\hyperref[Radon_Transform_ps]{(\ref*{Radon_Transform})} we have that
\phantomsection\label{Inverse_Radon_Transform_CAT_ps}\begin{align}
\log\left(\frac{I_0}{I}\right)&=\hspace{-.1cm}\int\limits_{\hspace{.23cm}0}^{\hspace{.45cm}l}\alpha\big(x(s),y(s)\big)\diff s\nonumber\\
&=\hspace{-.26cm}\int\limits_{\hspace{.5cm}\mathbb{R}^2}\hspace{-.19cm}\alpha(x,y)\delta(X-x\cos\theta-y\sin\theta)\diff x\diff y=\mathcal{R}\alpha(X,\theta).
\end{align}
If we change to polar coordinates in~\hyperref[Inverse_Radon_Transform_2_ps]{(\ref*{Inverse_Radon_Transform_2})}, this is,
\begin{align*}\label{Change_polar_coords}
  \mu&=k\cos\theta,\nonumber\\
  \nu&=k\sin\theta,
\end{align*}
and taking into account that
\phantomsection\label{homogeneity_tom_for_4_ps}
\begin{equation}\label{homogeneity_tom_for_4}
\mathcal{R}f(kX,k\mu,k\nu)=\frac{1}{|k|}\mathcal{R}f(X,\mu,\nu),
\end{equation}
because of the homogeneity condition of the delta function:
\phantomsection\label{delta_property_ps}\begin{equation}\label{delta_property}
\delta(kx)=\frac{1}{|k|}\delta(x),\qquad k\in\mathbb{C},
\end{equation}
we obtain:
\phantomsection\label{Inverse_Radon_Transform_CAT_2_ps}\begin{equation}\label{Inverse_Radon_Transform_CAT_2}
\alpha(x,y)=\frac{1}{(2\pi)^2}\hspace{-0.2cm}\int\limits_{\hspace{.04cm}-\infty}^{\hspace{.35cm}\infty}\hspace{-.3cm}\int\limits_{\hspace{.15cm}0}^{\hspace{.35cm}\infty}\hspace{-.32cm}\int\limits_{\hspace{.15cm}0}^{\hspace{.35cm}2\pi}\hspace{-.06cm}\mathcal{R}\alpha(X,\theta)\e^{-ik(X-x\cos\theta-y\sin\theta)}k\hspace{0.02cm}\diff\theta\diff k\diff X.
\end{equation}

Finally, to reconstruct the image, the absorption function $\alpha$ usually is plotted in a grey scale, and because the absorption coefficient at point $(x,y)$ is proportional to the quantity of matter in that point, this plot will represent the distribution of matter in the interior of the body.

\section{Reconstruction of signals in classical telecommunications}\label{section_telecomm}

Physicists in Quantum Optics realized \hyperref[ps_Au09]{{\color{red}[\citen*{Au09}, {sec.\,15.4}]}} that they could implement a way for reconstructing the matrix elements of the matrix representation of a quantum state of a light source by a clever use of the Radon Transform described in the \hyperref[section_Radon_Transform]{{\color{black}\textbf{first section}}}. The way for doing that is inspired on how signals are sent and detected in classical telecommunications \RED{[}\citen{Ru87}\RED{,}\hspace{0.05cm}\citen{Si01}\RED{]}.

In classical telecommunications, signals are emitted by modulating a high frequency signal, usually called the \textit{carrier signal}, with the signal that we want to transmit, which is a low frequency signal called the \textit{modulating signal}.

There are two main modulations:
\begin{itemize}
\item AM (Amplitude Modulation).

\item Angular Modulation:
\begin{itemize}
\item FM (Frequency Modulation).
\item PM (Phase Modulation).
\end{itemize}
\end{itemize}

We will only consider here AM because there are a lot of similitudes between it and the ``quantization'' of an electric field.

Amplitude Modulation, as its name tells, consists on modulating the amplitude of the carrier signal with the signal carrying the information, then the envelope of the carrier will vary with the same modulating signal frequencies. Maybe, the reader is wondering why is necessary to modulate a signal for transmitting the information. The prize we have to pay for transmitting modulated signals is that the attenuation of radiation is greater as the frequency grows (that produces losses in the signal) and that the power emission necessary for transmitting increases as the frequency does, then the cost in energy to transmit a modulated signal is greater. However, the rest of arguments are advantages, let us see it.

\begin{figure}[htbp]
\centering
\includegraphics{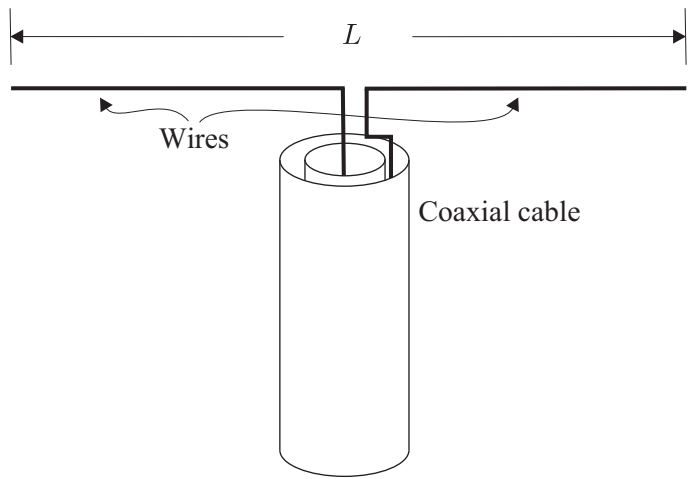}
\caption{\hfil\MYGREEN{Figure 1.3.1}: Half-wavelength simple dipole.\hfil}
\label{Half-wave}
\end{figure}

A reception antenna is simply a resonance circuit. The resonance frequencies greatly depend on the length of the antenna, however they depend also on the design: half-wavelength dipole, inverted V-dipole, vertical monopole, etc., and they also depend on the impedance and other circuit factors \cite{Ba08}. A simple half-wavelength dipole is an antenna composed by two wires of same length $
\,$ connected to a coaxial cable,  {\changeurlcolor{mygreen}\hyperref[Half-wave]{Figure \ref*{Half-wave}}}.
                                                                                                                                                                                                                                                                                                    
\vspace{-0.2cm}

The current induced in the wires has two nodes in the extremes, then the resonance wavelengths are \hspace{0.02cm}$%
\,,$ \hspace{0.02cm}for\hspace{0.04cm} that, this\hspace{0.04cm} dipole\hspace{0.04cm} is

\noindent known as \textit{half-wavelength dipole}. The addition of an impedance to the antenna may reduce the length of the antenna in multiples of 2, $L=%
\,,\ldots,$ without changing the fundamental wavelength $\lambda$.                                                                                                                                                                                                                                                                                           \begin{figure}[htbp]
\centering
\includegraphics{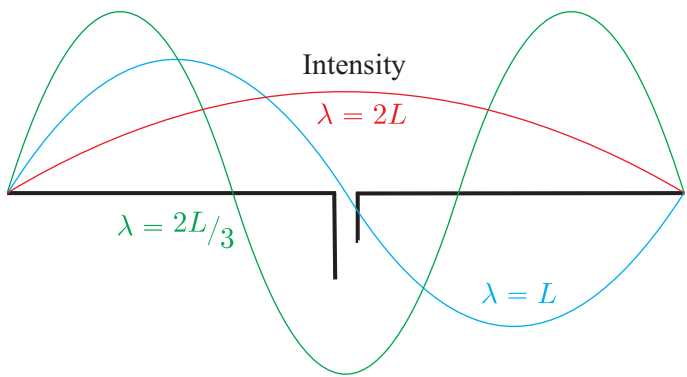}
\caption{\hfil\MYGREEN{Figure 1.3.2}: Current in a half-wavelength simple dipole.\hfil}
\label{Current_half}
\end{figure}
                                                                                                                                                                                                                                                                                                  
\noindent Thanks to that, we see that the length of the antenna needed to transmit or receive signals decreases the higher is the frequency. For example, digital television in Spain transmits with frequencies between 400\,Mhz and 800\,Mhz, and the length needed for an ideal half-wavelength dipole is:
\begin{equation*}\label{Wave-length_dipole}
L=\frac{\lambda}{2}=\frac{c}{2f},\qquad 18.75\,\textrm{cm}\leq L\leq 37.5\,\textrm{cm},
\end{equation*}
where $c$ is the velocity of light.

The previous argument is an important reason for justifying the use of modulation, however the most important reason for doing this is that the carrier wave allows to sort the information in channels making possible to send at the same time different information if the frequencies among their carriers are separated enough for non superposing.

Let us consider first a modulated signal of a simple tone. Let $x_m(t)=V_m\cos(\omega_mt)$ be the simple monotone modulating signal (or message) and $x_c(t)=V_c\cos(\omega_ct)$ the carrier signal. The modulated signal is obtained by adding to the carrier the product of the two signals with a \textit{modulating factor} $\beta$:
\phantomsection\label{AM_ps}\begin{equation}\label{AM}
x(t)=\big(1+\beta x_m(t)\big)x_c(t).
\end{equation}

\begin{figure}[htbp]
\centering
\includegraphics{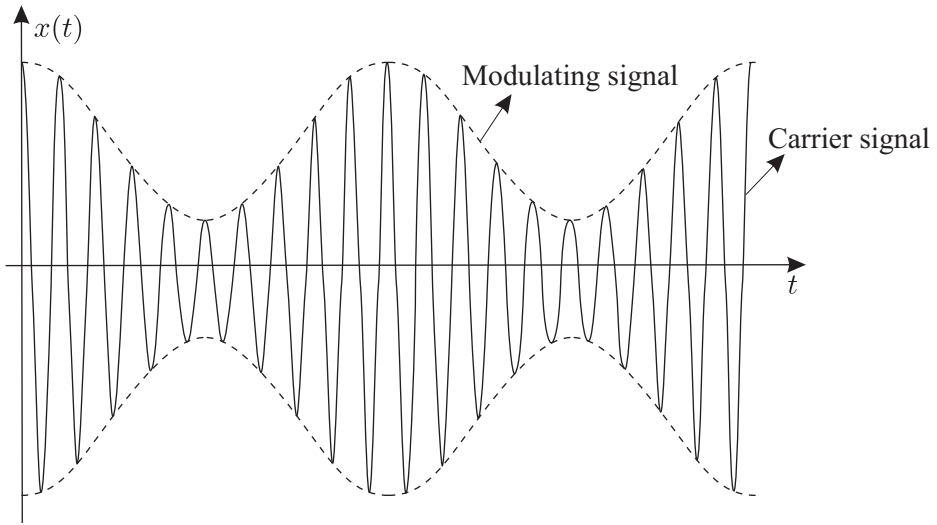}
\caption{\hfil\MYGREEN{Figure 1.3.3}: Signal modulated to 57\%.\hfil}
\label{mod_57}
\end{figure}

The factor $\beta$ depends on the modulator system we use. If $|\beta V_m|\leq 1$, the maximum and minimum values of the modulated signal are:
\begin{align*}
  x_{max}(t)&=(1+\beta V_m)V_c,\\
  x_{min}(t)&=(1-\beta V_m)V_c.
\end{align*}
In this case, the signal is well modulated and it is possible to obtain the modulating signal from the envelope of the signal, {\changeurlcolor{mygreen}\hyperref[mod_57]{Figure \ref*{mod_57}}}.
\vspace{0cm}

\begin{figure}[h]
\centering
\includegraphics{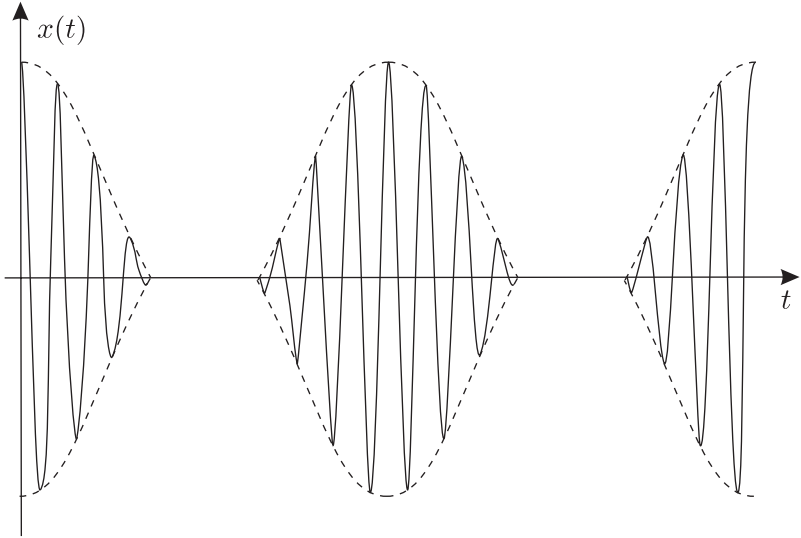}
\caption{\MYGREEN{Figure 1.3.4}: Signal overmodulated to 170\%. This plot corresponds to the output of a real circuit. Notice that differs to what we would have got from eq.\,\hyperref[AM_ps]{(\ref*{AM})}.}
\label{overmod_170}
\end{figure}

If $|\beta V_m|>1$, the signal is overmodulated what produces a distortion in the envelope, and for that, a loss of information at the time of recovering the modulating signal, {\changeurlcolor{mygreen}\hyperref[overmod_170]{Figure \ref*{overmod_170}}}.

\begin{figure}[htbp]
\centering
\includegraphics{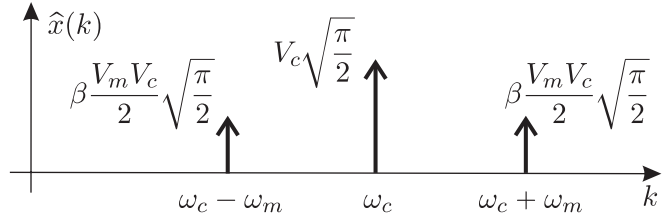}
\caption{\hfil\MYGREEN{Figure 1.3.5}: Fourier Transform of AM monotone signal.\hfil}
\label{FT_AM_mono}
\end{figure}

If we compute the Fourier Transform of \hyperref[AM_ps]{(\ref*{AM})}, we get:
\phantomsection\label{FT_AM_monotone_ps}\begin{multline}\label{FT_AM_monotone}
  \widehat{x}(k)=V_c\sqrt{\frac{\pi}{2}}\left[\delta(\omega_c-k)+\delta(\omega_c+k)+\beta\frac{V_m}{2}\big(\delta(\omega_c+\omega_m-k)\right.\\
  +\delta(\omega_c+\omega_m+k)+\delta(\omega_c-\omega_m-k)+\delta(\omega_c-\omega_m+k)\big)\bigg].
\end{multline}
In {\changeurlcolor{mygreen}\hyperref[FT_AM_mono]{Figure \ref*{FT_AM_mono}}}, we show the amplitude of this Fourier Transform with respect to the frequency. There, we can identify three resonance frequencies:
\begin{itemize}
\item At $\omega_c$ with amplitude $\displaystyle{A(\omega_c)=V_c\sqrt{\frac{\pi}{2}}}$ .
\item At $\omega_c\pm\omega_m$ with amplitude $\displaystyle{A(\omega_c\pm\omega_m)=\beta \frac{V_m V_c}{2}\sqrt{\frac{\pi}{2}}}$ .
\end{itemize}

\vspace{-0.3cm}

\noindent From the medium peak, we can identify the voltage and the frequency of the carrier, and using this information, we can obtain the frequency of the modulating signal and the factor $\beta V_m$ from one of the other resonances.

If we have a multitone signal, because the signals we generate are periodic, we can express them as a Fourier series,
\phantomsection\label{multitone_AM_ps}\begin{equation*}\label{multitone_AM}
x_m(t)=\hspace{-.1cm}\sum_{m=1}^\infty\hspace{-0.05cm} V_m(\omega_m)\e^{i\omega_mt}.
\end{equation*}
If we modulate this signal with a carrier signal, as saw in~\hyperref[AM_ps]{(\ref*{AM})}, the Fourier Transform of the modulated signal becomes:
\phantomsection\label{FT_AM_multione_ps}\begin{multline}\label{FT_AM_multione}
\widehat{x}(k)=V_c\sqrt{\frac{\pi}{2}}\,\bigg[\delta(\omega_c-k)+\delta(\omega_c+k)\\
+\beta\hspace{-.1cm}\sum_{m=1}^\infty\hspace{-.05cm} V_m(\omega_m)\big(\delta(\omega_m+\omega_c-k)+\delta(\omega_m-\omega_c-k)\big)\bigg].
\end{multline}

\begin{figure}[htbp]
\centering
\includegraphics{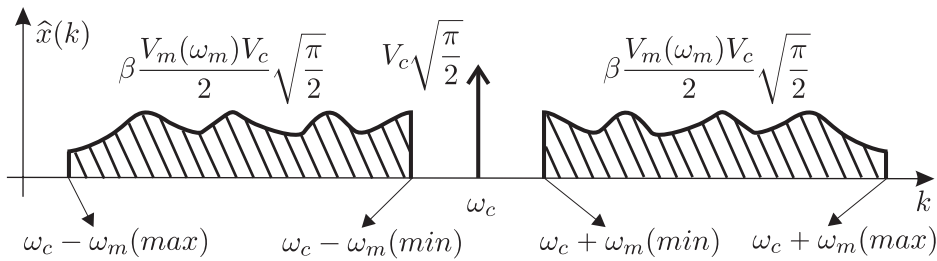}
\caption{\hfil \MYGREEN{Figure 1.3.6}: Fourier Transform of AM multitone signal.\hfil}
\label{FT_AM_multi}
\end{figure}

Plotting this result we obtain, instead of two delta functions, two bands in both sides of the peak at $\omega_c$, because of that, standard AM is usually called \textit{Double Side Band Amplitude Modulation} (DSB--AM), {\changeurlcolor{mygreen}\hyperref[FT_AM_multi]{Figure \ref*{FT_AM_multi}}}. The recovering of the modulating signal can be done in the same way as it was done for the monotone case, however instead of one peak we have as many peaks as frequencies compose the modulating signal.

We have shown how one can reconstruct the modulating signal by analyzing the spectrum of the modulated signal. By the way, the modulating signal can be obtained in the output of a demodulator. For an AM signal, the demodulation can be done using several devices: envelope detectors, homodyne detectors, heterodyne detectors, etc. The choice among them depends on the frequency, simplicity or efficiency of the antenna, among other variables. Following, we will describe homodyne and heterodyne detectors because that kind of detection can be easily adapted to optical devices in Quantum Mechanics.

These two types of detectors consist on multiplying the signal received in the antenna by a local oscillator to demodulate the carrier signal, and after mixing the signal with the local oscillator, the modulating signal is recovered by using a low-pass filter \cite{Ki78}.

\phantomsection
\label{ph_section_local}
A local oscillator is an oscillating circuit together with an amplifier and a feedback circuit to counteract the softening of the oscillations due to losses of energy because of Joule effect and other factors \hyperref[Ru87_ps]{{\color{red}[\citen*{Ru87}, ch.\,5]}}. The simplest oscillating circuit we can build consists of a capacitor and an inductor in parallel.

\begin{figure}[htbp]
 \centering
 \subfigure[\hspace{-0.55cm}\label{Res_circ_a}]{\includegraphics{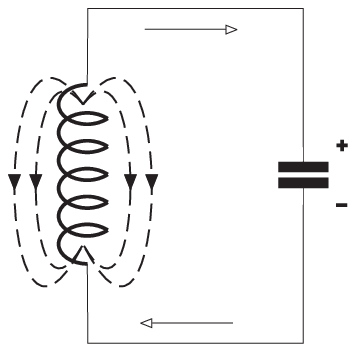}}\hspace{0.3cm}
 \subfigure[\hspace{-0.6cm}\label{Res_circ_b}]{\includegraphics{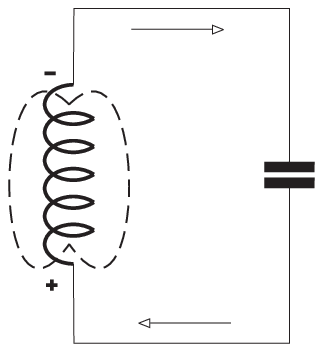}}\hspace{0.5cm}
 \subfigure[\label{Res_circ_c}]{\includegraphics{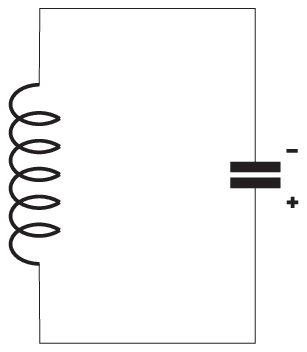}}
 \caption{\MYGREEN{Figure 1.3.7}: \MYGREEN{(a)} Discharge of the capacitor that induce a magnetic flux in the inductance. \MYGREEN{(b)} Electric field induced by the inductance. \MYGREEN{(c)} Charge of the capacitor with different polarity than \MYGREEN{(a)}.}\label{Res_circ}
 \end{figure}

If we feed the circuit during a small period of time, the capacitor will become charged, hence when we stop of feeding, the capacitor will  induce a current in the inductor, {\changeurlcolor{mygreen}\hyperref[Res_circ]{Figure~1.3.7(a)}}. When the capacitor is discharged, the magnetic flux will tend to disappear and producing an electric current in the same direction, {\changeurlcolor{mygreen}\hyperref[Res_circ]{Figure~1.3.7(b)}}, that will charge the capacitor with opposite polarity to {\changeurlcolor{mygreen}\hyperref[Res_circ]{(a)}}, {\changeurlcolor{mygreen}\hyperref[Res_circ]{Figure~1.3.7(c)}} and the process will repeat. 

\noindent This process produces a sinusoidal signal of frequency
\phantomsection\label{Frequency_Oscillator_ps}\begin{equation*}
\omega_0=\frac{1}{LC},
\end{equation*}
where $L$ is the inductance and $C$ the capacity.

\begin{figure}[htbp]
\centering
\includegraphics{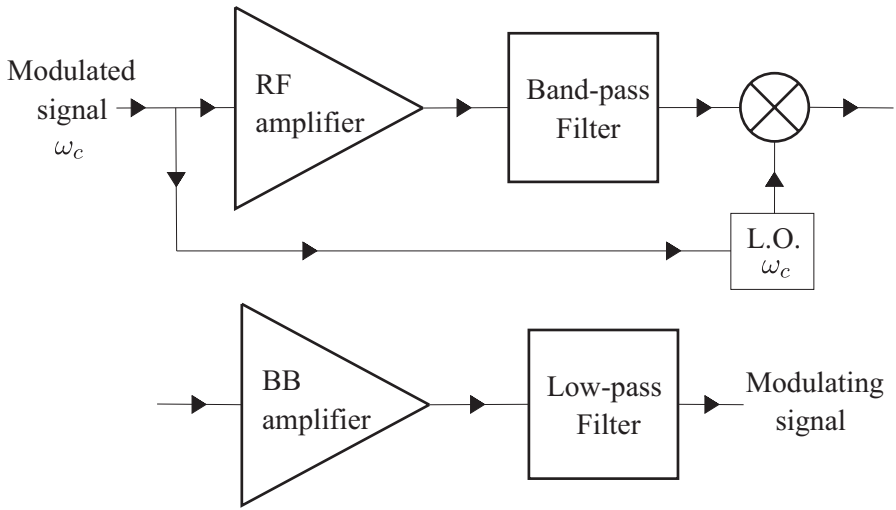}
\caption{\hfil\MYGREEN{Figure 1.3.8}: Homodyne detector scheme.\hfil}
\label{Homodyne_cl}
\end{figure}

\phantomsection
\label{ph_section_Homodyne_cl}
In the homodyne detector, first, the signal is amplified in Radio-Frequen-cy and next, we pass the signal through a band-pass filter to remove noise and other unwanted contributions. After that, the signal is mixed with a local oscillator with the same frequency than the carrier signal to shift the signal to \textit{baseband}, this is lowering to frequency zero, then the signal is amplified in baseband and filtered with a low-pass filter to get the modulating signal, {\changeurlcolor{mygreen}\hyperref[Homodyne_cl]{Figure \ref*{Homodyne_cl}}}.

\begin{figure}[h]
\centering
\includegraphics{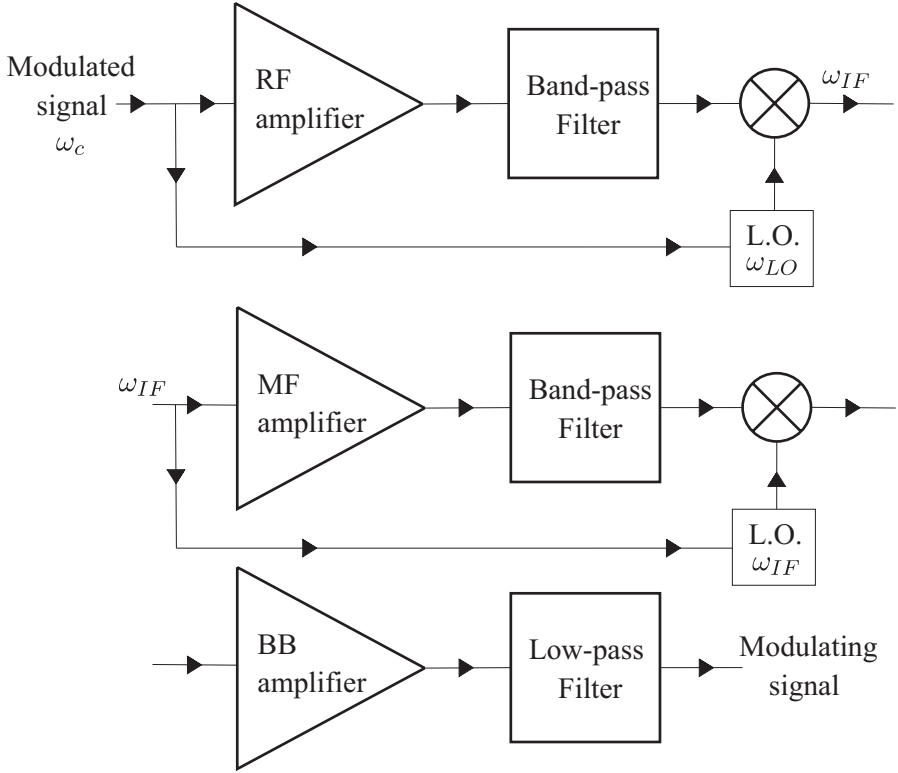}
\caption{\hfil\MYGREEN{Figure 1.3.9}: Heterodyne detector scheme.\hfil}
\label{Heterodyne_cl}
\end{figure}

The heterodyne detector is similar to the homodyne, the only difference is that we mix the signal in the output of the Radio-Frequency band-pass filter with a local oscillator with different frequency than the carrier, hence the name \textit{heterodyne} (\textit{other force} in classic greek), to shift the signal to an intermediate frequency at the level of Medium Frequencies $10^2-10^3$ kHz. Because of this, the homodyne detection is usually called \textit{zero intermediate frequency detection}. Following, the signal is treated the same way as in the homodyne detector, this is, the signal is amplified in Medium-Frequency, filtered with a band-pass filter and mixed with a local oscillator at the intermediate frequency and so on, {\changeurlcolor{mygreen}\hyperref[Heterodyne_cl]{Figure \ref*{Heterodyne_cl}}}.

When the modulated signal is mixed with the first oscillator, two bands appear, one at $\left|\omega-\omega_{LO}\right|$ and other at $\omega+\omega_{LO}$. Obviously, the intermediate frequency is the smallest frequency $\omega_{IF}=\left|\omega-\omega_{LO}\right|$. A problem comes from the fact that the intermediate frequency is different from zero. Usually, information is emitted in channels of different frequencies at the same time, then there are two frequencies, one smaller than $\omega_{LO}$ and other bigger, with the same intermediate frequency:
\phantomsection\label{Image_Frequency_ps}\begin{equation*}
\omega_{LO}-\omega_1=\omega_{IF}=\omega_2-\omega_{LO}.
\end{equation*}
The non desired frequency is called \textit{image frequency}:
\phantomsection\label{Image_Frequency_2_ref_ps}\begin{equation}\label{Image_Frequency_ref}
\omega_{image}=\left|\omega-2\omega_{LO}\right|.
\end{equation}
For that, the emission of information at the image frequency of every allowed channel is forbidden.

\newpage


\section{Classical Electromagnetic field}\label{section_Clas_EM}

In \hyperref[section_hom_det_clas]{{\color{black}\textbf{section~\ref*{section_hom_det_clas}}}}, we will adapt this treatment for demodulating signals modulated in AM to Quantum Optics. For that purpose, we will discuss the classical and quantum electromagnetic fields \hyperref[ps_Ba98]{{\color{red}[\citen*{Ba98}, {ch.\,19}]}} because we want to use the similarity between an AM signal and the explicit form of the quantum electromagnetic field.

Let us consider a manifold $\mathcal{M}=\mathbb{R}\times\mathcal{V}$, where $\mathbb{R}$ denotes the time part and $\mathcal{V}$ a domain in the spatial part that represents a cavity with surface $\partial\mathcal{V}$. The Maxwell's equations that describe the classical electric and magnetic fields in empty space are:
\phantomsection\label{Maxwell_ps}
\begin{alignat}{3}
\label{Maxwells_equation}\nabla\times \boldsymbol{E}&=-\frac{1}{c}\frac{\partial \boldsymbol{B}}{\partial t},& &\hspace{1cm}\mbox{\BLUE{(1.4.1\textrm{a})}}&\hspace{2.2cm} \nabla\cdot \boldsymbol{E}&=0,\tag{1.4.1b}\\
\label{Maxwells_equation_2}\nabla\times \boldsymbol{B}&=\frac{1}{c}\frac{\partial \boldsymbol{E}}{\partial t},& &\hspace{1cm}\mbox{\BLUE{(1.4.1\textrm{c})}}&\hspace{2.2cm}\nabla\cdot \boldsymbol{B}&=0,\tag{1.4.1d}
\end{alignat}
where $c$ is the velocity of light. From these equations, it is possible to see that the electric field satisfies the wave equation:
\setcounter{equation}{1}
\phantomsection\label{wave_equation_electric field_ps}\begin{equation}\label{wave_equation_electric field}
\nabla^2\boldsymbol{E}-\frac{1}{c^2}\frac{\partial^2 \boldsymbol{E}}{\partial t^2}=0.
\end{equation}

The solution of this equation can be written as the sum of mode functions:
\phantomsection\label{solution_Electric_field_ps}\begin{equation*}
\boldsymbol{E}(\boldsymbol{x},t)=\hspace{-0.1cm}\sum_{p=1}^\infty f_p(t)\boldsymbol{u}_p(\boldsymbol{x}),
\end{equation*}
where $f_p(t)$ satisfies
\phantomsection\label{Temporal_condition_EF_ps}\begin{equation}\label{Temporal_condition_EF}
\frac{\diff^2f_p(t)}{\diff t^2}+c^2k_p^2f_p(t)=0,
\end{equation}
and $u_p(\boldsymbol{x})$ satisfies the eigenvector equation for the Laplace operator:
\phantomsection\label{conditions_Electric_field_1_ps}\begin{equation}\label{conditions_Electric_field_1}
\nabla^2\boldsymbol{u}_p(\boldsymbol{x})+k_p\boldsymbol{u}_p(\boldsymbol{x})=0,\tag{1.4.4a}
\end{equation}
\vspace{-1.15cm}
\phantomsection
\label{conditions_Electric_field_3_ps}
\begin{align}
\label{conditions_Electric_field_2}\nabla\cdot \boldsymbol{u}_p(\boldsymbol{x})&=0\tag{1.4.4b}\\
\label{conditions_Electric_field_3}\hat{\boldsymbol{n}}\times \boldsymbol{u}_p(\boldsymbol{x})\hspace{-0cm}&=0\tag{1.4.4c}
\end{align}

\phantom{a}
\vspace{-2.35cm}\begin{equation*}
	\hspace{4cm}\left.\begin{matrix}
\phantom{a}\\
\phantom{a}
\end{matrix}\right\}\hspace{0.5cm}\mbox{\textrm{in}}\,\partial\mathcal{V}
\end{equation*}
where $\hat{\boldsymbol{n}}$ is the unit vector normal to the surface. The last condition must be imposed because the tangential component of the electric field must vanish on the conducting surface. Because it is a Hermitian eigenvalue problem, the functions $\boldsymbol{u}_p(\boldsymbol{x})$ are orthogonal:
\setcounter{equation}{4}
\phantomsection\label{orthogonality relation_EF_ps}\begin{equation}\label{orthogonality relation_EF}
\int\limits_{\hspace{0.3cm}\mathcal{V}}\boldsymbol{u}_p(\boldsymbol{x})\cdot\boldsymbol{u}_{p'}(\boldsymbol{x})\diff^3\boldsymbol{x}=\delta_{pp'}.
\end{equation}

The magnetic field can be determined from~\hyperref[Maxwell_ps]{(1.4.1a)}:
\phantomsection\label{magnetic_field_ps}\begin{equation*}
\boldsymbol{B}(\boldsymbol{x},t)=\hspace{-0.1cm}\sum_{p=1}^\infty g_p(t)\big(\nabla\times \boldsymbol{u}_p(\boldsymbol{x})\big),
\end{equation*}
with
\phantomsection\label{g_p_magnetic_field_ps}\begin{equation}\label{g_p_magnetic_field}
\frac{\diff g_p(t)}{\diff t}=-cf_p(t).
\end{equation}
From~\hyperref[conditions_Electric_field_3_ps]{(\ref*{conditions_Electric_field_3})}, we see that
\phantomsection\label{conditions_Magnetic_field_1_ps}\begin{equation}\label{conditions_Magnetic_field_1}
\hat{\boldsymbol{n}}\cdot\big(\nabla\times\boldsymbol{u}_p(\boldsymbol{x})\big)=0,
\end{equation}
hence, from~\hyperref[conditions_Electric_field_1_ps]{(\ref*{conditions_Electric_field_1})} and~\hyperref[conditions_Magnetic_field_1_ps]{(\ref*{conditions_Magnetic_field_1})}, we obtain the following orthogonality condition:
\phantomsection\label{orthogonality relation_MF_ps}\begin{equation}\label{orthogonality relation_MF}
\int\limits_{\hspace{0.3cm}\mathcal{V}}\big(\nabla\times\boldsymbol{u}_p(\boldsymbol{x})\big)\!\cdot\!\left(\nabla\times\boldsymbol{u}_{p'}(\boldsymbol{x})\right)\diff^3\boldsymbol{x}=k_p^2\delta_{pp'}.
\end{equation}
Notice that from Maxwell's equation~\hyperref[Maxwells_equation_2]{(1.4.1c)}, we see that $g_p(t)$ verifies the same equation as $f_p(t)$, eq.\,\hyperref[Temporal_condition_EF_ps]{(\ref*{Temporal_condition_EF})}:
\phantomsection\label{g_m_equal_f_m_ps}\begin{equation}\label{g_m_equal_f_m}
\frac{\diff^2g_p(t)}{\diff t^2}+c^2k_p^2g_p(t)=0.
\end{equation}

If we define the classical Hamiltonian of the electromagnetic field by computing its total energy:
\begin{equation*}
H_{cl}=\frac{1}{8\pi}\hspace{-0.13cm}\int\limits_{\hspace{0.3cm}\mathcal{V}}\hspace{-0.13cm}\left(|\boldsymbol{E}|^2+|\boldsymbol{B}|^2\right)\diff^3\boldsymbol{x}=\frac{1}{8\pi}\hspace{-0.1cm}\sum_{\stackrel{\scriptstyle{\hspace{0.08cm}p=1}}{p'\hspace{-0.06cm}=1}}^\infty f_p(t)\bar{f}_{p'}(t)\hspace{-0.19cm}\int\limits_{\hspace{0.3cm}\mathcal{V}}\hspace{-0.13cm}\boldsymbol{u}_p(\boldsymbol{x})\cdot\boldsymbol{u}_{p'}(\boldsymbol{x})\diff^3\boldsymbol{x}
\end{equation*}
\vspace{-0.5cm}\begin{equation*}
\hspace{-1.9cm}+\hspace{-0.14cm}\sum_{\stackrel{\scriptstyle{\hspace{0.08cm}p=1}}{p'\hspace{-0.06cm}=1}}^\infty g_p(t)\bar{g}_{p'}(t)\hspace{-0.19cm}\int\limits_{\hspace{0.3cm}\mathcal{V}}\hspace{-0.13cm}\left(\nabla\times\boldsymbol{u}_p(\boldsymbol{x})\right)\cdot\left(\nabla\times\boldsymbol{u}_{p'}(\boldsymbol{x})\right)\diff^3\boldsymbol{x}
\end{equation*}
\vspace{-0.9cm}\phantomsection\label{Hamiltonian_electromagnetic_field_ps}\begin{equation}\label{Hamiltonian_electromagnetic_field}
\hspace{5.2cm} =\frac{1}{8\pi}\hspace{-0.05cm}\sum_{p=1}^\infty\left(|f_p(t)|^2+k_p^2|g_p(t)|^2\right)\!,
\end{equation}
and if we think of $f_p(t)$ as the position of a particle $p$ at time $t$,
\phantomsection\label{Classical_Position_EMF_ps}\begin{equation}\label{Classical_Position_EMF}
Q_p=\frac{f_p}{2ck_p\sqrt{\pi}}\,,
\end{equation}
because the classical momentum is the derivative of the position with respect to time, from~\hyperref[g_p_magnetic_field_ps]{(\ref*{g_p_magnetic_field})} and~\hyperref[g_m_equal_f_m_ps]{(\ref*{g_m_equal_f_m})} we get that
\phantomsection\label{Classical_momentum_EMF_ps}\begin{equation}\label{Classical_momentum_EMF}
P_p=\frac{\textrm{d}Q_p}{\textrm{d}t}=\frac{k_pg_p(t)}{2\sqrt{\pi}}\,.
\end{equation}
Thus, denoting $\omega_p=ck_p$ and using the previous identifications, the Hamiltonian becomes:
\phantomsection\label{Hamiltonian_electromagnetic_field_2_ps}\begin{equation}\label{Hamiltonian_electromagnetic_field_2}
H_{cl}=\frac{1}{2}\sum_{p=1}^\infty\left(P_p^2+\omega_p^2Q_p^2\right)\!,
\end{equation}
which is the Hamiltonian of a set of infinite classical harmonic oscillators of frequencies $\omega_p$, $p=1,2,\ldots$\:.

\markboth{The birth of Quantum Tomography}{1.5. The\hspace{0.02cm} quantum\hspace{0.02cm} harmonic\hspace{0.02cm} oscillator\hspace{0.02cm} and\hspace{0.02cm} the\hspace{0.02cm} quantization\hspace{0.02cm} of\hspace{0.02cm} the\hspace{0.02cm} Electro-\newline{\color{white}.......\!\!} magnetic field}
\section{The quantum harmonic oscillator and the quantization of the Electromagnetic field}\label{section_Quantization_elec}
\markboth{The birth of Quantum Tomography}{1.5. The\hspace{0.02cm} quantum\hspace{0.02cm} harmonic\hspace{0.02cm} oscillator\hspace{0.02cm} and\hspace{0.02cm} the\hspace{0.02cm} quantization\hspace{0.02cm} of\hspace{0.02cm} the\hspace{0.02cm} Electro-\newline{\color{white}.......\!\!} magnetic field}

To obtain a proper quantum description of the Electromagnetic field, we have to provide a quantum description of the system given by the Hamiltonian in \hyperref[Hamiltonian_electromagnetic_field_2_ps]{(\ref*{Hamiltonian_electromagnetic_field_2})}.

First of all, let us define two operators that usually appear in Quantum Mechanics, the \textit{annihilation} and \textit{creation} operators. These two operators are defined as:
\phantomsection\label{creation_annihilation_ps}\begin{equation}\label{creation_annihilation}
a=\frac{1}{\sqrt{2\hbar}}\left(\sqrt{\omega}\textbf{Q}+\frac{i}{\sqrt{\omega}}\textbf{P}\right)\!,\quad
a^\dagger=\frac{1}{\sqrt{2\hbar}}\left(\sqrt{\omega}\textbf{Q}-\frac{i}{\sqrt{\omega}}\textbf{P}\right)\!,
\end{equation}
where $\hbar$ is the reduced Planck constant $\hbar=h/2\pi$ and $\omega$ the frequency. The operators $\textbf{Q}$ and $\textbf{P}$ are the operators on the Hilbert space $\mathcal{H}=L^2(\mathbb{R})$ corresponding to position and momentum respectively. The operator $\textbf{Q}$ is the multiplication operator and the operator $\textbf{P}$ is the operator $-i\hbar\nabla$:
\phantomsection\label{rep_Q_and_P_ps}
\begin{equation}\label{rep_Q_and_P}
\textbf{Q}\psi(\boldsymbol{q})=\boldsymbol{q}\psi(\boldsymbol{q}),\qquad\textbf{P}\psi(\boldsymbol{q})=-i\hbar\nabla\psi(\boldsymbol{q}).
\end{equation}
This framework is usually called \textit{coordinate representation}. It is easy to verify that the commutator of the position and momentum operators is $\left[\vphantom{a^\dagger}\textbf{Q},\textbf{P}\right]=i\hbar\mathds{1}$.

From here to the rest of this text, bold capital letters will denote operators on a Hilbert space. In cases in which may exist any confusion, we will use the circumflex accent \,$\widehat{\phantom{a}}$\, but, in general, this symbol will be reserved to denote the Fourier Transform.

The operator $a$ is the \textit{annihilation} operator and $a^\dagger$ is the \textit{creation} operator and they satisfy the canonical commutation relation:
\phantomsection\label{Commutator_creation_annihilation_ps}\begin{equation}\label{Commutator_creation_annihilation}
\left[a,a^\dagger\right]=\mathds{1}.
\end{equation}
The Hamiltonian of a quantum harmonic oscillator of frequency $\omega$ (and mass $m=1$) is given by:
\phantomsection\label{harmonic_oscillator_ps}\begin{equation}\label{harmonic_oscillator}
\textbf{H}=\frac{\textbf{P}^2}{2}+\frac{\omega^2}{2}\textbf{Q}^2.
\end{equation}
When one substitutes the momentum and position operators by the annihilation and creation operators \hyperref[creation_annihilation_ps]{(\ref*{creation_annihilation})}, the Hamiltonian becomes:
\phantomsection\label{harmonic_oscillator_2_ps}\begin{equation}\label{harmonic_oscillator_2}
\textbf{H}=\hbar\omega\Big(a^\dagger a+\frac{1}{2}\Big).
\end{equation}
The problem of finding the spectrum of $\textbf{H}$ is reduced to the problem of finding the spectrum of the \textit{number operator}:
\phantomsection\label{number_operator_ps}\begin{equation}\label{number_operator}
\textbf{N}=a^\dagger a,
\end{equation}
which satisfies the commutation relations:
\begin{equation}
\left[\vphantom{a^\dagger}\textbf{N},a\right]=-a,\qquad\left[\textbf{N},a^\dagger\right]=a^\dagger.
\end{equation}

To deal with the quantum system defined by \hyperref[harmonic_oscillator_2_ps]{(\ref*{harmonic_oscillator_2})} is better for the purposes of this work (see later \hyperref[chap_QFT]{{\color{black}\textbf{chapter~\ref*{chap_QFT}}}}) to consider an abstract realitation of the Hilbert space of states of the system called \textit{Fock space} $\mathcal{F}$.\newpage

\subsection{Canonical commutation relations and the Fock space}\label{can_section_fock_ch_1}

Let us consider now that the creation and annihilation operators \hyperref[creation_annihilation_ps]{(\ref*{creation_annihilation})} as abstract symbols, and consider the associative algebra generated by them with the commutation relation given by \hyperref[Commutator_creation_annihilation_ps]{(\ref*{Commutator_creation_annihilation})}. There is a natural representation of this algebra as operators on the Hilbert space $\mathcal{F}$ defined as the complex space generated by the eigenvectors $|n\rangle$ of the number operator \hyperref[number_operator_ps]{(\ref*{number_operator})}, with eigenvalues $n=0,1,\ldots$\,, completed with respect to the norm defined by them and where the symbols $a$ and $a^\dagger$ are realized as the operators (denoted with the same symbols): 
\phantomsection\label{creation_annihilation_number_eigenvectors_ps}\begin{align}\label{creation_annihilation_number_eigenvectors}
a^\dagger|n\rangle&=\sqrt{n+1}|n+1\rangle,\qquad n=0,1,\ldots,\nonumber\\
a|n\rangle&=\sqrt{n}|n-1\rangle ,\hspace{0.67cm}\qquad n=1,2,\ldots,\nonumber\\
a|0\rangle&=0.
\end{align}
Notice that
\phantomsection\label{vector_harmonic_oscillator_f_ps}
\begin{equation}\label{vector_harmonic_oscillator_f}
|n\rangle=\frac{(a^\dagger)^n}{\sqrt{n!}}|0\rangle.
\end{equation}

From \hyperref[creation_annihilation_number_eigenvectors_ps]{(\ref*{creation_annihilation_number_eigenvectors})}, it is easy to check that the operators $a$ and $a^\dagger$ are unbounded operators, which are adjoint to each other, and also that the spectrum of the positive self-adjoint operator $\textbf{N}\!=\!a^\dagger a$ is $n=0,1,\ldots$ with eigenvectors $|n\rangle$.

In this representation, we think of the state $|0\rangle$ as the fundamental state (or \textit{vacuum}) of the theory. The vector $|1\rangle=a^\dagger|0\rangle$ is the state representing of one ``particle'' and $|n\rangle$ is the state representing of $n$ particles. Because of the action of the operators $a^\dagger$ and $a$ over the vectors $|n\rangle$, showed in \hyperref[creation_annihilation_number_eigenvectors_ps]{(\ref*{creation_annihilation_number_eigenvectors})}, it is easy to understand why the name of \textit{creation} and \textit{annihilation}.  

If we have a multipartite system composed by $N$ particles vibrating at different frequencies $\omega_k$, the Hamiltonian of the total system is the sum of the Hamiltonians of every one:
\phantomsection\label{harmonic_oscillator_multipartite_ps}\begin{equation}\label{harmonic_oscillator_multipartite}
\textbf{H}=\hbar\sum_{k=1}^N\left(\omega_ka^\dagger_ka_k+\frac{1}{2}\omega_k\right)\!,
\end{equation}
where now, the canonical commutation relations are:
\phantomsection\label{Commutator_creation_annihilation_f_ps}\begin{equation}\label{Commutator_creation_annihilation_f}
\left[\vphantom{a^\dagger}\right.\hspace{-0.1cm}a_k,a_{k'}^\dagger\hspace{-0.1cm}\left.\vphantom{a^\dagger}\right]=\delta_{kk'},\quad\left[\vphantom{a^\dagger}\right.\hspace{-0.1cm}a_k,a_{k'}\hspace{-0.1cm}\left.\vphantom{a^\dagger}\right]=\left[\vphantom{a^\dagger}\right.\hspace{-0.1cm}a_k^\dagger,a_{k'}^\dagger\hspace{-0.1cm}\left.\vphantom{a^\dagger}\right]=0.
\end{equation}
The physical meaning of this commutation relations is that the creation and annihilation of particles of each mode does not affect the others.

Repeating the previous construction, we set now the Fock space $\mathcal{F}_N$ that will be generated by orthonormal vectors $|n_1,\ldots,n_N\rangle$ with $n_1,\ldots,n_N=0,1,\ldots$\,, and denoting the fundamental state $|0,\ldots,0\rangle$ by $|0\rangle$, we get:
\begin{equation}
a_k^\dagger|0\rangle=|0,\ldots,\overset{k}{1},\ldots,0\rangle,\qquad a_k|0\rangle=0.
\end{equation}
Notice that $\mathcal{F}_N=\mathcal{F}_1\otimes\cdots\otimes\mathcal{F}_1^{\hspace{-1.1cm}N}$\hspace{0.05cm}.

In addition to the abstract Fock space representation of the quantum harmonic oscillator, it is sometimes useful (see later on \hyperref[resume_heis]{{\color{black}\textbf{section~\ref*{resume_heis}}}}) to use the representation provided by the operators presented before in \hyperref[rep_Q_and_P_ps]{(\ref*{rep_Q_and_P})}. For that, consider the function
\begin{equation*}
\psi_0(x)=\sqrt[4]{\frac{\omega}{\pi\hbar}}\e
\,,
\end{equation*}
hence,
\begin{multline*}
a\psi_0(x)=\frac{1}{\sqrt{2\hbar}}\left(\sqrt{\omega}\textbf{Q}+\frac{i}{\sqrt{\omega}}\textbf{P}\right)\psi_0(x)\\
=\frac{1}{\sqrt{2\hbar}}\left(\sqrt{\omega}x+\frac{i}{\sqrt{\omega}}(i\omega x)\right)\psi_0(x)=0,
\end{multline*}
and
\begin{equation*}
\|\psi_0\|^2_{L^2}=\hspace{-.2cm}\int\limits_{\hspace{-.04cm}-\infty}^{\hspace{.45cm}\infty}\hspace{-0.1cm}|\psi_0(x)|^2\diff x=\sqrt{\frac{\omega}{\pi\hbar}}\hspace{-.2cm}\int\limits_{\hspace{-.04cm}-\infty}^{\hspace{.45cm}\infty}\hspace{-0.1cm}\e
\,\diff x=1.
\end{equation*}

Clearly the functions
\begin{equation*}
\psi_n(x)=\frac{(a^\dagger)^n}{\sqrt{n!}}\psi_0(x),
\end{equation*}
provide an orthonormal basis of $L^2(\mathbb{R})$ because they are the eigenfunctions of the self-adjoint operator $\textbf{H}$ in eq.\,\hyperref[harmonic_oscillator_2_ps]{(\ref*{harmonic_oscillator_2})}.

The map $\mathcal{F}_1\longrightarrow L^2(\mathbb{R})$ defined by $|n\rangle\rightsquigarrow\psi_n$ defines a unitary operator between both spaces providing the \textit{coordinate representation} of the quantum harmonic oscillator.

Expanding the functions $\psi_n$, we get \hyperref[ps_Ba98]{{\color{red}[\citen*{Ba98}, {page 156}]}}:
\begin{equation*}
\psi_n(x)=\sqrt[4]{\frac{\omega}{\pi\hbar}}\sqrt{\frac{1}{2^nn!}}H_n\left(\sqrt{\frac{\omega}{\hbar}}x\right)\e
\,,
\end{equation*}
where $H_n(x)$ are the Hermite Polynomials
\phantomsection\label{Hermite_pols_ps}
\begin{equation}\label{Hermite_pols}
H_n(x)=(-1)^n\e^{x^2}\frac{\diff^n}{\diff x^n}\e^{-x^2}.
\end{equation}

We see that the difference between the Hamiltonians of the classical harmonic oscillator \hyperref[Hamiltonian_electromagnetic_field_2_ps]{(\ref*{Hamiltonian_electromagnetic_field_2})} and of the quantum harmonic oscillator \hyperref[harmonic_oscillator_ps]{(\ref*{harmonic_oscillator})} is the substitution of the classical position and momentum $Q$ and $P$ by the operators $\textbf{Q}$ and $\textbf{P}$ respectively. This is a very extended way to generalize classical results to the quantum case and it is commonly called \textit{canonical quantization}:
\vspace{-0.2cm}
\phantomsection
\label{ps_Classical_to_Quantum_EMF}
\phantomsection\label{Classical_to_Quantum_EMF_ps}\begin{align*}
&\hspace{-2.43cm}Q_p\longrightarrow\textbf{Q}_p\nonumber\\
&\hspace{-2.37cm}P_p\longrightarrow\textbf{P}_p\nonumber\\
\end{align*}

\vspace{-1.3cm}\begin{equation}\label{Classical_to_Quantum_EMF}
\hspace{2.35cm}H_{cl}\longrightarrow\textbf{H}=\frac{1}{2}\sum_{p=1}^\infty\left(m_p\textbf{P}_p^2+\omega_p^2\textbf{Q}_p^2\right).
\end{equation}

\vspace{-0.2cm}\subsection{Canonical Quantization of the Electromagnetic field}

Applying the canonical quantization scheme \hyperref[ps_Classical_to_Quantum_EMF]{(\ref*{Classical_to_Quantum_EMF})} to the E.M. field and writing it in terms of the creation and annihilation operators~\hyperref[creation_annihilation_ps]{(\ref*{creation_annihilation})}, the classical electric and magnetic fields become:
\begin{align*}
\boldsymbol{E}(\boldsymbol{x},t)&\longrightarrow\widehat{\textbf{E}}(\boldsymbol{x},t)=\sqrt{2\pi\hbar}\sum_{p=1}^\infty \sqrt{\omega_p}\left(a_p^\dagger(t)+a_p(t)\right)\boldsymbol{u}_p(\boldsymbol{x}),\\
\boldsymbol{B}(\boldsymbol{x},t)&\longrightarrow\widehat{\textbf{B}}(\boldsymbol{x},t)=ic\sqrt{2\pi\hbar}\sum_{p=1}^\infty \frac{1}{\sqrt{\omega_p}}\left(a_p^\dagger(t)-a_p(t)\right)\!\big(\nabla\times\boldsymbol{u}_p(\boldsymbol{x})\big),
\end{align*}
where
\phantomsection\label{annihilation_time_dependence_ps}\begin{equation*}
a_p(t)=a_p\e^{-i\omega_pt}.
\end{equation*}
Therefore, the electric and magnetic fields can be written finally as:
\phantomsection\label{Classical_to_Quantum_EF_3_ps}
\begin{align}\label{Classical_to_Quantum_EF_3}
\widehat{\textbf{E}}(\boldsymbol{x},t)&=\sqrt{2\pi\hbar}\sum_{p=1}^\infty \sqrt{\omega_p}\left(a_p^\dagger\e^{i\omega_pt}+a_p\e^{-i\omega_pt}\right)\boldsymbol{u}_p(\boldsymbol{x}),\nonumber\\
\widehat{\textbf{B}}(\boldsymbol{x},t)&=ic\sqrt{2\pi\hbar}\hspace{-0.0cm}\sum_{p=1}^\infty \frac{1}{\sqrt{\omega_p}}\left(a_p^\dagger\e^{i\omega_pt}-a_p\e^{-i\omega_pt}\right)\!\big(\nabla\times\boldsymbol{u}_p(\boldsymbol{x})\big).
\end{align}

Notice that the Electromagnetic field operators can be separated in two parts, one that creates photons of frequencies $\omega_p$ and other that destroys them:
\phantomsection\label{Classical_to_Quantum_EF_4_ps}\vspace{-0.5cm}\begin{align}\label{Classical_to_Quantum_EF_4}
\widehat{\textbf{E}}(\boldsymbol{x},t)&=\widehat{\textbf{E}}^{-}(\boldsymbol{x},t)+\widehat{\textbf{E}}^{+}(\boldsymbol{x},t),\nonumber\\
\widehat{\textbf{B}}(\boldsymbol{x},t)&=\widehat{\textbf{B}}^{-}(\boldsymbol{x},t)+\widehat{\textbf{B}}^{+}(\boldsymbol{x},t),
\end{align}
where
\begin{align*}
\widehat{\textbf{E}}^{+}(\boldsymbol{x},t)&=\sqrt{2\pi\hbar}\sum_{p=1}^\infty \sqrt{\omega_p}a_p\e^{-i\omega_pt}\boldsymbol{u}_p(\boldsymbol{x}),\\
\widehat{\textbf{B}}^{+}(\boldsymbol{x},t)&=-ic\sqrt{2\pi\hbar}\sum_{p=1}^\infty \frac{1}{\sqrt{\omega_p}}a_p\e^{-i\omega_pt}\left(\nabla\times\boldsymbol{u}_p(\boldsymbol{x})\right),
\end{align*}
is the part that destroys them, and
\begin{align*}
\widehat{\textbf{E}}^{-}(\boldsymbol{x},t)&=\widehat{\textbf{E}}^{+}(\boldsymbol{x},t)^\dagger,\\
\widehat{\textbf{B}}^{-}(\boldsymbol{x},t)&=\widehat{\textbf{B}}^{+}(\boldsymbol{x},t)^\dagger,
\end{align*}
the part that creates them.

If we come back to \hyperref[section_telecomm]{{\color{black}\textbf{section~\ref*{section_telecomm}}}}, we remember that in Amplitude Modulation a signal is composed by a function, that only depends on the message, multiplied by a sinusoidal function that corresponds to the frequency in which the message is sent~\hyperref[AM_ps]{(\ref*{AM})}. If we emit different messages at different frequencies, the total signal is a linear superposition of all the modulated signals, hence we can write it as:
\phantomsection\label{AM_multiplex_ps}\begin{equation}\label{AM_multiplex}
x(t)=\hspace{-0.1cm}\sum_{k=1}^{\infty}\big(1+x_{m,k}(t)\big)\e^{-i\omega_{c,k} t}.
\end{equation}

If we compare this formula with eq.\,\hyperref[Classical_to_Quantum_EF_3_ps]{(\ref*{Classical_to_Quantum_EF_3})}, we see that are similar, however the role of the modulating signal is played by
\phantomsection\label{Quantum_modulating_signal_ps}\begin{equation}\label{Quantum_modulating_signal}
a^\dagger_p|n_1,n_2,\ldots,n_p,\ldots\,\rangle=|n_1,n_2,\ldots,a_p^\dagger n_p,\ldots\,\rangle,
\end{equation}
for that, experimentalists thought that the way for reconstructing the quantum state $|n_p\rangle$ should be an adaptation of the process saw in \hyperref[Homodyne_cl]{{\color{black}\textbf{section~\ref*{section_telecomm}}}} for demodulating signals in telecommunications. However, we cannot use the same devices that were shown in \hyperref[Homodyne_cl]{{\color{black}\textbf{section~\ref*{section_telecomm}}}}, because the \textit{quantum modulating signal} is an operator on $\mathcal{B}(\mathcal{H})$, therefore here is where take into action the photodetection process.

Before involving in this task, let us comment briefly the notation of state we have described before. Recall that a quantum state $|n\rangle$ of the harmonic oscillator is given by \hyperref[vector_harmonic_oscillator_f_ps]{(\ref*{vector_harmonic_oscillator_f})}. However, physically, states differing in phases are equivalent, so it is convenient to consider the projector operators instead. Moreover, when dealing with ensembles of systems, their states are statistical mixtures of such pure states, i.e.,
\begin{equation}
\boldsymbol{\rho}=\sum_{i=1}^N \,p_i|n_i\rangle\langle n_i|,\qquad p_i\geq 0,\qquad \sum_{i=1}^N\,p_i=1.
\end{equation}
We will call such states \textit{mixed states}. Thus, a mixed state for the quantum E.M. field will be an operator $\boldsymbol{\rho}$ of the form:
\begin{equation}
\boldsymbol{\rho}=\hspace{-0.2cm}\sum_{i_1,\ldots i_N}\hspace{-0.1cm} p_{i_1,\ldots,i_N}|n_{i_1},\ldots,n_{i_N}\rangle\langle n_{i_1},\ldots,n_{i_N}|
\end{equation}
where $p_{i_1,\ldots i_N}\geq 0$ and $\hspace{-0.1cm}\displaystyle{\sum_{i_1,\ldots i_N}\hspace{-0.1cm}p_{i_1,\ldots,i_N}=1}$.

Typically, mixed states for the E.M. field consist of a finite number of excitations, i.e., the sum above is finite. Hence, we may think of it as a mixed state for a finite ensemble of harmonic oscillators. This is the point of view we will adopt in what follows.

Let us point out that mixed states are also called \textit{density operators} and a formal treatment of them will be given in \hyperref[chap_tom_qu]{{\color{black}\textbf{chapter~\ref*{chap_tom_qu}}}}.

\phantomsection\label{section_QM_ps}
\section{Reconstruction of matrix elements of quantum density operators}\label{section_real_QM_ps}

As it was commented \hyperref[section_telecomm]{{\color{black}\textbf{at the beginning of section~\ref*{section_telecomm}}}}, experimentalists in Quantum Optics wondered how to measure the matrix elements of the representation of the quantum state of a light source, and the answer they found was to measure the Radon Transform of a quantity
\noindent that is directly related with these matrix elements, the \textit{Wigner's function}\footnote{More information about Wigner's function and its relation with the reconstruction of quantum density matrices can be found in modern texts due to Giuseppe Marmo \textit{et al.}, as for example \RED{[}\citen{Er07}\RED{,}\hspace{0.05cm}\citen{Ca08}\RED{,}\hspace{0.05cm}\citen{As15}\RED{]}.}, \cite{Wi32}:
\phantomsection
\label{section_Wigner}
\phantomsection\label{Wigner_function_ps}\begin{equation}\label{Wigner_function}
\boldsymbol{\rho}_w(\boldsymbol{q},\boldsymbol{p})=\frac{1}{(2\pi\hbar)^n}\hspace{-.23cm}\int\limits_{\hspace{.5cm}\mathbb{R}^n}\hspace{-.19cm} \left\langle \boldsymbol{q}-\frac{\boldsymbol{y}}{2}\right|\boldsymbol{\rho}\left|\boldsymbol{q}+\frac{\boldsymbol{y}}{2}\right\rangle\e^{ipy/\hbar}\diff^n \boldsymbol{y},
\end{equation}
where $\boldsymbol{\rho}$ is the operator representing a mixed state (see eq.\,\hyperref[definition_state_rho_ps]{(\ref*{definition_state_rho})}), $n$ the number of excited modes describing such states, and
\begin{equation*}
\boldsymbol{q}=(q_1,q_2,\ldots,q_n),\qquad\boldsymbol{p}=(p_1,p_2,\ldots,p_n).
\end{equation*}

To express the inner product on the Hilbert space $\mathcal{H}$, we have used the Bra-ket notation introduced by Dirac where $|\cdot\rangle$ denotes a vector on $\mathcal{H}$ and $\langle\cdot|$ denotes its dual.

The matrix elements of $\boldsymbol{\rho}$, in coordinate representation, can be recovered from the Fourier Transform of the Wigner's function:
\phantomsection\label{matrix_element_rho_ps}\begin{equation}\label{matrix_element_rho}
\boldsymbol{\rho}(\boldsymbol{q},\boldsymbol{q}')=\hspace{-.3cm}\int\limits_{\hspace{.5cm}\mathbb{R}^n}\hspace{-.19cm}\e^{-i\boldsymbol{p}\cdot(\boldsymbol{q}'-\boldsymbol{q})/\hbar}\boldsymbol{\rho}_w\left(\frac{\boldsymbol{q}'+\boldsymbol{q}}{2},\boldsymbol{p}\right)\diff^n \boldsymbol{p}.
\end{equation}
An equivalent formula can be obtained in \textit{momentum representation}.

The \textit{eigenvectors} of momentum and position operators form an \textit{orthogonal} basis that satisfy:
\phantomsection\label{Position_Momentum_relation_ps}\begin{equation}\label{Position_Momentum_relation}
\langle \boldsymbol{q}|\boldsymbol{p}\rangle=\frac{1}{(2\pi\hbar)^{n/2}}\e
,
\end{equation}
and to pass from one representation to other, we use the Fourier Transform:
\phantomsection\label{Fourier_Transform__ps}\begin{equation}\label{Fourier_Transform_}
\psi(\boldsymbol{p})=\langle \boldsymbol{p}|\psi\rangle=\hspace{-.3cm}\int\limits_{\hspace{.5cm}\mathbb{R}^n}\hspace{-.19cm}\langle \boldsymbol{p}|\boldsymbol{q}\rangle\langle \boldsymbol{q}|\psi\rangle\diff^n\boldsymbol{q}=\frac{1}{(2\pi\hbar)^{n/2}}\hspace{-.23cm}\int\limits_{\hspace{.5cm}\mathbb{R}^n}\hspace{-.19cm}\e%
\,\psi(\boldsymbol{q})\diff^n\boldsymbol{q}.
\end{equation}

A function $\boldsymbol{\rho}_{ps}(\boldsymbol{q},\boldsymbol{p})$ is a probability distribution in phase space if there exists a mixed state $\boldsymbol{\rho}$ such that:
\phantomsection
\label{Phase-space_distribution1_ps}
\begin{align}
\label{Phase-space_distribution1}\langle \boldsymbol{q}|\boldsymbol{\rho}|\boldsymbol{q}\rangle&=\hspace{-.3cm}\int\limits_{\hspace{.5cm}\mathbb{R}^n}\hspace{-.19cm}\boldsymbol{\rho}_{ps}(\boldsymbol{q},\boldsymbol{p})\diff^n \boldsymbol{p},\tag{1.6.5a}\\
\label{Phase-space_distribution2}\langle \boldsymbol{p}|\boldsymbol{\rho}|\boldsymbol{p}\rangle&=\hspace{-.3cm}\int\limits_{\hspace{.5cm}\mathbb{R}^n}\hspace{-.19cm}\boldsymbol{\rho}_{ps}(\boldsymbol{q},\boldsymbol{p})\diff^n \boldsymbol{q},\tag{1.6.5b}\\
\label{Phase-space_distribution3}\boldsymbol{\rho}_{ps}&(\boldsymbol{q},\boldsymbol{p})\geq 0.\tag{1.6.5c}
\end{align}
\setcounter{equation}{5}

Wigner proved that there is not a function that satisfy these three conditions, however he found a function, that is not a probability distribution because is not bigger or equal than zero, which satisfies the marginal probability conditions~\hyperref[Phase-space_distribution1_ps]{(\ref*{Phase-space_distribution1})} and~\hyperref[Phase-space_distribution2]{(\ref*{Phase-space_distribution2})}. That function is the Wigner's function $\boldsymbol{\rho}_w(\boldsymbol{q},\boldsymbol{p})$ defined previously in~\hyperref[Wigner_function_ps]{(\ref*{Wigner_function})}:
\phantomsection\label{Wigner_state_square}
\begin{align*}
\int\limits_{\hspace{.5cm}\mathbb{R}^n}\hspace{-.19cm}\boldsymbol{\rho}_w(\boldsymbol{q},\boldsymbol{p})\diff^n \boldsymbol{p}&=\frac{1}{(2\pi\hbar)^n}\hspace{-.33cm}\int\limits_{\hspace{.5cm}\mathbb{R}^{2n}}\hspace{-.23cm}\left\langle \boldsymbol{q}-\frac{\boldsymbol{y}}{2}\right|\boldsymbol{\rho}\left|\boldsymbol{q}+\frac{\boldsymbol{y}}{2}\right\rangle\e
\,\diff^n \boldsymbol{y}\diff^n \boldsymbol{p}\\
 &=\hspace{-.2cm}\int\limits_{\hspace{-.04cm}-\infty}^{\hspace{.45cm}\infty}\hspace{-.04cm}\left\langle \boldsymbol{q}-\frac{\boldsymbol{y}}{2}\right|\boldsymbol{\rho}\left|\boldsymbol{q}+\frac{\boldsymbol{y}}{2}\right\rangle\delta^{(n)}(\boldsymbol{y})\diff^n y=\langle \boldsymbol{q}|\boldsymbol{\rho}|\boldsymbol{q}\rangle,\tag*{\hyperref[Phase-space_distribution1_ps]{(\ref*{Phase-space_distribution1})}}
 \end{align*}
 \begin{equation*}
 \int\limits_{\hspace{.5cm}\mathbb{R}^n}\hspace{-.19cm}\boldsymbol{\rho}_w(\boldsymbol{q},\boldsymbol{p})\diff^n \boldsymbol{q}=\langle \boldsymbol{p}|\boldsymbol{\rho}|\boldsymbol{p}\rangle,\tag*{\hyperref[Phase-space_distribution2]{(\ref*{Phase-space_distribution2})}}
 \end{equation*}
where $\delta^{(n)}(\boldsymbol{x})$ is the $n$-dimensional analogue of Dirac's delta function \hyperref[delta_function_ps]{(\ref*{delta_function})} with integral representation given by:
\phantomsection\label{delta_function_n_dim_ps}\begin{equation}\label{delta_function_n_dim}
\delta^{(n)}(\boldsymbol{x})=\frac{1}{(2\pi)^n}\hspace{-.23cm}\int\limits_{\hspace{.5cm}\mathbb{R}^n}\hspace{-.19cm} \e^{i\boldsymbol{k}\cdot\boldsymbol{x}}\diff^n \boldsymbol{k}.
\end{equation}
The Wigner's function, although is not a probability distribution, is normalized:
\phantomsection\label{Wigner_function_normalization_ps}\begin{equation}\label{Wigner_function_normalization}
\int\limits_{\hspace{.5cm}\mathbb{R}^{2n}}\hspace{-.19cm}\boldsymbol{\rho}_w(\boldsymbol{q},\boldsymbol{p})\diff^n\boldsymbol{q}\diff^n\boldsymbol{p}=\hspace{-.3cm}\int\limits_{\hspace{.5cm}\mathbb{R}^n}\hspace{-.19cm}\langle\boldsymbol{p}|\boldsymbol{\rho}|\boldsymbol{p}\rangle\diff^n\boldsymbol{p}=\Tr(\boldsymbol{\rho})=1.
\end{equation} 

For the following computations, is important to define first functions of operators on a Hilbert space. Let $\textbf{A}$ be a self-adjoint operator on a Hilbert space $\mathcal{H}$. Let us define $f(\textbf{A})$ as a regular enough function on $\textbf{A}$:
\phantomsection\label{Functions of operator_ps}\begin{equation}\label{Functions of operator}
f(\textbf{A})\coloneq\hspace{-.2cm}\int\limits_{\hspace{-.04cm}-\infty}^{\hspace{.45cm}\infty}\hspace{-.1cm}f(a)E_{\textbf{A}}(\diff a),
\end{equation}
where $E_{\textbf{A}}(\diff a)$ is the spectral measure associated to the self-adjoint operator $\textbf{A}$ (see for instance \hyperref[Re80_ps]{\RED{[\citen*{Re80}, ch.\,7]}}). From this, we can define the delta function of an operator on a Hilbert space as a distribution with values in operators (see \hyperref[delta_to_section_1_ps]{{\color{black}\textbf{section~\ref*{section_tomograms_group}}}}). 

However, if $X$ is an eigenvalue of $\textbf{A}$, $\textbf{A}|X\rangle=X|X\rangle$, we can see that the delta function $\delta(X\mathds{1}-\textbf{A})$ is nothing but the projector over the eigenvector $|X\rangle$:

\phantom{a}
\vspace{-1.3cm}\phantomsection\label{delta_function_operator_ps}\begin{multline}\label{delta_function_operator}
\delta(X\mathds{1}-\textbf{A})=\frac{1}{2\pi}\hspace{-.2cm}\int\limits_{\hspace{-.04cm}-\infty}^{\hspace{.45cm}\infty}\hspace{-0.1cm}\e^{ik(X\mathds{1}-\textbf{A})}\diff k=\hspace{-.3cm}\int\limits_{\hspace{.5cm}\mathbb{R}^{2}}\hspace{-.19cm}\e^{ik(X-a)}E_{\textbf{A}}(\diff a)\diff k\\
=\hspace{-.2cm}\int\limits_{\hspace{-.04cm}-\infty}^{\hspace{.45cm}\infty}\delta(X-a)E_{\textbf{A}}(\diff a)=|X\rangle\langle X|.
\end{multline}
Hence, if we compute the mean value of the delta function of the operator $X\mathds{1}-\textbf{A}$ (with $X$ an eigenvalue of $\textbf{A}$) over a state $\boldsymbol{\rho}$, we get:
\phantomsection\label{delta_function_operator_mean_value_ps}\begin{equation}\label{delta_function_operator_mean_value}
\langle\delta(X\mathds{1}-\textbf{A})\rangle_{\boldsymbol{\rho}}=\Tr\!\big(\boldsymbol{\rho}\delta(X\mathds{1}-\textbf{A})\big)=\Tr(\boldsymbol{\rho}|X\rangle\langle X|)=\langle X|\boldsymbol{\rho}|X\rangle .
\end{equation}

To implement the reconstruction setting of the matrix elements of a state $\boldsymbol{\rho}$, we need to define new position and momentum variables via a rotation of angle $\phi$:
\phantomsection\label{Position_momentum_rotation_ps}\begin{equation}\label{Position_momentum_rotation}
\begin{pmatrix}
  q_\phi \\
  p_\phi \\
\end{pmatrix}=\begin{pmatrix}
                \cos\phi & \sin\phi \\
                -\sin\phi & \cos\phi \\
              \end{pmatrix}\begin{pmatrix}
  q \\
  p \\
\end{pmatrix}.
\end{equation}
The marginal probability distribution of $q_\phi$ is
\phantomsection\label{Marginal_probability_q_phi_ps}\begin{multline}\label{Marginal_probability_q_phi}
P_\phi(q_\phi)=\langle q_\phi|\boldsymbol{\rho}|q_\phi\rangle\\
=\hspace{-0.15cm}\int\limits_{\hspace{-.04cm}-\infty}^{\hspace{.45cm}\infty}\boldsymbol{\rho}_{w}(q_\phi\cos\phi-p_\phi\sin\phi,q_\phi\sin\phi+p_\phi\cos\phi)\diff p_\phi,
\end{multline}
where $|q_\phi\rangle$ is the eigenvector of $\textbf{Q}_\phi=\textbf{Q}\cos\phi+\textbf{P}\sin\phi$ with eigenvalue $q_\phi$.
We have put $n=1$ for simplicity, however this result can be generalized for any $n$ by changing the momentum and position variables by $n$-dimensional vectors.

Let us prove now eq.\,\hyperref[Marginal_probability_q_phi_ps]{(\ref*{Marginal_probability_q_phi})}. From~\hyperref[delta_function_operator_mean_value_ps]{(\ref*{delta_function_operator_mean_value})}, we have that
\phantomsection\label{Marginal_probability_q_phi_1_ps}\begin{equation}\label{Marginal_probability_q_phi_1}
P_\phi(q_\phi)=\langle q_\phi|\boldsymbol{\rho}|q_\phi\rangle=\Tr\!\big(\boldsymbol{\rho}\delta(q_\phi\mathds{1}-\textbf{Q}\cos\phi-\textbf{P}\sin\phi)\big).
\end{equation}
Hence, using the exponential representation of the delta function, we get:
\begin{multline*}\label{Marginal_probability_q_phi_2}
\Tr\!\big(\boldsymbol{\rho}\delta(q_\phi\mathds{1}-\textbf{Q}\cos\phi-\textbf{P}\sin\phi)\big)\\
=\frac{1}{2\pi}\Tr\!\left(\boldsymbol{\rho}\hspace{-.2cm}\int\limits_{\hspace{-.04cm}-\infty}^{\hspace{.45cm}\infty}\hspace{-0.1cm}\e^{ikq_\phi\mathds{1}}\e^{-ik(\textbf{Q}\cos\phi+\textbf{P}\sin\phi)}\diff k\right),
\end{multline*}
and applying the Baker--Campbell--Hausdorff formula
\phantomsection\label{BCH_ps}\begin{equation}\label{BCH}
\e^{\textbf{A}+\textbf{B}}=\e^{\textbf{A}}\e^{\textbf{B}}\e^{-1/2[\textbf{A},\textbf{B}]},
\end{equation}
for $\textbf{A}$ and $\textbf{B}$ satisfying $\left[\vphantom{a^\dagger}\textbf{A},[\textbf{A},\textbf{B}]\right]=\left[\vphantom{a^\dagger}\textbf{B},[\textbf{A},\textbf{B}]\right]=0$, we obtain that
\begin{align*}\label{Marginal_probability_q_phi_3}
\frac{1}{2\pi}&\Tr\!\left(\boldsymbol{\rho}\hspace{-.2cm}\int\limits_{\hspace{-.04cm}-\infty}^{\hspace{.45cm}\infty}\hspace{-.1cm}\e^{ikq_\phi\mathds{1}}\e^{-ik(\textbf{Q}\cos\phi+\textbf{P}\sin\phi)}\diff k\right)\nonumber\\
&\hspace{-0.1cm}=\frac{1}{2\pi}\Tr\!\Bigg(\boldsymbol{\rho}\hspace{-0.05cm}\int\limits_{\hspace{-.26cm}-\infty}^{\hspace{.15cm}\infty}\hspace{-.1cm}\int\limits_{\hspace{-.08cm}-\infty}^{\hspace{.15cm}\infty}\e^{ikq_\phi\mathds{1}}\e
\e^{-ikq\cos\phi}\langle q|\boldsymbol{\rho}|q+k\sin\phi\rangle\diff q\diff k.
\end{align*}
Making the following change of variable in $q$:
\begin{equation*}\label{change_variables_marginal_distribution}
\tilde{q}=q+\frac{k}{2}\sin\phi,
\end{equation*}
and using eq.\,\hyperref[matrix_element_rho_ps]{(\ref*{matrix_element_rho})}, we obtain that
\begin{align*}
\frac{1}{2\pi}\hspace{-0.05cm}\int\limits_{\hspace{-.26cm}-\infty}^{\hspace{.15cm}\infty}\hspace{-.1cm}\int\limits_{\hspace{-.08cm}-\infty}^{\hspace{.15cm}\infty}&\e^{ikq_\phi}\e
\e^{-ikq\cos\phi}
\langle q|\boldsymbol{\rho}|q+k\sin\phi\rangle\diff q\diff k\\
&=\frac{1}{2\pi}\hspace{-0.05cm}\int\limits_{\hspace{-.26cm}-\infty}^{\hspace{.15cm}\infty}\hspace{-.1cm}\int\limits_{\hspace{-.08cm}-\infty}^{\hspace{.15cm}\infty}\e^{ik(q_\phi-\tilde{q}\cos\phi)}\langle\tilde{q}-\frac{k}{2}\sin\phi|\boldsymbol{\rho}|\tilde{q}+\frac{k}{2}\sin\phi\rangle\diff\tilde{q}\diff k\\
&=\frac{1}{2\pi}\hspace{-0.05cm}\int\limits_{\hspace{-.34cm}-\infty}^{\hspace{.15cm}\infty}\hspace{-.08cm}\int\limits_{\hspace{-.15cm}-\infty}^{\hspace{.15cm}\infty}\hspace{-.08cm}\int\limits_{\hspace{-.04cm}-\infty}^{\hspace{.15cm}\infty}\e^{ik(q_\phi-\tilde{q}\cos\phi-p\sin\phi)}\boldsymbol{\rho}_{w}(\tilde{q},p)\diff p\diff\tilde{q}\diff k\nonumber\\
&=\hspace{-0.05cm}\int\limits_{\hspace{-.26cm}-\infty}^{\hspace{.15cm}\infty}\hspace{-.1cm}\int\limits_{\hspace{-.08cm}-\infty}^{\hspace{.15cm}\infty}\boldsymbol{\rho}_{w}(\tilde{q},p)\delta(q_\phi-\tilde{q}\cos\phi-p\sin\phi)\diff p\diff\tilde{q}.
\end{align*}
\\
\\
Then, we have shown that
\phantomsection\label{Tomogram_quantum_to_classic_4_ps}\begin{multline}\label{Tomogram_quantum_to_classic_4}
\Tr\!\big(\boldsymbol{\rho}\delta(q_\phi\mathds{1}-\textbf{Q}\cos\phi-\textbf{P}\sin\phi)\big)\\
=\hspace{-0.05cm}\int\limits_{\hspace{-.26cm}-\infty}^{\hspace{.15cm}\infty}\hspace{-.1cm}\int\limits_{\hspace{-.08cm}-\infty}^{\hspace{.15cm}\infty}\boldsymbol{\rho}_{w}(q,p)\delta(q_\phi-q\cos\phi-p\sin\phi)\diff q\diff p.
\end{multline}
From~\hyperref[Position_momentum_rotation_ps]{(\ref*{Position_momentum_rotation})}, we have that
\begin{equation*}\label{Position_momentum_rotation_inverse}
p=q_\phi\sin\phi+p_\phi\cos\phi,
\end{equation*}
hence, finally we have:
\begin{equation*}
\Tr\!\big(\boldsymbol{\rho}\delta(q_\phi\mathds{1}-\textbf{Q}\cos\phi-\textbf{P}\sin\phi)\big)=\hspace{-.2cm}\int\limits_{\hspace{-.04cm}-\infty}^{\hspace{.45cm}\infty}\hspace{-0.15cm}\boldsymbol{\rho}_{w}\left(\frac{q_\phi}{\cos\phi}-p\tan\phi,p\right)\frac{\diff p}{|\cos\phi|}
\end{equation*}

\phantom{a}
\vspace{-0.3cm}\phantomsection\label{Marginal_probability_q_phi_5_ps}\vspace{-1.3cm}\begin{align*}
&=\hspace{-.2cm}\int\limits_{\hspace{-.04cm}-\infty}^{\hspace{.45cm}\infty}\hspace{-0.15cm}\boldsymbol{\rho}_{w}\bigg(\frac{q_\phi}{\cos\phi}(1-\sin^2\phi)-p_\phi\sin\phi, q_\phi\sin\phi+p_\phi\cos\phi\bigg)\diff p_\phi \\
&=\hspace{-.2cm}\int\limits_{\hspace{-.04cm}-\infty}^{\hspace{.45cm}\infty}\hspace{-0.15cm}\boldsymbol{\rho}_{w}(q_\phi\cos\phi-p_\phi\sin\phi,q_\phi\sin\phi+p_\phi\cos\phi)\diff p_\phi.
\end{align*}

\vspace{-.6cm}{\hfill\hyperref[Marginal_probability_q_phi_ps]{{\color{black}$\blacksquare$}}}

Because the probability distribution $P_\phi(q_\phi)$ is an average of the Wigner's function over the plane $q_\phi-q\cos\phi-p\sin\phi=0$, people working in Quantum Optics decided to call it a \textit{Tomogram}. 

Applying now the Inverse Radon Transform in polar coordinates, as in~\hyperref[Inverse_Radon_Transform_CAT_2_ps]{(\ref*{Inverse_Radon_Transform_CAT_2})}, to the tomograms $P_\phi(q_\phi)$, we can recover the Wigner's function $\boldsymbol{\rho}_{w}$:

\phantom{a}
\vspace{-1.4cm}\phantomsection\label{Inverse_Radon_Transform_Wigner_ps}\begin{equation}\label{Inverse_Radon_Transform_Wigner}
\boldsymbol{\rho}_{w}(q,p)=\frac{1}{(2\pi)^2}\hspace{-0.05cm}\int\limits_{\hspace{-.12cm}-\infty}^{\hspace{.35cm}\infty}\hspace{-.3cm}\int\limits_{\hspace{.15cm}0}^{\hspace{.35cm}\infty}\hspace{-.33cm}\int\limits_{\hspace{.15cm}0}^{\hspace{.35cm}2\pi}\hspace{-.06cm}P_\phi(X)\e^{-ik(X-q\cos\phi -p\sin\phi)}k\hspace{0.02cm}\diff\phi\diff k\diff X,
\end{equation}
and finally, from~\hyperref[matrix_element_rho_ps]{(\ref*{matrix_element_rho})}, we can reconstruct easily the matrix elements of the density operator $\boldsymbol{\rho}(q,q')$:
\phantomsection\label{Inverse_Radon_Transform_Wigner_ps}\begin{multline}\label{Inverse_Radon_Transform_Wigner}
\boldsymbol{\rho}(q,q')=\frac{1}{(2\pi)^2}\hspace{-0.05cm}\int\limits_{\hspace{-.26cm}-\infty}^{\hspace{.15cm}\infty}\hspace{-.1cm}\int\limits_{\hspace{-.08cm}-\infty}^{\hspace{.15cm}\infty}\hspace{-.23cm}\int\limits_{\hspace{.15cm}0}^{\hspace{.35cm}\infty}\hspace{-.33cm}\int\limits_{\hspace{.15cm}0}^{\hspace{.35cm}2\pi}\hspace{-.06cm}P_\phi(X)\e
k\hspace{0.02cm}\diff\phi\diff k\diff X\diff p.
\end{multline}

\section{Photodetection}\label{section_photodetection}

Now, we will focus in how to measure the tomograms of a quantum state of a radiation source. To get that, we will adapt the homodyne and heterodyne detection described in \hyperref[ph_section_Homodyne_cl]{{\color{black}\textbf{section~\ref*{section_telecomm}}}} for quantum devices \cite{Ar03}.
In this setting, the equivalent to the mixing of two electric fields is more complicated because now, they are operators on a Hilbert space, and instead of a simple multiplication, we have a tensorial product. The devices that let us mix two signals in Quantum Optics are beam-splitters and photodetectors.

\begin{figure}[htbp]
\centering
\includegraphics[]{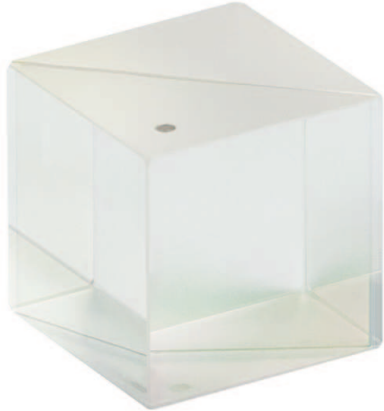}
\caption{\hfil\MYGREEN{Figure 1.7.1}: Beam-splitter.\hfil}
\label{Bs}
\end{figure}

\noindent{\MYBROWN{a) Beam-splitters}}:

\vspace{0.1cm}A beam-splitter (see {\changeurlcolor{mygreen}\hyperref[Bs]{Figure \ref*{Bs}}}) is an optical device that separates a ray in two components. Generally, it is composed by two triangular prisms sticken together forming a cube. It reflects part of the ray and it transmits the other part. In the classical picture, let us suppose that we have a beam-splitter of reflectance $R$ and transmittance $T$. If the beam-splitter is lossless, then $|R|^2+|T|^2=1$.

\begin{figure}[htbp]
\centering
\includegraphics{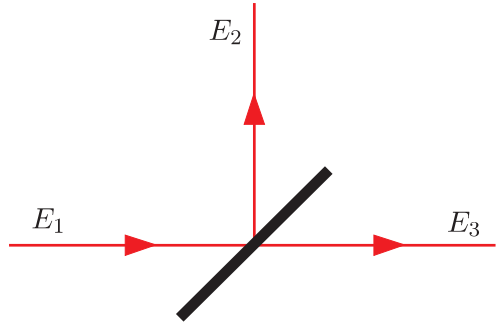}
\caption{\hfil\MYGREEN{Figure 1.7.2}: Classical beam-splitting.\hfil}
\label{Cl_bsing}
\end{figure}

If a ray enters with electric field $E_1$, the electric field reflected will be $E_2=RE_1$ and the part transmitted will be $E_3=TE_1$, see {\changeurlcolor{mygreen}\hyperref[Cl_bsing]{Figure \ref*{Cl_bsing}}}.

Let us consider the quantum case. Here, the electric field is an operator with form~\hyperref[Classical_to_Quantum_EF_3_ps]{(\ref*{Classical_to_Quantum_EF_3})}, then because it can be separated in the sum of an operator and its Hermitean conjugate~\hyperref[Classical_to_Quantum_EF_4_ps]{(\ref*{Classical_to_Quantum_EF_4})}, to obtain the output of an electric field operator $\widehat{\textbf{E}}_1$, we only need to see how the annihilation operator $a_1$ is transmitted and reflected. Notice that the annihilation and creator operators satisfy the commutation relations~\hyperref[Commutator_creation_annihilation_f_ps]{(\ref*{Commutator_creation_annihilation_f})}, then if we had a system similar to {\changeurlcolor{mygreen}\hyperref[Cl_bsing]{Figure \ref*{Cl_bsing}}}, with $a_2=Ra_1$ and $a_3=Ta_1$, the commutation relations would not be verified, so it would be necessary to add something to them.

\begin{figure}[htbp]
\centering
\includegraphics{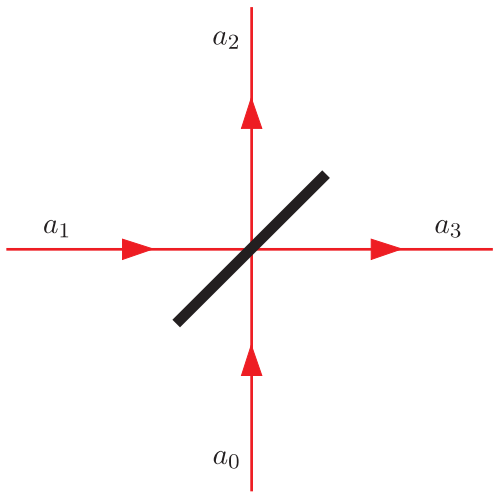}
\caption{\hfil\MYGREEN{Figure 1.7.3}: Quantum beam-splitting.\hfil}
\label{Q_bsing}
\end{figure}

In Quantum Mechanics and Quantum Field Theory, the vacuum plays an important role and in the previous discussion we have forgot it. If we suppose that our initial state is a tensor product of two states $\boldsymbol{\rho}_0\otimes\boldsymbol{\rho}_1$, where $\boldsymbol{\rho}_0$ is the vacuum state $|0\rangle\langle 0|$ and $\boldsymbol{\rho}_1$ is the state in which $\widehat{\textbf{E}}_1$ acts, we have to add a second input $a_0$ at the same frequency which acts on the vacuum, see {\changeurlcolor{mygreen}\hyperref[Q_bsing]{Figure \ref*{Q_bsing}}}.
\vspace{0cm}

In Quantum Mechanics, two states differing in a phase are physically equivalent, then let us suppose that for the second input we have a reflectance $\tilde{R}$ and transmittance $\tilde{T}$ with $|R|=|\tilde{R}|$ and $|T|=|\tilde{T}|$. Now, the new outputs in the beam-splitter will be:
\phantomsection\label{Outputs_beam_splitter_ps}\begin{equation}\label{Outputs_beam_splitter}
a_2=Ra_1+\tilde{T}a_0,\qquad a_3=Ta_1+\tilde{R}a_0.
\end{equation}
The operators $a_0$ and $a_1$ satisfy the commutation relations~\hyperref[Commutator_creation_annihilation_f_ps]{(\ref*{Commutator_creation_annihilation_f})}, then $a_2$ and $a_3$ will satisfy them too if:
\phantomsection\label{Outputs_beam_splitter_2_ps}\begin{align*}
|R|^2+|T|^2=1,\qquad RT^*+\tilde{R}^*\tilde{T}=0.
\end{align*}
We are assuming that the beam-splitter is lossless, hence the energy is conserved. From~\hyperref[harmonic_oscillator_multipartite_ps]{(\ref*{harmonic_oscillator_multipartite})} and because $a_0$ and $a_1$ act at the same frequency (or in the same mode), we have that $a_0^\dagger a_0+a_1^\dagger a_1=a_2^\dagger a_2+a_3^\dagger a_3$, then it must be verified that
\phantomsection\label{Outputs_beam_splitter_3_ps}\begin{equation*}
\tilde{R}T^*+R^*\tilde{T}=0.
\end{equation*}
Then, the reflectance and transmittance in the two outputs can be written as:
\phantomsection\label{Outputs_beam_splitter_4_ps}\begin{equation*}
R=\sin\theta\e^{i\phi},\quad \tilde{R}=\sin\theta\e^{i\tilde{\phi}},\quad T=\cos\theta\e^{i\psi},\quad \tilde{T}=\cos\theta\e^{i\tilde{\psi}},
\end{equation*}
with
\begin{equation*}
\e^{i\phi}=-\e^{i(\psi+\tilde{\psi}-\tilde{\phi})},
\end{equation*}
hence, we have:
\phantomsection\label{Beam_splitter_transformation_ps}\begin{equation*}
\begin{pmatrix}
  a_3 \\
  a_2 \\
\end{pmatrix}=\begin{pmatrix}
                \cos\theta\e^{i\psi} & \sin\theta\e^{i\tilde{\phi}} \\
                -\sin\theta\e^{i(\psi+\tilde{\psi}-\tilde{\phi})} & \cos\theta\e^{i\tilde{\psi}} \\
              \end{pmatrix}\begin{pmatrix}
  a_1 \\
  a_0 \\
\end{pmatrix},
\end{equation*}
and if we choose all phases $0$, we have that
\phantomsection\label{Beam_splitter_transformation_2_ps}\begin{equation}\label{Beam_splitter_transformation_2}
U_{bs}=\begin{pmatrix}
                \cos\theta & \sin\theta \\
                -\sin\theta & \cos\theta \\
              \end{pmatrix}.
\end{equation}
In the configurations that we will show later, we will use $50/50$ beam-splitters with $\theta=\pi/4$, that is, beam-splitters that reflect and transmit at $50\%$:
\vspace{0.1cm}
\phantomsection\label{Outputs_beam_splitter_50_50_ps}\begin{equation}\label{Outputs_beam_splitter_50_50}
a_2=\frac{1}{\sqrt{2}}(a_0-a_1),\qquad a_3=\frac{1}{\sqrt{2}}(a_0+a_1).
\end{equation}

\phantomsection\label{res_foto}
\noindent{\MYBROWN{b) Photodetectors}}:

\vspace{0.1cm}A photodetector is a device that generates an electric current by photoelectric effect when photons reach it. The transition probability of absorbing $n$ photons from an initial state $|i\rangle$ to a final state $|f\rangle$ is:
\phantomsection\label{probability_transition_ps}\begin{equation}\label{probability_transition}
T_{if}=|\langle f|\widehat{\textbf{E}}^{+}(\boldsymbol{x},t)|i\rangle|^2.
\end{equation}

Thus, the probability of absorbing any photon (or average intensity of the electric field) will be the sum of all the probabilities of absorbing $0,1,\ldots$ photons:
\phantomsection\label{Average_intensity_ps}\begin{multline*}
I(\boldsymbol{x},t)=\sum_fT_{if}=\sum_f\langle i|\widehat{\textbf{E}}^{-}(\boldsymbol{x},t)|f\rangle\cdot\langle f|\widehat{\textbf{E}}^{+}(\boldsymbol{x},t)|i\rangle\\
=\langle i|\widehat{\textbf{E}}^{-}(\boldsymbol{x},t)\cdot\widehat{\textbf{E}}^{+}(\boldsymbol{x},t)|i\rangle.
\end{multline*}
If our initial state is a mixed state $\displaystyle{\boldsymbol{\rho}=\sum_ip_i|i\rangle\langle i|}$ and $\displaystyle{\sum_ip_i=1}$, $p_i\geq 0$, the
\phantom{a}

\vspace{-0.67cm}\noindent average intensity will be:

\vspace{-0.4cm}\phantomsection\label{Average_intensity_2_ps}\begin{equation*}
I(\boldsymbol{x},t)=\sum_ip_i\langle i|\widehat{\textbf{E}}^{-}(\boldsymbol{x},t)\cdot\widehat{\textbf{E}}^{+}(\boldsymbol{x},t)|i\rangle\\
=\Tr\!\left(\boldsymbol{\rho}\widehat{\textbf{E}}^{-}(\boldsymbol{x},t)\cdot\widehat{\textbf{E}}^{+}(\boldsymbol{x},t)\right).
\end{equation*}

The intensity of the electric field measures the number of photons per unit of time. In the classical case, the probability $\textrm{d}p$ that a photodetector counts one photon in a time $\diff t$ will be:
\phantomsection\label{Probability_count_photon_ps}\begin{equation}\label{Probability_count_photon}
\diff p(t)=\alpha I_{cl}(t)\diff t,
\end{equation}
where the parameter $\alpha$ measures the sensitivity of the photodetector and $I_{cl}(t)$ is the classical current. Notice that this formula is equivalent to the formula~\hyperref[Photon_absorption_ps]{(\ref*{Photon_absorption})}, hence the probability that a count does not occur in the interval $[t,t+T]$ is similar to eq.\,\hyperref[Photon_absorption_2_ps]{(\ref*{Photon_absorption_2})}:
\phantomsection\label{Probability_no_count_ps}\begin{equation*}
P_0(t,t+T)=\exp\bigg(\hspace{-0.1cm}-\alpha\hspace{-0.5cm}\int\limits_{\hspace{.14cm}t}^{\hspace{.75cm}t+T}\hspace{-.4cm}I_{cl}(t')\diff t'\bigg),
\end{equation*}
and by induction, we can get the probability of counting $n$ photons:
\phantomsection\label{Probability_count_n_ps}\begin{equation}\label{Probability_count_n}
P_n(t,t+T)=\frac{1}{n!}\bigg[\alpha\hspace{-0.5cm}\int\limits_{\hspace{.14cm}t}^{\hspace{.75cm}t+T}\hspace{-.4cm}I_{cl}(t')\diff t'\bigg]^n\hspace{-.1cm}\exp\hspace{-.05cm}\bigg(\hspace{-.1cm}-\alpha\hspace{-0.5cm}\int\limits_{\hspace{.14cm}t}^{\hspace{.75cm}t+T}\hspace{-.4cm}I_{cl}(t')\diff t'\bigg).
\end{equation}
\vspace{-0cm}\begin{figure}[h]
\centering
\includegraphics{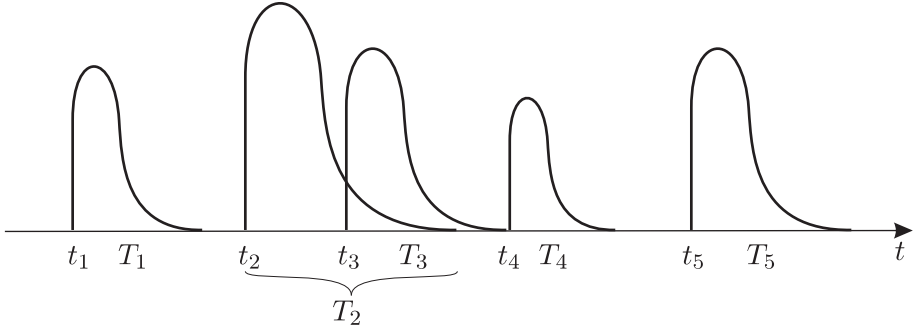}
\caption{\hfil\MYGREEN{Figure 1.7.4}: Process of absorption of photons in a photodetector.\hfil}
\label{Ph_abs}
\end{figure}

It can be shown (see \cite{Wa94}) that, in the quantum case, formula \hyperref[Probability_count_n_ps]{(\ref*{Probability_count_n})} becomes:
\phantomsection\label{Probability_count_n_quantum_ps}\begin{multline}\label{Probability_count_n_quantum}
P_n(t,t+T)=\frac{1}{n!}\Tr\!\Bigg(\boldsymbol{\rho}:\hspace{-.1cm}\bigg[\alpha\hspace{-0.5cm}\int\limits_{\hspace{.14cm}t}^{\hspace{.75cm}t+T}\hspace{-.4cm}\widehat{\textbf{E}}^{-}(x,t')\cdot\widehat{\textbf{E}}^{+}(x,t')\diff t'\bigg]^n\\
\cdot\exp\hspace{-.05cm}\bigg(\hspace{-.1cm}-\alpha\hspace{-0.5cm}\int\limits_{\hspace{.14cm}t}^{\hspace{.75cm}t+T}\hspace{-.4cm}\widehat{\textbf{E}}^{-}(x,t')\cdot\widehat{\textbf{E}}^{+}(x,t')\diff t'\bigg)\hspace{-.1cm}:\!\!\Bigg),
\end{multline}
where $:\!\!*\!\!:$ denotes the normal ordering of the operators inside, that is, the creation operators to the left and the annihilation operators to the right. The statistics of absorption of photons in the detector \cite{Ou95} is the probability of the independent events of absorbing $n_1$ photons in the interval $[t_1,t_1+T_1]$, $n_2$ in $[t_2,t_2+T_2]$, $\ldots$ and $n_k$ in $[t_k,t_k+T_k]$, (see {\changeurlcolor{mygreen}\hyperref[Ph_abs]{Figure~\ref*{Ph_abs}}}):
\phantomsection\label{Probability_count_nk_quantum_ps}\begin{align*}
&\hspace{-0cm}P_{n_1,\ldots,n_k}\big([t_1,t_1+T_1],\ldots,t_k,[t_k+T_k]\big)\\
&\hspace{2cm}=\Tr\Bigg(\boldsymbol{\rho}:\hspace{-.1cm}\prod_{i=1}^k\frac{1}{n_i!}\hspace{0cm}\cdot\bigg[\alpha\hspace{-0.6cm}\int\limits_{\hspace{.17cm}t_i}^{\hspace{.78cm}t_i+T_i}\hspace{-.47cm}\widehat{\textbf{E}}^{-}(x,t')\cdot\widehat{\textbf{E}}^{+}(x,t')\diff t'\bigg]^{n_i}\hspace{-.2cm}
\end{align*}
\begin{equation}\label{Probability_count_nk_quantum}
\hspace{-4.7cm}\cdot\exp\hspace{-.05cm}\bigg(\hspace{-.1cm}-\alpha\hspace{-0.6cm}\int\limits_{\hspace{.17cm}t_i}^{\hspace{.78cm}t_i+T_i}\hspace{-.47cm}\widehat{\textbf{E}}^{-}(x,t')\cdot\widehat{\textbf{E}}^{+}(x,t')\diff t'\bigg)\hspace{-.1cm}:\!\!\Bigg).
\end{equation}

\begin{figure}[h]
\centering
\includegraphics{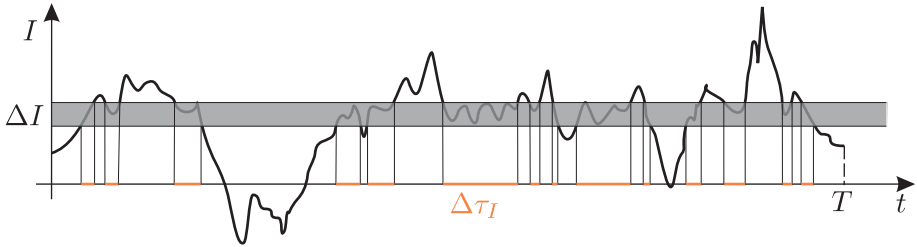}
\caption{\hfil\MYGREEN{Figure 1.7.5}: Intensity statistics.\hfil}
\label{Intensity_photons}
\end{figure}

The photodetector produces an output that is the average of the intensity of the beam over the state $\boldsymbol{\rho}$. From here, we can obtain the probability distribution of the intensity operator $\textbf{I}=\widehat{\textbf{E}}^{-}\cdot\widehat{\textbf{E}}^{+}$ by means of
\phantomsection\label{probability_distribution_intensity_ps}\begin{equation}\label{probability_distribution_intensity}
P(\textbf{I})=\frac{\Delta\tau_I}{T\Delta I}\,,
\end{equation}
where $\Delta\tau_I$ is the total time the detector has been measuring values in the interval $\Delta I$, and $T$ is the time it has been taking measures, {\changeurlcolor{mygreen}\hyperref[Intensity_photons]{Figure \ref*{Intensity_photons}}}.

Notice that from~\hyperref[Marginal_probability_q_phi_1_ps]{(\ref*{Marginal_probability_q_phi_1})}, we see that the probability distribution of $\textbf{I}$ is nothing but its tomogram:
\phantomsection\label{tomogram_intensity_ps}\begin{equation}\label{tomogram_intensity}
P(\textbf{I})=\Tr\!\big(\boldsymbol{\rho}\delta(I-\textbf{I})\big),
\end{equation}
where $I$ are the eigenvalues of $\textbf{I}$.

For a single mode field, the average of the field is proportional to $\langle a^\dagger a\rangle$, then suitable configurations of beam-splitters and photodetectors can be used to mix quantum electric fields in an analogue way as we saw in \hyperref[ph_section_Homodyne_cl]{{\color{black}\textbf{sec-}}} \hyperref[ph_section_Homodyne_cl]{{\color{black}\textbf{tion~\ref*{section_telecomm}}}} for signals in telecommunications.\newpage

\section{Homodyne and heterodyne detection in Quantum Optics}\label{section_hom_det_clas}

If we write the tomogram~\hyperref[Marginal_probability_q_phi_1_ps]{(\ref*{Marginal_probability_q_phi_1})} in terms of creation and annihilation operators, we have:
\phantomsection\label{tomogram_creation_annihilation_ps}\begin{equation}\label{tomogram_creation_annihilation}
P_{\phi}(X)=\Tr\!\big(\boldsymbol{\rho}\delta(X\mathds{1}-\overline{w}a-wa^\dagger)\big),
\end{equation}
where
\phantomsection\label{tomogram_creation_annihilatio_ps}\begin{equation*}
w=\sqrt{\frac{\hbar}{2}}\left(\frac{\cos\phi}{\sqrt{\omega}}+i\sqrt{\omega}\sin\phi\right).
\end{equation*}
Our aim in this section will be show how to measure this tomogram.

\subsection{Homodyne detection}

This first kind of detection is called homodyne because the radiation source is mixed with a strong laser beam of the same frequency. A strong laser beam is a radiation source which emits light in a highly excited coherent state\footnote{A more detailed description specifying the most important properties of coherent states will be presented in \hyperref[holomorphic_quantum_1_sec]{{\color{black}\textbf{subsection~\ref*{holomorphic_quantum_1_sec}}}}.}, i.e., $|z|\rightarrow\infty$:
\phantomsection\label{coherent_state_ps}\begin{equation}\label{coherent_state}
|z\rangle=\e
\hspace{-0.1cm}\sum_{n=0}^\infty\frac{z^n}{\sqrt{n!}}|n\rangle,\qquad a|z\rangle=z|z\rangle.
\end{equation}
The answer to the question, why do we mix the signal with a strong laser beam, is that in the limit $|z|\rightarrow\infty$, the laser beam behaves as a semiclassical source, then the action of the electric field of the laser over the state of the radiation source will be only a change of phase, as we will see \hyperref[Output_homodyne_4_ps]{{\color{black}\textbf{later}}}.

Let our input be a single mode field:
\phantomsection\label{Input_homodyne_ps}\begin{equation*}
\widehat{\textbf{E}}^{+}(\boldsymbol{x},t)=\sqrt{2\pi\hbar\omega}\,a\e^{-i\omega t}\boldsymbol{u}(\boldsymbol{x}),
\end{equation*}
that emits light at a state $\boldsymbol{\rho}$, and a strong laser beam:
\phantomsection\label{Laser_homodyne_ps}\begin{equation*}
\widehat{\textbf{E}}_L^{+}(\boldsymbol{x},t)=\sqrt{2\pi\hbar\omega}\,b\e^{-i\omega t}\boldsymbol{u}_L(\boldsymbol{x}),
\end{equation*}
that emits at the same frequency with state $|z\rangle\langle z|$, where $\displaystyle{z=|z|\e^{i\theta}}$ and $|z|^2\!\gg\!\langle a^\dagger a\rangle_{\boldsymbol{\rho}}$. Let mix both fields in a $50/50$ beam-splitter, hence the two outputs will be:
\vspace{-0.4cm}\phantomsection\label{Output_beam_splitter_homodyne_ps}\begin{align}\label{Output_beam_splitter_homodyne}
c=\frac{1}{\sqrt{2}}(a-b),\qquad d=\frac{1}{\sqrt{2}}(a+b).
\end{align}
\vspace{-0.3cm}\begin{figure}[h]
\centering
\includegraphics{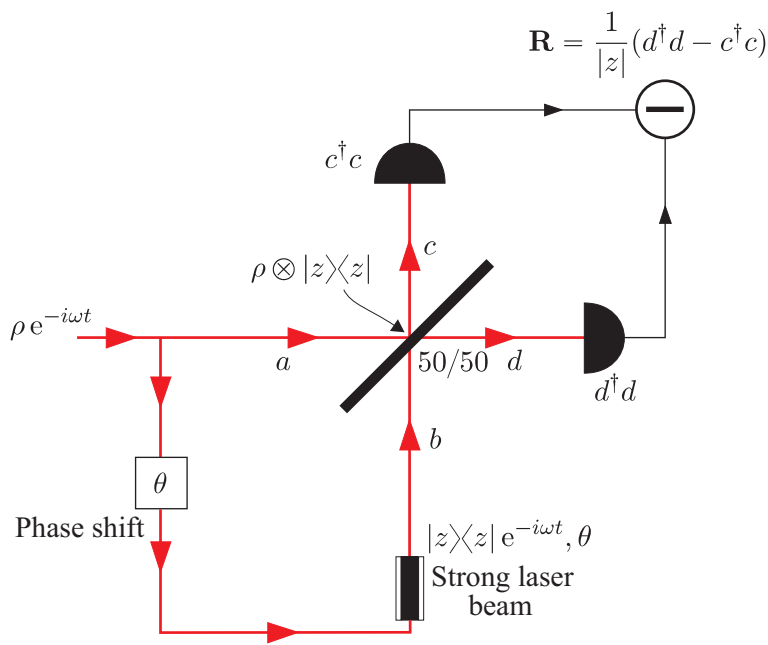}
\caption{\hfil\MYGREEN{Figure 1.8.1}: Quantum homodyne detection.\hfil}
\label{Q_homodyne}
\end{figure}

If we put two photodetectors in each output, substract both results, as it can be seen in {\changeurlcolor{mygreen}\hyperref[Q_homodyne]{Figure \ref*{Q_homodyne}}}, and divide it by $|z|$, we get the statistics of the operator:
\vspace{0.2cm}\phantomsection\label{Output_homodyne_ps}\begin{equation}\label{Output_homodyne}
\textbf{R}=\frac{1}{|z|}(d^\dagger d-c^\dagger c)=\frac{1}{|z|}(a^\dagger b+b^\dagger a),
\end{equation}
\\
with tomogram
\phantomsection\label{Output_homodyne_tom_ps}\begin{equation}\label{Output_homodyne_tom}
\mathcal{W}_\theta^\textrm{hom}(R)=\hspace{.1cm}\frac{1}{2\pi}\hspace{-.2cm}\int\limits_{\hspace{-.04cm}-\infty}^{\hspace{.45cm}\infty}\hspace{-0.15cm}\e^{ikR}\Tr\!\Big(\boldsymbol{\rho}\otimes|z\rangle\langle z|\e
\Big)\diff k.
\end{equation}

The annihilation operators $a$ and $b$ satisfy eq.\,\hyperref[Commutator_creation_annihilation_f_ps]{(\ref*{Commutator_creation_annihilation_f})}:
\phantomsection\label{a_b_relations_ps}\begin{equation}\label{a_b_relations}
\left[\vphantom{a^\dagger}a,a^\dagger\right]=\left[\vphantom{a^\dagger}b,b^\dagger\right]=\mathds{1},\qquad \left[\vphantom{a^\dagger}a,b\right]=\left[\vphantom{a^\dagger}a,b^\dagger\right]=0,
\end{equation}
\phantomsection\label{ph_algebra_su2}

\vspace{-0.4cm}\noindent hence, the operators $a^\dagger b$, $b^\dagger a$ and $2N_0=a^\dagger a-b^\dagger b$ satisfy the commutation relations of the $\mathfrak{su}(2)$ algebra:
\phantomsection\label{ab_su2_algebra_ps}\begin{align}\label{ab_su2_algebra}
&\left[a^\dagger b,b^\dagger a\right]=a^\dagger a-b^\dagger b=2N_0,\nonumber\\
&\hspace{-1cm}\left[a^\dagger b,N_0\right]=-a^\dagger b,\qquad \left[b^\dagger a,N_0\right]=b^\dagger a,
\end{align}
where $N_0$ corresponds to the $z$-component of the angular momentum and $a^\dagger b$, $b^\dagger a$ are ladder operators.

Therefore, if we use the Baker--Campbell--Hausdorff formula for the $SU(2)$ group \RED{[}\citen{Wo85}\RED{,}\hspace{0.05cm}\citen{Ar92}\RED{]}, we have:
\phantomsection\label{BCH_SU2_ps}\begin{equation}\label{BCH_SU2}
\e^{\xi ab^\dagger-\hspace{0.03cm}\bar{\xi}a^\dagger b}=\e^{\zeta b^\dagger a}\e

\e^{-\bar{\zeta} a^\dagger b},\qquad \zeta=\frac{\xi}{|\xi|}\tan|\xi|.
\end{equation}
If we insert this result in~\hyperref[Output_homodyne_tom_ps]{(\ref*{Output_homodyne_tom})}, we get:
\phantomsection\label{Output_homodyne_2_ps}\begin{align*}
\frac{1}{2\pi}&\hspace{-.2cm}\int\limits_{\hspace{-.04cm}-\infty}^{\hspace{.45cm}\infty}\hspace{-0.15cm}\e^{ikR}\Tr\!\Big(\boldsymbol{\rho}\otimes|z\rangle\langle z|\e%
 \Big)\exp\hspace{-0.1cm}\left(\frac{2|z|^2}{\cos\left(\frac{k}{|z|}\right)}\sin^2\left(\frac{k}{2|z|}\right)\right)\diff k.
\end{align*}
Using the standard BCH formula~\hyperref[BCH_ps]{(\ref*{BCH})} again, and the identity
\phantomsection\label{expXexpY_ps}\begin{equation}\label{expXexpY}
\e^X\e^Y=\e^{\exp(s)Y}\e^{X},
\end{equation}
whenever $\left[\vphantom{a^\dagger}X,Y\right]=sY$, we obtain:
\phantomsection\label{Output_homodyne_3_ps}\begin{multline*}
\mathcal{W}_\theta^\textrm{hom}(R)=\frac{1}{2\pi}\hspace{-.2cm}\int\limits_{\hspace{-.04cm}-\infty}^{\hspace{.45cm}\infty}\hspace{-0.15cm}\e^{ikR}
\left.\Tr\!\Big(\boldsymbol{\rho}\e
\right.\Big)\diff k.
\end{multline*}
And finally, in the limit $|z|^2\!\gg\!\langle a^\dagger a\rangle_{\boldsymbol{\rho}}$, we get:
\phantomsection\label{Output_homodyne_4_ps}\begin{multline}\label{Output_homodyne_4}
\mathcal{W}_\theta^\textrm{hom}(R)=\frac{1}{2\pi}\hspace{-.2cm}\int\limits_{\hspace{-.04cm}-\infty}^{\hspace{.45cm}\infty}\hspace{-0.15cm}\e^{ikR}\Tr\!\Big(\boldsymbol{\rho}\e^{-ik\e^{i\theta}a^\dagger}\e%
\e^{-ik\e^{-i\theta}a}\Big)\diff k\\
=\frac{1}{2\pi}\hspace{-.2cm}\int\limits_{\hspace{-.04cm}-\infty}^{\hspace{.45cm}\infty}\hspace{-0.15cm}\e^{ikR}\Tr\!\Big(\boldsymbol{\rho}\e^{-ik(\e^{i\theta}a^\dagger+\e^{-i\theta}a)}\Big)\diff k.
\end{multline}

If we use the creation and annihilation formulas \hyperref[creation_annihilation_ps]{(\ref*{creation_annihilation})} to write the tomogram $\mathcal{W}_\theta^\textrm{hom}(R)$ in terms of the operators $\textbf{Q}$ and $\textbf{P}$ to compare it with the Radon Transform formula \hyperref[Radon_Transform_ps]{(\ref*{Radon_Transform})} via the correspondence with the average integral over the Wigner function $\boldsymbol{\rho}_w$ given in \hyperref[Tomogram_quantum_to_classic_4_ps]{(\ref*{Tomogram_quantum_to_classic_4})}, we get that
\begin{equation}
\mathcal{W}_\theta^\textrm{hom}(R)=\sqrt{\frac{\hbar}{2}}\mathcal{R}\boldsymbol{\rho}_w\left(\sqrt{\frac{\hbar}{2}}R,\sqrt{\omega}\cos\theta,\frac{\sin\theta}{\sqrt{\omega}}\right).
\end{equation}
That shows that the quantum tomogram $\mathcal{W}^{\textrm{hom}}_\theta(R)$ measured in the homodyne detection device, {\changeurlcolor{mygreen}\hyperref[Q_homodyne]{Figure \ref*{Q_homodyne}}}, is just the Radon Transform of the Wigner's function of the state $\boldsymbol{\rho}$ (in the strong laser limit).

Hence, using a rescaled version of the change of variables that we used to get the formula \hyperref[Inverse_Radon_Transform_CAT_2_ps]{(\ref*{Inverse_Radon_Transform_CAT_2})}:
\begin{align*}
  \mu&=k\sqrt{\omega}\cos\theta,\nonumber\\
  \nu&=\frac{k}{\sqrt{\omega}}\sin\theta,
\end{align*}
we get the following formula to reconstruct the Wigner's function:
\phantomsection\label{Inverse_Radon_Transform_Wigner_homod_ps}
\begin{multline}\label{Inverse_Radon_Transform_Wigner_homod}
\boldsymbol{\rho}_{w}(q,p)=\frac{1}{(2\pi)^2}\hspace{-0.15cm}\int\limits_{\hspace{-.12cm}-\infty}^{\hspace{.35cm}\infty}\hspace{-.3cm}\int\limits_{\hspace{.15cm}0}^{\hspace{.35cm}\infty}\hspace{-.33cm}\int\limits_{\hspace{.15cm}0}^{\hspace{.35cm}2\pi}\hspace{-.06cm}\mathcal{W}_\theta^\textrm{hom}(R)\\
\cdot\e
k\diff\theta\diff k\diff R,
\end{multline}
and the matrix elements of the state $\boldsymbol{\rho}$ are recovered by means of the Fourier Transform \hyperref[matrix_element_rho_ps]{(\ref*{matrix_element_rho})}.

\subsection{Heterodyne detection}

As in the classical case in telecommunications, the difference between homodyne and heterodyne detection is that in the heterodyne case the frequency of the local oscillator is different to the frequency of the signal. 

We saw in \hyperref[Image_Frequency_2_ref_ps]{{\color{black}\textbf{section~\ref*{section_telecomm}}}} that heterodyne detection has the inconvenient that there are frequencies, that we called image frequencies, in which we cannot emit signals. In quantum optical detection, information is emitted at every frequency because even if we are not at an excited state, the state is the vacuum. However, we will see that this is not an inconvenient in the final result. We will show that it gives only an extra contribution because of the nature of the expected value over a state composed by a tensor product.

\begin{figure}[htbp]
\centering
\includegraphics{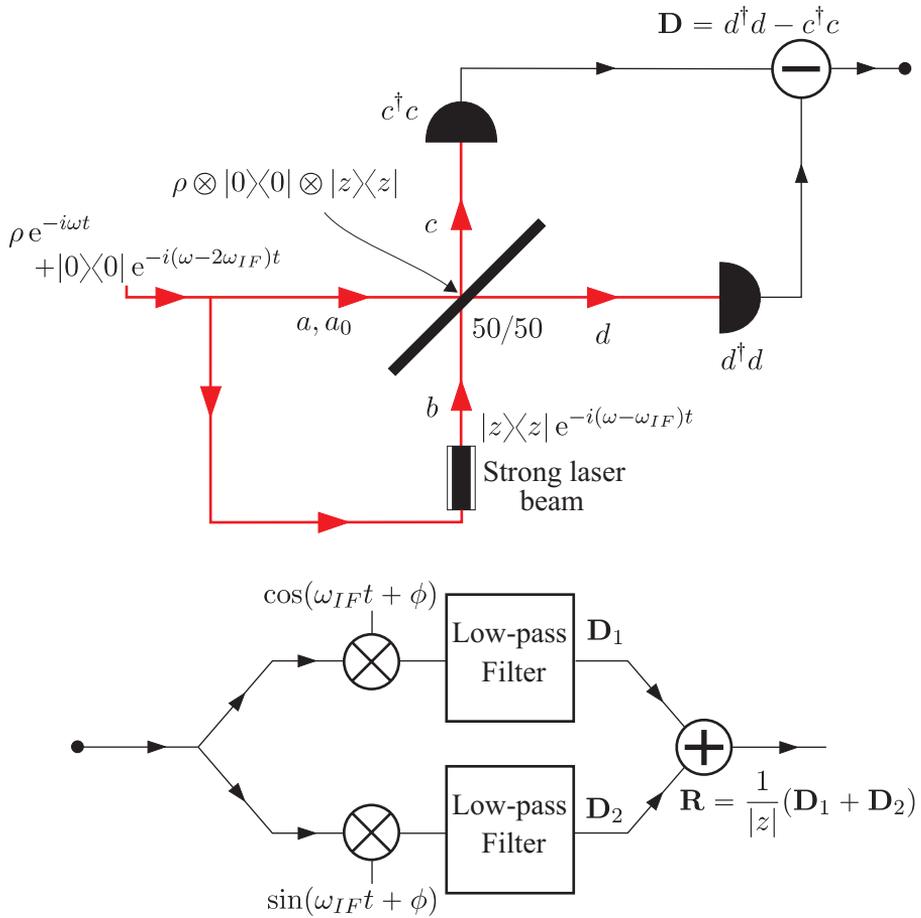}
\caption{\hfil\MYGREEN{Figure 1.8.2}: Quantum heterodyne detection.\hfil}
\label{Q_heterodyne}
\end{figure}

Let the input signal be:
\phantomsection\label{Input_Qheterodyne_ps}\begin{equation*}
\widehat{\textbf{E}}^{+}(\boldsymbol{x},t)=\sqrt{2\pi\hbar}\left(\sqrt{\omega}\,a\e^{-i\omega t}\boldsymbol{u}_{\boldsymbol{\rho}}(\boldsymbol{x})+\sqrt{\omega-2\omega_{IF}}a_0\e^{-i(\omega-2\omega_{IF}) t}\boldsymbol{u}_0(\boldsymbol{x})\right)\!,
\end{equation*}
acting on the state $\boldsymbol{\rho}\otimes|0\rangle\langle 0|$, where $a$ acts on $\boldsymbol{\rho}$ and $a_0$ on $|0\rangle\langle 0|$. We do not write in this formula the rest of frequencies that also act on the vacuum because they will not contribute to the final result. Let also be a strong laser beam:
\phantomsection\label{Laser_heterodyne_ps}\begin{equation*}
\widehat{\textbf{E}}_L^{+}(\boldsymbol{x},t)=\sqrt{2\pi\hbar(\omega-\omega_{IF})}\,b\e^{-i(\omega-\omega_{IF}) t}\boldsymbol{u}_L(\boldsymbol{x}),
\end{equation*}
that emits light in a coherent state $|z\rangle\langle z|$ with phase $\theta=0$, $z=|z|$ and $|z|^2\gg\langle (a\bar{\zeta}+a_0\zeta)^\dagger(a\bar{\zeta}+a_0\zeta)\rangle_{\boldsymbol{\rho}\otimes |0\rangle\langle 0|}$ with $\zeta\in\mathbb{C}$ and $|\zeta|=1$. The heterodyne detection process is the following (see {\changeurlcolor{mygreen}\hyperref[Q_heterodyne]{Figure \ref*{Q_heterodyne}}}):

First, we mix the two fields in a $50/50$ beam-splitter to get:
\phantomsection\label{Output_heterodyne_ps}\begin{align}\label{Output_heterodyne}
c&=\frac{1}{\sqrt{2}}\left(a\e^{-i\omega t}+a_0\e^{-i(\omega-2\omega_{IF}) t}-b\e^{-i(\omega-\omega_{IF})t}\right),\nonumber\\
d&=\frac{1}{\sqrt{2}}\left(a\e^{-i\omega t}+a_0\e^{-i(\omega-2\omega_{IF}) t}+b\e^{-i(\omega-\omega_{IF})t}\right).
\end{align}

Second, we put two photodetectors in each output and substract the result to obtain:
\phantomsection\label{Output_heterodyne_2_ps}\begin{equation}\label{Output_heterodyne_2}
\textbf{D}=\frac{1}{2}\left((a^\dagger b+b^\dagger a_0)\e^{i\omega_{IF}t}+(b^\dagger a+a_0^\dagger b)\e^{-i\omega_{IF}t}\right).
\end{equation}

Third, we divide the signal in two parts and multiply one by $\cos(\omega_{IF}t+\phi)$ and the other by $\sin(\omega_{IF}t+\phi)$ and we pass the results through a low-pass filter to get:
\phantomsection\label{Output_heterodyne_3_ps}\begin{align*}
\textbf{D}_1&=\textbf{D}\cdot\cos(\omega_{IF}t+\phi)*\textrm{F}_\textrm{low}(t)=\frac{1}{2}\left((a^\dagger b+b^\dagger a_0)\e^{-i\phi}+(b^\dagger a+a_0^\dagger b)\e^{i\phi}\right)\!,\nonumber\\
\textbf{D}_2&=\textbf{D}\cdot\sin(\omega_{IF}t+\phi)*\textrm{F}_\textrm{low}(t)=\frac{i}{2}\left((a^\dagger b+b^\dagger a_0)\e^{-i\phi}-(b^\dagger a+a_0^\dagger b)\e^{i\phi}\right).
\end{align*}
The symbol $*$ indicates the convolution product. The low-pass filters are necessary for removing the terms of other frequencies that appear during the process.

Finally, if we sum $\textbf{D}_1$ and $\textbf{D}_2$ and divide it by $|z|$, we will get the operator whose statistics we want to obtain:
\phantomsection\label{Output_heterodyne_4_ps}\begin{align}\label{Output_heterodyne_4}
\textbf{R}&=\frac{1}{|z|}(\textbf{D}_1+\textbf{D}_2)=\frac{1}{2|z|}\left((a^\dagger b+b^\dagger a_0)(1+i)\e^{-i\phi}\hspace{-0.02cm}+(b^\dagger a+a_0^\dagger b)(1-i)\e^{i\phi}\right)\nonumber\\
&\hspace{0cm}=\frac{1}{|z|\sqrt{2}}\left(\big(a^\dagger\e^{i\theta}+a_0^\dagger\e^{-i\theta}\big)b+b^\dagger\big(a\e^{-i\theta}+a_0\e^{i\theta}\big)\right)\!,
\end{align}
with
\begin{equation}
\theta=\frac{\pi}{4}-\phi.
\end{equation} 

Notice that the operator
\phantomsection\label{New_annihilation_ps}\begin{equation*}
\textbf{A}=\frac{1}{\sqrt{2}}(a\e^{-i\theta}+a_0\e^{i\theta}),
\end{equation*}
where
\phantomsection\label{New_annihilation_2_ps}\begin{equation}\label{New_annihilation_2}
\left[a,a^\dagger\right]=\left[\hspace{-0.1cm}\vphantom{a^\dagger}\right.a_0,a_0^\dagger\hspace{-0.1cm}\left.\vphantom{a^\dagger}\right]=\mathds{1},\qquad\left[a_0,a^\dagger\right]=0,
\end{equation}
verifies:
\phantomsection\label{New_annihilation_3_ps}\begin{equation*}
\left[\vphantom{a^\dagger}\textbf{A},\textbf{A}^\dagger\right]=\mathds{1},
\end{equation*}
hence, the operators $\textbf{A}^\dagger b$, $b^\dagger \textbf{A}$ and $2N_0=\textbf{A}^\dagger\textbf{A}-b^\dagger b$ determine again a $\mathfrak{su}(2)$ algebra as in~\hyperref[ab_su2_algebra_ps]{(\ref*{ab_su2_algebra})}. Because of this, we can repeat the computations made in the homodyne case and then, the tomogram
\begin{equation*}
\mathcal{W}_\theta^\textrm{het}(R)=\hspace{.1cm}\frac{1}{2\pi}\hspace{-.2cm}\int\limits_{\hspace{-.04cm}-\infty}^{\hspace{.45cm}\infty}\hspace{-0.15cm}\e^{ikR}\Tr\!\Big(\boldsymbol{\rho}\otimes|0\rangle\langle 0|\otimes|z\rangle\langle z|\e
\Big)\diff k,
\end{equation*}
in the limit $|z|^2\!\gg\!\langle \textbf{A}^\dagger \textbf{A}\rangle_{\boldsymbol{\rho}\otimes|0\rangle\langle 0|}$ reads as:
\phantomsection\label{Output_Heterodyne_Tom_final_ps}\begin{multline}\label{Output_Heterodyne_Tom_final}
\mathcal{W}_\theta^\textrm{het}(R)=\hspace{.1cm}\frac{1}{2\pi}\hspace{-.2cm}\int\limits_{\hspace{-.04cm}-\infty}^{\hspace{.45cm}\infty}\hspace{-0.15cm}\e^{ikR}\Tr\!\Big(\boldsymbol{\rho}\otimes|0\rangle\langle 0|\e^{-ik(\textbf{A}^\dagger+\textbf{A})}\Big)\diff k\\
\hspace{.0cm}=\frac{1}{2\pi}\hspace{-.2cm}\int\limits_{\hspace{-.04cm}-\infty}^{\hspace{.45cm}\infty}\hspace{-0.15cm}\e^{ikR}\Tr\!\left(\boldsymbol{\rho}\otimes|0\rangle\langle 0|\e
\right)\diff k.
\end{multline}
\begin{figure*}[htbp]
\centering
\includegraphics{Grafico17col}
\end{figure*}

Taking into account that $a+a_0^\dagger$ commutes with its adjoint and because $a$ and $a_0$ commute each other, we can split the contribution of the vacuum using the BCH formula \hyperref[BCH_ps]{(\ref*{BCH})}:
\phantomsection\label{Output_Heterodyne_Tom_final_2_ps}\begin{multline*}
\Tr\!\left(\boldsymbol{\rho}\otimes|0\rangle\langle 0|\e
|0\rangle=\langle 0|0\rangle=1,
                                                                                                                                                                                                                                                                                                  $$                                                                                                                                                                                                                                                                                               we have that
\phantomsection\label{Trace_het_ps}                               
\begin{multline}\label{Trace_het}
\Tr\!\left(\boldsymbol{\rho}\otimes|0\rangle\langle 0|\e%
\Big).
\end{multline}
Hence, we have that the heterodyne tomogram is the Fourier Transform over the variable $k$ of expression \hyperref[Trace_het_ps]{(\ref*{Trace_het})}:
\phantomsection\label{Output_Heterodyne_Tom_final_2_ps}\begin{equation}\label{Output_Heterodyne_Tom_final_2}
\mathcal{W}_\theta^\textrm{het}(R)=\frac{1}{2\pi}\hspace{-.2cm}\int\limits_{\hspace{-.04cm}-\infty}^{\hspace{.45cm}\infty}\hspace{-0.15cm}\e^{ikR}\Tr\!\Big(\boldsymbol{\rho}\e%
\Big)\diff k.
\end{equation}

From the tomogram \hyperref[Output_Heterodyne_Tom_final_2_ps]{(\ref*{Output_Heterodyne_Tom_final_2})}, we cannot obtain the Wigner's function $\boldsymbol{\rho}_w(q,p)$ as in the homodyne case, however we can obtain the matrix elements of $\boldsymbol{\rho}$ in the coherent basis $|z\rangle$ by means of the Husimi distribution \cite{Hu37}:
\begin{equation}
\boldsymbol{\rho}_Q(z,\bar{z})=\frac{1}{\pi}\langle z|\boldsymbol{\rho}|z\rangle.
\end{equation}
The way to do it is by noticing that the Husimi function $\boldsymbol{\rho}_Q(z,\bar{z})$ and the anti-normal ordered characteristic function
\begin{equation*}
\chi_A(\xi,\bar{\xi})=\Tr\!\left(\boldsymbol{\rho}\e^{-\bar{\xi}a}\e^{\xi a^\dagger}\right)
\end{equation*}
are related by a two-dimesional Fourier Transform. Let see it using the properties of the coherent states that will be exposed in \hyperref[holom_chap_2]{{\color{black}\textbf{subsec.\,\ref*{holom_chap_2}}}}:
\phantomsection\label{char_husimi_ps}
\begin{align}\label{char_husimi}
\chi_A(\xi,\bar{\xi})&=\frac{1}{\pi}\Tr\Bigg(\hspace{-0.3cm}\int\limits_{\hspace{.5cm}\mathbb{R}^{2}}\hspace{-.19cm}\boldsymbol{\rho}\e^{-\bar{\xi}a}|z\rangle\langle z|\e^{\xi a^\dagger}\diff^2 z\Bigg)\nonumber\\
&=\frac{1}{\pi}\Tr\Bigg(\hspace{-0.3cm}\int\limits_{\hspace{.5cm}\mathbb{R}^{2}}\hspace{-.19cm}\boldsymbol{\rho}\e^{-\bar{\xi}z}|z\rangle\langle z|\e^{\xi\bar{z}}\diff^2 z\Bigg)=\frac{1}{\pi}\hspace{-0.3cm}\int\limits_{\hspace{.5cm}\mathbb{R}^{2}}\hspace{-.19cm}\langle z|\boldsymbol{\rho}|z\rangle\e^{-\bar{\xi}z}\e^{\xi\bar{z}}\diff^2 z.
\end{align}

Because of eq.\,\hyperref[char_husimi_ps]{(\ref*{char_husimi})}, the Husimi function $\boldsymbol{\rho}_Q(z,\bar{z})$ is obtained from the Inverse Fourier Transform in the variables $\xi$ and $\bar{\xi}$:
\begin{equation}
\boldsymbol{\rho}_Q(z,\bar{z})=\frac{1}{\pi^2}\hspace{-0.3cm}\int\limits_{\hspace{.5cm}\mathbb{R}^{2}}\hspace{-.19cm}\chi_A(\xi,\bar{\xi})\e^{\bar{\xi}z}\e^{-\xi\bar{z}}\diff^2\xi,
\end{equation}
hence, writing this expression in terms of the heterodyne tomogram \hyperref[Output_Heterodyne_Tom_final_2_ps]{(\ref*{Output_Heterodyne_Tom_final_2})}, we finally obtain the inversion formula:
\begin{multline}\label{Inverse_Radon_Transform_Wigner_homod}
\boldsymbol{\rho}_{Q}(z,\bar{z})=\frac{1}{2\pi^2}\hspace{-0.15cm}\int\limits_{\hspace{-.12cm}-\infty}^{\hspace{.35cm}\infty}\hspace{-.3cm}\int\limits_{\hspace{.15cm}0}^{\hspace{.35cm}\infty}\hspace{-.33cm}\int\limits_{\hspace{.15cm}0}^{\hspace{.35cm}2\pi}\hspace{-.06cm}\mathcal{W}_\theta^\textrm{het}(R)\\
\cdot\e                                                                                                                                                                                                                                                                                                  

k\diff\theta\diff k\diff R.
\end{multline}

Then, if we express the coherent factor $z$ in terms of the position and momentum variables $q$ and $p$:
\begin{equation}
z=\frac{1}{\sqrt{2\hbar}}\Big(\sqrt{\omega}q+\frac{i}{\sqrt{\omega}}p\Big),
\end{equation}
we can write the Husimi function $\boldsymbol{\rho}_Q(z,\bar{z})$ as the convolution of the Wigner's function $\boldsymbol{\rho}_w(q,p)$ with a Gaussian filter \cite{Le15}:
\begin{equation}
\boldsymbol{\rho}_Q(z,\bar{z})=\frac{1}{\pi\hbar}\hspace{-0.3cm}\int\limits_{\hspace{.5cm}\mathbb{R}^{2}}\hspace{-.19cm}\boldsymbol{\rho}_w(q',p')\e                                                                                                                                                                                                                                                                                                  
\,\diff q'\diff p',
\end{equation}
where $\sigma_q$ and $\sigma_p$ are the variances of the Gaussian wave-packet which satisfy the Heisenberg minimum uncertainty relation
\begin{equation*}
\sigma_q\sigma_p=\frac{\hbar}{2}\,.
\end{equation*}

%% file: Tesischap2.tex
\chapter{The tomographic picture of quantum systems}\label{chap_tom_qu}

The ideas presented in the \hyperref[chap_birth]{{\color{black}\textbf{introduction}}} can be extended and formalized by considering with more care the role of the observables of the system in the construction of the tomograms $\mathcal{W}$.

The description of a physical system involves always the selection of its algebra of observables $\mathcal{A}$ and a family of states $\mathcal{S}$. The outputs of measuring a given observable $a\in\mathcal{A}$, when the system is in the state $\rho\in\mathcal{S}$, are described by a probability measure $\mu_{a,\rho}$ on the real line such that $\mu_{a,\rho}(\Delta)$ is the probability that the output of $a$ belongs to the subset $\Delta\subset\mathbb{R}$. Thus, a measure theory (or better a theory of measurement) for the physical system under consideration is a pairing $\langle\rho,a\rangle$ between states $\rho$ and observables $a$ that assigns to pairs of them probability measures $\mu_{a,\rho}$. Then, the expected value of the observable $a$ in the state $\rho$ is given by:
$$
\langle a\rangle_\rho=\hspace{-0.1cm}\int\limits_{\hspace{.3cm}\mathbb{R}}\lambda\diff\mu_{a,\rho}(\lambda).
$$
\phantomsection\label{introoduction_bold_states_ps}
Such picture applies equally to both, classical and quantum systems. For closed quantum systems, the observables are usually described as a family of self-adjoint operators on a Hilbert space $\mathcal{H}$ while states are described by density operators $\boldsymbol{\rho}$ acting on such Hilbert space, that is, positive self-adjoint operators $\boldsymbol{\rho}\geq 0$, $\boldsymbol{\rho}=\boldsymbol{\rho}^\dagger$ such that $\Tr\boldsymbol{\rho}=1$. The pairing $\langle \rho,a\rangle$ written before is provided by the assignment $\mu_{a,\rho}=\Tr(\boldsymbol{\rho} E_{a})$, where $E_a$ denotes the projector-valued spectral measure associated to the Hermitean operator $a$.

This picture of quantum systems can somehow be enhanced by using a more algebraic presentation. The rest of this chapter will be devoted to this. A general discussion of Quantum Tomography in the setting of $C^*$--algebras will be analyzed and a number of applications, including Group Tomography, will be discussed. 

A \textit{Picture} of Quantum Mechanics is a mathematical representation of quantum systems. There are three main standard pictures \cite{Ga90} that differ in the role of time regarding states and observables:

\begin{itemize}

\item Schr\"{o}dinger picture: Where the time evolution is carried by the state and the observables are considered static.

\item Heisenberg picture: Where the time evolution is carried by the observables and the states are independent of time.

\item Dirac picture: In which both, states and observables, are time dependent.

\end{itemize}

These pictures together with Wigner--Moyal representation constitute the most common mathematical descriptions of quantum mechanical systems. The representation we will use here, by means of $C^*$--algebras, embraces all the previous ones in a clear mathematical way and it is the one we will consider here.

Moreover, already in \hyperref[section_QM_ps]{{\color{black}\textbf{section~\ref*{section_real_QM_ps}}}}, it was presented a way for reconstructing matrix elements of density operators by means of certain measurements of observables. This is the so called \textit{tomographic picture} of Quantum Mechanics. In this sense, \textit{Quantum Tomography} would not be considered a picture of Quantum Mechanics by itself because derives from Heisenberg's picture, however it was shown by Ibort \textit{et al.} \cite{Ib10} that it is truly equivalent to the standard ones. 

\phantomsection\label{resumen_C_alg}
\section{$C^{*}$--algebras and Quantum Tomography}\label{section_C-algebras}

The algebra of bounded operators $\mathcal{B}(\mathcal{H})$ on a Hilbert space $\mathcal{H}$ is usually considered as the algebra of (bounded) observables of the system, however, as it was proposed by von Neumann, it is possible to generalize that and consider more general algebras. In this Thesis, we will present a tomographic picture of Quantum Mechanics in which observables are elements of a $C^*$--algebra $\mathcal{A}$.

Let us recall that a $*$--algebra $\mathcal{A}$ \cite{Pe79} is a complex Banach algebra with a norm $\|\hspace{-0.02cm}\cdot\hspace{-0.02cm}\|$ and an \textit{involution} operation $^*$ satisfying:

\begin{enumerate}

\item[\MYBROWN{(a)}] $(a^*)^*=a$,

\item[\MYBROWN{(b)}] $(ab)^*=b^*a^*$,

\item[\MYBROWN{(c)}] $(a+\lambda b)^*=a^*+\bar{\lambda} b^*$,

\end{enumerate}
for all $a,b\in\mathcal{A}$ and $\lambda\in\mathbb{C}$. A $C^*$--algebra is a $*$--algebra $\mathcal{A}$ such that 
$$
\|a^*a\|=\|a\|^2,\qquad\forall a\in\mathcal{A}.
$$
We will also ask for the algebras considered here to be unital in the sense that there exists a neutral element $\mathds{1}$ such that $\mathds{1}a=a\mathds{1}$ for all $a\in\mathcal{A}$.

\phantomsection
\label{ph_section_observables}
An element will be called self-adjoint if $a^*=a$. The subspace of all self-adjoint elements is denoted by $\mathcal{A}_{sa}$ and constitutes the Lie--Jordan Banach algebra of observables of the corresponding quantum system (see \cite{Fa13} for more details).

In particular, we can consider the $C^*$--algebra $\mathcal{B}(\mathcal{H})$ equipped with the operator norm and the involution defined by the adjoint operation, that is, $\textbf{A}^*\coloneq\textbf{A}^\dagger$ where $\textbf{A}^\dagger$ denotes the adjoint operator of $\textbf{A}\in\mathcal{B}(\mathcal{H})$.

\phantomsection\label{resumen_states}
The states of the theory are normalized positive functionals on $\mathcal{A}$, that is, linear maps $\rho:\mathcal{A}\rightarrow\mathbb{C}$ such that
\begin{equation}
\rho(\mathds{1})=1,\qquad \rho(a^*a)\geq 0,\qquad\forall a\in\mathcal{A}.
\end{equation}

\phantomsection
\label{section_Gleason}
In the case in which $\mathcal{A}=\mathcal{B}(\mathcal{H})$, because of Gleason's theorem \cite{Gl57}, states are in one-to-one correspondence with normalized non-negative Hermitean operators $\boldsymbol{\rho}$ acting on the Hilbert space $\mathcal{H}$:
\phantomsection\label{definition_state_rho_ps}
\begin{equation}\label{definition_state_rho}
\textrm{Tr}(\boldsymbol{\rho})=1,\qquad\boldsymbol{\rho}^\dagger=\boldsymbol{\rho},\qquad\boldsymbol{\rho}\geq 0,
\end{equation}
which are the density operators presented in \hyperref[section_QM_ps]{{\color{black}\textbf{section~\ref*{section_real_QM_ps}}}} and in the \hyperref[introoduction_bold_states_ps]{{\color{black}\textbf{intro-}}} \hyperref[introoduction_bold_states_ps]{{\color{black}\textbf{duction to this chapter}}}.

The relation between states $\rho$ of the $C^*$--algebra $\mathcal{B}(\mathcal{H})$ and density operators $\boldsymbol{\rho}$ on the Hilbert space $\mathcal{H}$ is given by the formula:
\phantomsection\label{expected_value_c_star_ps}\begin{equation}\label{expected_value_c_star}
\rho(\textbf{A})=\textrm{Tr}(\boldsymbol{\rho}\textbf{A}),\qquad\forall\textbf{A}\in\mathcal{B}(\mathcal{H}).
\end{equation}

The space of states of a given $C^*$--algebra $\mathcal{A}$ will be denoted by $\mathcal{S}(\mathcal{A})$ and it is a convex weak$^*$-compact subset of the topological dual $\mathcal{A}^\prime$ of $\mathcal{A}$ \cite{Al78}.

Notice that according to the physical interpretation of the $C^*$--algebra $\mathcal{A}$ as the algebra of observables of a given physical system, when the algebra is commutative it will be describing a classical system, whereas non-commutativity will correspond to ``genuine'' quantum systems.

\phantomsection\label{evol_state_chap_4_ps}
A state $\rho$ of the $C^*$--algebra $\mathcal{A}$ represents the state of the physical system under consideration and the number $\rho(a)$, for a given $a\in\mathcal{A}$, is interpreted as the expected value of the observable $a$ measured in the state $\rho$, consequently it is also denoted as:
\begin{equation}\label{mean_value_conclusion}
\langle a\rangle_\rho\coloneq\rho(a).
\end{equation}
In this sense, eq.\,\hyperref[expected_value_c_star_ps]{(\ref*{expected_value_c_star})} represents the expected value of the observable described by the operator $\textbf{A}$ when the system is in the state given by the density operator $\boldsymbol{\rho}$.

Each self-adjoint element $a\in\mathcal{A}_{sa}$ defines a continuous affine function $\widecheck{a}: \mathcal{S}(\mathcal{A})\rightarrow\mathbb{R}$,
\begin{equation}
\widecheck{a}\coloneq\rho(a).
\end{equation}
A theorem by Kadison \cite{Ka51} states that the correspondence $a\rightarrow\widecheck{a}$ is an isometric isomorphism from the self-adjoint part of $\mathcal{A}$ onto the space of all continuous affine functions from $\mathcal{S}(\mathcal{A})$ into $\mathbb{R}$. Thus, the self-adjoint part of the algebra of observables can be recovered directly from the space of states and its complexification provides the whole algebra \cite{Fa13}.\newpage

\subsection{The GNS construction}\label{ps_GNS}

The Hilbert space picture is recovered by means of the called GNS construction \RED{[}\citen{Ge43}\RED{,}\hspace{0.05cm}\citen{Se47}\RED{]} named in honor of Gel'fand, Naimark and Segal. 

Given a state $\rho$ of a $C^*$--algebra $\mathcal{A}$, we can construct a representation $\pi_\rho$ of $\mathcal{A}$ in the $C^*$--algebra of bounded operators of a Hilbert space $\mathcal{H}_\rho$ canonically associated to it. The Hilbert space $\mathcal{H}_\rho$ is constructed as the completion of the inner product space $\mathcal{A}/\mathcal{J}_\rho$ where
\begin{equation}
\mathcal{J}_\rho=\left\{a\in\mathcal{A}|\,\rho(a^*a)=0\right\}
\end{equation}
is the Gel'fand's ideal of null elements for $\rho$, and the inner product is defined as:
\begin{equation}
\langle[a],[b]\rangle_\rho\coloneq\rho(a^*b),\qquad a,b\in\mathcal{A},
\end{equation}
where $[a]$ denotes the class $a+\mathcal{J}_\rho$ in the quotient space. The representation $\pi_\rho$ is defined as:
\phantomsection\label{pi_rep_ps}\begin{equation}\label{pi_rep}
\pi_\rho(a)[b]\coloneq[ab],\qquad\forall a,b\in\mathcal{A}.
\end{equation}

The GNS construction provides a cyclic representation of $\mathcal{A}$ with the cyclic vector corresponding to the unit element $\mathds{1}$. Such vector will be called the \textit{vacuum vector} of $\mathcal{H}_\rho$ and denoted by $|0\rangle$. Moreover, we get that the state $\rho$ is also described by:
\phantomsection\label{sampling_go_back}
\begin{equation}
\rho(a)=\langle0|\pi_\rho(a)|0\rangle,\qquad a\in\mathcal{A}.
\end{equation}
In addition, given any element $a\in\mathcal{A}$, we have the associated vector $\pi_\rho(a)|0\rangle=[a\mathds{1}]=[a]$. In what follows, we will denote by $|a\rangle$ the vectors $[a]\in\mathcal{H}_\rho$, thus:
\begin{equation}
\pi_\rho(a)|0\rangle=|a\rangle.
\end{equation}

By duality, $\mathcal{A}$ acts on the space of states $\mathcal{S}(\mathcal{A})$, i.e., $(a\cdot\rho)(b)=\rho(ab)$. Thus, if we fix the state $\rho$, then the orbit of $\mathcal{A}$ through $\rho$ can be identified with the Hilbert space $\mathcal{H}_\rho$. Now, each unit vector $|a\rangle\in\mathcal{H}_\rho$ defines a state on $\mathcal{A}$ by means of
\begin{equation}
\rho_a(b)=\langle a|\pi_\rho(b)|a\rangle=\rho(a^*ba).
\end{equation}
Such states will be called vector states of the representation $\pi_\rho$. More general states can be defined by means of density operators $\boldsymbol{\sigma}$ in $\mathcal{B}(\mathcal{H}_\rho)$ by the formula:
\phantomsection\label{folium_rho_ps}\begin{equation}\label{folium_rho}
\sigma(a)=\Tr\!\big(\boldsymbol{\sigma}\pi_\rho(a)\big),\qquad\forall a\in\mathcal{A}.
\end{equation}
Notice that
\phantomsection\label{folium_ps}\begin{equation}\label{folium}
\sigma(\mathds{1})=1,\qquad\sigma(a^*a)=\langle a|\boldsymbol{\sigma}|a\rangle\geq 0,\quad\forall a\in\mathcal{A}.
\end{equation}
The family of states given by~\hyperref[folium_rho_ps]{(\ref*{folium_rho})} is called a \textit{folium} of the representation $\pi_\rho$ (see for instance \hyperref[Ha96_ps]{\RED{[\citen*{Ha96}, page 124]}}).

\begin{figure}[h]
\centering
\includegraphics{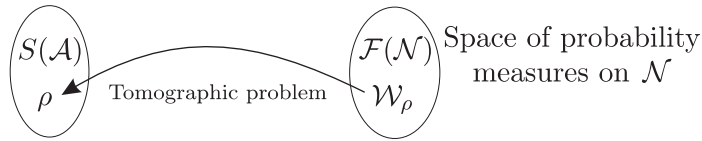}
\caption{\hfil\MYGREEN{Figure 2.1.1}: Tomographic problem.\hfil}
\label{Tom_problem_fig}
\end{figure}

The tomographic description of states $\rho$ consists on assigning to this state a probability density $\mathcal{W}_\rho$ in some auxiliary space $\mathcal{N}$, in such a way that given $\mathcal{W}_\rho$ the state $\rho$ can be reconstructed unambiguously \cite{Ib09} (see figure {\changeurlcolor{mygreen}\hyperref[Tom_problem_fig]{Figure \ref*{Tom_problem_fig}}}).

There is not a single ``tomographic theory'' neither a standard way to construct $\mathcal{W}_\rho$ out of $\rho$. In what follows, we will show that it is possible to construct the tomograms $\mathcal{W}_\rho$ using the following two tools: a \textit{Generalized Sampling Theory} and a \textit{Generalized Positive Transform}. We will discuss these two basic ingredients in the \hyperref[section_Sampling_C]{{\color{black}\textbf{following sections}}} as well as the \hyperref[section_equivariant]{{\color{black}\textbf{equiv-}}} \hyperref[section_equivariant]{{\color{black}\textbf{ariant version}}} of them. Finally, we will provide a particular instance of the theory based on harmonic analysis in groups.\newpage

\section{Sampling theory on $C^*$--algebras}\label{section_Sampling_C}

We consider a family of elements $U(x)$ in $\mathcal{A}$ parametrized by an index $x$ which can be discrete or continuous. This family can be described by a map $U:\MM\rightarrow\mathcal{A}$ where $\MM$ is the space labeling the elements $U(x)$.
\begin{figure}[htbp]
\centering
\includegraphics{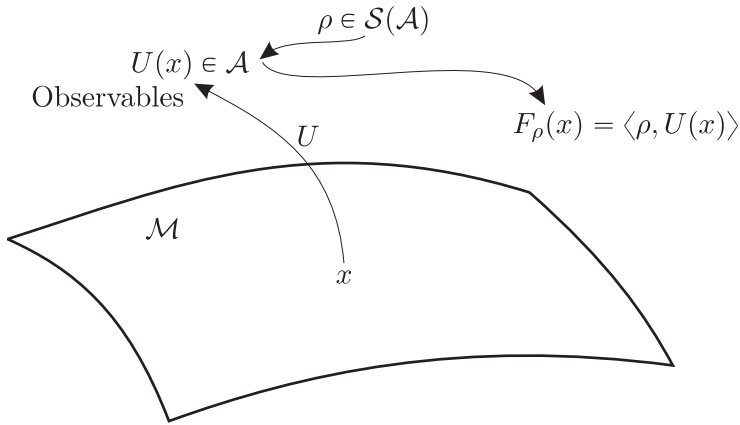}
\caption{\hfil\MYGREEN{Figure 2.2.1}: Tomographic map $U$.\hfil}
\label{Tom_set}
\end{figure}

Given a state $\rho$ and a set $U(x)$, we will call the function $F_\rho:\mathcal{M}\rightarrow\mathbb{C}$ defined as
\phantomsection\label{resumen_sampling_fun}
\begin{equation}
F_\rho(x)=\langle\rho,U(x)\rangle=\rho\big(U(x)\big),\qquad x\in\MM,
\end{equation}
the \textit{sampling function} of $\rho$ with respect to $U$, {\changeurlcolor{mygreen}\hyperref[Tom_set]{Figure \ref*{Tom_set}}}. In what follows, we will use indistinctly the notation $\rho(a)$ or $\langle\rho,a\rangle$ to denote the evaluation of the state $\rho$ in the element $a\in\mathcal{A}$. 

We will assume now that the map $U$ \textit{separates states}, i.e., given two states $\rho$ and $\tilde{\rho}$ there exist $x\in\MM$ such that $F_\rho(x)\neq F_{\tilde{\rho}}(x)$. In such case, sometimes the map $U$ is called a \textit{tomographic map} and its range $\left\{\vphantom{a^\dagger}U(x)\,|\,x\in\MM\right\}$ a \textit{tomographic set}.

Let us consider that $\MM$ is a measurable space with $\sigma$--algebra $\Sigma$ (if $\MM$ is a topological space, $\Sigma$ will be the Borelian $\sigma$--algebra on $\MM$) and let $\mu:\Sigma\rightarrow\mathbb{R}^+$ be a positive measure. We will also assume that the map $U$ is measurable (and continuous in the topological setting) and integrable in the sense that for any $\rho\in\mathcal{S}(\mathcal{A})$, the sampling function $F_\rho$ is integrable, that is, $F_\rho\in L^1(\MM,\mu)$. 

We will consider now the special case where there exists another map $D:\MM\rightarrow\mathcal{A}^\prime$, measurable and integrable in the sense that for any $a\in\mathcal{A}$ the function
\begin{equation}
G_a(x)=\langle D(x),a\rangle
\end{equation}
is integrable, and such that
\phantomsection\label{biortho_ps}\begin{equation}\label{biortho}
\langle D(x),U(y)\rangle=\delta(x,y),\qquad x,y\in\MM,
\end{equation}
where $\delta(x,y)$ is the delta distribution on $\MM$ with integral representation:
\phantomsection\label{deltaxxp_ps}
\begin{equation}
\phi(x)=\hspace{-.26cm}\int\limits_{\hspace{.3cm}\MM}\hspace{-0.1cm}\delta(x,y)\phi(y)\diff\mu(y),
\end{equation}
where $\phi$ is any test function on $\MM$. We will call the set $D(x)$ a \textit{dual tomographic set}.

If such map $D$ exists, we will say that $U$ and $D$ are \textit{biorthogonal}. These two maps $U$ and $D$ define what could be called a \textit{Generalized Fourier Transform} because its resemblance to the standard Fourier Transform, that is, if we denote the sampling function by
\begin{equation}
\widecheck{\rho}(x)\coloneq\langle\rho,U(x)\rangle
\end{equation}
and define:
\begin{equation}
\widehat{\phi}=\hspace{-.26cm}\int\limits_{\hspace{.3cm}\MM}\hspace{-0.1cm}\phi(x)D(x)\diff\mu(x)
\end{equation}
(notice that this integral is well-defined), we have two maps:
\begin{equation*}
\widecheck{\phantom{a}}:\,\mathcal{A}^\prime\longrightarrow L^{1}(\MM,\mu)
\end{equation*}

\phantom{a}
\vspace{-1cm}
\noindent and
\begin{equation}
\hspace{0.1cm}\widehat{\phantom{a}}:\,L^{1}(\MM,\mu)\longrightarrow \mathcal{A}^\prime,
\end{equation}
where the map $\widehat{\phantom{a}}$ \,is a left-inverse of the map $\widecheck{\phantom{a}}$\,. We will write formally this fact in the following theorem.\footnote{The formalism involving a tomographic map $U$ and a tomographic dual map $D$ has been widely used by Marmo \textit{et al.} (see for instance \cite{As15_2}) for applications in different settings and it was introduced by G. Marmo and V. Man'ko under the name of the ``quantizer-dequantizer'' formalism. Today, it is common to call the functions $F_\rho(x)$ ``tomographic symbols'' of the state $\rho$. }

{\MYBROWN{\begin{theorem}\label{GFT}\color{black}{
Let $U$ be a tomographic map in the $C^*$--algebra $\mathcal{A}$, and $D:\MM\rightarrow\mathcal{A}^\prime$ be an integrable map such that $U$ and $D$ are biorthogonal, then the map\; $\widehat{\phantom{a}}:L^1(\MM,\mu)\rightarrow\mathcal{A}^\prime$ given by
$$
\widehat{\phi}=\hspace{-.26cm}\int\limits_{\hspace{.3cm}\MM}\hspace{-0.1cm}\phi(x)D(x)\diff\mu(x)
$$
is the left-inverse of the tomographic map\; $\widecheck{\phantom{a}}:\mathcal{A}^\prime\rightarrow L^1(\MM,\mu)$ given by $\widecheck{\rho}=F_\rho$ if the function $G_a(x)=\langle D(x),a\rangle$ is in $L^\infty$ for any $a\in\AM$.}
\end{theorem}}}

\phantomsection\label{proof_GFT_ps}
\noindent \MYBROWN{\textbf{Proof}}: Let us consider $\widecheck{\rho}=F_\rho$, then, we will see that the element $\widehat{\widecheck{\rho}}$ is equal to $\rho$. For that, we will check now that the following functional:
\begin{equation*}
\tilde{\rho}(a)=\hspace{-.26cm}\int\limits_{\hspace{.3cm}\MM}\hspace{-0.1cm} F_\rho(x)\langle D(x),a\rangle\diff\mu(x),\qquad a\in\mathcal{A},
\end{equation*}
is continuous. Notice that this functional satisfies
\begin{equation*}
\|\tilde{\rho}\|\leq\|F_\rho\|_{L^1}\|\langle D(\cdot),a\rangle\|_{L^\infty},
\end{equation*}
but $D(x)\in\mathcal{A}^\prime$, therefore this means that $|\langle D(x),a\rangle|\leq K\|a\|$ for some $K$ independent of $x$, hence $\tilde{\rho}$ is continuous.

To show that $\rho=\tilde{\rho}$, we will prove that $\tilde{\rho}\big(U(x)\big)=\rho\big(U(x)\big)$ for all $x\in\MM$, hence because $U$ separates states, we will have that $\rho=\tilde{\rho}$\,:
\begin{equation*}
\langle\tilde{\rho},U(x)\rangle=\hspace{-.26cm}\int\limits_{\hspace{.3cm}\MM}\hspace{-0.1cm} F_\rho(y)\langle D(y),U(x)\rangle\diff\mu(y)=F_\rho(x)=\langle\rho,U(x)\rangle.
\end{equation*}

Conversely, we may compute first the\; $\widehat{\phantom{a}}$\, map and later the $\widecheck{\phantom{a}}$\hspace{0.01cm} map on $F_\rho$. If we apply the first map, we have that
\begin{equation*}
\widehat{F}_\rho=\hspace{-.26cm}\int\limits_{\hspace{.3cm}\MM}\hspace{-0.1cm}F_\rho(x)D(x)\diff\mu(x),
\end{equation*}
then, if we apply the $\widecheck{\phantom{a}}$\hspace{0.01cm} map, we get what we expected:
\begin{equation*}
\widecheck{\widehat{F}}_\rho(x)=\langle \widehat{F}_\rho,U(x)\rangle=\hspace{-.26cm}\int\limits_{\hspace{.3cm}\MM}\hspace{-0.1cm} F_\rho(y)\langle D(y),U(x)\rangle\diff\mu(y)=F_\rho(x).
\end{equation*}

\vspace{-.3cm}\hfill\hyperref[proof_GFT_ps]{{\color{black}{$\blacksquare$}}}

It is also noticeable that
\begin{equation*}
\int\limits_{\hspace{.3cm}\MM}\hspace{-0.1cm} F_\rho(x)\diff\mu(x)=\hspace{-.26cm}\int\limits_{\hspace{.3cm}\MM}\hspace{-0.1cm}\langle\rho,U(x)\rangle\diff\mu(x)=\rho\Bigg(\hspace{-.16cm}\int\limits_{\hspace{.3cm}\MM}\hspace{-0.1cm} U(x)\diff\mu(x)\Bigg),
\end{equation*}
thus, if $U$ is normalized, that is:
\begin{equation}
\int\limits_{\hspace{.3cm}\MM}\hspace{-0.1cm} U(x)\diff\mu(x)=\mathds{1},
\end{equation}
it is clear that $F_\rho$ is normalized too:
\begin{equation}
\int\limits_{\hspace{.3cm}\MM}\hspace{-0.1cm} F_\rho(x)\diff\mu(x)=1.
\end{equation}

We may also define another sampling function, but this time depending on two arguments as follows:
\begin{equation}
F_\rho(x,y)=\langle\rho,U(x)^*U(y)\rangle,\qquad\mbox{for any }x,y\in\MM.
\end{equation}
We will say that a function $F:\MM\times\MM\rightarrow\mathbb{C}$ is of positive type or semidefinite positive if for all $N\in\mathbb{N}$, $\xi_i\in\mathbb{C}$ and any $x_i\in\MM$, $i=1,\ldots,N$, it satisfies that
\phantomsection\label{positive_function_ps}\begin{equation}\label{positive_function}
\sum_{i,j=1}^N\bar{\xi}_i\xi_j F(x_i,x_j)\geq 0.
\end{equation}
This notion of positivity implies the following theorem.

{\MYBROWN{\begin{theorem}\label{tom_positive}\color{black}{
Given a state $\rho$ and a tomographic set $U:\MM\rightarrow\mathcal{A}$ in a $C^*$--algebra $\mathcal{A}$, then the sampling function $F_\rho(x,y)=\langle\rho,U(x)^*U(y)\rangle$, $x,y\in\MM$ is of positive type.}
\end{theorem}}}

\phantomsection\label{tom_positive_ps}
\noindent\MYBROWN{\textbf{Proof}}: It is a straightforward computation:

\phantom{a}

\vspace{-1.2cm}
\begin{multline*}
\sum_{i,j=1}^N\bar{\xi}_i\xi_jF_\rho(x_i,x_j)=\hspace{-0.1cm}\sum_{i,j=1}^N\bar{\xi}_i\xi_j\langle\rho,U(x_i)^*U(x_j)\rangle\\
=\langle\rho,\hspace{-0.1cm}\sum_{i,j=1}^N\bar{\xi}_i\xi_jU(x_i)^*U(x_j)\rangle=\langle\rho,\left(\sum_{i=1}^N\xi_iU(x_i)\right)^{\hspace{-0.1cm}*}\hspace{-0.1cm}\left(\sum_{j=1}^N\xi_jU(x_j)\right)\rangle\geq 0.
\end{multline*}

\vspace{-.5cm}\hfill\hyperref[tom_positive_ps]{{{\color{black}$\blacksquare$}}}

We will take advantage of this property later on when dealing with tomography in groups. We will conclude this section by establishing the notion of equivalence of tomographic sets.

Given two tomographic sets $U:\MM\rightarrow\mathcal{A}$ and $\widetilde{U}:\hspace{0.14cm}\widetilde{\phantom{U}}\hspace{-0.43cm}\mathcal{M}\rightarrow\AM$, we will say that they are equivalent if there exists an invertible measure preserving map $\varphi:\MM\rightarrow\hspace{0.14cm}\widetilde{\phantom{U}}\hspace{-0.43cm}\mathcal{M}$ such that $\widetilde{U}=U\circ\varphi^{-1}$. Clearly, if $U$ and $\widetilde{U}$ are equivalent, then the sampling functions corresponding to a given state are related by $\widetilde{F}_\rho=F_\rho\circ\varphi^{-1}$:
\begin{multline}
\widetilde{F}_\rho(\tilde{x})=\langle\rho, \widetilde{U}(\tilde{x})\rangle\\
=\langle\rho,(U\circ\varphi^{-1})(\tilde{x})\rangle=F_\rho\big(\varphi^{-1}(\tilde{x})\big)=(F_\rho\circ\varphi^{-1})(\tilde{x}).
\end{multline}
Consider that $\varphi$ is a measure preserving map $\tilde{\mu}(\hspace{0.01cm}\widetilde{\Delta}\hspace{0.01cm})=\mu\big(\varphi^{-1}(\hspace{0.01cm}\widetilde{\Delta}\hspace{0.01cm})\big)$, for any measurable set $\widetilde{\Delta}\subset\hspace{0.14cm}\widetilde{\phantom{U}}\hspace{-0.43cm}\mathcal{M}$, hence if $U$ and $\widetilde{U}$ are equivalent and $D:\MM\rightarrow\AM^\prime$ is biorthogonal to $U$, the pair $(\hspace{0.02cm}\widetilde{U},\widetilde{D}\hspace{0.02cm})$ is biorthogonal too, with $\widetilde{D}=D\circ\varphi$:
\begin{multline}
\langle \widetilde{D}(\tilde{x}),\widetilde{U}(\tilde{y})\rangle=\langle D\big(\varphi^{-1}(\tilde{x})\big),U\big(\varphi^{-1}(\tilde{y})\big)\rangle\\
=\delta\big(\varphi^{-1}(\tilde{x}),\varphi^{-1}(\tilde{y})\big)\hspace{-0.02cm}=\delta(\tilde{x},\tilde{y}).
\end{multline}

\begin{figure}[h]
\centering
\includegraphics{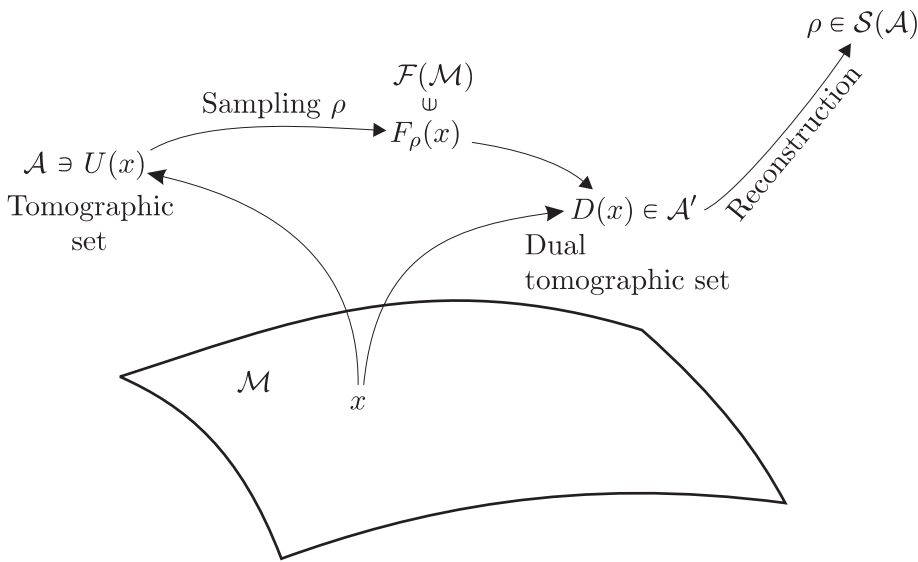}
\caption{\hfil\MYGREEN{Figure 2.2.2}: Sampling diagram.\hfil}
\label{Samp_diagr}
\end{figure}

The theory that we have sketched in this section, which consists basically on reconstructing the functional $\rho$ by means of a set of samples $F_\rho(x)$:
\phantomsection\label{reconstruction_state_D_ps}\begin{equation}\label{reconstruction_state_D}
\rho=\hspace{-.26cm}\int\limits_{\hspace{.3cm}\MM}\hspace{-0.1cm} F_\rho(x)\langle D(x),\;\cdot\;\rangle\diff\mu(x),
\end{equation}
could be called a \textit{Generalized Sampling Theory} on $C^*$--algebras, {\changeurlcolor{mygreen}\hyperref[Samp_diagr]{Figure}} {\changeurlcolor{mygreen}\hyperref[Samp_diagr]{\ref*{Samp_diagr}}}.

Recently, there has been some results trying to extend the classical theory of sampling to quantum systems (see for instance \cite{Fe15} and references therein). In this sense, notice that the tomographic map and its dual generalize the notion of frame (and its dual coframe).

The problem we are dealing with now is how to get the samples $F_\rho(x)$. In principle, it is not possible to measure directly the sampling function $F_\rho$ because, in general, they are complex numbers and as we have seen in \hyperref[Intensity_photons]{{\color{black}\textbf{section~\ref*{section_photodetection}}}}, what we can measure in the laboratory are probability distributions. For that reason, we need to include another tool which will allow us to obtain the sampling function $F_\rho$ from probability distributions. We will call this second tool a \textit{Generalized Positive Transform} and we will describe it in the \hyperref[ps_GPT]{{\color{black}\textbf{following section}}}.
\vspace{0.0cm}
\newpage


\section{A Generalized Positive Transform}\label{ps_GPT}

The second tool in our programme is the choice of a Generalized Positive Transform. \hyperref[section_Radon_Transform]{{\color{black}\textbf{At the beginning of this text}}}, we introduced the Radon Transform which maps probability distributions into probability distributions. We will try to generalize this concept in what follows.

To offer an abstract presentation of this transform, we will consider a second auxiliary space $\NN$ that parametrizes a family of elements in the dual space $\mathcal{D}(\MM)\subset\mathcal{F}(\MM)^\prime$ of the space of continuous functions on $\MM$\footnote{$\MM$ will be assumed to be a topological space in what follows and, consequently, a Borelian measurable space.}. If we denote by $y\in\NN$ the elements of $\NN$, such family of elements will have the form $\left\{\vphantom{a^\dagger}R(y)\,|\,y\in\NN\right\}$. Thus, $R$ will be a map $R:\NN\rightarrow\mathcal{D}(\MM)$ and it will allow us to define a transform of any continuous function on $\MM$ by means of
\phantomsection\label{ps_conclusions_s_map}
\begin{equation}
\mathcal{R}(F)(y)\coloneq\langle R(y),F\rangle,
\end{equation}
where $\langle\cdot,\cdot\rangle$ denotes the natural pairing between $\mathcal{D}(\MM)$ and $\mathcal{F}(\MM)$. We will say that this map $F\rightsquigarrow\mathcal{R}(F)$ is a \textit{Generalized Positive Transform} if it maps functions of positive type on $\MM$ into non-negative functions on $\NN$, i.e., if $F:\MM\times\MM\rightarrow\mathbb{C}$ is of positive type, then
\begin{equation}
\mathcal{R}(F)(y)=\langle R(y),F\rangle\geq 0,\qquad\forall y\in\NN.
\end{equation}

Again, if $\NN$ is a measure space with measure $\sigma$, we will assume that $\mathcal{R}$ is integrable in the sense that the function $\mathcal{R}(F)=\langle R(\cdot),F\rangle$ is $\sigma$--integrable for any $F$ integrable.

We will say that $\mathcal{R}$ is normalized if
\begin{equation}
\hspace{-.26cm}\int\limits_{\hspace{.3cm}\NN}\hspace{-0.1cm}\mathcal{R}(F)(y)\diff\sigma(y)=1
\end{equation}
for any function $F$ on $\MM$ such that
\begin{equation*}
\hspace{-.26cm}\int\limits_{\hspace{.3cm}\MM}\hspace{-0.1cm}F(x)\diff\mu(x)=1.
\end{equation*}

Notice that in this case, $\mathcal{R}:L^1(\MM,\mu)\rightarrow L^1(\NN,\sigma)$ is a continuous map and we will say that $\mathcal{R}$ is non-degenerate if it has a left inverse, i.e., if there exists a map $\mathcal{R}^{-1}:L^1(\NN,\sigma)\rightarrow L^1(\MM,\mu)$ such that
\begin{equation}
\mathcal{R}^{-1}\circ\mathcal{R}=\textrm{id}_{L^1(\MM,\mu)}.
\end{equation}

\begin{figure}[h]
\centering
\includegraphics{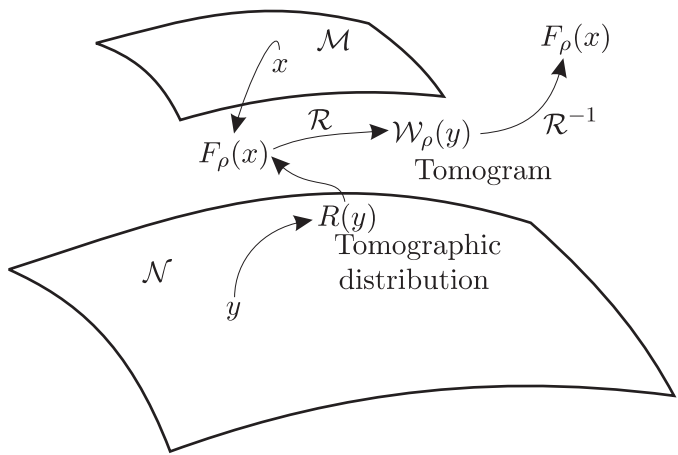}
\caption{\hfil\MYGREEN{Figure 2.3.1}: Generalized Positive Transform diagram.\hfil}
\label{Pos_diagr}
\end{figure}

Under this rather long list of conditions, we will conclude by noticing that if $\rho$ is a state and $U$ is a normalized tomographic map, then $F_\rho$ will be a normalized function of positive type, \hyperref[tom_positive]{\MYBROWN{\textbf{Thm.\,\ref*{tom_positive}}}}, and in consequence $\mathcal{R}(F_\rho)$ will be a normalized non-negative function on $\NN$:
\begin{equation}
\hspace{-.26cm}\int\limits_{\hspace{.3cm}\NN}\hspace{-0.1cm}\mathcal{R}(F_\rho)(y)\diff\sigma(y)=1.
\end{equation}\phantomsection\label{resumen_tomogram}
Moreover, if we know $\mathcal{R}(F_\rho)$, we could obtain $F_\rho$ by applying a left-inverse map $\mathcal{R}^{-1}$, i.e., $F_\rho=\mathcal{R}^{-1}\circ\mathcal{R}(F_\rho)$. The function $\mathcal{R}(F_\rho)$ will be called the \textit{tomogram} of the state $\rho$ and we will denote it by $\mathcal{W}_\rho$, {\changeurlcolor{mygreen}\hyperref[Pos_diagr]{Figure \ref*{Pos_diagr}}}:
\phantomsection\label{ps_tomogram_conclusion}
\begin{equation}\label{tomogram_conclusion}
\mathcal{W}_\rho(y)=\langle R(y),F_\rho\rangle.
\end{equation}
Notice again that the tomogram $\mathcal{W}_\rho(y)$ satisfies that it is a probability distribution related with the state $\rho$:
\begin{equation}
\mathcal{W}_\rho\geq 0,\qquad\hspace{0cm}\int\limits_{\hspace{.3cm}\NN}\hspace{-0.1cm}\mathcal{W}_\rho(y)\diff\sigma(y)=1.
\end{equation}

A particular instance of this setting is obtained when the tomographic set is trivial, i.e., $\MM=\mathcal{A}$ and $U=\textrm{id}_\mathcal{A}$. Then, we may assume that $R$ is a map $R:\NN\rightarrow\mathcal{A}\subset\mathcal{A}^{''}$, and in that case, the tomogram of the state $\rho$ will be obtained directly from:
\begin{equation}
\mathcal{W}_\rho(y)=\langle\rho,R(y)\rangle.
\end{equation}
This is just the situation for the Classical Radon Transform presented \hyperref[section_Radon_Transform]{{\color{black}\textbf{at}}} \hyperref[section_Radon_Transform]{{\color{black}\textbf{the beginning of this Thesis}}}, where now $\mathcal{A}$ can be taken to be the algebra of continuous functions on a compact domain $\Omega$ in $\mathbb{R}^2$, $\NN$ will be the set of lines on $\mathbb{R}^2$ and $R:\NN\rightarrow\mathcal{A}^\prime$, $L\rightsquigarrow R(L)=\delta_L$:
$$
\mathcal{W}_\rho(L)=\langle\rho,R(L)\rangle=\hspace{-0.1cm}\int\limits_{\hspace{.3cm}L}\hspace{-0.1cm}\rho\big(x(s),y(s)\big)\diff s,
$$
which is just the Radon Transform \hyperref[Radon_Transform_ps]{(\ref*{Radon_Transform})}.

\section{Equivariant tomographic theories on $C^*$--algebras}\label{section_equivariant}

In many situations of interest, there is a group present in the system whose states we want to describe tomographically. Such group could be, for instance, a group of symmetries of the dynamics or a group which is describing the background of the theory (as the Poincar\'e group in \hyperref[section_WS_axioms]{{\color{black}\textbf{chapter}}} \hyperref[section_WS_axioms]{{\color{black}\textbf{\ref*{chap_QFT}}}}). In any of these circumstances, we will assume that the group $G$ is a Lie group acting on the $C^*$--algebra $\mathcal{A}$, i.e., that there is a continuous map $T:G\rightarrow\textrm{Aut}(\mathcal{A})$:
\begin{equation}
T_e=\mathds{1},\qquad T_{g_1}T_{g_2}=T_{g_1g_2},\qquad\forall g_1,g_2\in G.
\end{equation}
In such case, in order to obtain a reasonable theory, we will assume that the group acts on the auxiliary spaces used to construct the tomographic description. Thus, the group $G$ will act on $\MM$ and $\NN$ and such actions will be simply denoted by $x\rightsquigarrow g\cdot x$ and $y\rightsquigarrow g\cdot y$, $g\in G$, $x\in\MM$ and $y\in\NN$.

The natural compatibility condition for a tomographic map $U:\mathcal{M}\rightarrow\mathcal{A}$ to be equivariant is that
\phantomsection\label{equivariant_U_ps}\begin{equation}\label{equivariant_U}
U(g\cdot x)=T_g U(x),\qquad\forall x\in\MM\quad\mbox{and}\quad\forall g\in G.
\end{equation}
This could be interpreted by saying that if $\tilde{x}=g\cdot x$, then the two obser-vables $U(x)$ and $U(\tilde{x})$ are equivariant with respect to $G$, that is:
\begin{equation}
U(\tilde{x})=T_gU(x).
\end{equation}
Under these conditions, it is easy to conclude that the sampling function $F_\rho$ corresponding to the state $\rho$ satisfies the following condition:
\phantomsection\label{property_equivariant_rho_ps}
\begin{equation}\label{property_equivariant_rho}
F_\rho(g\cdot x)=F_{g^*\rho}(x),
\end{equation}
because
\begin{equation*}
F_\rho(g\cdot x)=\langle\rho,U(g\cdot x)\rangle=\langle\rho,T_g U(x)\rangle=\langle T_g^*\rho, U(x)\rangle=F_{g^*\rho}(x),
\end{equation*}
where $g^*\rho=T_g^*\rho$ is the adjoint action of $G$ on $\mathcal{A}^\prime$. Notice that if $\rho$ is an invariant state, $T_g^*\rho=\rho$, then the corresponding sampling function will be invariant too:
\begin{equation}
F_\rho(g\cdot x)=F_\rho(x),\qquad\forall g\in G,\quad x\in\MM.
\end{equation}
As indicated before, we will also consider that the group $G$ acts on the auxiliary space $\NN$ used to define the Generalized Positive Transform. The map $R:\NN\rightarrow\mathcal{D}(\MM)$ is said to be equivariant if
\begin{equation}
R(g^{-1}\cdot y)=g_*R(y),
\end{equation}
where $g_*$ indicates now the natural action induced on the space $\mathcal{D}(\MM)\subset\mathcal{F}(\MM)^\prime$ given by the action of $G$ on $\MM$, more explicitly:
\begin{equation*}
\langle g_*R(y),F\rangle=\langle R(y),g^*F\rangle\quad\mbox{and}\quad g^*F(x)=F(g\cdot x).
\end{equation*}
If $R$ is actually a map that induces a Generalized Positive Transform and $\mathcal{W}_\rho$ the tomogram of the state $\rho$, we will have that:
\begin{multline}
\mathcal{W}_\rho(g^{-1}\cdot y)=\langle R(g^{-1}\cdot y),F_\rho\rangle=\langle g_* R(y),F_\rho\rangle\\
=\langle R(y),g^* F_\rho\rangle=\langle R(y),F_{g^*\rho}\rangle=\mathcal{W}_{g^*\rho}(y).
\end{multline}
Therefore, we will conclude this discussion by observing that under the conditions stated before, if $\rho$ is an invariant state, its tomogram is invariant too:
\begin{equation}
\mathcal{W}_\rho(g\cdot y)=\mathcal{W}_\rho(y),\qquad\forall g\in G.
\end{equation}

\markboth{\textbf{The tomographic picture of quantum systems}}{\textbf{2.5. A particular instance of Quantum Tomography: Quantum Tomography\newline{\color{white}........\!}with groups}}
\section{A particular instance of Quantum Tomography: Quantum Tomography with groups}\label{section_particular group}
\markboth{\textbf{The tomographic picture of quantum systems}}{\textbf{2.5. A particular instance of Quantum Tomography: Quantum Tomography\newline{\color{white}........\!}with groups}}

We will discuss now a particular instance of the tomographic programme where a group $G$ plays a paramount role. Such situation happens, for example, in Spin Tomography \cite{Ma97} where the group $G$ is $SU(N)$ (see \hyperref[section_Spin_Tomography]{{\color{black}\textbf{section~\ref*{section_Spin_Tomography}}}}), in the standard tomography of quantum states presented in \hyperref[Marginal_probability_q_phi_1_ps]{{\color{black}\textbf{chapter~\ref*{chap_birth}}}} with $G$ being the Heisenberg--Weyl group (see \hyperref[resume_heis]{{\color{black}\textbf{section~\ref*{resume_heis}}}}) and other physical situations that will show up later on.

In this setting, we will assume that the auxiliary space $\MM$ is a Lie group $G$ and the tomographic map $U:G\rightarrow\mathcal{A}$ is provided by a continuous unitary representation of $G$ on $\mathcal{A}$, this is:
\phantomsection\label{product_U_g_ps}
\begin{equation*}
U(g)^*=U(g)^{-1}=U(g^{-1}),\qquad\forall g\in G,
\end{equation*}

\phantom{a}
\vspace{-0.7cm}\noindent and
\begin{equation}\label{product_U_g}
U(g_1g_2)=U(g_1)U(g_2),\qquad\forall g_1,g_2\in G.
\end{equation}
Notice that because of \hyperref[product_U_g_ps]{(\ref*{product_U_g})}, then $U(e)=\mathds{1}$.

If we denote by $T_g:G\rightarrow\textrm{Aut}(\mathcal{A})$ the action of $G$ on $\mathcal{A}$ given by
\begin{equation}
T_g(a)=U(g)^*a\,U(g),
\end{equation}
with $a\in\mathcal{A}$ and $g\in G$, then we see immediately that
\begin{equation}
U(g^{-1}hg)=U(g)^*U(h)\hspace{0.02cm}U(g)=T_gU(h),\qquad\forall g,h\in G,
\end{equation}
which is the equivariant property \hyperref[equivariant_U_ps]{(\ref*{equivariant_U})} for the adjoint action of $G$ on itself, $h\rightsquigarrow g^{-1}\hspace{-0.03cm}\cdot h\cdot g$.

The sampling function corresponding to the state $\rho$ is given by
\phantomsection\label{sampling_go_back_2}
\begin{equation}
F_\rho(g)=\langle\rho,U(g)\rangle,
\end{equation}
and we may check that the map $F_\rho:G\rightarrow\mathbb{C}$ is of positive type because the function $F_\rho(g,h)=F_\rho(g^{-1}h)$ satisfies \hyperref[tom_positive]{\MYBROWN{\textbf{Thm.\,\ref*{tom_positive}}}}:
\phantomsection\label{positivity_functions_2_ps}\begin{equation}\label{positivity_functions_2}
\sum_{i,j=1}^N\bar{\xi}_i\xi_jF_\rho(g_i^{-1}g_j)\geq 0,
\end{equation}
for all $N\in\mathbb{N}$, $\xi_i\in\mathbb{C}$, $g_i\in G$ with $i=1,\ldots,N$. Moreover, it satisfies property \hyperref[property_equivariant_rho_ps]{(\ref*{property_equivariant_rho})}.

In the case in which $\mathcal{A}=\mathcal{B}(\mathcal{H})$, because of the one-to-one correspondence between states and density operators, the sampling function $F_\rho$ can be written as:
\phantomsection\label{smeared_character_chap_2_ps}\begin{equation}\label{smeared_character_chap_2}
F_\rho(g)=\Tr\!\big(\boldsymbol{\rho}U(g)\big),
\end{equation}
hence, because the character of a group representation is defined as:
\phantomsection\label{character_group_ps}\begin{equation}\label{character_group}
\chi(g)=\Tr\!\big(U(g)\big),
\end{equation}
we will denote the sampling function $\chi_\rho(g)\coloneq F_\rho(g)$ and we will call it a \textit{smeared character} of the representation $U$ with respect to the state $\rho$. Let us notice that if $\mathcal{H}$ has finite dimension $n$ and the state is the trivial one, $\boldsymbol{\rho}=\frac{1}{n}\mathds{1}$, the smeared character is just the standard character~\hyperref[character_group_ps]{(\ref*{character_group})} divided by $n$.

Consider again the strongly continuous action $T_g$ of $G$ on $\mathcal{A}$ and notice that the map $f_a(g)=\|T_g(a)\|$ is continuous for all $a\in\mathcal{A}$. The \hyperref[ps_GNS]{{\color{black}\textbf{GNS}}} \hyperref[ps_GNS]{{\color{black}\textbf{construction}}} described at the beginning of this chapter (\hyperref[ps_GNS]{{\color{black}\textbf{section \ref*{ps_GNS}}}}) provides, given a state $\rho$, a representation $\pi_\rho$ of $\mathcal{A}$ in $\mathcal{H}_\rho$ and then, we get a strongly continuous unitary representation of the group $G$ by means of
\phantomsection\label{U_rho_GNS_ps}
\begin{equation}\label{U_rho_GNS}
U_\rho(g)=\pi_\rho\big(U(g)\big).
\end{equation}
Notice that $U_\rho(g)$ is actually a unitary operator on the Hilbert space $\mathcal{H}_\rho$ because:
\begin{equation*}
\langle U_\rho(g)[a],U_\rho(g)[b]\rangle_\rho=\rho\big((U(g)a)^*(U(g)b)\big)=\rho(a^*b)=\langle[a],[b]\rangle_\rho,
\end{equation*}
for all $g\in G$, $[a],[b]\in\mathcal{H}_\rho$.

Now, the \hyperref[sampling_go_back_2]{{\color{black}\textbf{sampling function}}} of a representation $U$ corresponding to a state $\rho$ can be written as:
\begin{equation}
F_\rho(g)=\langle 0|U_\rho(g)|0\rangle,
\end{equation}
where $|0\rangle$ is the fundamental vector of $\mathcal{H}_\rho$. Fixed the state $\rho$, the smeared character of $U$ with respect to any other state $\sigma$ in the folium of $\rho$, \hyperref[folium_rho_ps]{(\ref*{folium_rho})}, will be given by
\phantomsection\label{character_folium_ps}\begin{equation}\label{character_folium}
\chi_\sigma(g)=\langle\sigma,U(g)\rangle=\Tr\!\big(\boldsymbol{\sigma}\pi_\rho\big(U(g)\big)\big)=\Tr\!\big(\boldsymbol{\sigma}U_\rho(g)\big).
\end{equation}

We can conclude this section by stating the following characterization of states.
{\MYBROWN{\begin{theorem}\label{prop_positive}\color{black}{Let $\rho\colon \mathcal{A} \to \mathbb{C}$ be a linear function and consider the sampling function $F_\rho (g)=\langle\rho,U(g)\rangle$ where $U$ is a completely reducible strongly continuous unitary representation of the Lie group $G$ on $\mathcal{A}$.  Then, $\rho$ is a state iff $F_\rho$ is a positive type function on $G$ and $F_\rho (e) = 1$.}
\end{theorem}}}

\phantomsection\label{prop_positive_proof_ps}
\noindent\MYBROWN{\textbf{Proof}}: We have seen before in~\hyperref[positivity_functions_2_ps]{(\ref*{positivity_functions_2})} that $F_\rho$ is of positive type if $\rho$ is a state, and $F_\rho(e)=1$ because of the normalization of $\rho$. Conversely, if $F_\rho$ is a positive type function on $G$, because of Naimark's theorem \cite{Na64}, there exist a complex separable Hilbert space $\mathcal{H}$ supporting a strongly continuous unitary representation $U$ of $G$, and a vector $|0\rangle\in\mathcal{H}$ such that
$$
F_\rho(g)=\langle 0|U(g)| 0\rangle.
$$
Now, because $U$ is completely reducible, then $U$ can be written as a direct sum of irreducible representations: 
$$
U=\bigoplus_{\alpha}U^{\alpha},
$$
and any $a\in\mathcal{A}$ can be written as:
$$
a=\bigoplus_{\alpha}a^\alpha,
$$
where $a^\alpha\in \overline{\textrm{span}}\left\{\vphantom{a^\dagger}U^\alpha(g)\tilde{a}\right\}$ for some $\tilde{a}\in\mathcal{A}$. Hence, we can restrict to the subspaces $\mathcal{A}^\alpha=\overline{\textrm{span}}\left\{\vphantom{a^\dagger}U^\alpha\tilde{a}\right\}$ where $U^\alpha$ is irreducible. Once we have that we can restrict to the subspaces $\mathcal{A}^\alpha$, we can proceed similarly to the proof made for finite groups in \cite{Ib11} generalizing it to any Lie group $G$. 

\vspace{-.0cm}\hfill{\hyperref[prop_positive_proof_ps]{{\color{black}$\blacksquare$}}}

\vspace{-.3cm}\section{Quantum tomograms associated to group representations}\label{section_tomograms_group}

We are ready now to introduce the notion of quantum tomogram of a given state $\rho$ associated to a unitary group representation  $(G,U)$.

Given an element $\xi$ in the Lie algebra $\mathfrak{g}$ of the Lie group $G$, we can consider the space $\mathfrak{g}\times\mathbb{R}$ and the extended exponential map $\exp:\frak{g}\times\mathbb{R}\rightarrow G$ given by $\exp(t,\xi)\coloneq\exp(t\xi)$, where $\exp:\mathfrak{g}\rightarrow G$ is the ordinary exponential map. Notice that if $G$ is a matrix Lie group, then: 
\begin{equation}
\exp(t\xi)=\sum_{n=0}^\infty\frac{t^n}{n!}\xi^n.
\end{equation}

Also, we can consider the one-parameter group of unitary operators in the Hilbert space $\mathcal{H}_\rho$, $U_\rho\big(\exp(t\xi)\big)$, obtained using the \hyperref[ps_GNS]{{\color{black}\textbf{GNS construc-}}} \hyperref[ps_GNS]{{\color{black}\textbf{tion}}}, eq.\hyperref[U_rho_GNS_ps]{(\ref*{U_rho_GNS})}, with $t\in\mathbb{R}$. Because of Stone's theorem \cite{St32}, there exists a self-adjoint operator $\boldsymbol{\xi}$ on $\mathcal{H}_\rho$ such that
\phantomsection\label{selfadjoint_exp_ps}\begin{equation}\label{selfadjoint_exp}
\e^{it\boldsymbol{\xi}}=U_\rho\big(\exp(t\xi)\big).
\end{equation}
Notice that the element $\xi$ in the Lie algebra and the operator $\boldsymbol{\xi}$ have opposite symmetry because of the $i$ factor in the exponent, that is, if $G$ is a matrix Lie group, then $\xi\in\mathfrak{g}$ is skew-Hermitian while $\boldsymbol{\xi}$ is Hermitean.

Let us denote by $\Theta$ the canonical left-invariant Cartan $1$-form on $G$ that has the tautological definition $\Theta(\xi)=\xi$, for any left-invariant vector field $\xi$ in $G$. Let also $\boldsymbol{\Theta}$ be the ``quantization'' of that $1$-form, i.e., $\boldsymbol{\Theta}$ is a left-invariant 1-form on $G$ with values in self-adjoint operators on $\mathcal{H}_\rho$, and is defined as:
\begin{equation}
\langle\boldsymbol{\Theta},\xi\rangle=\boldsymbol{\xi},\qquad\forall\xi\in\mathfrak{g}.
\end{equation}
Using that Cartan $1$-form, we can see that the operators $\boldsymbol{\xi}$ provide a representation of $\mathfrak{g}$ in $\mathcal{H}_\rho$, this is:
\begin{equation}
[\boldsymbol{\xi},\boldsymbol{\zeta}]=i\langle\boldsymbol{\Theta},[\xi,\zeta]\rangle,\qquad\forall\xi,\zeta\in\mathfrak{g}.
\end{equation}
To prove it, notice that because
$$
\e^{t\xi}\e^{t\eta}\e^{-t\xi}\e^{-t\eta}=\mathds{1}+t^2[\xi,\eta]+\cdots,
$$
we have:
\begin{multline*}
i\langle\boldsymbol{\Theta},[\xi,\zeta]\rangle=\left.\frac{\diff}{\diff t}U_\rho\big(\exp({t[\xi,\eta]})\big)\right|_{t=0}\hspace{-0.4cm}=\left.\frac{\diff}{\diff t}U_\rho\left(\e^{\sqrt{t}\xi}\e^{\sqrt{t}\eta}\e^{-\sqrt{t}\xi}\e^{-\sqrt{t}\eta}\right)\right|_{t=0}\\
=\left.\frac{\diff}{\diff t}\left(\e^{i\sqrt{t}\boldsymbol{\xi}}\e^{i\sqrt{t}\boldsymbol{\eta}}\e^{-i\sqrt{t}\boldsymbol{\xi}}\e^{-i\sqrt{t}\boldsymbol{\eta}}\right)\right|_{t=0}\hspace{-0.4cm}=[\boldsymbol{\xi},\boldsymbol{\eta}].
\end{multline*}

We may use now the spectral theorem \hyperref[Re80_ps]{\RED{[\citen*{Re80}, ch.\,7]}} to write each operator $\boldsymbol{\xi}$ on $\mathcal{H}_\rho$ as follows:
\phantomsection\label{spectral_thm_ch_2_ps}
\begin{equation}\label{spectral_thm_ch_2}
\boldsymbol{\xi}=\hspace{-.25cm}\int\limits_{\hspace{-.04cm}-\infty}^{\hspace{.45cm}\infty}\hspace{-0.15cm}\lambda E_{\xi}(\diff\lambda),
\end{equation}
where $E_{\xi}$ denotes the spectral measure of $\boldsymbol{\xi}$, and using~\hyperref[selfadjoint_exp_ps]{(\ref*{selfadjoint_exp})}, we can write:
\phantomsection\label{spectral_measure_ps}\begin{equation}\label{spectral_measure}
U_\rho\big(\exp(t\xi)\big)=e^{it\boldsymbol{\xi}}=\hspace{-.25cm}\int\limits_{\hspace{-.04cm}-\infty}^{\hspace{.45cm}\infty}\hspace{-0.15cm}\e^{it\lambda} E_{\xi}(\diff\lambda).
\end{equation}

Now, let $\sigma$ be a state on the folium of $\rho$, i.e., $\boldsymbol{\sigma}$ is a density operator on $\mathcal{H}_\rho$ defined by eq.\,\hyperref[folium_rho_ps]{(\ref*{folium_rho})}, then let us consider the measure $\mu_{\sigma,\xi}(\diff\lambda)=\textrm{Tr}\big(\boldsymbol{\sigma}E_{\xi}(\diff\lambda)\big)$. In other words, if $\Delta$ is a Borel set in $\mathbb{R}$, we have:
\phantomsection\label{physical_int_measure_ps}\begin{equation}\label{physical_int_measure}
P(\boldsymbol{\xi},\sigma;\Delta)=\hspace{-0.15cm}\int\limits_{\hspace{.3cm}\Delta}\hspace{-0.1cm}\mu_{\sigma,\xi}(\diff\lambda)=\mu_{\sigma,\xi}(\Delta)=\textrm{Tr}\big(\boldsymbol{\sigma}E_{\xi}(\Delta)\big).
\end{equation}

Notice that the physical interpretation of the measure $\mu_{\sigma,\xi}(\Delta)$ associated to the state $\sigma$ and the observable $\boldsymbol{\xi}$, as in the \hyperref[chap_tom_qu]{{\color{black}\textbf{introduction of this}}} \hyperref[chap_tom_qu]{{\color{black}\textbf{chapter}}}, is that the number $P(\boldsymbol{\xi},\sigma;\Delta)$ in eq.\,\hyperref[physical_int_measure_ps]{(\ref*{physical_int_measure})} is the probability of getting the output of measuring the observable $\boldsymbol{\xi}$ in the set $\Delta$ when the system is in the state $\sigma$. Then, obviously, we see that $\mu_{\sigma,\xi}(\mathbb{R})=1$. Moreover, if the measure $\mu_{\sigma,\xi}(\diff\lambda)$ is absolutely continuous with respect to the Lebesgue measure $\diff X$, then there will exist a function $\mathcal{W}_\sigma(X;\xi)$ in $L^1(\mathbb{R},\diff X)$ such that for all measurable $\Delta$:
\phantomsection\label{absolutely_tom_ps}\begin{equation}\label{absolutely_tom}
\int\limits_{\hspace{.3cm}\Delta}\hspace{-0.1cm}\mu_{\sigma,\xi}(\diff\lambda)=\hspace{-0.15cm}\int\limits_{\hspace{.3cm}\Delta}\hspace{-0.1cm}\mathcal{W}_\sigma(X;\xi)\diff X\geq 0.
\end{equation}
In general, this will not be true if the measure $\mu_{\sigma,\xi}(\diff\lambda)$ have singular part, for instance, if $\boldsymbol{\xi}$ has eigenvalues.  

{\MYBROWN{\begin{definition}\color{black}{Given a state $\sigma$ in the folium of $\rho$ and a unitary representation $U$ of a Lie group $G$ on the unital $C^*$--algebra $\mathcal{A}$, we will call the quantum tomogram family of $\sigma$ the family of Borelian probability measures $\mu_{\sigma,\xi}(\diff\lambda)=\Tr\big(\boldsymbol{\sigma}E_{\xi}(\diff\lambda)\big)$ on $\mathbb{R}$, $\xi\in\mathfrak{g}$ and $X\in\mathbb{R}$. The absolutely continuous part of them define a function $\mathcal{W}_\sigma:\mathfrak{g}\times\mathbb{R}\rightarrow\mathbb{R}$ given by eq.\,$\hyperref[absolutely_tom_ps]{(\ref*{absolutely_tom})}$, which is commonly called the quantum tomogram of $\sigma$, in other words, $\mathcal{W}_\sigma(X;\xi)$ is the Radon--Nikodym derivative of the measure $\mu_{\sigma,\xi}(\diff X)$ with respect to the Lebesgue measure $\diff X$:
\phantomsection\label{Radon_Nik_ps}
\begin{equation}\label{Radon_Nik}
\mathcal{W}_\sigma(X;\xi)=\frac{\delta\mu_{\sigma,\xi}(\diff X)}{\delta X}\,.
\end{equation}}
\end{definition}}}
Notice that \hyperref[Radon_Nik_ps]{(\ref*{Radon_Nik})} is another way of rewriting \hyperref[absolutely_tom_ps]{(\ref*{absolutely_tom})} and recall that if $\mathcal{W}_\sigma$ is continuous, then necessarily $\mathcal{W}_\sigma$ is non-negative, $\mathcal{W}_\sigma\geq 0$.

From~\hyperref[character_folium_ps]{(\ref*{character_folium})} and~\hyperref[spectral_measure_ps]{(\ref*{spectral_measure})}, we get immediately:
\phantomsection\label{smeared_character_Fourier_ps}\begin{equation}\label{smeared_character_Fourier}
\chi_\sigma\big(\hspace{-0.04cm}\exp(t\xi)\big)=\hspace{-.25cm}\int\limits_{\hspace{-.04cm}-\infty}^{\hspace{.45cm}\infty}\hspace{-0.15cm}\e^{itX}\mu_{\sigma,\xi}(\diff X),
\end{equation}
i.e., $\chi_\sigma\big(\exp(t\xi)\big)$ is the Inverse Fourier Transform of the measure $\mu_{\sigma,\xi}(\diff X)$, hence if the measure had only continuous part, we would have that
\phantomsection\label{Fourier_Transform_Tom_chap_2_ps}\begin{equation}\label{Fourier_Transform_Tom_chap_2}
\mathcal{W}_\sigma(X;\xi)=\frac{1}{2\pi}\hspace{-.25cm}\int\limits_{\hspace{-.04cm}-\infty}^{\hspace{.45cm}\infty}\hspace{-0.15cm}\e^{-itX}\chi_\sigma\big(\hspace{-0.04cm}\exp(t\xi)\big)\diff t.
\end{equation}

{\MYBROWN{\begin{proposition}\label{tom_conds_prop}\color{black}{
Under the conditions stated above, the quantum tomogram $\mathcal{W}_\sigma$ is non-negative and:
\begin{enumerate}

{\MYBROWN{\item \color{black}{\hspace{4.1cm}$\displaystyle{\hspace{-.25cm}\int\limits_{\hspace{-.04cm}-\infty}^{\hspace{.45cm}\infty}\hspace{-0.15cm}\mathcal{W}_\sigma(X;\xi)\diff X=1}$.}}}

{\MYBROWN{\item \color{black}{\hspace{2.8cm}$\displaystyle{\mathcal{W}_\sigma(sX;s\xi)=\frac{1}{s}\mathcal{W}_\sigma(X;\xi)},\quad s>0$.}}}

\end{enumerate}}
\end{proposition}}}

We will obtain now a representation of the quantum tomogram $\mathcal{W}_\sigma$, or more properly a representation of the measure $\mu_{\sigma,\xi}(\diff X)$, in a form that it will put the notion of quantum tomogram introduced in \hyperref[Fourier_Transform_Tom_chap_2_ps]{(\ref*{Fourier_Transform_Tom_chap_2})} in perfect parallelism with the Radon Transform discussed in \hyperref[section_Radon_Transform]{{\color{black}\textbf{chapter~\ref*{chap_birth}}}}. This will justify that such expression could be called the \textit{Quantum Radon Transform} of a given state.

{\MYBROWN{\begin{theorem}\label{tom_cart}\color{black}{
Given a state $\rho$ in a unital $C^*$--algebra $\mathcal{A}$, then the quantum tomogram $\mathcal{W}_{\sigma}(X;\xi)$ of any state $\sigma$ in the folium of $\rho$ associated to the unitary representation $U$ of the Lie group $G$ on $\mathcal{A}$ is given by
\begin{equation*}
\mathcal{W}_{\sigma}(X;\xi)=\Tr\!\big(\boldsymbol{\sigma}\delta(X\mathds{1}-\langle\boldsymbol{\Theta},\xi\rangle)\big),\qquad\forall\xi\in\mathfrak{g},\, X\in\mathbb{R}.
\end{equation*}}
\end{theorem}}}

\phantomsection\label{delta_to_section_1_ps}
Inside the trace, the delta function of an operator on $\mathcal{H}_\rho$ appears. We have already introduced the concept of delta function of an operator in \hyperref[chap_birth]{{\color{black}\textbf{chapter~\ref*{chap_birth}}}} in~\hyperref[delta_function_operator_ps]{(\ref*{delta_function_operator})}, however is convenient to consider it again and comment a few aspects of it. The delta function of a bounded operator $\textbf{T}$ on $\mathcal{H}_\rho$ is defined as the operator-valued distribution given by:
\begin{equation}
\delta(\textbf{T})=\frac{1}{2\pi}\hspace{-.2cm}\int\limits_{\hspace{-.04cm}-\infty}^{\hspace{.45cm}\infty} \hspace{-.1cm}\e^{ik\textbf{T}}\diff k,
\end{equation}
and for any test function $\phi(\lambda)$ in the Schwartz space $\mathscr{S}(\mathbb{R})$, it follows:
$$
\langle\delta(\textbf{T}),\phi\rangle=\frac{1}{2\pi}\hspace{-.05cm}\int\limits_{\hspace{-.26cm}-\infty}^{\hspace{.15cm}\infty}\hspace{-.1cm}\int\limits_{\hspace{-.08cm}-\infty}^{\hspace{.15cm}\infty}\e^{ik\lambda}\phi(\lambda)E_{\textbf{T}}(\diff\lambda)\diff k,
$$
where $E_{\textbf{T}}$ is the spectral measure defined by $\textbf{T}$. Notice that the previous integral is well-defined and notice also that if $\textbf{T}$ is self-adjoint and if $\phi$ is real, then the operator $\langle\delta(\textbf{T}),\phi\rangle$ is self-adjoint too.

Thus, in our case we have that
\begin{equation}
\langle\delta(X\mathds{1}-\langle\boldsymbol{\Theta},\xi\rangle),\phi\rangle=\frac{1}{2\pi}\hspace{-.05cm}\int\limits_{\hspace{-.26cm}-\infty}^{\hspace{.15cm}\infty}\hspace{-.1cm}\int\limits_{\hspace{-.08cm}-\infty}^{\hspace{.15cm}\infty}\e^{ik(X\mathds{1}-\langle\boldsymbol{\Theta},\xi\rangle)}\phi(\lambda)E_{\textbf{T}}(\diff\lambda)\diff k.
\end{equation}

In \hyperref[tom_conds_prop]{\MYBROWN{\textbf{Prop.\,\ref*{tom_conds_prop}}}}, we wrote the normalization and homogeneity conditions that the quantum tomogram $\mathcal{W}_\sigma(X;\xi)$ satisfies and before, in \hyperref[absolutely_tom_ps]{(\ref*{absolutely_tom})}, we wrote the non-negativity condition. But we know that a tomogram must satisfy that it is a probability distribution, therefore let us see that $\mathcal{W}_\sigma(X;\xi)$ is also real:
\begin{align}
\overline{\mathcal{W}_\sigma(X;\xi)}&=\overline{\Tr\!\big(\boldsymbol{\sigma}\delta(X\mathds{1}-\langle\boldsymbol{\Theta},\xi\rangle)\big)}=\Tr\!\big(\boldsymbol{\sigma}\delta(X\mathds{1}^\dagger-\langle\boldsymbol{\Theta},\xi\rangle^\dagger)\big)\nonumber\\
&=\Tr\!\big(\boldsymbol{\sigma}\delta(X\mathds{1}-\langle\boldsymbol{\Theta},\xi\rangle)\big)=\mathcal{W}_\sigma(X;\xi).
\end{align}

\section{Reconstruction of states sampled with compact Lie groups}\label{section_recosntruction_states_groups}

In \hyperref[section_tomograms_group]{{\color{black}\textbf{previous section}}}, we have discussed how to define tomograms using a group representation. In this one, we will discuss how to recover the state $\sigma$ in the folium of a state $\rho$ from its tomograms. 

Recall that because~\hyperref[smeared_character_Fourier_ps]{(\ref*{smeared_character_Fourier})}, we can obtain the smeared character $\chi_\sigma$ as the Inverse Fourier Transform of the tomogram $\mathcal{W}_\sigma$:
\phantomsection\label{smeared_character_Fourier_new_ps}\begin{equation}\label{smeared_character_Fourier_new}
\Tr\!\big(\boldsymbol{\sigma}U_\rho\big(\hspace{-0.04cm}\exp(t\xi)\big)\big)=\chi_\sigma\big(\hspace{-0.04cm}\exp(t\xi)\big)=\hspace{-.25cm}\int\limits_{\hspace{-.04cm}-\infty}^{\hspace{.45cm}\infty}\hspace{-0.15cm}\e^{itX}\mathcal{W}_\sigma(X;\xi)\diff X,
\end{equation}
then, what we need is to recover the state $\sigma$ from the smeared characters $\chi_\sigma\big(\hspace{-0.04cm}\exp(t\xi)\big)$.

In \hyperref[section_Sampling_C]{{\color{black}\textbf{section~\ref*{section_Sampling_C}}}}, we have explained that we need a tomographic set $U(x)$ and a dual tomographic set $D(x)$ to reconstruct a state of a system. To get the dual tomographic map $D$, we need a notion of orthogonality in our theory. 

In group theory, there exist natural orthogonality relations associated to representations of finite groups or compact Lie groups. We will concentrate in these two cases because they appear commonly in many quantum systems. Furthermore, there are other situations in which we can find orthogonality relations that allow us to reconstruct the desired states, for instance, the cases of Heisenberg--Weyl and Poincar\'e groups that will be considered later on.

\phantomsection\label{resumen_rep}
Let $(\mathcal{H},U)$ be a unitary representation, $U:G\rightarrow\mathcal{U}(\mathcal{H})$. The representation is \textit{irreducible} if there are no proper invariant subspaces of $\mathcal{H}$ under the action of the representation on the Hilbert space $\mathcal{H}$. Let us suppose that the Hilbert space $\mathcal{H}$ is finite dimensional with $n=\textrm{dim}(\mathcal{H})$ and let $e_i$, $i=1,\ldots,n$, be a given orthonormal basis on such space. We will denote by $U_{ij}(g)$ the elements of the unitary matrix associated to $U(g)$, $g\in G$, in the previous basis, i.e.,
\begin{equation}
U(g)e_i=\sum_{i=1}^nU_{ji}(g)e_j.
\end{equation}
In what follows, $U(g)$ will refer to its associated matrix too assuming that an orthonormal basis has been fixed.

Suppose that $G$ is finite, then Schur's orthogonality theorem (see for instance \cite{Jo97}) asserts that given two unitary irreducible representations $U^{(a)}$ and $U^{(b)}$ of dimensions $n_a$ and $n_b$ respectively of a finite group $G$ of order $|G|$, then:
\phantomsection\label{schur_ortho_ps}
\begin{equation}\label{schur_ortho}
\frac{n_a}{|G|}\sum_{g\in G} U_{ij}^{(a)}(g)^\dagger U_{rs}^{(b)}(g)=\delta_{ab}\delta_{ir}\delta_{js}.
\end{equation}
Therefore, if we choose as the dual tomographic map the Hermitean conjugate of $U(g)$, $D(g)=U(g)^\dagger$, we will get the biorthogonality condition~\hyperref[biortho_ps]{(\ref*{biortho})}:
\phantomsection\label{biortho_finite_groups_ps}\begin{equation}\label{biortho_finite_groups}
\frac{n}{|G|}\sum_{g\in G} D_{ij}(g) U_{rs}(g)=\delta_{ir}\delta_{js}.
\end{equation}
Hence, if $\boldsymbol{\rho}$ is the corresponding density operator related to the state $\rho$ and $U$ is irreducible, the reconstruction formula~\hyperref[reconstruction_state_D_ps]{(\ref*{reconstruction_state_D})} becomes:
\phantomsection\label{rec_states_schur_ps}\begin{equation}\label{rec_states_schur}
\boldsymbol{\rho}=\frac{n}{|G|}\sum_{g\in G}\chi_\rho(g)U(g)^\dagger.
\end{equation}

In the case of finite groups, the tomogram of the state $\rho$ can be obtained by using the discrete version of formula~\hyperref[Fourier_Transform_Tom_chap_2_ps]{(\ref*{Fourier_Transform_Tom_chap_2})}, \cite{Ib11}. Let us transform $U(g)$ in a diagonal matrix $d_g$ by means of a unitary matrix $V_g$:
\begin{equation}
U(g)=V_gd_gV_g^\dagger,\qquad d_g=\textrm{diag}\left[\e^{i\theta_1(g)},\ldots,\e^{i\theta_n(g)}\right],
\end{equation}
then, let us compute the smeared character of $U(g)$:
\begin{equation}
\chi_\rho(g)=\Tr\!\left(\boldsymbol{\rho} V_gd_gV_g^\dagger\right)=\Tr\!\left(d_gV_g^\dagger\boldsymbol{\rho} V_g\right)=\sum_{m=1}^n\e^{i\theta_m(g)}\left(V_g^\dagger\boldsymbol{\rho} V_g\right)_{mm}.
\end{equation}
Therefore, the tomograms of the state are given by: 
\begin{equation}
\mathcal{W}_\rho(m;g)\coloneq\left(V_g^\dagger\boldsymbol{\rho} V_g\right)_{mm}.
\end{equation}
These tomograms are, by definition, a stochastic vector, i.e.,
\begin{equation}
\sum_{m=1}^n\mathcal{W}_\rho(m;g)=\sum_{m=1}^n\left(V_g^\dagger\boldsymbol{\rho} V_g\right)_{mm}=\Tr(\boldsymbol{\rho})=1
\end{equation}
and
\begin{equation}
0\leq\mathcal{W}_\rho(m;g)\leq 1,\qquad m=1,\ldots,n,\qquad\forall g\in G.
\end{equation}
The proof of the last statement is a direct consequence of Schur's inequalities (see for instance \cite{Bh97}).

Therefore, we see that the smeared characters of the state $\rho$ can be obtained as a Discrete Fourier Transform of the tomograms defined before:
\begin{equation}
\chi_\rho(g)=\sum_{m=1}^n\e^{i\theta_m(g)}\mathcal{W}_\rho(m;g).
\end{equation}

If the group $G$ is now a compact Lie group, we have the same orthogonality relation~\hyperref[biortho_finite_groups_ps]{(\ref*{biortho_finite_groups})} by making the obvious substitutions:
\begin{equation}
\frac{1}{|G|}\sum_{g\in G}\rightsquigarrow\hspace{-0.1cm}\int\limits_{\hspace{.3cm}G}\hspace{-0.1cm}\diff\mu(g),
\end{equation}
with $\mu(g)$ the normalized Haar measure on the group:
\begin{equation}
\int\limits_{\hspace{.3cm}G}\hspace{-0.1cm}\diff\mu(g)=1.
\end{equation}
Then, if $U$ is an irreducible representation of dimension $n$, the state is reconstructed with the formula:
\phantomsection\label{rec_states_compact_schur_ps}\begin{equation}\label{rec_states_compact_schur}
\boldsymbol{\rho}=n\hspace{-0.15cm}\int\limits_{\hspace{.3cm}G}\hspace{-0.1cm}\chi_\rho(g)U(g)^\dagger\diff\mu(g),
\end{equation}
where $n=\textrm{dim}\mathcal{H}$ is the dimension of the irreducible representation and $D(g)=nU(g)^\dagger$ is the dual tomographic set. The tomograms are defined with the formula~\hyperref[Fourier_Transform_Tom_chap_2_ps]{(\ref*{Fourier_Transform_Tom_chap_2})}.

Let us consider now a subgroup $H\subset G$ of a finite or compact Lie group $G$. The restriction of the representation $U$ to the subgroup $H$, sometimes denoted by $U\!\downarrow\! H$ and called the subduced representation of $U$ to $H$, will be, in general, reducible even if $U$ is irreducible.

Let us suppose that the state $\rho$ satisfies the following orthogonality relations:
\phantomsection\label{cond_adapted_states_ps}\begin{equation}\label{cond_adapted_states}
\Tr\!\big(\boldsymbol{\rho} U(g)\big)=0,\qquad g\in G\setminus H,
\end{equation}
that is, the inner products with the unitary operators corresponding to the elements of $G$ not in the subgroup $H$ vanish. Therefore, in this case, we have similar formulas to~\hyperref[rec_states_schur_ps]{(\ref*{rec_states_schur})} and~\hyperref[rec_states_compact_schur_ps]{(\ref*{rec_states_compact_schur})} even if the representation $U\!\downarrow\! H$ is reducible:
\phantomsection
\label{adapted_state_a_ps}
\begin{equation}
\boldsymbol{\rho}=\frac{n}{|G|}\sum_{g\in H}\chi_\rho(g)U(g)^\dagger,\tag{2.7.16a}\label{adapted_state_a}
\end{equation}
in the finite case and
\begin{equation}
\boldsymbol{\rho}=n\hspace{-0.15cm}\int\limits_{\hspace{.3cm}H}\hspace{-0.1cm}\chi_\rho(g)U(g)^\dagger\diff\mu(g),\tag{2.7.16b}\label{adapted_state_b}
\end{equation}
in the compact situation.

Such states will be said to be \textit{adapted states} to the subgroup $H$ and they will constitute the main tool of the numerical algorithm to decompose reducible representations that will be presented in the \hyperref[chap_smily]{{\color{black}\textbf{following chapter}}}. Let us summarize this results by writing the following theorem.

{\MYBROWN{\begin{theorem}\label{reconstruction_thm}\color{black}{
Let $G$ be a compact Lie group and $(\mathcal{A},U)$ a unitary representation of $G$ on a $C^*$--algebra $\mathcal{A}$. Given an adapted state $\sigma$ in the folium of the state $\rho$, the density operator $\boldsymbol{\sigma}$ in $\mathcal{B}(\mathcal{H}_\rho)$ can be obtained by means of:
$$
\boldsymbol{\sigma}=n\hspace{-0.15cm}\int\limits_{\hspace{.3cm}G}\hspace{-0.1cm}\chi_\sigma(g)U_\rho(g)^\dagger\diff\mu(g),
$$
where $U_\rho(g)=\pi_\rho\big(U(g)\big)$, $\mathcal{H}_\rho$ and $\pi_\rho$ are the unitary representation, the Hilbert space and representation of $\mathcal{A}$ obtained with the \hyperref[ps_GNS]{{\color{black}\textbf{GNS construc-}}} \hyperref[ps_GNS]{{\color{black}\textbf{tion}}}, and $n=\dim\mathcal{H}_\rho$.}
\end{theorem}}}

\noindent\MYBROWN{\textbf{Proof}}: The proof follows immediately from the arguments stated before to obtain the formula \hyperref[rec_states_compact_schur_ps]{(\ref*{rec_states_compact_schur})} and from the definition of adapted states \hyperref[cond_adapted_states_ps]{(\ref*{cond_adapted_states})} that allows to use the formula \hyperref[rec_states_compact_schur_ps]{(\ref*{rec_states_compact_schur})} even if the unitary representation $U$ is not irreducible.

\vspace{-.0cm}\hfill{\hyperref[biortho_finite_groups_ps]{{\color{black}$\blacksquare$}}}

Another case in which an orthogonality relation can be defined is when we consider the regular representation of a group. The regular representation of a group $G$ is the unitary representation obtained from the action of the group $G$ on the Hilbert space of square integrable functions on the group, $\mathcal{H}=L^{2}(G,\mu)$, where $\mu$ denotes the left(right)-invariant Haar measure by left (right) translations.

Thus, the left regular representation $U^{reg}_L(h)$ is defined as follows:
\setcounter{equation}{16}
\begin{equation}
\big(U^{reg}_L(h)\psi\big)(g)=\psi(h^{-1}g),\qquad \psi\in L^{2}(G,\mu).
\end{equation}
The right regular representation is defined analogously.

If $G$ is finite, it is clear that $L^{2}(G)$ is isometrically isomorphic with the group algebra $\mathbb{C}[G]$:
\begin{equation}
\mathcal{H}\cong\mathbb{C}[G]=\Big\{|\alpha\rangle=\hspace{-0.05cm}\sum_{g\in G}\alpha_g|g\rangle\,\big|\,\alpha_g\in\mathbb{C}\Big\},
\end{equation}
with inner product $\displaystyle{\langle\alpha,\beta\rangle=\sum_{g\in G}\overline{\alpha}_g\beta_g}$. The action of the group is given by:
\begin{equation}
U_L^{reg}(h)|\alpha\rangle=\hspace{-0.05cm}\sum_{g\in G}\alpha_{h^{-1}g}|g\rangle=\hspace{-0.05cm}\sum_{g'\in G}\alpha_{g'}|hg'\rangle,
\end{equation}
then, we can interpret the left regular representation $U_L^{reg}$ as:
\phantomsection\label{left_regular_rep_ps}\begin{equation}\label{left_regular_rep}
U_L^{reg}(h)|g\rangle=|hg\rangle,\qquad\forall g,h\in G.
\end{equation}

From the orthogonality relation satisfied by the regular representation:
\begin{equation}
\Tr\!\left(U_L^{reg}(g)^\dagger U_L^{reg}(g')\right)=n\delta_{g^{-1}g'},
\end{equation}
the character of the representation is easily computed:
\phantomsection\label{regular_character_ps}\begin{equation}\label{regular_character}
\chi^{reg}(g)=n\delta_g=\left\{\begin{matrix}
n& g=e,\\
0&\mbox{otherwise\,,}
\end{matrix}\right.
\end{equation}
with $n=\dim\mathcal{H}$. In that case the reconstruction formula of $\boldsymbol{\rho}$ is \hyperref[adapted_state_a_ps]{(\ref*{adapted_state_a})}. 

For compact groups, we have similar results, however the character $\chi^{reg}$ is now a Dirac's delta distribution:
\begin{equation}
\chi^{reg}(g)=\delta(g),\qquad g\in G,
\end{equation}
and the theorem of Harish--Chandra \cite{Ar88} allows to extend the result in eq.\,\hyperref[regular_character_ps]{(\ref*{regular_character})} to semisimple Lie groups. In that case, the reconstruction formula is \hyperref[adapted_state_a_ps]{(\ref*{adapted_state_b})} with $n=1$.\newpage

\subsection{Spin Tomography}\label{section_Spin_Tomography}

\begin{figure}[h]
\centering
\includegraphics{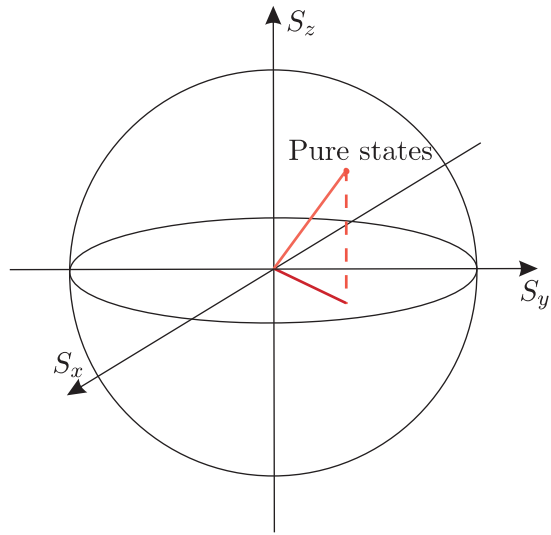}
\caption{\hfil\MYGREEN{Figure 2.7.1}: Bloch's sphere representing states of a particle of spin $1/2$.\hfil}
\label{B_sphere}
\end{figure}

The $SU(2)$ group is the group that underlies the description of the states of a particle with spin $1/2$\,, \hyperref[Ga90_ps]{\RED{[\citen*{Ga90}, ch.\,5]}}. Is very common in Quantum Optics to work with photons or qubits that are particles with this symmetry group. The states of a particle with spin $1/2$ may be represented in the so called \textit{Bloch's sphere}, {\changeurlcolor{mygreen}\hyperref[B_sphere]{Figure \ref*{B_sphere}}}.

These states can be parametrized as:
\begin{equation}
\boldsymbol{\rho}=\frac{1}{2}\begin{pmatrix}
1+r\cos\theta & \sin\theta\e^{-i\phi}\\
\sin\theta\e^{i\phi} & 1-r\cos\theta
\end{pmatrix},
\end{equation}
with
$$
0\leq r\leq 1 ,\qquad 0\leq\theta\leq\pi,\qquad 0\leq\phi\leq 2\pi.
$$
The states in the surface, i.e., the states with $r=1$, are the pure states of the system.

The $SU(2)$ group is a compact Lie group, hence we can reconstruct a given state $\boldsymbol{\rho}$ by means of~\hyperref[rec_states_compact_schur_ps]{(\ref*{rec_states_compact_schur})}. The irreducible representation of this group for spin one-half can be written in terms of the exponential of the elements of the Lie algebra $\mathfrak{su}(2)$:
\begin{equation}
U(s_x,s_y,s_z)=\e^{i(s_x\textbf{S}_x+s_y\textbf{S}_y+s_z\textbf{S}_z)},
\end{equation}
where the operators corresponding to the spin in the axis $x,y,z$ are:
\phantomsection\label{spin_operators_tom_ps}\begin{equation}\label{spin_operators_tom}
\textbf{S}_i=\frac{\hbar}{2}\sigma_i,\qquad i=x,y,z,
\end{equation}
and where the $\sigma_i$ are the Pauli matrices:
\phantomsection\label{pauli_matrices_ps}\begin{equation}\label{pauli_matrices}
\sigma_x=\begin{pmatrix}
0 & 1\\
1 & 0
\end{pmatrix},\quad\sigma_y=\begin{pmatrix}
0 & -i\\
i & 0
\end{pmatrix},\quad\sigma_z=\begin{pmatrix}
1 & 0\\
0 & -1
\end{pmatrix}.
\end{equation}

We will define the tomograms of the state $\rho$ of this system as:
\begin{equation}
\mathcal{W}_\rho(X,s_x,s_y,s_z)=\Tr\!\Big(\boldsymbol{\rho}\delta\big(X\mathds{1}-\frac{s_x}{2}\sigma_x-\frac{s_y}{2}\sigma_y-\frac{s_z}{2}\sigma_z\big)\Big).
\end{equation}
Instead of using the exponential representation of the delta function, we will use the result obtained in~\hyperref[delta_function_operator_mean_value_ps]{(\ref*{delta_function_operator_mean_value})}, i.e., we will use the interpretation of the delta function of an operator as a projector over the eigenstates corresponding to the eigenvalues equal to $X$:
\begin{equation*}
\delta(X\mathds{1}-\textbf{A})=|X\rangle\langle X|,\qquad \textbf{A}|X\rangle=X|X\rangle,
\end{equation*} 
to write the tomogram as
\begin{equation}
\mathcal{W}_\rho(X,s_x,s_y,s_z)=\langle X|\boldsymbol{\rho}| X\rangle,
\end{equation}
where $\displaystyle{\textbf{A}=\frac{s_x}{2}\sigma_x+\frac{s_y}{2}\sigma_y+\frac{s_z}{2}\sigma_z}$.

The eigenvalues of this operator are $\lambda=\pm|\boldsymbol{s}|/2$, and its corresponding eigenvectors are
\begin{equation*}
|v_{+}\rangle=\frac{1}{\sqrt{2|\boldsymbol{s}|(|\boldsymbol{s}|-s_z)}}
.
\end{equation*}
Thus, the tomogram can be written as:
\begin{equation*}
\mathcal{W}_\rho(X,s_x,s_y,s_z)=\delta(X-%
)\langle v_{-}|\boldsymbol{\rho}|v_{-}\rangle,
\end{equation*}
and therefore, we finally get:
\begin{align}
\mathcal{W}_\rho(X,s_x,s_y,s_z)=&\nonumber\\
&\hspace{-0.9cm}\frac{1}{2}\left(\Big(1+\frac{s_z}{|\boldsymbol{s}|}r\cos\theta\Big)\right.\nonumber\\
&\hspace{-0.2cm}\left.+\sin\theta\Big(\frac{s_x}{|\boldsymbol{s}|}\cos\phi+\frac{s_y}{|\boldsymbol{s}|}\sin\phi\Big)\right)\delta(X-%
).
\end{align}
\section{Tomography with the Heisenberg--Weyl group}\label{resume_heis}

To finish this discussion, let us mention a case in which the group is neither finite or compact but we know how to reconstruct their states. We do not need to go far away to find it because it was the main topic in \hyperref[section_Quantization_elec]{{\color{black}\textbf{previous chapter}}}. We are talking about the quantum harmonic oscillator. 

In that \hyperref[section_real_QM_ps]{{\color{black}\textbf{chapter}}}, we found a reconstruction formula for the matrix elements of a density operator $\boldsymbol{\rho}$, eq.\,\hyperref[Inverse_Radon_Transform_Wigner_ps]{(\ref*{Inverse_Radon_Transform_Wigner})}, so it is natural to think that we could find a reconstruction formula for the operator $\boldsymbol{\rho}$.

The position and momentum operators, which satisfy the commutation relation $[\textbf{Q},\textbf{P}]=i\mathds{1}$, are a realization of the Lie algebra of the Heisenberg--Weyl group which may be presented as the group of triples of real numbers with the composition law:
\phantomsection\label{HW_composition_law_ps}
\begin{equation}\label{HW_composition_law}
(\mu,\nu,t)\circ(\mu',\nu',t')=\big(\mu+\mu',\nu+\nu',t+t'-\frac{1}{2}(\mu\nu'-\nu\mu')\big),
\end{equation}
and an irreducible representation of it is provided by:
\begin{equation}
U(\mu,\nu,t)=\e^{i(\mu\textbf{Q}+\nu\textbf{P})}\e^{it\mathds{1}}.
\end{equation}

\phantomsection\label{HW_conclusions}
The symmetry group of the harmonic oscillator is the projective subgroup of Heisenberg--Weyl group with $t=0$, then its representation is:
\begin{equation}\label{HW_conclusions_formula}
U(\mu,\nu)\coloneq U(\mu,\nu,0)=\e^{i(\mu\textbf{Q}+\nu\textbf{P})}.
\end{equation}
If we compute the trace of the composition of two elements of the group, we obtain:
\begin{align}
\Tr\!\big(U(\mu,\nu)U(\mu',\nu')^\dagger\big)&\nonumber\\
&\hspace{-3cm}=\e
\hspace{-0.1cm}\int\limits_{\hspace{-.26cm}-\infty}^{\hspace{.15cm}\infty}\hspace{-.1cm}\int\limits_{\hspace{-.08cm}-\infty}^{\hspace{.15cm}\infty}\hspace{0.1cm}\e^{i(\mu-\mu')q}\e^{i(\nu-\nu')p}\diff q\diff p\nonumber\\
&\hspace{3cm}=2\pi\delta(\mu-\mu')\delta(\nu-\nu'),                                                                                                                                                                                                                                                                       
\end{align}
where we have used the BCH formula~\hyperref[BCH_ps]{(\ref*{BCH})} and the inner product of momentum and position eigenvectors~\hyperref[Position_Momentum_relation_ps]{(\ref*{Position_Momentum_relation})}. Finally, the reconstruction formula for the state $\rho$ reads as:
\phantomsection\label{reconstruction_rho_HW_ps}\begin{multline}\label{reconstruction_rho_HW}
\boldsymbol{\rho}=\frac{1}{2\pi}\hspace{-0.1cm}\int\limits_{\hspace{-.17cm}-\infty}^{\hspace{.15cm}\infty}\hspace{-.14cm}\int\limits_{\hspace{.04cm}-\infty}^{\hspace{.15cm}\infty}\chi_\rho(\mu,\nu)U(\mu,\nu)^\dagger\diff\mu\diff\nu\\
=\frac{1}{2\pi}\hspace{-0.3cm}\int\limits_{\hspace{.5cm}\mathbb{R}^3}\hspace{-0.17cm}\mathcal{W}_\rho(X,\mu,\nu)\e^{i(X\mathds{1}-\mu\textbf{Q}-\nu\textbf{P})}\diff X\diff\mu\diff\nu.
\end{multline}

If we compare this equation with the classical one obtained in \hyperref[chap_birth]{{\color{black}\textbf{chap-}}} \hyperref[chap_birth]{{\color{black}\textbf{ter~\ref*{chap_birth}}}} in~\hyperref[Inverse_Radon_Transform_2_ps]{(\ref*{Inverse_Radon_Transform_2})}, we see that are similar only by substituting the density probability $f(q,p)$ by the density operator $\boldsymbol{\rho}$ and the classical position and momentum $q$ and $p$ by its corresponding operators $\textbf{Q}$ and $\textbf{P}$, however instead of $(2\pi)^2$ in the denominator here we have $2\pi$. The factor $(2\pi)^2$ is due to the Fourier Transform of classical position and momentum, hence for each one we have a factor $2\pi$. But here, position and momentum are operators on a Hilbert space and they are related by means of a Fourier Transform, then that fact makes that in formula~\hyperref[reconstruction_rho_HW_ps]{(\ref*{reconstruction_rho_HW})} only one factor $2\pi$ appears.

\subsection{The holomorphic representation of an ensemble of quantum harmonic oscillators}\label{holomorphic_quantum_1_sec}\label{holom_chap_2}

The choice of the complex coordinates \hyperref[complex_holom_ps]{(\ref*{complex_holom})} in \hyperref[section_Toms_q_h_o]{{\color{black}\textbf{subsection \ref*{section_Toms_q_h_o}}}}, that will be used to facilitate the computation of the tomograms of an ensemble of quantum harmonic oscillators of a given state, is not just a convenient mathematical transformation because it provides another realization of the Fock space $\mathcal{F}_N$ that will be fundamental when dealing with tomograms of quantum fields (see \hyperref[resumen_quantization]{{\color{black}\textbf{chapter~\ref*{chap_QFT}}}}).

Let us consider first a quantum harmonic oscillator with Hamiltonian:
$$
\textbf{H}=\hbar\omega\Big(a^\dagger a+\frac{1}{2}\Big).
$$
This quantum system is nicely described in the Fock space $\mathcal{F}_1$ discussed in \hyperref[harmonic_oscillator_2_ps]{{\color{black}\textbf{section~\ref*{section_Quantization_elec}}}}.

It is easy to check that the coherent states presented in eq.\,\hyperref[coherent_state_ps]{(\ref*{coherent_state})},
\phantomsection\label{coherent_new_2_ps}
\begin{equation}\label{coherent_new_2}
|z\rangle=\e
\hspace{-0.1cm}\sum_{n=0}^\infty\frac{z^n}{n!}(a^\dagger)^n|0\rangle\in\mathcal{F}_1,
\end{equation}
are eigenvectors of the creation operator $a$ with eigenvalue $z$, i.e.:
\begin{equation}
a|z\rangle=z|z\rangle,\qquad\forall z\in\mathbb{C},
\end{equation}
and similarly $a^\dagger|\bar{z}\rangle=\bar{z}|\bar{z}\rangle$. Such states have the beautiful property that their evolution in time mimics the classical solution of the system, this is:
\begin{equation}
U_t|z_0\rangle=\e^{-it\textbf{H}}|z_0\rangle=|z(t)\rangle,
\end{equation}
with $z(t)=z_0\e^{-i\omega t}$ (this is easily checked by noticing that the evolution of $a$ is given by $U^\dagger_taU_t=a\e^{-i\omega t}$).

The expression \hyperref[coherent_new_2_ps]{(\ref*{coherent_new_2})} for $|z\rangle$ can be written in compact form as:
\begin{equation*}
|z\rangle=\e
\e^{za^\dagger}|0\rangle,
\end{equation*}
hence, as in other situations, taking advantage of the BCH formula \hyperref[BCH_ps]{(\ref*{BCH})}, it verifies the normalization of the coherent states:
\begin{equation*}
\langle z|z\rangle=\e^{-|z|^2}\langle 0|\e^{\bar{z}a}\e^{za^\dagger}|0\rangle=\langle 0|\e^{za^\dagger}\e^{\bar{z}a}|0\rangle=\langle 0|0\rangle=1.
\end{equation*}

The displacement operator $D(z)$, that is nothing but the unitary representation of the Heisenberg--Weyl group \hyperref[HW_conclusions]{(\ref*{HW_conclusions_formula})} written in terms of annihilation and creation operators:
\begin{equation}
D(z)=\e^{za^\dagger-\bar{z}a},
\end{equation}
satisfies
\begin{equation}
D(z)|0\rangle=|z\rangle
\end{equation}
and consequently, because of the composition law \hyperref[HW_composition_law_ps]{(\ref*{HW_composition_law})}, we have that
\begin{equation}
D(z)|w\rangle=|w+z\rangle\quad\mbox{and}\quad D(z)^\dagger|w\rangle=|w-z\rangle.
\end{equation}
It is important to remark that the coherent states are not orthogonal:
\phantomsection\label{prod_coherent_ps}
\begin{equation}\label{prod_coherent}
\langle z|w\rangle=\e                                                                                                                                                                                                                                                                                                  
,
\end{equation}
but they obey the following completeness relation:
\begin{align}
\frac{1}{\pi}\hspace{-0.3cm}\int\limits_{\hspace{.5cm}\mathbb{R}^{2}}\hspace{-.19cm}|z\rangle\langle z|\diff^2 z&=\frac{1}{\pi}\hspace{-0.3cm}\int\limits_{\hspace{.5cm}\mathbb{R}^{2}}\hspace{-.19cm}\e^{-|z|^2}\hspace{-0.2cm}\sum_{n,m=0}^\infty\hspace{-0.1cm}|n\rangle\langle m|\frac{z^n\bar{z}^m}{\sqrt{n!m!}}\diff^2 z\nonumber\\
&=\frac{1}{\pi}\hspace{-.15cm}\int\limits_{\hspace{.15cm}0}^{\hspace{.35cm}\infty}\hspace{-.33cm}\int\limits_{\hspace{.15cm}0}^{\hspace{.35cm}2\pi}\hspace{-.06cm}\e^{-r^2}\hspace{-0.2cm}\sum_{n,m=0}^\infty\hspace{-0.1cm}|n\rangle\langle m|\frac{r^{n+m}}{\sqrt{}n!m!}\e^{i(n-m)\theta}r\diff\theta\diff r\nonumber\\
&=2\sum_{n=0}^\infty|n\rangle\langle n|\hspace{-.15cm}\int\limits_{\hspace{.15cm}0}^{\hspace{.35cm}\infty}\e^{-r^2}\frac{r^{2n+1}}{n!}\diff r=\sum_{n=0}^\infty|n\rangle\langle n|=\mathds{1}.
\end{align}

Let us remark also, that the projectors $\textbf{E}_z=|z\rangle\langle z|$, providing a resolution of the identity, are not selft-adjoint $\textbf{E}_z^\dagger=\textbf{E}_{\bar{z}}\neq \textbf{E}_z$ neither orthogonal to each other: 
$$
\textbf{E}_z\textbf{E}_w=|z\rangle\langle z|w\rangle\langle w|=\e                                                                                                                                                                                                                                                                                                  
|z\rangle\langle w|\neq 0\quad\mbox{if}\quad z\neq w.
$$

\subsection{The Bargmann--Segal Hilbert space of entire functions}\label{B_S_chap_2}

Consider an entire function $\psi(z)$ with proper series expansion
\begin{equation}
\psi(z)=\sum_{n=0}^\infty c_nz^n.
\end{equation}
To such function, we may associate the vector $|\psi\rangle\in\mathcal{F}_1$ given by:
\phantomsection\label{psi_entire_ps}
\begin{equation}\label{psi_entire}
|\psi\rangle=\sum_{n=0}^\infty c_n\sqrt{n!}|n\rangle=\sum_{n=0}^\infty c_n(a^\dagger)^n|0\rangle,
\end{equation}
and then, we get that
\begin{multline}
\langle \bar{z}|\psi\rangle=\hspace{-0.2cm}\sum_{n,m=0}^\infty c_n\sqrt{n!}\frac{1}{\sqrt{m!}}z^m\langle m|n\rangle\e
\psi(z).
\end{multline}
Thus, if we consider the space of entire functions $\psi(z)$ such that
$$
\int\limits_{\hspace{.5cm}\mathbb{R}^{2}}\hspace{-.19cm}\e^{-|z|^2}|\psi(z)|^2\diff^2 z<\infty,
$$
we have that the map which assigns to each function $\psi(z)$ the expression $|\psi\rangle$ in \hyperref[psi_entire_ps]{(\ref*{psi_entire})} is well-defined on $\mathcal{F}_1$ because:
$$
\langle\psi|\psi\rangle=\frac{1}{\pi}\hspace{-0.3cm}\int\limits_{\hspace{.5cm}\mathbb{R}^{2}}\hspace{-.19cm}\langle\psi|z\rangle\langle z|\psi\rangle\diff^2 z=\frac{1}{\pi}\hspace{-0.3cm}\int\limits_{\hspace{.5cm}\mathbb{R}^{2}}\hspace{-.19cm}\e^{-|z|^2}|\psi(z)|^2\diff^2 z<\infty.
$$

Hence, we define the Bargmann--Segal space $\mathcal{F}_{\textrm{BS}}$ as the Hilbert space of entire functions $\psi$ such that
\begin{equation}
\|\psi\|_{\textrm{BS}}^2=\frac{1}{\pi}\hspace{-0.3cm}\int\limits_{\hspace{.5cm}\mathbb{R}^{2}}\hspace{-.19cm}\e^{-|z|^2}|\psi(z)|^2\diff^2 z<\infty,
\end{equation}  
with inner product:
\begin{equation}
\langle\psi,\phi\rangle_{\textrm{BS}}=\frac{1}{\pi}\hspace{-0.3cm}\int\limits_{\hspace{.5cm}\mathbb{R}^{2}}\hspace{-.19cm}\e^{-|z|^2}\overline{\psi(z)}\phi(z)\diff^2 z.
\end{equation}

Clearly, the space $\mathcal{F}_{\textrm{BS}}$ is unitarily equivalent to the Fock space $\mathcal{F}_1$ by the following map:
\begin{alignat*}{2}
T\hspace{-0.05cm}:&\hspace{0.1cm}\mathcal{F}_{\textrm{BS}}\hspace{0.1cm}\longrightarrow\hspace{0.2cm}\mathcal{F}_1\\
&\hspace{0.3cm}\psi\hspace{0.32cm}\rightsquigarrow\hspace{0.1cm}|\psi\rangle=T\psi.
\end{alignat*}
Moreover, the creation and annihilation operators become:
\begin{equation}
a_{\textrm{BS}}\psi=T^\dagger aT\psi=\frac{\partial}{\partial z}\psi,\qquad a_{\textrm{BS}}^\dagger\psi=T^\dagger a^\dagger T\psi=z\psi,
\end{equation}
then, the ground state of the theory is just:
\begin{equation}
\psi_0=T^\dagger|0\rangle=1,\qquad(\|\psi_0\|_{\textrm{BS}}^2=1),
\end{equation}
and
\begin{equation}
\textbf{N}_{\textrm{BS}}=a_{\textrm{BS}}^\dagger a_{\textrm{BS}}=z\frac{\partial}{\partial z},
\end{equation}
which is the (complex) \textit{Euler operator}.\newpage

\subsection{Tomograms of an ensemble of quantum harmonic oscillators}\label{section_Toms_q_h_o}

We will show the explicit form of the tomograms of a pure state of an ensemble of quantum harmonic oscillators using the holomorphic representation described before in \hyperref[holom_chap_2]{{\color{black}\textbf{subsection \ref*{holom_chap_2}}}}. Let us recall the Hamiltonian of an ensemble of harmonic oscillators given in~\hyperref[harmonic_oscillator_multipartite_ps]{(\ref*{harmonic_oscillator_multipartite})}:
\begin{equation*}
\textbf{H}=\sum_{k=1}^n\omega_ka^\dagger_ka_k+\frac{1}{2}\sum_{k=1}^n\omega_k,
\end{equation*}
where we have put $\hbar=1$ to simplify the notation of the following results. And also let us recall the canonical commutation relations of the annihilation and creation operators given in \hyperref[Commutator_creation_annihilation_f_ps]{(\ref*{Commutator_creation_annihilation_f})}:
\begin{equation*}
\left[\vphantom{a^\dagger}\right.\hspace{-0.1cm}a_k,a_{k'}^\dagger\hspace{-0.1cm}\left.\vphantom{a^\dagger}\right]=\delta_{kk'},\quad\left[\vphantom{a^\dagger}\right.\hspace{-0.1cm}a_k,a_{k'}\hspace{-0.1cm}\left.\vphantom{a^\dagger}\right]=\left[\vphantom{a^\dagger}\right.\hspace{-0.1cm}a_k^\dagger,a_{k'}^\dagger\hspace{-0.1cm}\left.\vphantom{a^\dagger}\right]=0.
\end{equation*}
Let $\boldsymbol{\rho}$ be the pure state corresponding to the system in which each particle has momentum $k_i$, $i=1,\ldots,n$:
\begin{equation}
\boldsymbol{\rho}=|1_{k_1},\ldots,1_{k_n}\rangle\langle1_{ k_1},\ldots,1_{k_n}|.
\end{equation}
Recall that the annihilation and creation operators act on the ground state $|0,\ldots,0\rangle$ this way:
\begin{equation}
a_j^\dagger|0,\ldots,0\rangle=|0,\ldots,\overset{j}{1},\ldots,0\rangle,\qquad a_j|0,\ldots,0\rangle=0.
\end{equation}
The \textit{center of mass} tomogram has the following form:
\begin{equation}
\mathcal{W}_{cm}(X,\boldsymbol{\mu},\boldsymbol{\nu})=\Tr\!\big(\boldsymbol{\rho}\delta(X-\boldsymbol{\mu}\cdot\textbf{Q}-\boldsymbol{\nu}\cdot\textbf{P})\big),
\end{equation}
and introducing the holomorphic variables
\phantomsection\label{complex_holom_ps}
\begin{equation}\label{complex_holom}
w_j=\frac{\mu_j+i\nu_j}{\sqrt{2}}\,,
\end{equation}
we have:
\phantomsection
\label{ps_tom_q_h_o}
\begin{align}
&\mathcal{W}_{cm}(X,w,\overline{w})=\Tr\!\big(\boldsymbol{\rho}\delta(X-\overline{\boldsymbol{w}}\cdot\boldsymbol{a}-\boldsymbol{w}\cdot\boldsymbol{a}^\dagger)\big)\nonumber\\
&\hspace{1.2cm}=\vphantom{\sum^n}\langle 0|a_1\cdots a_n|\delta(X-\overline{\boldsymbol{w}}\cdot\boldsymbol{a}-\boldsymbol{w}\cdot\boldsymbol{a}^\dagger)|a_n^\dagger\cdots a_1^\dagger|0\rangle\nonumber\\
&\hspace{1.2cm}=\frac{1}{2\pi}\hspace{-.15cm}\int\limits_{\hspace{-.04cm}-\infty}^{\hspace{.45cm}\infty}\hspace{-0.15cm}\e^{ikX}\e
\langle 0|a_1\cdots a_n|\e^{-ik(\overline{\boldsymbol{w}}\cdot\boldsymbol{a}+\boldsymbol{w}\cdot\boldsymbol{a}^\dagger)}|a_n^\dagger\cdots a_1^\dagger|0\rangle\diff k.
\end{align}

From the canonical commutation relations \hyperref[Commutator_creation_annihilation_f_ps]{(\ref*{Commutator_creation_annihilation_f})} and the property of the Lie bracket:
\begin{equation}
\left[\vphantom{a^\dagger}\right.\hspace{-0.1cm}\textbf{A}\textbf{B},\textbf{C}\hspace{-0.1cm}\left.\vphantom{a^\dagger}\right]=\textbf{A}\left[\vphantom{a^\dagger}\right.\hspace{-0.1cm}\textbf{B},\textbf{C}\hspace{-0.1cm}\left.\vphantom{a^\dagger}\right]+\left[\vphantom{a^\dagger}\right.\hspace{-0.1cm}\textbf{A},\textbf{C}\hspace{-0.1cm}\left.\vphantom{a^\dagger}\right]\textbf{B},
\end{equation}
for any operators $\textbf{A}$, $\textbf{B}$ and $\textbf{C}$ on $\mathcal{H}$, by recurrence we have that
\phantomsection\label{commutation_a_n_a_dagger_ps}
\begin{equation}\label{commutation_a_n_a_dagger}
\left[\vphantom{a^\dagger}\right.\hspace{-0.1cm}a^n_k,a^\dagger_j\hspace{-0.1cm}\left.\vphantom{a^\dagger}\right]=na^{n-1}_k\delta_{kj},
\end{equation}
therefore, we get the following:
\begin{align}
\langle 0|a_k\e^{-ikw_ia^\dagger_j}&=\langle 0|(a_k-ikw_i\delta_{kj}),\nonumber\\
\e^{-ik\overline{w}_ia_j}a^\dagger_k|0\rangle&=(a^\dagger_k-ik\overline{w}_i\delta_{kj})|0\rangle.
\end{align}
Hence, using this result and the BCH formula~\hyperref[BCH_ps]{(\ref*{BCH})}, we get:
\begin{align}\label{Tomogram_q_h_oscillator}
\mathcal{W}_{cm}(X,\boldsymbol{w},\overline{\boldsymbol{w}})&=\frac{1}{2\pi}\hspace{-.15cm}\int\limits_{\hspace{-.04cm}-\infty}^{\hspace{.45cm}\infty}\hspace{-0.15cm}\e^{ikX}\e
(1-k^2|w_1|^2)\cdots(1-k^2|w_n|^2)\diff k\nonumber\\                                                                                                                                                                                                                                                              
                                                                                                                                                                                                                                                                                                  &=\frac{1}{\sqrt{\pi(\boldsymbol{\mu}^2+\boldsymbol{\nu}^2)}}\left(1+\alpha_1\frac{\diff^2}{\diff X^2}+\cdots\right.\nonumber\\
                                                                                                                                                                                                                                                                                                  &\hspace{3cm}\left.+\alpha_{n}\frac{\diff^{2n}}{\diff X^{2n}}\right)\exp\left(-\frac{X^2}{\boldsymbol{\mu}^2+\boldsymbol{\nu}^2}\right),                                                                                                                                                                                                                                                                                                                                                                                                                                                                                                                                                                       
\end{align}
where
\begin{align*}
&\alpha_1=\sum_{i_1=1}^n|w_{i_1}|^2=|\boldsymbol{w}|^2=2^{-1}(\boldsymbol{\mu}^2+\boldsymbol{\nu}^2),\\
&\hspace{-0.1cm}\alpha_2\hspace{0.1cm}=\hspace{-0.3cm}\sum_{\hspace{0.2cm}i_1,\,i_2>i_1}^n\hspace{-0.33cm}|w_{i_1}|^2|w_{i_2}|^2=2^{-2}\hspace{-0.4cm}\sum_{\hspace{0.2cm}i_1,\,i_2>i_1}^n\hspace{-0.3cm}(\mu_{i_1}^2+\nu_{i_1}^2)(\mu_{i_2}^2+\nu_{i_2}^2),\\
&\hspace{0.63cm}\vdots\hspace{3.03cm}\\
&\hspace{0.63cm}\vdots\hspace{3.03cm}\\
&\hspace{-0.5cm}\alpha_{n-1}\hspace{0.1cm}=\hspace{-0.4cm}\sum_{\substack{i_1,\,i_2>i_1,\ldots,\\i_{n-1}>\cdots>i_1}}^n\hspace{-0.5cm}|w_{i_1}|^2\cdots|w_{i_n}|^2=2^{-(n-1)}\hspace{-0.6cm}\sum_{\substack{i_1,\,i_2>i_1,\ldots,\\i_{n-1}>\cdots>i_1}}\hspace{-0.25cm}(\mu_{i_1}^2+\nu_{i_1}^2)\cdots(\mu_{i_n}^2+\nu_{i_n}^2),\\
&\alpha_n=|w_1|^2\cdots|w_n|^2=2^{-n}(\mu_1^2+\nu_1^2)\cdots(\mu_n^2+\nu_n^2).
\end{align*}
Thus, using the formula \hyperref[Hermite_pols_ps]{(\ref*{Hermite_pols})} of Hermite polynomials, we finally obtain:
\phantomsection\label{NDimqharm_ps}
\vspace{-0.41cm}
\begin{multline}\label{NDimqharm}
\mathcal{W}_{cm}(X,\boldsymbol{\mu},\boldsymbol{\nu})=\frac{1}{\sqrt{\pi(\boldsymbol{\mu}^2+\boldsymbol{\nu}^2)}}\left[1+\frac{\alpha_1}{\boldsymbol{\mu}^2+\boldsymbol{\nu}^2}H_2\left(\frac{X}{\sqrt{\boldsymbol{\mu}^2+\boldsymbol{\nu}^2}}\right)+\right.\\
\cdots+\left.\frac{\alpha_{n}}{(\boldsymbol{\mu}^2+\boldsymbol{\nu}^2)^n}H_{2n}\left(\frac{X}{\sqrt{\boldsymbol{\mu}^2+\boldsymbol{\nu}^2}}\right)\right]\exp\left(-\frac{X^2}{\boldsymbol{\mu}^2+\boldsymbol{\nu}^2}\right).
\end{multline}

%% file: Tesischap3.tex
\chapter{A numerical algorithm to reduce unitary representations}\label{chap_smily}

\section{The Clebsh--Gordan problem}\label{CG_problem_beg_section}

In the \hyperref[chap_tom_qu]{{\color{black}\textbf{previous chapter}}}, we have discussed the tomographic problem of reconstructing a state $\rho$ of a quantum system from a family of probability distributions. We have seen that in the case in which the auxiliary space $\mathcal{M}$ is a compact Lie group, a reconstruction formula for recovering the state $\rho$, using a unitary representation $U$ of that group, can be obtained.

In this chapter, we will deal with a sort of converse problem where we will try to determine a unitary representation $U$ from the properties of a family of states. We will show here that the adapted states presented in \hyperref[section_recosntruction_states_groups]{{\color{black}\textbf{section~\ref*{section_recosntruction_states_groups}}}}, eqs.\,\hyperref[adapted_state_a]{$(2.7.16)$} play a paramount role in the description of the proper invariant subspaces under the action of the representation $U$. 

More precisely, let $G$ be a Lie group and $(\mathcal{H},U)$ a finite dimensional irreducible unitary representation (\textit{irrep} in what follows) of it. Let us consider now a closed subgroup $H \subset G$.  The restriction of $U$ to $H$ will define, in general, a reducible unitary representation of $H$.   

If we denote by $\widehat{H}$ the family of equivalence classes of irreps of $H$ (recall that two unitary representations of $H$, $V \colon H\rightarrow U(E)$ and $V^\prime \colon H \to U(E')$ are equivalent if there exists a unitary map $T \colon E \to \displaystyle{E'}$ such that $\displaystyle{V}'(h) \circ T = T \circ V(h)$ for all $h \in H$), because $\mathcal{H}$ is finite dimensional, then:
 \phantomsection\label{decomposition_ps}\begin{equation}\label{decomposition}
 \mathcal{H} = \bigoplus_{\alpha \in \widehat{H}} \mathcal{L}^\alpha \, ,\qquad \mathcal{L}^\alpha=c_\alpha\mathcal{H}^\alpha\, ,
 \end{equation}
where $c_\alpha$ denotes a non-negative integer,
$\alpha$ labels a subset in the class of irreps of the group $H$, that is, each $\alpha$ actually denotes a finite dimensional irrep $(\mathcal{H}^\alpha, U^\alpha)$ of $H$, and $c_\alpha \mathcal{H}^\alpha$ denotes the direct sum of the linear space $\mathcal{H}^\alpha$ with itself $c_\alpha$ times.  

Thus, the family of non-negative integer numbers $c_\alpha$ denotes the multiplicity of the irrep $\mathcal{H}^\alpha$ in $\mathcal{H}$ and it obviously satisfies $n = \sum_{\alpha} c_\alpha n_\alpha$ where $n_\alpha = \dim \mathcal{H}^\alpha$.  Notice that similarly, the unitary operator $U(h)$ will have the block structure:
\phantomsection\label{decompositionU_ps}\begin{equation}\label{decompositionU}
U(h) = \bigoplus_{\alpha \in \widehat{H}} c_\alpha U^\alpha(h) \, , \qquad \forall h \in H \, ,
\end{equation}
where $\displaystyle{U}^\alpha (h ) = U(h) \mid_{\mathcal{H}^\alpha} $.

\phantomsection\label{resumen_CGP}
The problem of determining an orthonormal basis of $\mathcal{H}$ adapted to the decomposition given in eq.\,\hyperref[decomposition_ps]{(\ref*{decomposition})} will be called the Clebsch--Gordan problem of $(\mathcal{H},U)$ with respect to the subgroup $H$, i.e., find an orthonormal basis $\{ u_{a,k}^\alpha \}$, $\alpha \in \widehat{H}$, $a = 1,\ldots, c_\alpha$ and $k = 1, \ldots, n_\alpha$, of $\mathcal{H}$ such that each family $\{ u_{a,k}^\alpha \}$ with given $\alpha, a$ defines an orthonormal basis of $\mathcal{H}^\alpha$.  Thus, if we are given an arbitrary orthonormal basis $\{ u_j \}_{j = 1, \ldots, n} \subset \mathcal{H}$, we could compute the $n\times n$ unitary matrix $C$ such that:
\phantomsection\label{CG_matrix_ps}\begin{equation}\label{CG_matrix}
u_l = \sum_{\alpha, a, k} C_{a,kl}^\alpha u_{a,k}^\alpha \, ,  \qquad  \alpha \in \widehat{H}, \quad a = 1,\ldots, c_\alpha, \quad k,l = 1, \ldots, n_\alpha \, .
\end{equation}
The coefficients of the matrix $C$ are usually expressed as the symbol $C_{a,kl}^\alpha=(l \mid \alpha, a, k)$ and are called the Clebsch-Gordan coefficients of the decomposition \hyperref[decompositionU_ps]{(\ref*{decompositionU})}.

The original Clebsh--Gordan problem (CGP for short) arises from the composition of two quantum systems possessing the same symmetry group (see for instance \hyperref[Ga90_ps]{\RED{[\citen*{Ga90}, ch.\,5]}}).

If $\mathcal{H}_A$ and $\mathcal{H}_B$ denote Hilbert spaces corresponding to two quantum systems $A$ and $B$ respectively, and both support irreps $U_A$ and $U_B$ of a Lie group $G$, then the composite system, whose Hilbert space is $\mathcal{H} = \mathcal{H}_A \widehat{\otimes} \mathcal{H}_B$, supports an irrep of the product group $G \times G$.    The interaction between both systems causes, typically, that the composite system just possesses $G$ as a symmetry group (considered as the diagonal subgroup $G \subset G\times G$ of the product group). The tensor product representation $U_A \otimes U_B$ will no longer be irreducible with respect to the subgroup $G \subset G \times G$ and we will be compelled to consider its decomposition in irrep components, i.e.:
\begin{equation}
U_A\otimes U_B=\bigoplus_{\alpha \in \widehat{H}}c_\alpha U^\alpha.
\end{equation}

A considerable effort has been put in computing the CG matrix for various situations of physical interest. In particular, the groups $SU(N)$ have been widely discussed (see for instance \RED{[}\citen{Gl07}\RED{,}\hspace{0.05cm}\citen{Al11}\RED{]} and references therein) because in such cases, for instance when considering the groups $SU(3)$ and $SU(2)$, the CG matrix provides the multiplet structure and the spin components of a composite system of particles with various spins \RED{[}\citen{Wi94}\RED{,}\hspace{0.05cm}\citen{Ro97}\RED{]}.  However all these results depend critically on the algebraic structure of the underlying group $G$ (and of the subgroup $H$) and no algorithm is known that will allow the efficient computation of the CG matrix for a general subgroup $H \subset G$.

On the other hand the problem of determining the decomposition of an irreducible representation with respect to a given subgroup has not been addressed from a numerical point of view.    The general theory asserts that the multiplicity of a given irreducible representation $(\mathcal{H}^\alpha, U^\alpha)$ of the compact group $G$ in the finite-dimensional representation $(\mathcal{H}, U)$ is given by the inner product in $L^2(G)$:
$$
c_\alpha = \langle \chi^\alpha, \chi \rangle_{L^2(G)},
$$
where $\chi^\alpha$ and $\chi$ are the characters of $U^\alpha$ and $U$ respectively and $\langle \cdot , \cdot \rangle$ the standard inner product of central functions in $G$ with respect to the (left-invariant) Haar measure. Hence, if the characters $\chi^\alpha$ of the irreducible representations of $G$ were known, the computation of the multiplicity would become, in principle, a simple task. Moreover, given the characters $\chi^\alpha$ of the irreducible representations, the \textit{projection method} \hyperref[Tu85_ps]{\RED{[\citen*{Tu85}, ch.\,4]}} would allow to construct the CG matrix explicitly.  However, there is not an easy way of determining the multiplicities $c_\alpha$ if the irreducible representations are not known in advance or are not explicitly described.    

Again, in principle, the computation of the irreducible representations of a finite group could be achieved by constructing its character table, i.e., a $n_C\times n_C$ unitary matrix where $n_C$ is the number of conjugacy classes of the group describing the characters of its irreps, however there is not a general numerical algorithm for doing that till now.

\section{The SMILY algorithm}\label{section_smily_alg}

Let $G$ be a compact Lie group (or a finite group) and $H\subset G$ a closed subgroup of it. Let us remember that an adapted state is a state \hyperref[adapted_state_a]{$(2.7.16)$} of the form:
\begin{equation*}
\boldsymbol{\rho}=\frac{n}{|G|}\sum_{g\in H}\chi_\rho(g)U(g)^\dagger,
\end{equation*}
or
\begin{equation*}
\boldsymbol{\rho}=n\hspace{-0.15cm}\int\limits_{\hspace{.3cm}H}\hspace{-0.1cm}\chi_\rho(g)U(g)^\dagger\diff\mu(g),
\end{equation*}
depending whether the group $G$ is finite or compact. Clearly, adapted states satisfy~\hyperref[cond_adapted_states_ps]{(\ref*{cond_adapted_states})}:
\begin{equation*}
\Tr\big(\boldsymbol{\rho} U(g)\big)=0,\qquad g\not\in H.
\end{equation*}

\phantomsection\label{resumen_trans}
The main idea of the algorithm is that the structure of proper invariant subspaces for the representation $U(h)$, $ \forall h\in H$, is the same as that for generic adapted states of the form written above, i.e., adapted states such that their eigenvalues have the smallest possible degeneracy. Then, the unitary matrix $C$ that diagonalizes in blocks any matrix representation of generic adapted states $\boldsymbol{\rho}$, \hyperref[adapted_state_a]{$(2.7.16)$}, will diagonalize in blocks the matrix representation $U(h)$ of $H$ too, and each block will correspond to an irrep of $H$. 

Thus, we will get that if we find a unitary matrix $C$ such that $\boldsymbol{\rho}$ transforms as:
 \phantomsection\label{structure_rho_ps}\begin{equation}\label{structure_rho}
C^\dagger \boldsymbol{\rho} C=\begin{pmatrix}
                   \mathds{1}_{c_1}\otimes  \boldsymbol{\sigma}^1 &  &   \\
                    &  \hspace{-1.4cm}\mathds{1}_{c_2}\otimes  \boldsymbol{\sigma}^2 &  &  &  \\
                    &  & \hspace{-0.6cm}\diagdots[-47]{1.4em}{.12em} &  & \mbox{\Huge{0}}&   \\
                    &  &  & \hspace{-0.4cm}\diagdots[-47]{1.4em}{.12em}  &  &  \\
                    & \hspace{-0.8cm}\mbox{\Huge{0}} & & & \hspace{-0.1cm}\diagdots[-47]{1.4em}{.12em}  &   \\
                    &  &  &  &  &  \hspace{-1.15cm}\mathds{1}_{c_N}\otimes  \boldsymbol{\sigma}^N
\end{pmatrix},
\end{equation}
where $N$ is the number of irreps decomposing $U$ in \hyperref[decompositionU_ps]{(\ref*{decompositionU})}, and $\boldsymbol{\sigma}^\alpha$ satisfy
$$
\boldsymbol{\sigma}^\alpha={\boldsymbol{\sigma}^{\alpha}}^\dagger,\qquad\boldsymbol{\sigma}^\alpha\geq 0 \, ,
$$
then, it will follow that:
\phantomsection\label{resumen_diag}
\begin{equation*}
C^\dagger D(h) C=\begin{pmatrix}
                   \mathds{1}_{c_1}\otimes D^1(h) &  &   \\
                    &  \hspace{-1.9cm}\mathds{1}_{c_2}\otimes D^2(h) &  &  &  \\
                    &  &\hspace{-1.7cm}\diagdots[-47]{1.4em}{.12em}  &  & \hspace{-0.4cm}\mbox{\Huge{0}}&   \\
                    &  &  &  \hspace{-1.5cm}\diagdots[-47]{1.4em}{.12em} &  &  \\
                    & \hspace{-2.0cm}\mbox{\Huge{0}} & & & \hspace{-0.9cm}\diagdots[-47]{1.4em}{.12em} &   \\
                    &  &  &  &  &  \hspace{-1.35cm}\mathds{1}_{c_N}\otimes D^N(h)
\end{pmatrix}.
\end{equation*}
Notice that because of the form of the matrix \hyperref[structure_rho_ps]{(\ref*{structure_rho})}, the state $\boldsymbol{\rho}$ will be generic if every eigenvalue of $\boldsymbol{\sigma}^\alpha$ has multiplicity one and the eigenvalues of all matrices $\boldsymbol{\sigma}^\alpha$ are different.

The algorithm will start from a generic adapted state $\boldsymbol{\rho}_1$. Consider a unitary matrix $V_1$ that diagonalizes the state $\boldsymbol{\rho}_1$.  Then, using a second generic adapted state $\boldsymbol{\rho}_2$, we will obtain several unitary transformations that will lead to the desired CG matrix.

The \hyperref[SMILY_step_1]{{\color{mybrown}{SMILY}}} algorithm is decomposed in eight steps:\newpage
\begin{enumerate}

\phantomsection\label{SMILY_step_1}
\item[\MYBROWN{1.}]  \MYBROWN{Generating the adapted states}: Create two generic independent adap-ted states $\boldsymbol{\rho}_1$ and $\boldsymbol{\rho}_2$. The states must be independent in the sense that they must have different eigenvectors. To create them, we generate first two random vectors $\boldsymbol{\varphi}_1$, $\boldsymbol{\varphi}_2$, with no zero components, of size $|H|$ and afterwards multiply their elements by the matrices $D(h)$, $\forall h\in H$:
\phantomsection\label{rho_tilde_ps}\begin{equation}\label{rho_tilde}
\tilde{\boldsymbol{\rho}}_{1,2}=\sum_{j=1}^{|H|}\varphi_{1,2}(j)D(h_j).
\end{equation}
After that, we construct the Hermitian matrices:
\begin{equation*}
\tilde{\boldsymbol{\rho}}^\prime_{1,2}=\tilde{\boldsymbol{\rho}}_{1,2}+\tilde{\boldsymbol{\rho}}_{1,2}^\dagger,
\end{equation*}
and finally, after shifting them by their spectral radius and dividing by their traces, we get two normalized positive definite matrices:
\phantomsection\label{adapted_state_ps}\begin{equation}\label{adapted_state}
\tilde{\boldsymbol{\rho}}''_{1,2}=\tilde{\boldsymbol{\rho}}_{1,2}^\prime+s_{\textrm{radius}}(\tilde{\boldsymbol{\rho}}^\prime_{1,2})\mathds{1},\qquad \boldsymbol{\rho}_{1,2}=\frac{\tilde{\boldsymbol{\rho}}''_{1,2}}{\mathrm{Tr}(\tilde{\boldsymbol{\rho}}''_{1,2})}\,.
\end{equation}

\item[\MYBROWN{2.}] \MYBROWN{Diagonalizing the first state}: Compute a matrix $V_1$ that diagonalizes the state $\boldsymbol{\rho}_1$:
$$
V_1=\begin{pmatrix}
|&| & &|\\
V_1^1&V_1^2&\cdots&V_1^n\\
|&| & &|
\end{pmatrix},
$$
where $V_1^j$, $j=1,\ldots,n$, are the eigenvectors of $\boldsymbol{\rho}_1$. Notice that because the matrix $\boldsymbol{\rho}_1$ is Hermitean, it is unitary diagonalizable.

\item[\MYBROWN{3.}] \MYBROWN{Reorganizing}: Construct the matrix $V_1^{sort_1}$\! by reordering the columns of $V_1$ grouping the eigenvectors corresponding to the same proper subspace $\mathcal{L}^\alpha$. The following routine will be used:
\\
\\
\phantom{aa}\BLUE{for} $j$ \BLUE{from} 1 \BLUE{to} $n$ \BLUE{do}\\
\phantom{aa}\phantom{aaaa}\BLUE{for} $k\neq j$ \BLUE{from} 1 \BLUE{to} $n$ \BLUE{do}\\
\phantom{aa}\phantom{aaaaaaaa}$\epsilon_{jk}={V_1^j}^\dagger\rho_2V_1^k$,\\
\phantom{aa}\phantom{aaaaaaaa}\BLUE{if} $\epsilon_{jk}\neq 0$ \BLUE{then}\\
\phantom{aa}\phantom{aaaaaaaaaaaa} $V_1^j$ and $V_1^k$ belong to the same proper subspace.\\
\phantom{aa}\phantom{aaaaaaaa}\BLUE{end}\\
\phantom{aa}\phantom{aaaa}\BLUE{end}\\
\phantom{aa}\BLUE{end}

Hence,
\[
\begin{array}{cccc}
  V_1^{sort_1} =\left[\begin{array}{c|c|c|c}
W_1\, & \,W_2\, & \cdots & \,W_N\end{array}\right],\vspace{-0.3cm} \\
\hspace{-1.62cm}\underbrace{\phantom{a}}_{c_1n_1} & \hspace{-3.55cm}\underbrace{\phantom{a}}_{c_2n_2} & \hspace{0.5cm}\underbrace{\phantom{aaa}}_{c_Nn_N}
 \end{array}
 \]
where the columns of $W_\alpha$ are the eigenvectors of $\boldsymbol{\rho}_1$ that belongs to the same proper subspace $\mathcal{L}_\alpha$.

\phantomsection\label{Sorting_step_4}
\item[\MYBROWN{4.}] \MYBROWN{Sorting}: Sort the columns of the matrices $W_\alpha$ in increasing or decreasing order according to their eigenvalues to group the eigenvectors corresponding to same eigenvalues. The matrix we obtain after this reordering will be denoted as $V_1^{sort_2}$:
\phantomsection\label{V1sort2_ps}\begin{equation}\label{V1sort2}
V_1^{sort_2} =\left[\begin{array}{c|c|c|c}
W_1^{sort} & W_2^{sort} & \cdots & W_N^{sort}\end{array}\right].
\end{equation}

A few comments are in order here. Already at this step, we can get the multiplicities $c_\alpha$ and the dimensions $n_\alpha$ of $\mathcal{H}^\alpha$. Actually, the multiplicity of the eigenvalues of $W_\alpha^{sort}$ will be the multiplicity $c_\alpha$ of the irrep $\alpha$. Then, the dimensions $n_\alpha$ are obtained immediately because the number of columns of  $W_\alpha^{sort}$ is equal to $c_\alpha n_\alpha$.

Notice that after applying  $W_\alpha^{sort}$ to $\boldsymbol{\rho}_1$, we get a diagonal matrix with the eigenvalues ordered, that is, a matrix of the form:
$$
{W_\alpha^{sort}}^\dagger\boldsymbol{\rho}_1W_\alpha^{sort}=\begin{pmatrix}
                   \lambda_{1}^\alpha\mathds{1}_{c_\alpha} &  &   \\
                    &  \hspace{-0.8cm}\lambda_{2}^\alpha\mathds{1}_{c_\alpha} &  &  &  \\
                    &  & \hspace{-0.5cm}\diagdots[-47]{1.4em}{.12em} &  & \mbox{\Huge{0}}&   \\
                    &  &  & \hspace{-0.26cm}\diagdots[-47]{1.4em}{.12em} &  &  \\
                    & \mbox{\Huge{0}} & & & \hspace{-0cm}\diagdots[-47]{1.4em}{.12em} &   \\
                    &  &  &  &  &  \hspace{-0.6cm}\lambda_{n_\alpha}^\alpha\mathds{1}_{c_\alpha}
\end{pmatrix},
$$
where $\lambda_{k}^\alpha$, $k=1,\ldots,n_\alpha$, are the eigenvalues of $\boldsymbol{\rho}_1$ corresponding to the subspaces $\mathcal{L}^\alpha$. Therefore, counting the multiplicity of the eigenvalues of these matrices, we get the multiplicities $c_\alpha$, and $n_\alpha$ are obtained by dividing the dimensions of these blocks by $c_\alpha$.

At this point, it would also be possible to obtain the characters of the irreps in the decomposition of $D(h)$ by computing:
$$
\chi(h)=\frac{1}{c_\alpha}\mathrm{Tr}\big({W_\alpha^{sort}}^\dagger D(h)W_\alpha^{sort}\big).
$$

\item[\MYBROWN{5.}] \MYBROWN{Diagonalizing in blocks the second state}: Transform $\boldsymbol{\rho}_2$ with $V_1^{sort_2}$. The resulting matrix ${V_1^{sort_2}}^\dagger\boldsymbol{\rho}_2V_1^{sort_2}$ will have the following structure:
\phantomsection\label{rho_2structure_ps}\begin{equation}\label{rho_2structure}
\hspace{1.3cm}\left(
\begin{array}{ccccc}
\left.\begin{array}{|cc|}\hline
 \phantom{a} &  \phantom{a}\\
 \phantom{a} &  \phantom{a}\\\hline
\end{array}\right\}c_1n_1
 & &  &  &\\
 &\hspace{-1.44cm} \left.\begin{array}{|cccc|}\hline
 \phantom{a}& \phantom{a} & \phantom{a} &  \phantom{a}\\
  \phantom{a}&  \phantom{a}&  \phantom{a}& \phantom{a} \\
  \phantom{a}& \phantom{a} & \phantom{a} & \phantom{a} \\
  \phantom{a}& \phantom{a} & \phantom{a} & \phantom{a}\\\hline
\end{array}\right\} c_2n_2 &   & &\\
& &\hspace{-2.25cm}\diagdots[-40]{1.4em}{.12em} & &\\
& & &\hspace{-1.85cm}\diagdots[-40]{1.4em}{.12em} &\\
& & & &\hspace{-2.4cm}c_Nn_N\left\{\begin{array}{|ccc|}\hline
 \phantom{a} &  \phantom{a}& \phantom{a}\\
 \phantom{a} &  \phantom{a}& \phantom{a}\\
 \phantom{a} &  \phantom{a}& \phantom{a}\\\hline
\end{array}\right.
 \end{array}
\right),\begin{array}{c}
\hspace{-6.75cm}\vspace{4.32cm}\boldsymbol{\Sigma}^1
\end{array}
\phantom{a}
\begin{array}{c}
\hspace{-5.22cm}\vspace{1.4cm}\boldsymbol{\Sigma}^2
\end{array}
\phantom{a}
\begin{array}{c}
\hspace{-2.37cm}\vspace{-4.0cm}\boldsymbol{\Sigma}^N
\end{array}
\end{equation}
where
$$
\boldsymbol{\Sigma}^\alpha={\boldsymbol{\Sigma}^{\alpha}}^\dagger\qquad\mbox{and}\qquad\boldsymbol{\Sigma}^\alpha\geq 0.
$$

In the remaining steps, we will focus on decomposing in a diagonal block structure the blocks $\boldsymbol{\Sigma}^\alpha$ obtained in this step. Here, is where the key point of the algorithm appears and we see why only two adapted states are necessary to get the CG matrix that diagonalizes in blocks all the elements of the representation.\newpage

\phantomsection\label{Tensor_block_step_5}
\item[\MYBROWN{6.}] \MYBROWN{Getting the tensor block structure}: The matrices $\boldsymbol{\Sigma}^\alpha$ have the following block structure decomposition:
\phantom{a}
\\
\phantomsection\label{Sigma_structure_alg_ps}\begin{equation}\label{Sigma_structure_alg}
\boldsymbol{\Sigma}^\alpha=\begin{pmatrix}
 R^{11}_\alpha & \hspace{-0.3cm}R^{12}_\alpha&\diagdots[0]{1.4em}{.12em} &\diagdots[0]{1.4em}{.12em}& R^{1n_\alpha}_\alpha \\
 R^{21}_\alpha  &\hspace{-0.3cm} R^{22}_\alpha&\diagdots[0]{1.4em}{.12em} &\diagdots[0]{1.4em}{.12em}&R^{2n_\alpha}_\alpha\\
\hspace{-0.2cm} \diagdots[90]{1.2em}{.12em} & \hspace{-0.5cm}\diagdots[90]{1.2em}{.12em}&\hspace{-0.6cm}\diagdots[-32]{1.4em}{.12em}& &\hspace{-0.4cm}\diagdots[90]{1.2em}{.12em}   \\
\hspace{-0.2cm} \diagdots[90]{1.2em}{.12em} & \hspace{-0.5cm} \diagdots[90]{1.2em}{.12em}&\hspace{0.8cm}\diagdots[-32]{1.4em}{.12em}&   &\hspace{-0.4cm}\diagdots[90]{1.2em}{.12em}\\
 \hspace{-0.2cm}\diagdots[90]{1.2em}{.12em} & \hspace{-0.5cm} \diagdots[90]{1.2em}{.12em} & &\hspace{0.26cm}\diagdots[-32]{1.4em}{.12em}& \hspace{-0.4cm}\diagdots[90]{1.2em}{.12em}\\
   R^{n_\alpha 1}_\alpha&  \hspace{-0.3cm}R^{n_\alpha 2}_\alpha&\diagdots[0]{1.4em}{.12em} &\diagdots[0]{1.4em}{.12em} &R^{n_\alpha n_\alpha}_\alpha
\end{pmatrix},
\end{equation}
where the blocks $R^{ij}_\alpha$, $i,j=1,\ldots,n_\alpha$, are square matrices of size $c_\alpha$. The blocks in the diagonal are the identity, $R^{ii}_\alpha=\mathds{1}_{c_\alpha}$ (we will verify this fact in the \hyperref[sec:proof]{{\color{black}\textbf{next section}}}).

Now, let us take the matrices of the first column (we can choose any column but we will do it with the first one) and divide each matrix by its norm:
\begin{equation}
\tilde{R}^{i1}_\alpha=\frac{R^{i1}_\alpha}{\|R^{i1}_\alpha\|},\qquad i=1,\ldots,n_\alpha.
\end{equation}
Create the following block diagonal matrix:
\begin{equation}
Y^\alpha=\begin{pmatrix}
                   \mathds{1}_{c_\alpha} &  &   \\
                    &  \hspace{-0.6cm}\tilde{R}^{21}_\alpha&  &  &  \\
                    &  & \hspace{-0.3cm}\diagdots[-47]{1.4em}{.12em}&  & \mbox{\Huge{0}}&   \\
                    &  &  & \hspace{-0.15cm}\diagdots[-47]{1.4em}{.12em} &  &  \\
                    & \mbox{\Huge{0}} & & & \hspace{0.1cm}\diagdots[-47]{1.4em}{.12em} &   \\
                    &  &  &  &  &  \hspace{-0.3cm}\tilde{R}^{n_\alpha1}_\alpha
\end{pmatrix}.
\end{equation}
Therefore, if we transfom the matrices $\boldsymbol{\Sigma}^\alpha$ with $Y^\alpha$ we get:
\begin{equation}
{Y^\alpha}^\dagger\boldsymbol{\Sigma}^\alpha Y^\alpha=\begin{pmatrix}
  \tilde{s}_{11}^\alpha \mathds{1}_{c_\alpha} & \hspace{-0.3cm} \tilde{s}_{12}^\alpha \mathds{1}_{c_\alpha}&\diagdots[0]{1.4em}{.12em} &\diagdots[0]{1.4em}{.12em}&  \tilde{s}_{1n_\alpha}^\alpha \mathds{1}_{c_\alpha} \\
  \tilde{s}_{21}^\alpha \mathds{1}_{c_\alpha}  &\hspace{-0.3cm}  \tilde{s}_{22}^\alpha \mathds{1}_{c_\alpha}&\diagdots[0]{1.4em}{.12em} &\diagdots[0]{1.4em}{.12em}& \tilde{s}_{2n_\alpha}^\alpha \mathds{1}_{c_\alpha}\\
\hspace{-0.2cm} \diagdots[90]{1.2em}{.12em} & \hspace{-0.5cm}\diagdots[90]{1.2em}{.12em}&\hspace{-1.2cm}\diagdots[-25]{1.4em}{.12em}& &\hspace{-0.4cm}\diagdots[90]{1.2em}{.12em}   \\
\hspace{-0.2cm} \diagdots[90]{1.2em}{.12em} & \hspace{-0.5cm} \diagdots[90]{1.2em}{.12em}&\hspace{0.8cm}\diagdots[-25]{1.4em}{.12em}&   &\hspace{-0.4cm}\diagdots[90]{1.2em}{.12em}\\
 \hspace{-0.2cm}\diagdots[90]{1.2em}{.12em} & \hspace{-0.5cm} \diagdots[90]{1.2em}{.12em} & &\hspace{0.56cm}\diagdots[-24.5]{1.4em}{.12em}& \hspace{-0.4cm}\diagdots[90]{1.2em}{.12em}\\
    \tilde{s}_{n_\alpha 1}^\alpha \mathds{1}_{c_\alpha}&  \hspace{-0.3cm} \tilde{s}_{n_\alpha 2}^\alpha \mathds{1}_{c_\alpha}&\diagdots[0]{1.4em}{.12em} &\diagdots[0]{1.4em}{.12em} & \tilde{s}_{n_\alpha n_\alpha}^\alpha \mathds{1}_{c_\alpha}
\end{pmatrix}
\end{equation}
where $\tilde{s}_{ij}$, $i,j=1,\ldots,n_\alpha$, are complex numbers which are the matrix elements of the Hermitean non-negative matrix
\begin{equation}
[\tilde{\boldsymbol{\sigma}}^\alpha]_{ij}=\tilde{s}_{ij},
\end{equation}
and ${Y^\alpha}^\dagger\boldsymbol{\Sigma}^\alpha Y^\alpha=\tilde{\boldsymbol{\sigma}}^\alpha\otimes\mathds{1}_{c_\alpha}$. (See \hyperref[sec:proof]{{\color{black}\textbf{following section}}} for the proof of these facts).

\phantomsection\label{switching_step_6}
\item[\MYBROWN{7.}] \MYBROWN{Switching the tensor block structure}: Let us define now the $n_\alpha c_\alpha\times n_\alpha c_\alpha$ row shift matrix $S_\alpha$:
\phantomsection\label{shift_matrix_alg_ps}\begin{equation}\label{shift_matrix_alg}
S_\alpha=

\end{equation}
\\

If we have to matrices $A$ and $B$, the matrix elements of the Kronecker product $A\otimes B$ are equal to the elements of $B\otimes A$ up to a rearrangement. Then, there exist two permutation matrices $P$ and $F$ \cite{He81}, such that
$$
F (A\otimes B) P= B\otimes A.
$$
However, if $A$ and $B$ are square matrices of size $n$ and $m$ respectively, $P$ can be chosen such that $P=F^\dagger$, where $F$ only depends on $n$ and $m$:
$$
F_{nm} (A_n\otimes B_m) F^\dagger_{nm}= B_m\otimes A_n.
$$

In our case, we want to find the permutation matrix which transforms $\tilde{\boldsymbol{\sigma}}^\alpha\otimes\mathds{1}_{c_\alpha}$ into $\mathds{1}_{c_\alpha}\otimes\tilde{\boldsymbol{\sigma}}^\alpha$:
\begin{equation}
F_{n_\alpha c_\alpha} (\tilde{\boldsymbol{\sigma}}^\alpha\otimes\mathds{1}_{c_\alpha}) F^\dagger_{n_\alpha c_\alpha}= \mathds{1}_{c_\alpha}\otimes\tilde{\boldsymbol{\sigma}}^\alpha.
\end{equation}

The permutation matrix $F_{n_\alpha c_\alpha}$ is constructed using the matrices defined previously in~\hyperref[shift_matrix_alg_ps]{(\ref*{shift_matrix_alg})} and~\hyperref[f_matrix_alg_ps]{(\ref*{f_matrix_alg})} as:
\begin{equation}
F_{n_\alpha c_\alpha}=\left[\begin{array}{c|c|c|c|c}
f_\alpha &S_\alpha f_\alpha & S_\alpha^2f_\alpha &\cdots&S_\alpha^{c_\alpha-1}f_\alpha\end{array}\right].
\end{equation}

\item[\MYBROWN{8.}] \MYBROWN{Getting the Clebsh--Gordan matrix}: Finally, we construct the following block diagonal matrices with the matrices obtained in steps \hyperref[Tensor_block_step_5]{{\color{mybrown}6}} and \hyperref[switching_step_6]{{\color{mybrown}7}}:
\phantomsection\label{Y_matrix_alg_ps}\begin{equation}\label{Y_matrix_alg}
\hat{Y}=\begin{pmatrix}
                   Y^1 &  &   \\
                    &  \hspace{-0.6cm}Y^2&  &  &  \\
                    &  & \hspace{-0.3cm}\diagdots[-47]{1.4em}{.12em}&  & \mbox{\Huge{0}}&   \\
                    &  &  & \hspace{-0.15cm}\diagdots[-47]{1.4em}{.12em} &  &  \\
                    & \mbox{\Huge{0}} & & & \hspace{0.1cm}\diagdots[-47]{1.4em}{.12em} &   \\
                    &  &  &  &  &  \hspace{-0.3cm}Y^N
\end{pmatrix}
\end{equation}
and
\phantomsection\label{F_matrix_alg_ps}\begin{equation}\label{F_matrix_alg}
\hat{F}=\begin{pmatrix}
                   F_{n_1c_1} &  &   \\
                    &  \hspace{-0.6cm}F_{n_2c_2}&  &  &  \\
                    &  & \hspace{-0.3cm}\diagdots[-47]{1.4em}{.12em}&  & \mbox{\Huge{0}}&   \\
                    &  &  & \hspace{-0.15cm}\diagdots[-47]{1.4em}{.12em} &  &  \\
                    & \mbox{\Huge{0}} & & & \hspace{0.1cm}\diagdots[-47]{1.4em}{.12em} &   \\
                    &  &  &  &  &  \hspace{-0.3cm}F_{n_Nc_N}
\end{pmatrix}
\end{equation}
then, the particular CG matrix $C$ which diagonalizes completely the adapted state $\boldsymbol{\rho}_1$ and that block diagonalizes any state $\boldsymbol{\rho}_2$ is constructed by making the product of the matrices $V_1^{sort_2}$, $\hat{Y}$ and $\displaystyle{{\hat{F}}^\dagger}$ in eqs.\,\hyperref[V1sort2_ps]{(\ref*{V1sort2})},~\hyperref[Y_matrix_alg_ps]{(\ref*{Y_matrix_alg})} and~\hyperref[F_matrix_alg_ps]{(\ref*{F_matrix_alg})}:
\phantomsection\label{CGmatrix_ps}\begin{equation}\label{CGmatrix}
C_1=V_1^{sort_2}\hat{Y}{\hat{F}}^\dagger.
\end{equation}

\end{enumerate}

\section{The proof of SMILY}\label{sec:proof}

In this section, we will provide a proof of the \hyperref[section_smily_alg]{{\color{black}\textbf{SMILY}}} algorithm presented in \hyperref[section_smily_alg]{{\color{black}\textbf{last section}}}. Let us start with the following lemma.

{\MYBROWN{\begin{lemma}\label{lemma_adapted_state}\color{black}{
Let $\boldsymbol{\rho}$ be a generic adapted state with respect to a closed subgroup $H\subset G$ and $(\mathcal{H}, U)$ an irreducible unitary representation of $G$. Let also $(\mathcal{H}^\alpha, U^\alpha)$, $\alpha\in\widehat{H}$, be the irreducible unitary representations into which the representation $(\mathcal{H}, U)$ is decomposed when restricted to $H$. Then, the proper invariant subspaces of $\mathcal{H}^\alpha$ are the same as the proper invariant subspaces of $\boldsymbol{\rho}$.}
\end{lemma}}}
\phantomsection\label{proof_Lemma_SMYLY}
\noindent\MYBROWN{\textbf{Proof}}: From the definition of adapted state \hyperref[adapted_state_a]{$(2.7.16)$}, if $C$ is the CG matrix that diagonalizes in blocks the matrix representation $D(h)$, because any adapted state is a linear combination of elements $D(h)$, $\forall h\in H$, $C$ will diagonalize in blocks any adapted state. 

Conversely, we have to prove that if $C$ is a matrix that diagonalizes in blocks a set of $|H|$ generic adapted states, then that matrix will diagonalize every element of the representation $D(h)$.

Let us write the smeared character~\hyperref[smeared_character_chap_2_ps]{(\ref*{smeared_character_chap_2})} of the representation $U$ with respect to the state $\boldsymbol{\rho}$ as a vector over the elements of the group:
$$
\boldsymbol{\chi}_\rho(H)=\begin{bmatrix}
\chi_\rho(e)&\chi_\rho(h_1)&\cdots &\chi_\rho(h_{r-1})
\end{bmatrix},
$$
where $r$ is the order of $H$. 

Clearly, two states are independent if their smeared character vectors are independent too. Then, let us consider the $r\times r$ matrix $\mathcal{Y}$ defined by $r$ independent  smeared character vectors, $[\mathcal{Y}]_{jk}=\chi_{\rho_j}(h_{k-1})$, with $j,k=1,\ldots r$, that is:
\[
\left[\begin{array}{c}
\boldsymbol{\rho}_1\\\hline
\boldsymbol{\rho}_2\\\hline
\vdots\\\hline
\boldsymbol{\rho}_r
\end{array}\right]=\mathcal{Y}\left[\begin{array}{c}
D(e)\\\hline
D(h_1)\\\hline
\vdots\\\hline
D(h_{r-1})
\end{array}\right].
\]
Therefore, because the rows of the matrix $\mathcal{Y}$ are independent, it is invertible. Thus, we can write the elements of the representation as linear combinations of adapted states:
\[
\left[\begin{array}{c}
D(e)\\\hline
D(h_1)\\\hline
\vdots\\\hline
D(h_{r-1})
\end{array}\right]=\mathcal{Y}^{-1}\left[\begin{array}{c}
\boldsymbol{\rho}_1\\\hline
\boldsymbol{\rho}_2\\\hline
\vdots\\\hline
\boldsymbol{\rho}_r
\end{array}\right].
\]
Then, we have proved that the unitary transformation that diagonalizes in blocks $r$ independent adapted states will diagonalize in blocks all the elements of $D(h)$, and that matrix is the CG matrix $C$. 

If the state $\boldsymbol{\rho}$ is generic and adapted, we can always construct a family of $r$ independent adapted states by taking permutations of the components of $\boldsymbol{\chi}_\rho(H)$, therefore the conclusion is reached.

\hfill{\hyperref[proof_Lemma_SMYLY]{{\color{black}$\blacksquare$}}}

Notice that for simplicity, we have supposed that the group $G$ is finite. In the case of compact groups, if we are considering finite-dimensional representations, the integral \hyperref[adapted_state_a_ps]{(\ref*{adapted_state_b})} can be approximated as well as we want by using appropriate quadrature rules. Then, it becomes a finite sum and the arguments written above can be repeated \textit{mutatis mutandis}.

\phantomsection\label{proof_secondpart_ps}
We will prove now that the matrix $V_1^{sort_2}$~\hyperref[V1sort2_ps]{(\ref*{V1sort2})} transforms every adapted state $\boldsymbol{\rho}_2$ in a block diagonal matrix~\hyperref[rho_2structure_ps]{(\ref*{rho_2structure})} where $\boldsymbol{\Sigma}^\alpha$ has the structure~\hyperref[Sigma_structure_alg_ps]{(\ref*{Sigma_structure_alg})}. For that, let us start by choosing an arbitrary CG matrix $C$ and transform with it the generic adapted state $\boldsymbol{\rho}_1$:
\begin{equation*}
C^\dagger\boldsymbol{\rho}_1 C=\begin{pmatrix}
                   \mathds{1}_{c_1}\otimes  \boldsymbol{\sigma}_1^1 &  &   \\
                    &  \hspace{-1.4cm}\mathds{1}_{c_2}\otimes  \boldsymbol{\sigma}_1^2 &  &  &  \\
                    &  & \hspace{-0.6cm}\diagdots[-47]{1.4em}{.12em} &  & \mbox{\Huge{0}}&   \\
                    &  &  & \hspace{-0.4cm}\diagdots[-47]{1.4em}{.12em}  &  &  \\
                    & \hspace{-0.8cm}\mbox{\Huge{0}} & & & \hspace{-0.1cm}\diagdots[-47]{1.4em}{.12em}  &   \\
                    &  &  &  &  &  \hspace{-1.15cm}\mathds{1}_{c_N}\otimes  \boldsymbol{\sigma}_1^N
\end{pmatrix}.
\end{equation*}
Next, we will diagonalize each block $\mathds{1}_{c_\alpha}\otimes \boldsymbol{\sigma}_1^\alpha$ to get the relation between $V_1^{sort_2}$ and $C$. 

Let $r^\alpha_j$ be the eigenvectors of $\boldsymbol{\sigma}_1^\alpha$, $j=1,\ldots,n_\alpha$:
\begin{equation}
\boldsymbol{\sigma}_{1}^\alpha r^{\alpha}_j=\lambda^{\alpha}_jr^{\alpha}_j,\qquad \langle r^{\alpha}_j,r^{\alpha}_k\rangle=\delta_{jk}.
\end{equation}
Because the state $\boldsymbol{\rho}_1$ is generic, then for $\alpha\neq\gamma$ we have:
\begin{equation*}
\lambda^{\alpha}_j\neq\lambda^{\gamma}_k,\qquad \alpha,\gamma=1,\ldots,N,\quad j=1,\ldots,n_\alpha,\quad k=1,\ldots,n_\gamma.
\end{equation*}

Let $\displaystyle{\left\{z_p^j\right\}_{p=1}^{c_\alpha}}$, $j=1,\ldots,n_\alpha$, be $n_\alpha$ arbitrary orthonormal basis of $\mathbb{C}^{c_\alpha}$. The eigenvectors of $\mathds{1}_{c_\alpha}\otimes \boldsymbol{\sigma}^{\alpha}_1$ will be $z_p^j\otimes r^{\alpha}_j$:
\begin{equation}
(\mathds{1}_{c_\alpha}\otimes \boldsymbol{\sigma}^{\alpha}_1)(z_p^j\otimes r^{\alpha}_j)=\lambda^\alpha_{j}z_p^j\otimes r^{\alpha}_j.
\end{equation}
If we construct a matrix such that its columns are the orthonormal vectors of the basis  $\displaystyle{\left\{z_p^j\right\}_{p=1}^{c_\alpha}}$:
\begin{equation}
Q_\alpha^j=
,
\end{equation}
the matrix that diagonalizes $\mathds{1}_{c_\alpha}\otimes \boldsymbol{\sigma}^{\alpha}_1$ will be:
\phantomsection\label{X_alpha_def_ps}\begin{equation}\label{X_alpha_def}
X_\alpha=\left[%
=\boldsymbol{\Lambda}^\alpha.
\end{equation}
If we define now the matrix
\begin{equation}
\hat{X}=%
,
\end{equation}
we have that any matrix that diagonalize $\boldsymbol{\rho}_1$, in which the eigenvalues are sorted in the same way as in $\boldsymbol{\Lambda}_{\alpha}$ in \hyperref[lambda_alpha_prove_ps]{(\ref*{lambda_alpha_prove})}, has the form $V_1^{sort_2}=C\hat{X}$:
\begin{equation*}
{V_1^{sort_2}}^\dagger\boldsymbol{\rho}_1 V_1^{sort_2}=\hat{X}^\dagger C^\dagger\boldsymbol{\rho}_1C\hat{X}=%
.
\end{equation*}
This factorization $V_1^{sort_2}=C\hat{X}$, when applied to a generic adapted state $\boldsymbol{\rho}_2$, gives the structure~\hyperref[rho_2structure_ps]{(\ref*{rho_2structure})}, in fact, we get:
\begin{equation*}
{V_1^{sort_2}}^\dagger\boldsymbol{\rho}_2 V_1^{sort_2}=\hat{X}^\dagger C^\dagger\boldsymbol{\rho}_2 C\hat{X}=\hat{X}^\dagger%
,
\end{equation}
where
\begin{equation*}\label{sigma_structure}
X_\alpha^\dagger(\mathds{1}_{c_1}\otimes \boldsymbol{\sigma}_2^\alpha)X_\alpha=\boldsymbol{\Sigma}^\alpha,
\end{equation*}
and now, it is easy to verify from the definition of $X_\alpha$~\hyperref[X_alpha_def_ps]{(\ref*{X_alpha_def})} that the matrix $\boldsymbol{\Sigma}^\alpha$ has the block structure \hyperref[Sigma_structure_alg_ps]{(\ref*{Sigma_structure_alg})} with each block $R_\alpha^{ij}$ given by:
\begin{equation}
R^{ij}_\alpha=\tilde{s}_{ij}^\alpha {Q^i_\alpha}^\dagger Q^j_\alpha
\end{equation}
with
\begin{equation}
\tilde{s}_{ij}^\alpha={r^{\alpha}_i}^\dagger\boldsymbol{\sigma}^{\alpha}_2 r^{\alpha}_j.
\end{equation}
Here, we can see that the dependence on the state $\boldsymbol{\rho}_2$ is only in $\tilde{s}_{ij}^\alpha$, because the matrices $Q^i_\alpha$ only depend on $\boldsymbol{\rho}_1$.

Finally, it is very easy to verify that the CG matrix $C_1=V_1^{sort_2}\hat{Y}\hat{F}^\dagger$ diagonalizes in blocks the state $\boldsymbol{\rho}_2$ and diagonalizes completely the state $\boldsymbol{\rho}_1$.

\vspace{0cm}\hfill{\hyperref[proof_secondpart_ps]{{\color{black}$\blacksquare$}}}\newpage

\section{The decomposition of the regular representation of a finite group}\label{decom_reg_group_arx}

The \hyperref[section_smily_alg]{{\color{black}\textbf{algorithm}}} we have presented in this chapter decomposes any finite dimensional unitary representation of any compact Lie group. In the case of finite groups, it is natural to apply it to the regular representation because it contains every irreducible representation with multiplicity equal to the dimension of the irreps, $c_\alpha=n_\alpha$ \hyperref[Se77_ps]{\RED{[\citen*{Se77}, ch.\,2]}}, thus:

\vspace{0.1cm}\begin{equation}
|G|=\sum_{\alpha=1}^Nn_\alpha^2.
\end{equation}
\vspace{0.1cm}

Let us remember that the left regular representation $U_L^{reg}$ introduced in \hyperref[section_recosntruction_states_groups]{{\color{black}\textbf{section~\ref*{section_recosntruction_states_groups}}}}, eq.\,\hyperref[left_regular_rep_ps]{(\ref*{left_regular_rep})}, is defined as

\vspace{0.1cm}$$
U_L^{reg}(h)|g\rangle=|hg\rangle,\qquad\forall g,h\in G,
$$

\vspace{0.3cm}\noindent when considered as the group $G$ acting on the group algebra $\mathbb{C}[G]$.\footnote{Then, we may consider that $G$ is a subgroup of the unitary group $U(|G|)$ acting on $L^2(G)=\mathbb{C}[G]$ by multiplication by unitary matrices, and apply to this situation the \hyperref[section_smily_alg]{{\color{black}\textbf{SMILY}}} algorithm.}

The matrix elements of the regular representation are obtained by computing the action of the group on the orthonormal basis $|g_i\rangle$, $i=1,\ldots,n$, of the Hilbert space $\mathcal{H}=\mathbb{C}[G]$:
\vspace{0.1cm}\begin{equation}
\big[U^{reg}_L\big]_{ij}(g)=\langle g_i| U_L^{reg}(g)| g_j\rangle=\langle g_i|gg_j\rangle=\delta_{g_ig_j^{-1}g^{-1}}.
\end{equation}

\vspace{0.1cm}\noindent Then, the matrix representation of the left regular representation $U_L^{reg}$ of the element $g_k$ can be easily computed from the table of the group written below (notice the inverse of the elements along the rows). The matrix $U^{reg}_L$ is obtained by constructing a matrix  with ones in the positions where $g_k$ appears in the table and zeros in the rest.\newpage

 \vspace{0cm}\phantomsection\label{group_table_ps}\begin{equation}\label{group_table}
 \begin{tabular}{c|ccccc}
\vspace{0.03cm}$T$&$e$&$g_1^{-1}$&\hspace{-0.1cm}$\cdots$ &\hspace{0.2cm}$\cdots$&$g_{n-1}^{-1}$\\\hline
\vspace{-0.4cm}$\phantom{a}$&$\phantom{a}$&$\phantom{a}$&\hspace{-0.1cm}$\phantom{a}$ &\hspace{0.2cm}$\phantom{a}$&$\phantom{a}$\\
$e$&$e$&$g_1^{-1}$&\hspace{-0.1cm}$\cdots$ &\hspace{0.2cm}$\cdots$&$g_{n-1}^{-1}$\\
\vspace{-0.42cm}$\phantom{a}$&$\phantom{a}$&$\phantom{a}$&\hspace{-0.1cm}$\phantom{a}$ &\hspace{0.2cm}$\phantom{a}$&$\phantom{a}$\\
$g_1$&$g_1$&$e$&\hspace{-0.1cm}$\cdots$ &\hspace{0.2cm}$\cdots$&$g_1g_{n-1}^{-1}$\\
$\vdots$&$\vdots$&$\vdots$&\hspace{-0.2cm}$\diagdots[-28]{1.4em}{.12em}$ &&$\vdots$\\
$\vdots$&$\vdots$&$\vdots$& &\hspace{0.4cm}$\diagdots[-28]{1.4em}{.12em}$&$\vdots$\\
$g_{n-1}$&$g_{n-1}$&$g_{n-1}g_1^{-1}$&\hspace{-0.1cm}$\cdots$ &\hspace{0.2cm}$\cdots$&$e$\\
 \end{tabular}\phantom{a}\begin{matrix}
\vspace{-2.93cm}\hspace{-0.2cm}.
 \end{matrix}
 \end{equation}
 \\
 
 The input of our algorithm will be the table relabeled by identifying $e$ with $1$ and $g_i$ with $i+1$. Once we get the table $T$ in the desired form, to create the matrices $\tilde{\boldsymbol{\rho}}$ in~\hyperref[rho_tilde_ps]{(\ref*{rho_tilde})} it is not necessary to write explicitly the regular representation of each element, we simply need to evaluate the random vectors $\boldsymbol{\varphi}$ on the elements of the table, i.e.:
\begin{equation}
[\tilde{\boldsymbol{\rho}}_{1,2}]_{ij}=\boldsymbol{\varphi}_{1,2}(T_{ij}),
\end{equation}
where $T_{ij}$ are the elements of the relabeled table~\hyperref[group_table_ps]{(\ref*{group_table})}.

To show the \hyperref[section_smily_alg]{{\color{black}\textbf{SMILY}}} algorithm in action, we will apply it to decompose the regular representation of two simple cases: the permutation group $S_3$ and the alternating group $A_4$.

\subsection{The decomposition of the left regular representation of the permutation group $S_3$}\label{dec_S3_arx}

 The $S_3$ group is the group of permutations of three elements and it has order six. The elements of this group can be generated with the set of transpositions $a_k=(k,k+1)$, $k=1,2$:
 \begin{equation}
 a_1^2=a_2^2=(a_1a_2)^3=e.
 \end{equation}\newpage
 
The CG matrix obtained with the \hyperref[section_smily_alg]{{\color{black}\textbf{SMILY}}} algorithm is the following:
 \vspace{-0cm}\begin{equation*}
\hspace{-0.6cm}C_1= \left(\begin{matrix}
  0.1997-0.2326i&-0.1408+0.4685i&-0.1120-0.4762i\\
 -0.3182+0.4508i&-0.0167+0.1692i&0.0063+0.1699i\\
 -0.2006-0.4508i&-0.2362-0.1847i&0.2471-0.1699i\\
 0.0302+0.1805i&-0.3809-0.3935i&-0.4041+0.3695i\\
 -0.2299+0.0521i&0.5217-0.0751i&0.5161+0.1068i\\
 0.5188-0.0000i&0.2529+0.0155i&-0.2534-0.0000i
\end{matrix} \right.
\end{equation*}
\vspace{-0.8cm}\begin{equation*}
\hspace{1.6cm}\left.\begin{matrix}
-0.1852-0.2444i&-0.4082-0.0000i&0.4082+0.0000i\\
-0.2901-0.4693i&0.4082+0.0000i&0.4082-0.0000i\\
-0.2277+0.4377i&0.4082+0.0000i&0.4082-0.0000i\\
-0.0411+0.1783i&-0.4082-0.0000i&0.4082-0.0000i\\
0.2263+0.0661i&-0.4082-0.0000i&0.4082+0.0000i\\
0.5178+0.0317i&0.4082&0.4082
\end{matrix}\right).
 \end{equation*}

\vspace{0.2cm}\hyperref[section_smily_alg]{{\color{black}\textbf{SMILY}}} decomposes the regular representation in two representations $\widehat{D}^1$ and $\widehat{D}^2$ of dimension one and multiplicity one, and another $\widehat{D}^3$ of dimension two and multiplicity two, exactly what it was expected\footnote{The notation $\widehat{D}$ used here is standard in numerical analysis and means that the corresponding object is the actual computed result of the algorithm.}. The representations obtained after applying the transformation $C_1$, written above, are the following:

$$
\hspace{0cm}\begin{array}{c!{\vrule width 1.5pt}c|c|}
S_3& \widehat{D}^1& \widehat{D}^2\\\noalign{\hrule height 1.5pt}
e&1.0000-0.0000i&1.0000-0.0000i\\\noalign{\hrule}
a_1&1.0000&-1.0000-0.0000i\\\noalign{\hrule}
a_2&1.0000&-1.0000+0.0000i\\\noalign{\hrule}
a_1a_2&1.0000-0.0000i,&1.0000-0.0000i\\\noalign{\hrule}
a_2a_1&1.0000+0.0000i&1.0000+0.0000i\\\noalign{\hrule}
a_2a_1a_2&1.0000-0.0000i&-1.0000\\\noalign{\hrule}
\end{array}
$$
\phantom{a}\\
\phantom{a}
$$
\begin{array}{c!{\vrule width 1.5pt}c|}
S_3& \widehat{D}^3\\\noalign{\hrule height 1.5pt}
\begin{matrix}
\phantom{a}\\
e\\
\phantom{a}
\end{matrix}&\begin{pmatrix}
1.0000+0.0000i&0.0000+0.0000i\\
0.0000-0.0000i&1.0000+0.0000i
\end{pmatrix}\\\noalign{\hrule}
\begin{matrix}
\phantom{a}\\
a_1\\
\phantom{a}
\end{matrix}&\begin{pmatrix}
-0.7501-0.0000i&0.6399-0.1671i\\
0.6399+0.1671i&0.7501+0.0000i
\end{pmatrix}\\\noalign{\hrule}
\begin{matrix}
\phantom{a}\\
a_2\\
\phantom{a}
\end{matrix}&\begin{pmatrix}
0.3542+0.0000i&-0.5615-0.7479i\\
-0.5615+0.7479i&-0.3542-0.0000i
\end{pmatrix}\\\noalign{\hrule}
\begin{matrix}
\phantom{a}\\
a_1a_2\\
\phantom{a}
\end{matrix}&\begin{pmatrix}
-0.5000+0.5723i&0.1945+0.6202i\\
-0.1945+0.6202i&-0.5000-0.5723i
\end{pmatrix}\\\noalign{\hrule}
\begin{matrix}
\phantom{a}\\
a_2a_1\\
\phantom{a}
\end{matrix}&\begin{pmatrix}
-0.5000-0.5723i&-0.1945-0.6202i\\
0.1945-0.6202i&-0.5000+0.5723i
\end{pmatrix}\\\noalign{\hrule}
\begin{matrix}
\phantom{a}\\
a_2a_1a_2\\
\phantom{a}
\end{matrix}&\begin{pmatrix}
0.3959-0.0000i&-0.0784+0.9149i\\
-0.0784-0.9149i&-0.3959
\end{pmatrix}\\\noalign{\hrule}
\end{array}
$$
\\

It is remarkable that these representations verify the table of the group with very good precision. 

\subsection{The decomposition of the left regular representation of the alternating group $A_4$}\label{dec_reg_A4_arx}

The alternating group $A_4$ is the group of even permutations of four elements. This group has twelve elements and it can be generated with three generators satisfying the relations:
\phantomsection\label{A4_relatios_ps}
\begin{equation}\label{A4_relatios}
a^2=b^2=c^3=(ab)^2=ac^2abc=bc^2ac=e.
\end{equation}

The left regular representation of this group has four irreducible representations, three of dimension one and one of dimension three, hence \hyperref[section_smily_alg]{{\color{black}\textbf{SMILY}}} will decompose the regular representation of this group in the three representations of dimension one with multiplicity one and in the representation of dimension three with multiplicity three. 

Below, we display the representation of dimension three obtained using \hyperref[section_smily_alg]{{\color{black}\textbf{SMILY}}}:

$$
\begin{array}{c!{\vrule width 1.5pt}c|}
A_4\;\;& \widehat{D}^4\\\noalign{\hrule height 1.5pt}
\begin{matrix}
\vspace{0.2cm}\phantom{a}\\
e\\
\vspace{0.2cm}\phantom{a}\end{matrix}&\begin{pmatrix}
1.0000    &        -0.0000 + 0.0000i & -0.0000 - 0.0000i\\
  -0.0000 - 0.0000i &  1.0000  &           0.0000 + 0.0000i\\
  -0.0000 + 0.0000i &  0.0000 - 0.0000i &  1.0000
\end{pmatrix}\\\noalign{\hrule}
\begin{matrix}
\vspace{0.2cm}\phantom{a}\\
a\\
\vspace{0.2cm}\phantom{a}\end{matrix}&\begin{pmatrix}
-0.9852     &       -0.0240 + 0.0941i &  0.1176 + 0.0789i\\
  -0.0240 - 0.0941i & -0.3653       &      0.3099 - 0.8724i\\
   0.1176 - 0.0789i &  0.3099 + 0.8724i &  0.3504
\end{pmatrix}\\\noalign{\hrule}
\begin{matrix}
\vspace{0.2cm}\phantom{a}\\
b\\
\vspace{0.2cm}\phantom{a}\end{matrix}&\begin{pmatrix}
0.6482       &     -0.2501 + 0.4766i&  -0.3940 - 0.3672i\\
  -0.2501 - 0.4766i&  -0.8242       &     -0.0464 + 0.1697i\\
  -0.3940 + 0.3672i & -0.0464 - 0.1697i & -0.8240 - 0.0000i
\end{pmatrix}\\\noalign{\hrule}
\begin{matrix}
\vspace{0.2cm}\phantom{a}\\
c\\
\vspace{0.2cm}\phantom{a}\end{matrix}&\begin{pmatrix}
-0.1137 - 0.4209i & -0.4113 - 0.2302i &  0.4649 - 0.6096i\\
  -0.0136 + 0.5419i &  0.0028 + 0.5742i  & 0.5988 - 0.1335i\\
  -0.6284 + 0.3482i  & 0.4483 - 0.4971i &  0.1110 - 0.1533i
\end{pmatrix}\\\noalign{\hrule}
\begin{matrix}
\vspace{0.2cm}\phantom{a}\\
c^2\\
\vspace{0.2cm}\phantom{a}\end{matrix}&\begin{pmatrix}
-0.1137 + 0.4209i & -0.0136 - 0.5419i & -0.6284 - 0.3482i\\
  -0.4113 + 0.2302i  & 0.0028 - 0.5742i  & 0.4483 + 0.4971i\\
   0.4649 + 0.6096i  & 0.5988 + 0.1335i &  0.1110 + 0.1533i
\end{pmatrix}\\\noalign{\hrule}
\end{array}
$$\newpage

$$
\begin{array}{c!{\vrule width 1.5pt}c|}
A_4\;\;& \widehat{D}^4\\\noalign{\hrule height 1.5pt}
\begin{matrix}
\vspace{0.2cm}\phantom{a}\\
ab\\
\vspace{0.2cm}\phantom{a}\end{matrix}&\begin{pmatrix}
-0.6631      &       0.2741 - 0.5707i &  0.2765 + 0.2883i\\
   0.2741 + 0.5707i &  0.1895       &     -0.2635 + 0.7028i\\
   0.2765 - 0.2883i & -0.2635 - 0.7028i & -0.5264
\end{pmatrix}\\\noalign{\hrule}
\begin{matrix}
\vspace{0.2cm}\phantom{a}\\
cb\\
\vspace{0.2cm}\phantom{a}\end{matrix}&\begin{pmatrix}
-0.0400 + 0.3917i &  0.4431 + 0.1902i & -0.4347 + 0.6508i\\
   0.0772 + 0.4789i & -0.3076 - 0.7107i & -0.3866 - 0.1247i\\
  -0.7438 + 0.2375i & -0.4095 + 0.0115i &  0.3475 + 0.3190i
\end{pmatrix}\\\noalign{\hrule}
\begin{matrix}
\vspace{0.2cm}\phantom{a}\\
ca\\
\vspace{0.2cm}\phantom{a}\end{matrix}&\begin{pmatrix}
0.1069 + 0.3505i  & 0.8684 + 0.3002i & -0.1455 + 0.0155i\\
   0.1273 - 0.6109i &  0.2504 + 0.2570i  & 0.6673 + 0.1914i\\
   0.5625 - 0.4001i & -0.0133 + 0.1634i & -0.3573 - 0.6075i
\end{pmatrix}\\\noalign{\hrule}
\begin{matrix}
\vspace{0.2cm}\phantom{a}\\
bc\\
\vspace{0.2cm}\phantom{a}\end{matrix}&\begin{pmatrix}
0.0468 - 0.3213i & -0.9002 - 0.2602i  & 0.1153 - 0.0567i\\
  -0.1908 - 0.4100i &  0.0544 - 0.1205i & -0.8795 + 0.0669i\\
   0.8097 - 0.1857i & -0.0255 + 0.3222i & -0.1013 + 0.4419i
\end{pmatrix}\\\noalign{\hrule}
\begin{matrix}
\vspace{0.2cm}\phantom{a}\\
bc^2\\
\vspace{0.2cm}\phantom{a}\end{matrix}&\begin{pmatrix}
-0.0400 - 0.3917i &  0.0772 - 0.4789i & -0.7438 - 0.2375i\\
   0.4431 - 0.1902i & -0.3076 + 0.7107i & -0.4095 - 0.0115i\\
  -0.4347 - 0.6508i & -0.3866 + 0.1247i &  0.3475 - 0.3190i
\end{pmatrix}\\\noalign{\hrule}
\begin{matrix}
\vspace{0.2cm}\phantom{a}\\
cbc\\
\vspace{0.2cm}\phantom{a}\end{matrix}&\begin{pmatrix}
0.1069 - 0.3505i &  0.1273 + 0.6109i  & 0.5625 + 0.4001i\\
   0.8684 - 0.3002i &  0.2504 - 0.2570i & -0.0133 - 0.1634i\\
  -0.1455 - 0.0155i  & 0.6673 - 0.1914i & -0.3573 + 0.6075i
\end{pmatrix}\\
\end{array}
$$
\\

Again, it is remarkable that the relations of the group \hyperref[A4_relatios_ps]{(\ref*{A4_relatios})} are verified with very good precision.\newpage

\section{Clebsch--Gordan coefficients for $SU(2)$}

Let $G$ be a compact Lie group and $H$ a closed subgroup (hence, compact too). Adapted states to $H$ will have the form:
\phantomsection\label{apapted_states_for_spins_ps}\begin{equation}\label{apapted_states_for_spins}
\boldsymbol{\rho}=\frac{1}{Z}\hspace{-0.15cm}\int\limits_{\hspace{.3cm}H}\hspace{-0.1cm}\chi_\rho(h)U(h)^\dagger\diff h,
\end{equation}
where $Z$ is the normalization factor
$$
Z=\hspace{-0.15cm}\int\limits_{\hspace{.3cm}H}\hspace{-0.1cm}\chi_\rho(h)\overline{\chi(h)}\diff h,
$$
and $\diff h$ denotes the invariant Haar measure on $H$.

Because our algorithm is numerical, we need to approximate the integral~\hyperref[apapted_states_for_spins_ps]{(\ref*{apapted_states_for_spins})} with a finite sum. Choosing a quadrature rule to approximate the integral \hyperref[apapted_states_for_spins_ps]{(\ref*{apapted_states_for_spins})} for a given $\boldsymbol{\rho}$ is equivalent to use another $\widehat{\boldsymbol{\rho}}$ such that $\chi_{\widehat{\rho}}(h)\neq 0$ only at a finite number of elements of the group. Then, the integral \hyperref[apapted_states_for_spins_ps]{(\ref*{apapted_states_for_spins})} for $\widehat{\boldsymbol{\rho}}$ reduces to a finite sum and the representation of $\widehat{\boldsymbol{\rho}}$ is exact. However, it could happen that the generic adapted states we obtain doing that do not have enough degrees of freedom, that is, it could happen that the block diagonal matrices of the representations would not be irreducible. But this problem can be solved by choosing a set of points large enough, for instance adaptive quadratures, because we know that the representations of a compact Lie group can be written in terms of a finite set of irreducible representations. 

The original CG problem was discussed at the \hyperref[CG_problem_beg_section]{{\color{black}\textbf{beginning of the chap-}}} \hyperref[CG_problem_beg_section]{{\color{black}\textbf{ter}}}. This problem consists on the reduction of a tensor product representation $U_A(g) \otimes U_B(g)$, $g\in G$, of two representations of the same group $G$ restricted to the diagonal subgroup of the product group. By associativity, this problem can be generalized to any number of tensor products $U_1(g) \otimes U_2(g)\otimes\cdots\otimes U_N(g)$.

The CG problem appeared for the first time studying the composition of two representations of the $SU(2)$ group related to the composition of angular momenta of two quantum systems.

The examples below, will show the reduction of a bipartite system of two spins with angular momenta $3/2$ and $1$ and the reduction of a tripartite system of three spins with momenta $1/2$, $1/2$ and $3/2$.

The angular momentum operators $ \textbf{J}_k$ satisfy the commutation relations:
\phantomsection\label{com_angular_momentum_ps}\begin{equation}\label{com_angular_momentum}
\left[\vphantom{a^\dagger}\right.\hspace{-0.1cm}\textbf{J}_k,\textbf{J}_l\hspace{-0.1cm}\left.\vphantom{a^\dagger}\right]=i\epsilon_{klm}\textbf{J}_m, \qquad k,l,m=x,y,z,
\end{equation}
and generate the Lie algebra $\mathfrak{su}(2)$ of the group $SU(2)$. Any element of a representation of $SU(2)$ can be written as:
\begin{equation}
U(\boldsymbol{\theta})=\textrm{e}^{i\boldsymbol{\theta}\cdot\textbf{J}},\qquad \theta_k\in[0,2\pi).
\end{equation}
The matrix representation of momentum $j$ of the angular momentum operators $\textbf{J}_k$ is usually written in a basis of eigenvectors of $\textbf{J}_z$:
\begin{equation}
\textbf{J}_z|j,m\rangle=m|j,m\rangle,\qquad m=j,j-1,\ldots,-j,
\end{equation}
and the representation of the operators $\textbf{J}_x$ and $\textbf{J}_y$ is usually obtained from the representation of the ladder operators $\textbf{J}_{\pm}=\textbf{J}_x\pm i\textbf{J}_y$:
\phantomsection\label{ladder_operator_ps}\begin{equation}\label{ladder_operator}
\langle j,m|\textbf{J}_{\pm}|j,m'\rangle=\sqrt{(j\mp m')(j\pm m'+1)}\,\delta_{mm'\pm 1}.
\end{equation}
For instance, if $ j=3/2$ we have:
$$
\textbf{J}_x=\begin{pmatrix}
0 &\frac{\sqrt{3}}{2}& 0& 0\\
\frac{\sqrt{3}}{2} &0 &1& 0\\
0&1& 0&\frac{\sqrt{3}}{2}\\
0 &0& \frac{\sqrt{3}}{2} & 0
\end{pmatrix},\qquad \textbf{J}_y=\begin{pmatrix}
0 &-i\frac{\sqrt{3}}{2}& 0& 0\\
i\frac{\sqrt{3}}{2} &0 &-i& 0\\
0&i& 0&-i\frac{\sqrt{3}}{2}\\
0 &0& i\frac{\sqrt{3}}{2} & 0
\end{pmatrix},
$$
$$
\textbf{J}_z=\begin{pmatrix}
\frac{3}{2}&0&0&0\\
0&\frac{1}{2}&0&0\\
0&0&-\frac{1}{2}&0\\
0&0&0&-\frac{3}{2}
\end{pmatrix},
$$
in the standard basis
\begin{alignat*}{2}
|3/2,3/2\rangle&=\begin{pmatrix}
1\\
0\\
0\\
0
\end{pmatrix},& \hspace{1.5cm}|3/2,1/2\rangle&=\begin{pmatrix}
0\\
1\\
0\\
0
\end{pmatrix},\\
|3/2,-1/2\rangle&=\begin{pmatrix}
0\\
0\\
1\\
0
\end{pmatrix},& \hspace{1.5cm}|3/2,-3/2\rangle&=\begin{pmatrix}
0\\
0\\
0\\
1
\end{pmatrix}.
\end{alignat*}

The standard CG matrix is constructed with eigenvectors of the total angular momentum operator ${\textbf{J}_T}$ with respect to the $z$ component:
\begin{multline}
{\textbf{J}_T}_{\!z}={\textbf{J}_1}_z\otimes\mathds{1}_2\otimes\cdots\otimes\mathds{1}_n+\mathds{1}_1\otimes{\textbf{J}_2}_z\otimes\cdots\otimes\mathds{1}_n+\cdots\\
+\mathds{1}_1\otimes\mathds{1}_2\otimes\cdots\otimes {\textbf{J}_N}_{\!z},
\end{multline}
where $N$ is the number of parts of the system. The eigenvectors of this operator are usually denoted by $|J,M\rangle$, where $J$ represents the total angular momentum and $M=J,J-1,\ldots,-J$:
\begin{equation}
{\textbf{J}_T}_{\!z}|J,M\rangle=M|J,M\rangle.
\end{equation}

The standard procedure to obtain the CG matrix consists in applying successively the ladder operator $\textbf{J}_-$ (or $\textbf{J}_+$) starting from the state of maximum (or minimum) momentum $M$, $|J_{max},M_{max}\rangle=|j_1+j_2,j_1+j_2\rangle$ (or $|J_{max},M_{min}\rangle=|j_1+j_2,-j_1-j_2\rangle$). Because of this, if we come back to the equation~\hyperref[ladder_operator_ps]{(\ref*{ladder_operator})}, because the matrix elements of the ladder operators are real, the Clebsh--Gordan coefficients are real.

Let us recall that the CG matrix provided by \hyperref[section_smily_alg]{{\color{black}\textbf{SMILY}}} is written in terms of the eigenvectors of the first adapted state $\boldsymbol{\rho}_1$. Thus, if we want to compare the Clebsh--Gordan coefficients obtained with our algorithm with the standard ones, we have to find a CG matrix $C_1$ which is conformed by eigenvectors of the operator ${\textbf{J}_T}_{\!z}$. To do that, first we will create two real adapted states. To create them, we will use that the operators $\textbf{J}_k$ verify:
\begin{equation}
\overline{\textbf{J}}_x=\textbf{J}_x,\quad \overline{\textbf{J}}_y=-\textbf{J}_y,\quad \overline{\textbf{J}}_z=\textbf{J}_z.
\end{equation}
Therefore, for any adapted state $\boldsymbol{\rho}$, $\overline{\boldsymbol{\rho}}$ is an adapted state too. To create a real adapted state from a complex one, we will add to the matrix $\tilde{\boldsymbol{\rho}}$ in~\hyperref[rho_tilde_ps]{(\ref*{rho_tilde})} its complex conjugate to obtain a real symmetric matrix and after that, we will multiply the result by its transpose to make it definite positive. Then finally, we will divide by the trace to normalize it:
\phantomsection\label{real_states_ps}\begin{equation}\label{real_states}
\boldsymbol{\tau}=\tilde{\boldsymbol{\rho}}+\overline{\tilde{\boldsymbol{\rho}}},\qquad \boldsymbol{\rho}_{real}=\frac{1}{\Tr(\boldsymbol{\tau}\boldsymbol{\tau}^t)}\boldsymbol{\tau}\boldsymbol{\tau}^t.
\end{equation}

Once we have two real adapted states $\boldsymbol{\rho}_{1\,real}$ and $\boldsymbol{\rho}_{2\,real}$, we apply our algorithm to get the real CG matrix $C_1$. After obtaining that real CG matrix $C_1$, we will transform the operator ${\textbf{J}_T}_{\!z}$ with $C_1$ to decompose it in irreducible representations:
\phantomsection\label{Jz_decomposition_ps}\begin{equation}\label{Jz_decomposition}
C_1^\dagger{\textbf{J}_T}_{\!z} C_1=\left(
\begin{array}{ccccc}
\begin{array}{|cc|}\hline
 * &  *\\
 * &  *\\\hline
\end{array}
 & &  &  &\\
 &\hspace{-0.342cm} \begin{array}{|cccc|}\hline
 *& * & * &  *\\
  *&  *&  *& * \\
  *& * & * & * \\
  *& * & * & *\\\hline
\end{array} &   & &\\
& &\hspace{-0.1cm}\diagdots[-40]{1.4em}{.12em} & &\\
& & &\hspace{0.0cm}\diagdots[-40]{1.4em}{.12em} &\\
& & & &\hspace{-0.2cm}\begin{array}{|ccc|}\hline
* & *& *\\
 * & *& *\\
 * & *& *\\\hline
\end{array}
 \end{array}
\right)
\end{equation}
and after that, we will diagonalize each block of this matrix transforming it with a block diagonal matrix $V_z$, $V_z^\dagger C_1^\dagger{\textbf{J}_T}_{\!z} C_1V_z$, which reorders the eigenvalues as follows:
\\
\\
\\
\\
\\
\phantomsection\label{order_eigs_Jzt_ps}\begin{equation}\label{order_eigs_Jzt}
\hspace{0.5cm}\begin{pmatrix}
j_1 & & & & & & & & &\\
& \hspace{-0.3cm}j_1-1 & & & & & & & &\\
& & \hspace{-0.4cm}\diagdots[-40]{1.4em}{.12em}  & & & & & & & \\
& & & \hspace{-0.4cm}-j_1 & & & & & & \\
& & & &\hspace{-0.2cm} j_2 & & & & & \\
& & & & & \hspace{-0.3cm}j_2-1 & & & &\\
& & & & & &\hspace{-0.6cm}\diagdots[-40]{1.4em}{.12em} & & & \\
& & & & &  & & \hspace{-0.5cm}-j_2 & & \\
& & & & &  & & &\hspace{-0.1cm}\diagdots[-40]{1.4em}{.12em}  & \\
& & & & &  & & & &\hspace{-2.8cm}\diagdots[-40]{1.4em}{.12em}   \\
& & & & &  & & & &\hspace{-1.7cm}j_N \\
& & & & &  & & & &\hspace{0.0cm}j_N-1 \\
& & & & &  & & & &\hspace{0.9cm}\diagdots[-40]{1.4em}{.12em}   \\
& & & & &  & & & &\hspace{2.4cm}\phantom{aaa} \\
\end{pmatrix}.\begin{matrix}
\vspace{-6.2cm}\hspace{-1.4cm}-j_N
\end{matrix}
\end{equation}
Therefore, the CG matrix whose columns are the eigenvectors of ${\textbf{J}_T}_{\!z}$ reordered in this way is given by:
\begin{equation}
C_z=C_1V_z.
\end{equation}

\subsection{Clebsh--Gordan coefficients for the spin system $3/2\otimes1$}\label{Clebsh_G_coeff_arx_1}

Suppose that we have a system of two particles in which the first particle has momentum $3/2$ and the second momentum $1$. It is well known \hyperref[Ga90_ps]{\RED{[\citen*{Ga90}, ch.\,5]}} that this system is decomposed in the direct sum of systems of momentum $5/2$, $3/2$ and $1/2$, each one with multiplicity one:
$$
3/2\otimes 1=5/2\oplus 3/2\oplus 1/2,
$$
or, in other words, that the representation of $SU(2)$ corresponding to the tensor product $3/2\otimes 1$ has irreducible representations $5/2$, $3/2$ and $1/2$ with multiplicity one.

To create the adapted states that will be the input of our algorithm, we have chosen $7$ random angles $\theta_{xi}$, $\theta_{yj}$, $\theta_{zk}$ for each axis $x$, $y$, $z$ and then, approximate the integral~\hyperref[apapted_states_for_spins_ps]{(\ref*{apapted_states_for_spins})} by the sum:
$$
\tilde{\boldsymbol{\rho}}_{1,2}=\hspace{-0.2cm}\sum\limits_{j,k,l=1}^{7}\hspace{-0.1cm}\varphi_{1,2}(j,k,l)U^{3/2}(\theta_{xi},\theta_{yj},\theta_{zk})\otimes U^{1}(\theta_{xi},\theta_{yj},\theta_{zk}),
$$
where $\varphi_{1,2}(j,k,l)$ are the elements of two random vectors of size $7^3$.

To represent the CG coefficients, we will use the following standard arrengement:
\begin{center}
\begin{figure*}[htbp]
\centering
\includegraphics{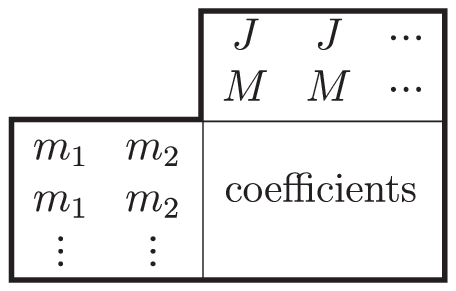}
\end{figure*}
\end{center}

\vspace{-0.5cm}The coefficients obtained for the system $3/2\otimes 1$ applying the \hyperref[section_smily_alg]{{\color{black}\textbf{SMILY}}} algorithm are:

\vspace{0.3cm}\begin{figure*}[htbp]
\centering
\includegraphics{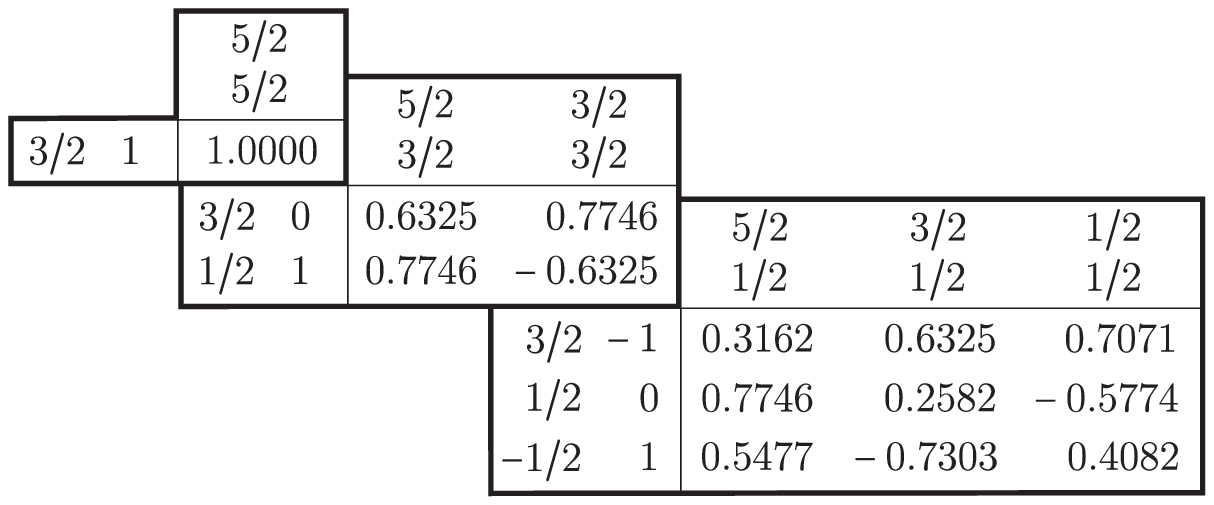}
\end{figure*}\newpage

\begin{figure}[htbp]
\centering
\includegraphics{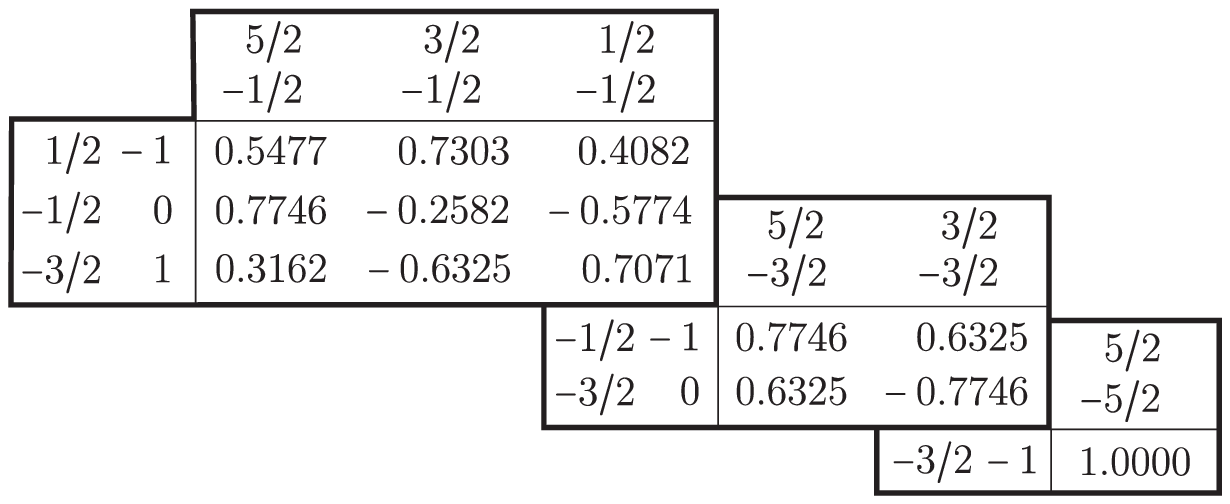}
\caption{\hfil\MYGREEN{Figure 3.5.1}: CG coefficients for $3/2\otimes 1$.\hfil}
\label{CG_table_1}
\end{figure}

Comparing this table with the standard one in \cite{Wi59}, we can see that the coefficients we have obtained are the same up to the precision of the numerical computations.

\subsection{Clebsh--Gordan coefficients for the spin system $1/2\otimes 1/2\otimes 3/2$}\label{Clebsh_G_coeff_arx_2}

To test the capabilities of the \hyperref[section_smily_alg]{{\color{black}\textbf{SMILY}}} algorithm, we will compute the CG coefficients of a system of  three particles with spin. These coefficients can be obtained from suitable choices of coefficients of products of two spins, however there are not tables for systems with more than two particles.

The standard procedure consists on reducing the representation of the first two particles, and after that, reducing the result with the next particle, and so on till we have finished. In this case, the product of three particles with spin $1/2$, $1/2$ and $3/2$ yields:
\phantomsection\label{product_3_spins_ps}\begin{equation}\label{product_3_spins}
1/2\otimes 1/2\otimes 3/2=(0\oplus 1)\otimes 3/2=3/2\oplus 5/2\oplus 3/2\oplus 1/2.
\end{equation}

In the first step, we diagonalize in blocks the first two spins:
\begin{equation*}
(C_{1/2\otimes1/2}\otimes\mathds{1}_4)^\dagger(U^{1/2}\otimes U^{1/2}\otimes U^{3/2})(C_{1/2\otimes1/2}\otimes\mathds{1}_4)
=\big(U^{0}\oplus U^{1}\big)\otimes U^{3/2}
\end{equation*}
and secondly, we diagonalize the result:\newpage

\phantom{a}
\vspace{-1.1cm}\begin{multline*}
\begin{pmatrix}
                   \mathds{1}_4 &0  \\
                    0 &C_{1\otimes 3/2}^\dagger
\end{pmatrix}\left((U^{0}\oplus U^{1})\otimes U^{3/2}\right)\begin{pmatrix}
                   \mathds{1}_4 &0  \\
                    0 &C_{1\otimes 3/2}
\end{pmatrix}\\
=U^{3/2}\oplus U^{5/2}\oplus U^{3/2}\oplus U^{1/2}.
\end{multline*}
Therefore, the CG matrix of this system is
\phantomsection\label{Clebsh_three_spins_ps}\begin{equation}\label{Clebsh_three_spins}
C_{1/2\otimes 1/2\otimes 3/2}=(C_{1/2\otimes 1/2}\otimes\mathds{1}_4)(\mathds{1}_4\oplus C_{1\otimes 3/2}).
\end{equation}

In this example, we see that for a multipartite system of spins, the multiplicities of the representations can be greater than one, as it can be seen in~\hyperref[product_3_spins_ps]{(\ref*{product_3_spins})}, then it may exist several eigenvectors with the same values of $J$ and $M$, so it is necessary to add another ``quantum number'', that we will denote by $c$, to differentiate them. This ``quantum number'' will be a label that indicates for which copy of the representation of multiplicity bigger than one belongs each eigenvector with the same $J$ and $M$. For that reason the symbol $c$ because it refers to the multiplicity and was used before for this purpose \hyperref[decompositionU_ps]{(\ref*{decompositionU})}. 

Using \hyperref[section_smily_alg]{{\color{black}\textbf{SMILY}}}, we do not need to group the system in groups of bipartite systems as in \hyperref[Clebsh_three_spins_ps]{(\ref*{Clebsh_three_spins})}, it can be done in one step. Again in this case, we have used in our algorithm 7 random angles for each axis to get the adapted states for the product of the three irreps. The coefficients will be represented in similar arrengements to the case of two spins but including the label $c$:
\begin{center}
\begin{figure*}[htbp]
\centering
\includegraphics{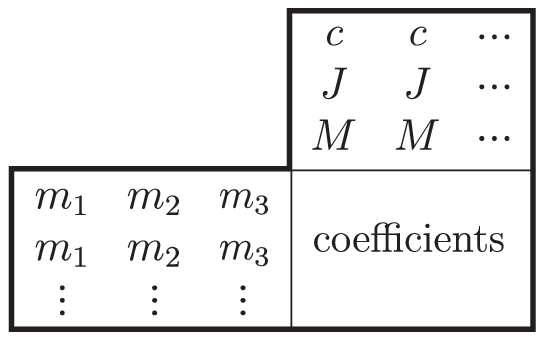}
\end{figure*}
\end{center}

\vspace{-0.5cm}The coefficients obtained for the tripartite system $1/2\otimes 1/2\otimes 3/2$ are the following:

\vspace{0cm}\begin{figure*}[htbp]
\centering
\includegraphics{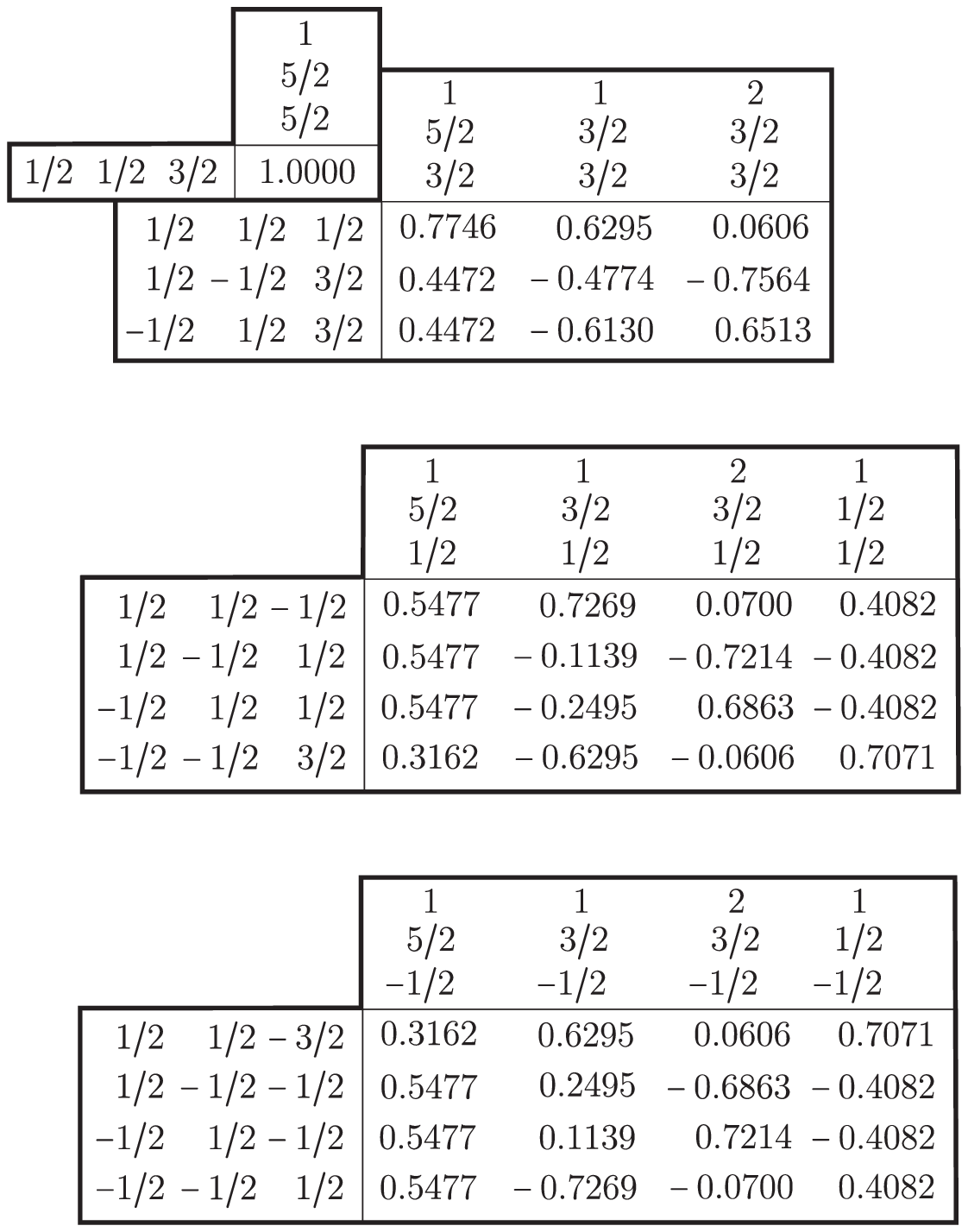}
\end{figure*}

\begin{figure}[htbp]
\centering
\includegraphics{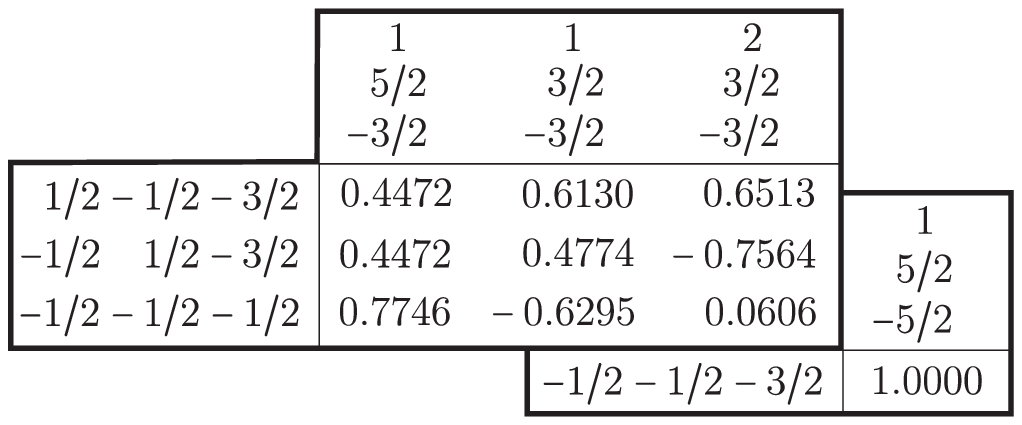}
\caption{\hfil\MYGREEN{Figure 3.5.2}: CG coefficients for $1/2\otimes 1/2\otimes 3/2$.\hfil}
\label{CG_table_2}
\end{figure}
\newpage

Let us recall that these coefficients are different to the coefficients that are obtained with the formula~\hyperref[Clebsh_three_spins_ps]{(\ref*{Clebsh_three_spins})} with the standard CG tables of two spins. The reason for this is that, in general, for systems of more than two spins, the choice of real Clebsh--Gordan coefficients of eigenvectors that only differ in the multiplicity $c$ is not unique. Because of that, there exists more than one linear combination that gives a valid CG matrix that diagonalizes ${\textbf{J}_{T}}_{\!z}$ with the eigenvalues reordered in the way given in \hyperref[order_eigs_Jzt_ps]{(\ref*{order_eigs_Jzt})}.

\section{A proposal for the experimental construction of adapted states}\label{conclusion_create_states}

We have shown that to implement \hyperref[section_smily_alg]{{\color{black}\textbf{SMILY}}} we only just need two generic adapted states of the form \hyperref[adapted_state_a]{$(2.7.16)$}. Numerically, as it was indicated in the step \hyperref[SMILY_step_1]{{\color{mybrown}{1}}} of the algorithm, we can create such states using just the unitary representation of the group we want to reduce, however thinking on the possible implementation of \hyperref[section_smily_alg]{{\color{black}\textbf{SMILY}}} in a quantum computer, we will discuss here an experimental setting to implement such states.

The proposed way to do it will be to measure the tomograms of certain clever configurations of coherent states and reconstruct the adapted states by means of formulas \hyperref[adapted_state_a]{$(2.7.16)$}, where the smeared characters $\chi_\rho(g)$ are recovered from the Inverse Fourier Transform of the tomograms $\mathcal{W}_\rho(X;\xi)$ (see \hyperref[section_recosntruction_states_groups]{{\color{black}\textbf{chapter~\ref*{chap_tom_qu}}}}, eq.\,\hyperref[smeared_character_Fourier_new_ps]{(\ref*{smeared_character_Fourier_new})}). 

To achieve it, we will use the Jordan--Schwinger map. The main idea behind it is that the Lie algebra of $n\times n$ complex matrices $\mathfrak{gl}(n,\mathbb{C})$ can be naturally represented in the Fock space $\mathcal{F}_n$ generated by $n$ creation and annihilation operators $a^\dagger_k,a_k$ (see sections \hyperref[can_section_fock_ch_1]{{\color{black}\textbf{\ref*{section_Quantization_elec}}}} and \hyperref[B_S_chap_2]{{\color{black}\textbf{\ref*{holom_chap_2}}}}). The map $S_J\!:\mathfrak{gl}(n,\mathbb{C})\rightarrow\mathfrak{L}(\mathcal{F}_n)$ given by:
\phantomsection\label{Jordan_schwinger_map_ch3_ps}
\begin{equation}\label{Jordan_schwinger_map_ch3}
S_J(\xi)=\tilde{\boldsymbol{\xi}}=\hspace{-0.1cm}\sum_{i,j=1}^na_i^\dagger\xi_{ij}a_j
\end{equation}
defines a Lie algebra homomorphism between the Lie algebra $\mathfrak{gl}(n,\mathbb{C})$ and the Lie algebra of operators on $\mathcal{F}_n$.

If we compute the commutator of two operators $\tilde{\boldsymbol{\xi}}$ and $\tilde{\boldsymbol{\zeta}}$, we get:
\begin{equation*}
\left[\vphantom{a^\dagger}\right.\hspace{-0.1cm}\tilde{\boldsymbol{\xi}},\tilde{\boldsymbol{\zeta}}\hspace{-0.1cm}\left.\vphantom{a^\dagger}\right]=\hspace{-0.1cm}\sum_{\substack{i,j=1\\r,s=1}}^n\hspace{-0.05cm}\xi_{ij}\zeta_{rs}\left[\vphantom{a^\dagger}\right.\hspace{-0.1cm}a_i^\dagger a_j,a_r^\dagger a_s\hspace{-0.1cm}\left.\vphantom{a^\dagger}\right],
\end{equation*}
then, using the commutation relations of the creation and annihilation operators~\hyperref[Commutator_creation_annihilation_f_ps]{(\ref*{Commutator_creation_annihilation_f})}, we get:
\phantomsection\label{correspondence_bosonic_Lie-algebra_ps}
\begin{equation}\label{correspondence_bosonic_Lie-algebra}
\left[\vphantom{a^\dagger}\right.\hspace{-0.1cm}\tilde{\boldsymbol{\xi}},\tilde{\boldsymbol{\zeta}}\hspace{-0.1cm}\left.\vphantom{a^\dagger}\right]=\hspace{-0.1cm}\sum_{\substack{i,j=1\\r,s=1}}^n\hspace{-0.05cm}\xi_{ij}\zeta_{rs}\left(a_i^\dagger a_s\delta_{jr}-a_r^\dagger a_j\delta_{is}\right)=\hspace{-0.1cm}\sum_{i,j=1}^n\hspace{-0.05cm}a_i^\dagger\left[\vphantom{a^\dagger}\right.\hspace{-0.1cm}\xi,\zeta\hspace{-0.1cm}\left.\vphantom{a^\dagger}\right]_{ij}a_j=\tilde{\boldsymbol{\eta}}.
\end{equation}
where $\left[\vphantom{a^\dagger}\right.\hspace{-0.1cm}\xi,\zeta\hspace{-0.1cm}\left.\vphantom{a^\dagger}\right]=\eta$.

Now, because of Ado's theorem \hyperref[Ja62_ps]{\RED{[\citen*{Ja62}, ch.\,6]}}, any finite dimensional Lie algebra $\mathfrak{g}$ can be considered as a subalgebra of the algebra $\mathfrak{gl}(n,\mathbb{C})$ for $n$ large enough, hence we can represent the Lie algebra $\mathfrak{g}$ by using the Jordan--Schwinger map $S_J$ as Hermitean operators on the Fock space $\mathcal{F}_n$.

Notice that this result does not depend directly on the commutation relations of the Heisenberg--Weyl algebra, it depends on the standard commutator
\begin{equation}
\left[\vphantom{a^\dagger}\right.\hspace{-0.1cm}a_i^\dagger a_j,a_r^\dagger a_s\hspace{-0.1cm}\left.\vphantom{a^\dagger}\right]=a_i^\dagger a_s\delta_{jr}-a_r^\dagger a_j\delta_{is},
\end{equation} 
hence, for any set of operators $X_{ij}$ satisfying
\begin{equation}
\left[\vphantom{a^\dagger}\right.\hspace{-0.1cm}X_{ij},X_{rs}\hspace{-0.1cm}\left.\vphantom{a^\dagger}\right]=X_{is}\delta_{jr}-X_{rj}\delta_{is},
\end{equation} 
the result~\hyperref[correspondence_bosonic_Lie-algebra_ps]{(\ref*{correspondence_bosonic_Lie-algebra})} holds.

Recall that we can obtain the tomogram of $a^\dagger a$ on a given state from the data gathered by a photodetector (see  \hyperref[Intensity_photons]{{\color{black}\textbf{section~\ref*{section_photodetection}}}}), therefore if we mix in a convenient way beam-splitters and photodetectors, we can get the tomogram of the desired combination of creation and annihilation operators $a_k^\dagger a_{k'}$ for any $k$ and $k^\prime$. 

To see how to implement this idea, we will show the configuration needed to measure the tomograms of a particle of spin one-half corresponding to the $x$, $y$ and $z$ components. Then, from linear combinations of them, we will be able to obtain the tomograms of the representations of the elements of the Lie algebra $\mathfrak{su}(2)$.

In \hyperref[section_Spin_Tomography]{{\color{black}\textbf{section~\ref*{section_Spin_Tomography}}}}, the spin operators $\textbf{S}_i$, eq.\,\hyperref[spin_operators_tom_ps]{(\ref*{spin_operators_tom})}, $i=x,y,z$, where introduced. It is obvious, because they are a realization of $\mathfrak{su}(2)$, that they satisfy the commutation relations of the angular momentum~\hyperref[com_angular_momentum_ps]{(\ref*{com_angular_momentum})}. 

Using the Jordan--Schwinger map, we can write them in terms of creation and annihilation operators:
\begin{align}
\tilde{\textbf{S}}_x&=\frac{1}{2}(a_1^\dagger a_2+a_2^\dagger a_1),\qquad\tilde{\textbf{S}}_y=-\frac{i}{2}(a_1^\dagger a_2-a_2^\dagger a_1),\nonumber\\
&\hspace{1.6cm}\tilde{\textbf{S}}_z=\frac{1}{2}(a_1^\dagger a_1-a_2^\dagger a_2).
\end{align}
Let us remember that we have already found the representation of the algebra $\mathfrak{su}(2)$ in the implementation of the homodyne and heterodyne detectors in \hyperref[ph_algebra_su2]{{\color{black}\textbf{section~\ref*{section_hom_det_clas}}}}. 

Thus, the configurations to get the tomograms associated to spin $x$ and $y$ are similar to the configuration of the homodyne detector, {\changeurlcolor{mygreen}\hyperref[Q_homodyne]{Figure \ref*{Q_homodyne}}}, the only difference is that now, the inputs are two laser beams with coherent factors $z_1$ and $z_2$ excited at the same frequency, instead of a radiation source with state $\boldsymbol{\rho}$ and a strong laser beam $|z\rangle\langle z|$.

To get the tomogram of $\tilde{\textbf{S}}_x$, we mix the laser beams with annihilation operators associated $a_1$ and $a_2$ in a $50/50$ beam-splitter, then the outputs at the beam-splitter give:
\begin{equation}
c_1=\frac{1}{\sqrt{2}}(a_1-a_2),\qquad c_2=\frac{1}{\sqrt{2}}(a_1+a_2).
\end{equation}
Thus, the photodetectors on each output will give $\langle c_1^\dagger c_1\rangle_1$ and $\langle c_2^\dagger c_2\rangle_2$, where the subindexes denote the mean value with respect to the states $|z_1\rangle\langle z_1|$ and $|z_2\rangle\langle z_2|$ respectively. Finally, substracting the measurements of both photodetectors and dividing by two the result, we get the statistics for the operator $\tilde{\textbf{S}}_x$, {\changeurlcolor{mygreen}\hyperref[Sx]{Figure \ref*{Sx}}}:
\begin{figure}[h]
\centering
\includegraphics{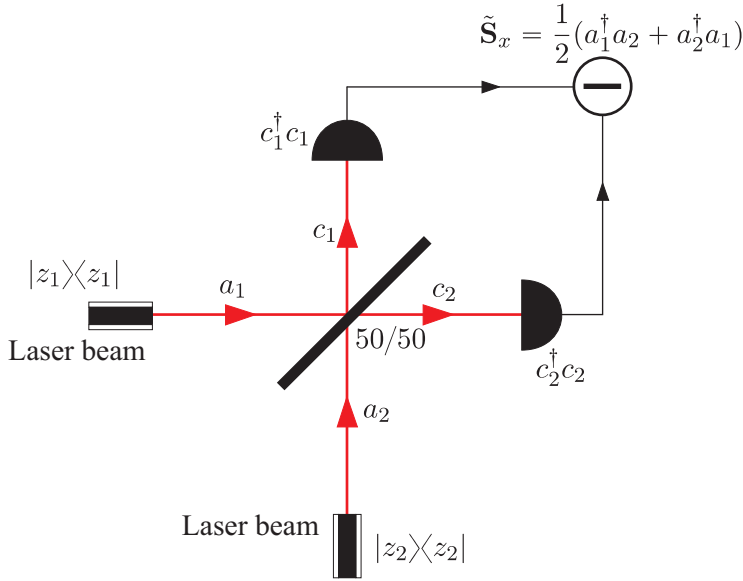}
\caption{\hfil\MYGREEN{Figure 3.6.1}: Configuration to obtain the tomogram of $\tilde{\textbf{S}}_x$.\hfil}
\label{Sx}
\end{figure}

The configuration to obtain the tomogram of $\tilde{\textbf{S}}_y$ is completely similar to the previous one, the only difference is that we have to add a phase factor of $
$ to the second laser beam, {\changeurlcolor{mygreen}\hyperref[Sy]{Figure \ref*{Sy}}}. Thus:
\begin{equation}
c_1=\frac{1}{\sqrt{2}}(a_1-ia_2),\qquad c_2=\frac{1}{\sqrt{2}}(a_1+ia_2).
\end{equation}

\phantom{a}                                                                                                                                                        
\vspace{-1cm}\begin{figure}[htbp]
\centering
\includegraphics{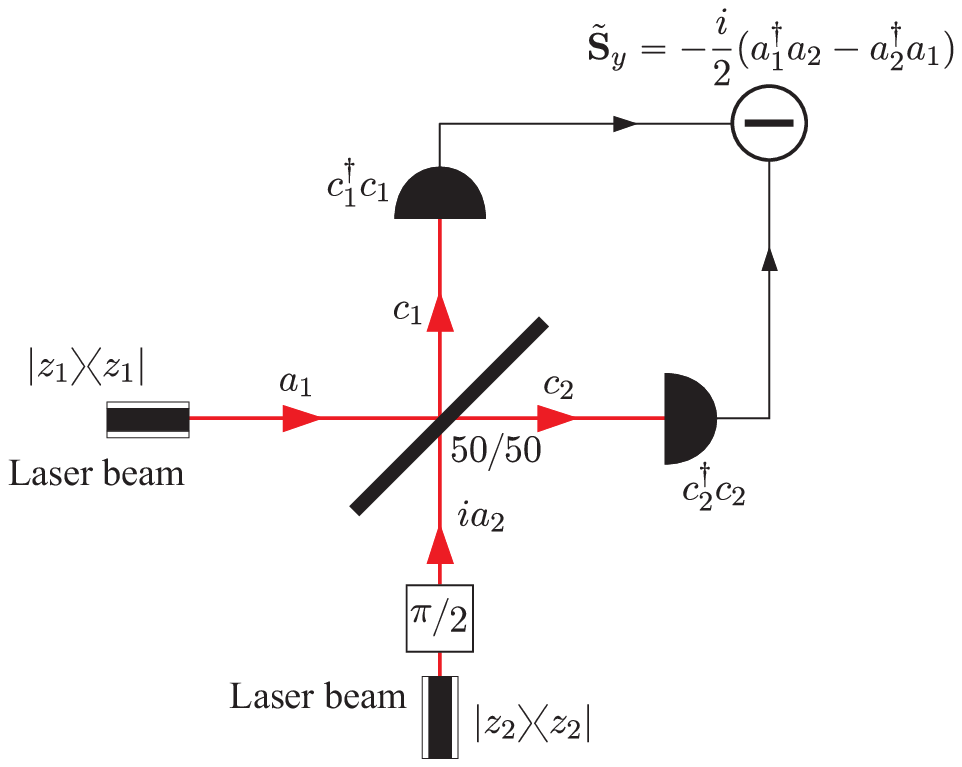}
\caption{\hfil\MYGREEN{Figure 3.6.2}: Configuration to obtain the tomogram of $\tilde{\textbf{S}}_y$.\hfil}
\label{Sy}
\end{figure}

Finally, the configuration to get the tomogram of $\tilde{\textbf{S}}_z$ is simpler than the previous ones, because we do not need a beam-splitter. We just need two photodetectors to measure the intensity of both laser beams $\langle a_1^\dagger a_1\rangle_1$, $\langle a_2^\dagger a_2\rangle_2$, substract them and divide the result by two, {\changeurlcolor{mygreen}\hyperref[Sz]{Figure \ref*{Sz}}}.


To conclude this section and chapter, we just make a comment about the experimental implementation of finite groups. We have shown that there is a correspondence between elements of the Lie algebra $\mathfrak{gl}(n,\mathbb{C})$ and the quadratic bosonic algebra $\mathfrak{L}(\mathcal{F}_n)$ generated by creation and annihilation operators. A finite group does not define a Lie algebra, however given a finite group, we can associate a Lie algebra to it. Thus, given the finite group $G$, consider its group algebra $\mathbb{C}[G]$ which is a finite dimensional Hilbert space of dimension $|G|$. Then, $G$ is faithfully represented on $\mathbb{C}[G]$ by unitary operators by using the left regular representation, hence we may consider $G$ as a subgroup of the group of unitary operators on $\mathbb{C}[G]$, $U(\mathbb{C}[G])\cong U(|G|)$.

\vspace{0.2cm}\begin{figure}[htbp]
\centering
\includegraphics{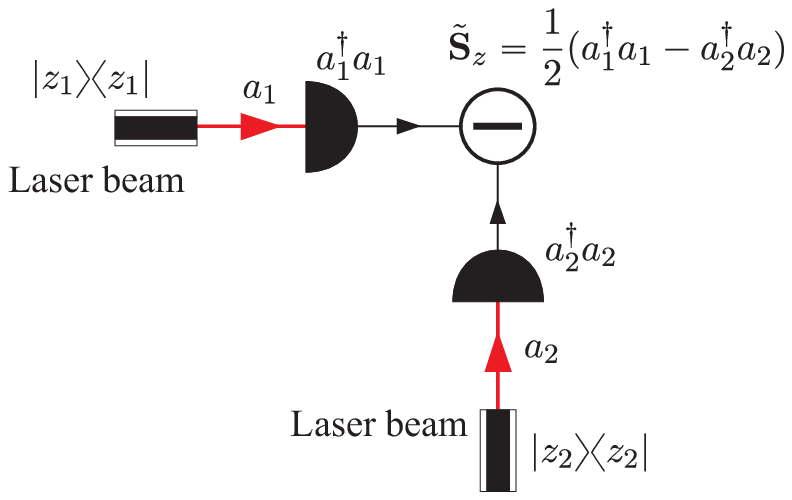}
\caption{\hfil\MYGREEN{Figure 3.6.3}: Configuration to obtain the tomogram of $\tilde{\textbf{S}}_z$.\hfil}
\label{Sz}
\end{figure}

Then, to get the elements of the ``Lie algebra'' of a finite group is very simple. We just have to consider that any unitary representation of a finite group can be written as the imaginary part of a Hermitian matrix by using the embedding of $G$ in $U(|G|)$ explained before:
\begin{equation}
U(g)=\e^{i\boldsymbol{\xi}},\qquad\boldsymbol{\xi}=\boldsymbol{\xi}^\dagger,
\end{equation}
hence, from the logarithm of the representation, we obtain the corresponding Hermitean operator:
\begin{equation}
\boldsymbol{\xi}=-i\log U(g),
\end{equation}
and finally, the elements of the bosonic algebra are obtained from the Jordan--Schwinger map $S_J$, eq.\,\hyperref[Jordan_schwinger_map_ch3_ps]{(\ref*{Jordan_schwinger_map_ch3})}.

%% file: Tesischap4.tex
\chapter{The tomographic picture of classical systems: finite and infinite dimensional}\label{chap_clas}
\markboth{The tomographic picture of classical systems}{}

\section{Classical Lagrangian and Hamiltonian systems}\label{section_Lagrangian}

The evolution of a large class of classical systems with a finite number of degrees of freedom can be obtained from the principle of least action. The action of a system is a functional $S:\mathcal{F}\rightarrow\mathds{R}$ on the space $\mathcal{F}$ of smooth curves $\boldsymbol{q}(t)=\big(q^1(t),\ldots,q^n(t)\big)$ in the configuration space $\mathcal{Q}$ of the system, which is a smooth manifold of dimension $n$ with local coordinates $q^i$, $i=1,\ldots,n$, and it is defined as the integral of the Lagrangian of the system $L(\boldsymbol{q}(t),\dot{\boldsymbol{q}}(t),t)$ in the interval of time $[t_0,t_1]$:
\begin{equation}
S[\boldsymbol{q}(t)]=\hspace{-0.2cm}\int\limits_{\hspace{0.4cm}t_0}^{\hspace{0.4cm}t_1}\hspace{-0.1cm}L(\boldsymbol{q}(t),\dot{\boldsymbol{q}}(t),t)\textrm{d}t.
\end{equation}

The \textit{principle of least action} (see for instance \cite{La69}) asserts that the evolution of a dynamical system follows a path in which the action is extremal, i.e.,
\phantomsection\label{Least_action_ps}\begin{equation}\label{Least_action}
\delta S=0,
\end{equation}
where $\delta$ in this context represents the differential of $S$.
\begin{figure}[h]
\centering
\includegraphics{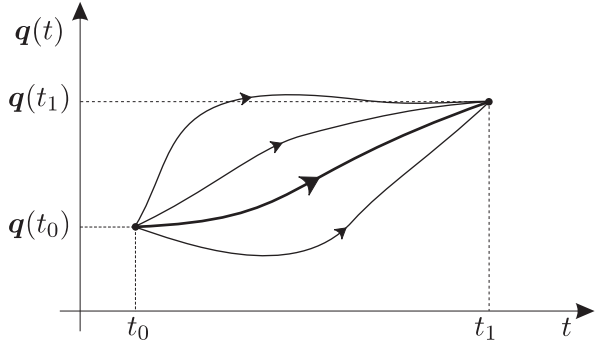}
\caption{\MYGREEN{Figure 4.1.1}: Various paths for $\boldsymbol{q}(t)$ with fixed end points at $t_0$ and $t_1$. The thick line indicates the path in which the action $S$ is extremal.}
\label{Least_action_fig}
\end{figure}

The differential of a functional $F[\phi]$, defined on functions $\phi(\boldsymbol{x})$ on a manifold $\mathcal{M}$, in the direction of the variation $\delta\phi(\boldsymbol{x})$ is defined as provided that the limit exists, as:
\begin{equation*}
\delta F_\phi[\delta\phi]=\lim_{\epsilon\rightarrow 0}\frac{F[\phi+\epsilon\delta\phi]-F[\phi]}{\epsilon}\,.
\end{equation*}
If $F$ is a local functional, it can be written as:
\phantomsection\label{functional_derivative_ps}\begin{equation}\label{functional_derivative}
\delta F_\phi[\delta\phi]=\hspace{-0.2cm}\int\limits_{\hspace{0.3cm}\mathcal{M}}\frac{\delta F}{\delta\phi(\boldsymbol{x})}\delta\phi(\boldsymbol{x})\diff\boldsymbol{x},
\end{equation}

\phantom{a}
\vspace{-0.3cm}\noindent with $\displaystyle{\frac{\delta F}{\delta\phi(\boldsymbol{x})}}$ a functional usually called the \textit{variational derivative} of $F$.

\phantom{a}
\vspace{-0.2cm}
\phantomsection
\label{section_motion}
The equations of motion for a Lagrangian system can be obtained applying this principle and are called \textit{Euler--Lagrange} equations:
\phantomsection\label{Euler-Lagrange_ps}\begin{equation}\label{Euler-Lagrange}
\frac{\delta S}{\delta q^i(t)}=\frac{\textrm{d}}{\textrm{d}t}\left(\frac{\partial L}{\partial\dot{q}^i}\right)-\frac{\partial L}{\partial q^i}=0,\qquad i=1,\ldots n.
\end{equation}

Often and for different reasons, it is more convenient to express the equations of motion in Hamiltonian form. The Hamiltonian of the system can be written in terms of the Lagrangian function as:
\phantomsection\label{Hamiltinian_ps}\begin{equation}\label{Hamiltinian}
H=\frac{\partial L}{\partial \dot{q}^i}\dot q^i-L.
\end{equation}
The \textit{canonical momentum} corresponding to the local coordinate $q^i$ is defined as:
\phantomsection\label{canonical_momentum_ps}\begin{equation}\label{canonical_momentum}
p_i=\frac{\partial L}{\partial\dot{q}^i}\,,
\end{equation}
thus, the equations of motion can be written in Hamiltonian form as follows:
\phantomsection\label{Hamilton_eq_ps}\begin{equation}\label{Hamilton_eq}
\dot{q}^i=\frac{\partial H}{\partial p_i},\qquad\dot{p}_i=-\frac{\partial H}{\partial q^i}.
\end{equation}

\section{The tomographic picture of classical Hamiltonian systems}\label{sec_tom_pic_clas}

The statistical states of a classical Hamiltonian system with finite degrees of freedom can be described by a probability density $\rho(\omega)$ on its phase space $\omega\in\Omega$, \RED{[}\citen{Is71}\RED{,}\hspace{0.05cm}\citen{Re98}\RED{]}. 

At the beginning of this Thesis, in \hyperref[section_computerized_axial_tom]{{\color{black}\textbf{section~\ref*{section_computerized_axial_tom}}}}, it was presented the Computerized Axial Tomography. CAT is an example of a classical system with finite degrees of freedom, in this case two, that is described by a state, in this case the absorption coefficient of a portion of matter $\alpha(x,y)$, {\changeurlcolor{mygreen}\hyperref[Section_CAT]{Figure~\ref*{Section_CAT}}}. In a more abstract way, the Radon Transform \hyperref[Radon_Transform_ps]{(\ref*{Radon_Transform})} can be considered too as an example of the tomographic analysis of the statistical states of a classical system with two degrees of freedom $(q,p)$. In this case, the domain $\Omega\subset\mathbb{R}^2$ represents the phase space of the system and the function $f:\Omega\rightarrow\mathbb{R}$ would be a probability density. In what follows, we will just call ``states'' to statistical states of classical systems.

The phase space carries a canonical measure, the Liouville's measure $\mu_{\textrm{Liouville}}$, that in canonical coordinates $(\boldsymbol{q},\boldsymbol{p})$ has the form $\textrm{d}\mu_{\textrm{Liouville}}(\boldsymbol{q},\boldsymbol{p})\!=\textrm{d}^n\boldsymbol{q}\textrm{d}^n\boldsymbol{p}$. In the case that $\Omega$ is a domain in $\mathbb{R}^{2n}$, the \textit{center of mass tomogram} $\mathcal{W}_{cm}$ of the probability density $\rho(\boldsymbol{q},\boldsymbol{p})$ is defined as the generalization to $n$ degrees of freedom of the Radon Transform~\hyperref[Radon_Transform_ps]{(\ref*{Radon_Transform})}:
\phantomsection\label{Radon_Transform_n_degrees_ps}\begin{equation}\label{Radon_Transform_n_degrees}
\mathcal{W}_{cm}(X,\boldsymbol{\mu},\boldsymbol{\nu})=\hspace{-.16cm}\int\limits_{\hspace{.3cm}\Omega}\hspace{0cm}\rho(\boldsymbol{q},\boldsymbol{p})\delta(X-\boldsymbol{\mu}\cdot\boldsymbol{q}-\boldsymbol{\nu}\cdot\boldsymbol{p})\textrm{d}^n\boldsymbol{q}\textrm{d}^n\boldsymbol{p},
\end{equation}
where $\boldsymbol{\mu}=(\mu_1,\ldots,\mu_n)$, $\boldsymbol{\nu}=(\nu_1,\ldots,\nu_n)$ belong to $\mathbb{R}^n$, and the equation $X-\boldsymbol{\mu}\cdot\boldsymbol{q}-\boldsymbol{\nu}\cdot\boldsymbol{p}=0$ determines a hyperplane $\Pi_X(\boldsymbol{\mu},\boldsymbol{\nu})$ in $\Omega$, that has the same geometrical interpretation than the line $L_X(q_0,p_0)$ in {\changeurlcolor{mygreen}\hyperref[Geometric_Radon]{Figure \ref*{Geometric_Radon}}} but in dimension $n$. Because of the definition of the $n$-dimensional Fourier Transform~\hyperref[Fourier_Transform_ps]{(\ref*{Fourier_Transform})}, repeating exactly the same analysis as in \hyperref[section_Radon_Transform]{{\color{black}\textbf{section~\ref*{section_Radon_Transform}}}}, but in a $n$-dimensional phase space, we have that the reconstruction formula~\hyperref[Inverse_Radon_Transform_2_ps]{(\ref*{Inverse_Radon_Transform_2})} for the classical state $\rho$ is
\phantomsection\label{Inverse_Radon_Transform_n-dimensional_ps}\begin{equation}\label{Inverse_Radon_Transform_n-dimensional}
\rho(\boldsymbol{q},\boldsymbol{p})=\frac{1}{(2\pi)^{2n}}\hspace{-.66cm}\int\limits_{\hspace{.84cm}\mathbb{R}^{2n+1}}\hspace{-.5cm}\mathcal{W}_{cm}(X,\boldsymbol{\mu},\boldsymbol{\nu})\e^{-i(X-\boldsymbol{\mu}\cdot\boldsymbol{q}-\boldsymbol{\nu}\cdot\boldsymbol{p})}\diff X\diff^n\boldsymbol{\mu}\diff^n\boldsymbol{\nu}.
\end{equation}

The description of a classical system, whose phase space is $\Omega$, can be easily established in the terms discussed in the introduction of \hyperref[chap_tom_qu]{{\color{black}\textbf{chapter \ref*{chap_tom_qu}}}} by considering as the algebra of operators $\mathcal{A}$ a class of functions on $\Omega$, and the states of the system as normalized positive functionals on $\mathcal{A}$. Let us remark that if $\mathcal{A}$ contains the algebra of continuous functions on $\Omega$, the states are probability measures on the phase space.

If we assume that the phase space is originally equipped with a measure $\mu$, for instance the Liouville's measure $\mu_{\textrm{Liouville}}$ in the case of mechanical systems, then we may restrict ourselves to the statistical states considered by Boltzmann corresponding to probability measures which are absolutely continuous with respect to the Liouville's measure determined by probability densities $\rho(\omega)$ on $\Omega$.

Given an observable $f(\omega)$ on $\Omega$, the pairing between states and observables will be realized by assigning to the observable $f$ its characteristic distribution $\rho_f$ with respect to the probability measure $\rho(\omega)\diff\mu(\omega)$, thus the probability of finding the measured value of the observable $f$ in the interval $\Delta$ is the following:
$$
\int\limits_{\hspace{.3cm}\Delta}\rho_f(\lambda)\diff\lambda=\hspace{-0.9cm}\int\limits_{\hspace{1.1cm}f^{-1}(\Delta)}\hspace{-0.8cm}\rho(\omega)\diff\mu(\omega),
$$
and the expected value of $f$ on the state $\rho$ will be given by:
$$
\langle f\rangle_\rho=\hspace{-0.1cm}\int\limits_{\hspace{.3cm}\mathbb{R}}\lambda\rho_f(\lambda)\diff\lambda.
$$

Sometimes, the center of mass tomogram $\mathcal{W}_{cm}$~\hyperref[Radon_Transform_n_degrees_ps]{(\ref*{Radon_Transform_n_degrees})} does not allow to cope with systems that can not be easily averaged over hyperplanes $X-\boldsymbol{\mu}\cdot\boldsymbol{q}-\boldsymbol{\nu}\cdot\boldsymbol{p}=0$ or, simply, it is more convenient to work with another parametrization. Hence, it is convenient to expand the scope of the formalism to make it more flexible for alternative and more general pictures. Thus, we can reproduce here the general discussion of tomographic theories for quantum systems but replacing the $C^*$--algebra $\mathcal{A}$ there by a commutative Banach algebra $\mathcal{A}$ containing the algebra of continuous functions $C(\Omega)$ ($\Omega$ will be assumed to be compact).

A general tomographic picture of a classical system can be given by starting with a family of elements in $\mathcal{A}$ parametrized by an index $x$ which can be discrete or continuous. Often, $x$ is a point on a finite dimensional manifold that we will denote by $\mathcal{M}$, thus $x\in\mathcal{M}$. The observables associated to the element $x$ will be denoted by $U_x$. Given a state $\rho$ of the classical system, the correspondence $x\rightarrow U_x$ allows to pull-back the observables $U_x$ to $\mathcal{M}$ by defining the function $F_\rho(x)$ on $\mathcal{M}$ associated to the state $\rho(\omega)$ by:
\begin{equation}
F_\rho(x)=\langle\rho,U_x\rangle\coloneq\hspace{-0.1cm}\int\limits_{\hspace{.3cm}\Omega}U_x(\omega)\rho(\omega)\diff\mu(\omega).
\end{equation}
The observables $U_x$ must be properly chosen so that the previous integral is well-defined. For instance, we could have chosen $\mathcal{M}=\Omega$ as in the definition of the center of mass tomogram~\hyperref[Radon_Transform_n_degrees_ps]{(\ref*{Radon_Transform_n_degrees})} and then consider $U_x(\omega)=\delta(\omega)$, thus the function $F_\rho$ associated to the state $\rho(\omega)$ will be again $\rho(\omega)$ itself. The original state $\rho(\omega)$ could be reconstructed from $F_\rho$ if and only if the family of observables $U_x$ separate states, i.e., given two different states $\rho\neq\tilde{\rho}$, there exist $x\in\mathcal{M}$ such that $\langle\rho,U_x\rangle\neq\langle\tilde{\rho},U_x\rangle$. Then, two states will be different if and only if the corresponding functions $F_\rho$ are different.

Clearly up to now, our construction does not discriminate the description between classical and quantum systems\footnote{The difference will appear only at the level of the product structure on the sampling functions $F_\rho$ as the Wigner--Weyl--Moyal approach shows.}. In the same way we introduced the Generalized Positive Transform in the quantum setting, let us introduce it now in this classical setting.

Let us consider, as it was in \hyperref[section_Sampling_C]{{\color{black}\textbf{chapter~\ref*{chap_tom_qu}}}}, the space of functions on the manifold $\mathcal{M}$, $\mathcal{F}(\mathcal{M})$ and its topological dual $\mathcal{F}({\mathcal{M}})^\prime$ (for that, we equip $\mathcal{F}(\mathcal{M})$ with the appropriate topology). And also consider a second auxiliary space $\mathcal{N}$ that parametrizes a certain subspace of smooth functions of compact support $\mathcal{D}(\mathcal{M})\subset\mathcal{F}({\mathcal{M}})^\prime$. In other words, for each $y\in\mathcal{N}$ there is an assignment $y\rightsquigarrow R(y)$ with $R(y)\in\mathcal{D}(\mathcal{M})$ a linear functional on the space of functions on $\mathcal{M}$. A \textit{Classical Generalized Positive Transform} is a map from $\mathcal{F}(\mathcal{M})$ to $\mathcal{F}(\mathcal{N})$ assigning to each $F\in\mathcal{F}(\mathcal{M})$:
$$
\mathcal{W}_F(y)=\langle R(y),F\rangle,
$$
and such that $\mathcal{W}_F$ is normalized and non-negative, $\mathcal{W}_F\geq 0$.

For instance, suppose that $\mathcal{N}$ parametrizes a family of submanifolds $S(y)$ of $\Omega$. If the submanifold $S(y)$ has the form $\Phi(\boldsymbol{q},\boldsymbol{p};X_1,\ldots,X_n)=X_0$, where $y=(X_0,X_1,\ldots,X_n)$ denotes a parametrization of $\mathcal{N}$, the corresponding Generalized Positive Transform would be written as:
\phantomsection\label{Classical_Generalized_Radon_Transform_ps}\begin{equation}\label{Classical_Generalized_Radon_Transform}
\mathcal{W}_\Phi(y)=\hspace{-.16cm}\int\limits_{\hspace{.3cm}\Omega}\hspace{0cm}\rho(\boldsymbol{q},\boldsymbol{p})\delta\big(X_0-\Phi(\boldsymbol{q},\boldsymbol{p};X_1,\ldots,X_n)\big)\textrm{d}^n\boldsymbol{q}\textrm{d}^n\boldsymbol{p},
\end{equation}
which is a generalization of the center of mass tomogram~\hyperref[Radon_Transform_n_degrees_ps]{(\ref*{Radon_Transform_n_degrees})}.

When the embedding is properly chosen, it turns out that $\mathcal{W}_F(y)$ is a fair probability distribution on $\mathcal{N}$, which we have constructed out of the state $\rho$. In the case in which $\Omega=\mathbb{R}^{2n}$ and $\mathcal{N}$ is the space of hyperplanes $\Pi_X$, let us recall the homogeneity condition \hyperref[homogeneity_tom_for_4_ps]{(\ref*{homogeneity_tom_for_4})} that the center of mass tomogram $\mathcal{W}_{cm}$ satisfies:
\phantomsection\label{homogeneity_Tomogram_1_ps}\begin{equation}\label{homogeneity_Tomogram_1}
\mathcal{W}_{cm}(\lambda X,\lambda\boldsymbol{\mu},\lambda\boldsymbol{\nu})=\frac{1}{|\lambda|}\mathcal{W}_{cm}(X,\boldsymbol{\mu},\boldsymbol{\nu}),
\end{equation}
then, we can derive the following relation for the center of mass tomogram by taking derivatives with respect to $\lambda$ in~\hyperref[homogeneity_Tomogram_1_ps]{(\ref*{homogeneity_Tomogram_1})} and evaluating in $\lambda=1$:
\phantomsection\label{homogeneity_relation_1_ps}\begin{equation}\label{homogeneity_relation_1}
\left[X\frac{\partial}{\partial X}+\boldsymbol{\mu}\cdot\frac{\partial}{\partial\boldsymbol{\mu}}+\boldsymbol{\nu}\cdot\frac{\partial}{\partial\boldsymbol{\nu}}+1\right]\mathcal{W}_{cm}(X,\boldsymbol{\mu},\boldsymbol{\nu})=0.
\end{equation}
Due to the homogeneity condition~\hyperref[homogeneity_relation_1_ps]{(\ref*{homogeneity_relation_1})}, $\mathcal{W}_{cm}$ depends effectively only on $2n$ variables instead of $2n+1$.

Similarly to \hyperref[Radon_Transform_n_degrees_ps]{(\ref*{Radon_Transform_n_degrees})}, we can introduce another kind of tomographic representation of the state $\rho$, the \textit{classical symplectic tomogram}:
\phantomsection\label{Classical_symplectic_tomogram_ps}\begin{equation}\label{Classical_symplectic_tomogram}
\mathcal{W}_{sym}(\boldsymbol{X},\boldsymbol{\mu},\boldsymbol{\nu})=\hspace{-.36cm}\int\limits_{\hspace{.6cm}\mathbb{R}^{2n}}\hspace{-0.23cm}\rho(\boldsymbol{q},\boldsymbol{p})\prod_{k=1}^n\delta(X_k-\mu_kq_k-\nu_kp_k)\textrm{d}^n\boldsymbol{q}\textrm{d}^n\boldsymbol{p}.
\end{equation}
Notice that in this case, we have taken $\mathcal{M}=\mathbb{R}^{2n}$ and $\mathcal{N}=\mathcal{N}_1\times\cdots\times\mathcal{N}_n$ with $\mathcal{N}_k$ the space of lines in $\mathbb{R}^{2}$, that is, the phase space of each individual degree of freedom of the physcal system under consideration. Thus, we have obtained a joint probability distribution of the $n$ random variables $(X_1,\ldots,X_n)=\boldsymbol{X}$. In contrast to the center of mass case, because of the presence of $n$ Dirac distributions, we find that the symplectic tomogram $\mathcal{W}_{sym}$ satisfies $n$ homogeneity conditions:
\phantomsection\label{homogeneity_relation_2_ps}\begin{equation}\label{homogeneity_relation_2}
\left[X_k\frac{\partial}{\partial X_k}+\mu_k\frac{\partial}{\partial\mu_k}+\nu_k\frac{\partial}{\partial\nu_k}+1\right]\mathcal{W}_{sym}(\boldsymbol{X},\boldsymbol{\mu},\boldsymbol{\nu})=0,
\end{equation}
$k=1,\ldots,n$. In other words, the classical symplectic tomogram $\mathcal{W}_{sym}$ depends effectively only on $2n$ variables instead of $3n$. In fact, one can show that the symplectic tomogram $\mathcal{W}_{sym}$ can be transformed into the center of mass tomogram $\mathcal{W}_{cm}$ of the same state $\rho$ and vice versa:
\phantomsection\label{Classical_symplectic_tomogram_vs_cm_tomogram_ps}\begin{multline}\label{Classical_symplectic_tomogram_vs_cm_tomogram}
\mathcal{W}_{sym}(\boldsymbol{X},\boldsymbol{\mu},\boldsymbol{\nu})=\frac{1}{(2\pi)^{2n}}\hspace{-.66cm}\int\limits_{\hspace{.84cm}\mathbb{R}^{4n+1}}\hspace{-0.5cm}\mathcal{W}_{cm}(X^\prime,\boldsymbol{\mu}^\prime,\boldsymbol{\nu}^\prime)\e^{-i(X^\prime-\boldsymbol{\mu}^\prime\cdot\boldsymbol{q}-\boldsymbol{\nu}^\prime\cdot\boldsymbol{p})}\\
\cdot\prod_{k=1}^n\delta(X_k-\mu_kq_k-\nu_kp_k)\diff X^\prime\diff^n\boldsymbol{\mu}^\prime\diff^n\boldsymbol{\nu}^\prime\textrm{d}^n\boldsymbol{q}\textrm{d}^n\boldsymbol{p}.
\end{multline}

Because the symplectic tomogram is composed by a product of delta distributions, we can obtain easily its reconstruction formula from the reconstruction formula of the center of mass tomogram~\hyperref[Inverse_Radon_Transform_n-dimensional_ps]{(\ref*{Inverse_Radon_Transform_n-dimensional})} for $n=1$:
\phantomsection\label{Reconstruction_symplectis_tomogram_ps}\begin{equation*}
\hspace{-6.5cm}\rho(\boldsymbol{q},\boldsymbol{p})=\frac{1}{(2\pi)^{2n}}\hspace{-.46cm}\int\limits_{\hspace{.64cm}\mathbb{R}^{3n}}\hspace{-.3cm}\mathcal{W}_{sym}(\boldsymbol{X},\boldsymbol{\mu},\boldsymbol{\nu})
\end{equation*}
\vspace{-0.5cm}\begin{equation}\label{Reconstruction_symplectis_tomogram}
\hspace{5cm}\cdot\e^{-i\sum\limits_{k=1}^n(X_k-\mu_kq_k-\nu_kp_k)}\diff^n \boldsymbol{X}\diff^n\boldsymbol{\mu}\diff^n\boldsymbol{\nu}.
\end{equation}

\section{Tomograms for states of an ensemble of classical oscillators}\label{section_Tomograms_clas_osc}

In contrast to what was done in \hyperref[section_Toms_q_h_o]{{\color{black}\textbf{chapter~\ref*{chap_tom_qu}}}}, where the tomograms of an ensemble of quantum harmonic oscillators were discussed, it is illuminating, for reasons that will be clear at the end of the computations, to do it again for their classical counterparts (see \cite{Ib12}).

\subsection{The canonical ensemble}\label{The canonical ensemble_sec_ps}

If we consider a family of $n$ independent one-dimensional oscillators with frequencies $\omega_k>0$, its phase space $\Omega$ will be $\mathbb{R}^{2n}$ with canonical coordinates $(q_k,p_k)$, $k=1,\ldots,n$. The Hamiltonian of the system will be (recall eq.\,\hyperref[Hamiltonian_electromagnetic_field_2_ps]{(\ref*{Hamiltonian_electromagnetic_field_2})}):
\phantomsection\label{Hamiltonian_classical_oscillator_chap_4_ps}\begin{equation}\label{Hamiltonian_classical_oscillator_chap_4}
H=\sum_{k=1}^n H_k(q_k,p_k),
\end{equation}
with $H_k$ the Hamiltonian of the oscillator of index $k$:
\phantomsection\label{Hamiltonian_classical_oscillator_chap_4_2_ps}\begin{equation*}
H_k=\frac{1}{2}(p_k^2+\omega_k^2q_k^2).
\end{equation*}
The dynamics of the system will be given by
\phantomsection\label{dynamics_oscillator_new_ps}\begin{equation}\label{dynamics_oscillator_new}
\dot{q}_k=p_k,\qquad \dot{p}_k=-\omega_k^2q_k,\qquad k=1,\ldots,n
\end{equation}
and the Liouville's measure will take again the form $\diff\mu_{\textrm{Liouville}}\!=\diff^n \boldsymbol{q}\diff^n\boldsymbol{p}$. Making the change of variables
\phantomsection\label{Cov_Oscillator_ps}\begin{equation}\label{Cov_Oscillator}
\xi_k=\frac{q_k}{\sqrt{\omega_k}},\qquad\eta_k=\sqrt{\omega_k}p_k,
\end{equation}
the dynamics is written in the ``symmetrical'' form
\phantomsection\label{sym_dynamics_ps}\begin{equation}\label{sym_dynamics}
\dot{\xi}_k=\omega_k\eta_k,\qquad\dot{\eta}_k=-\omega_k\xi_k,\qquad k=1,\ldots,n
\end{equation}
and the Hamiltonian becomes:
\phantomsection\label{Hamilonian_cov_ps}\begin{equation}\label{Hamilonian_cov}
H(\boldsymbol{\xi},\boldsymbol{\eta})=\sum_{k=1}^nH_k(\xi_k,\eta_k)=\frac{1}{2}\sum_{k=1}^n\omega_k\left(\xi_k^2+\eta_k^2\right).
\end{equation}

The state of a classical system can be used, instead of the equations of motion~\hyperref[Hamilton_eq_ps]{(\ref*{Hamilton_eq})}, to describe the dynamical evolution of a system in the phase space $\Omega$. Notice that Liouville's measure remains unchanged under the change of variables, $\diff\mu_{\textrm{Liouville}}\hspace{-0.03cm}=\diff^n \boldsymbol{q}\diff^n\boldsymbol{p}=\diff^n \boldsymbol{\xi}\diff^n\boldsymbol{\eta}$, and statistical states are described by probability densities $\rho(\boldsymbol{q},\boldsymbol{p})=\rho(\boldsymbol{\xi},\boldsymbol{\eta})$. 

Liouville's equation \hyperref[Re98_ps]{\RED{[\citen*{Re98}, ch.\,6]}} determines the evolution of the state:
\phantomsection\label{Liouville_eq_ps}\begin{equation}\label{Liouville_eq}
\frac{\diff}{\diff t}\rho=\left\{\vphantom{a^\dagger}\rho,H\right\},
\end{equation}
where the Poisson bracket $\{\vphantom{a^\dagger}\cdot,\cdot\}$ is defined by the canonical commutation relations:
\phantomsection\label{can_rel_Poisson_ps}\begin{equation}\label{can_rel_Poisson}
\left\{\vphantom{a^\dagger}q_k,p_l\right\}=\delta_{kl},\qquad\left\{\vphantom{a^\dagger}q_k,q_l\right\}=\left\{\vphantom{a^\dagger}p_k,p_l\right\}=0.
\end{equation}
In particular, the Gibbs state or canonical distribution is given by
\begin{equation}
\rho_{can}(\boldsymbol{q},\boldsymbol{p})=\frac{\e^{-\beta H}}{Z_0},
\end{equation}
(see \hyperref[Is71_ps]{\RED{[\citen*{Is71}, ch.\,2]}}), where the normalization constant $Z_0$ is easily evaluated:

\phantom{a}
\vspace{-0.6cm}
\phantomsection\label{Normalization_Gibbs_ps}\begin{equation*}
\hspace{-6.0cm}Z_0=\hspace{-.16cm}\int\limits_{\hspace{.3cm}\Omega}\hspace{0cm}\e^{-\beta H(\boldsymbol{q},\boldsymbol{p})}\diff\mu_{Liouville}(\boldsymbol{q},\boldsymbol{p})
\end{equation*}

\phantom{a}
\vspace{-1.5cm}\begin{equation*}
\hspace{3.5cm}=\hspace{-.26cm}\int\limits_{\hspace{.5cm}\mathbb{R}^{2n}}\hspace{-0.2cm}\e
\diff\boldsymbol{\xi}\diff\boldsymbol{\eta}=(2\pi)^n\prod_{k=1}^n(\beta\omega_k)^{-1},
\end{equation*}
where $\beta=(k_{B}T)^{-1}$ is the thermodynamic constant given by the inverse of the temperature of the system $T$ multiplied by the Boltzmann constant $k_B$.

For a given observable $f$, we have that its expected value over the Gibbs state $\rho_{can}$ is
\phantomsection\label{Expected_value_f_ps}\begin{equation}\label{Expected_value_f}
\langle f\rangle_{\rho_{can}}=\frac{1}{Z_0}\hspace{-.26cm}\int\limits_{\hspace{.5cm}\mathbb{R}^{2n}}\hspace{-0.2cm}f(\boldsymbol{\xi},\boldsymbol{\eta})\e
\diff\boldsymbol{\xi}\diff\boldsymbol{\eta}.
\end{equation}
More detailed information can be found in \cite{Kl68}.

Because of the form of the change of variables~\hyperref[Cov_Oscillator_ps]{(\ref*{Cov_Oscillator})}, we can perform the symplectic tomogram of a state $\rho(\boldsymbol{\xi},\boldsymbol{\eta})$ by means of:
\phantomsection\label{symplectic_cov_ps}\begin{equation}\label{symplectic_cov}
\mathcal{W}_{sym}(\boldsymbol{X},\boldsymbol{\mu},\boldsymbol{\nu})=\hspace{-.36cm}\int\limits_{\hspace{.6cm}\mathbb{R}^{2n}}\hspace{-0.23cm}\rho(\boldsymbol{\xi},\boldsymbol{\eta})\prod_{k=1}^n\delta(X_k-\mu_k\xi_k-\nu_k\eta_k)\textrm{d}^n\boldsymbol{\xi}\textrm{d}^n\boldsymbol{\eta},
\end{equation}
therefore, a simple computation shows that the Gibbs state tomogram reads as:
\phantomsection\label{Gibbs_Tomogram_ps}\begin{equation}\label{Gibbs_Tomogram}
\mathcal{W}_{can}(\boldsymbol{X},\boldsymbol{\mu},\boldsymbol{\nu})=\prod_{k=1}^n\sqrt{\frac{\beta\omega_k}{2\pi(\mu_k^2+\nu_k^2)}}\exp\left(-\frac{\beta\omega_kX_k^2}{2(\mu_k^2+\nu_k^2)}\right),
\end{equation}
and the state will be able to be reconstructed by means of eq.\,\hyperref[Reconstruction_symplectis_tomogram_ps]{(\ref*{Reconstruction_symplectis_tomogram})} only by changing $\boldsymbol{q}$ by $\boldsymbol{\xi}$ and $\boldsymbol{p}$ by $\boldsymbol{\eta}$ in that formula (compare with \hyperref[NDimqharm_ps]{(\ref*{NDimqharm})}).

An interesting family of states, which is the classical counterpart of quantum coherent states~\hyperref[coherent_state_ps]{(\ref*{coherent_state})}, can be introduced by means of the holomorphic representation of phase space:
\phantomsection\label{holomorphic_rep_ps}\begin{equation}\label{holomorphic_rep}
\zeta_k=\frac{1}{\sqrt{2}}(\xi_k+i\eta_k),\qquad k=1,\ldots,n.
\end{equation}
Hence, the phase space becomes into the complex space $\mathbb{C}^n$ with the Hermitean structure
\phantomsection\label{Hamiltonian_classical_coherent_ps}\begin{equation}\label{Hamiltonian_classical_coherent}
H(\boldsymbol{\zeta},\bar{\boldsymbol{\zeta}})=\sum_{k=1}^n\omega_k|\zeta_k|^2,
\end{equation}
and given a point $\boldsymbol{z}=(z_1,\ldots,z_n)\in\mathbb{C}^n$, we can construct the distribution
\phantomsection\label{Distribution_classical_coherent_ps}\begin{equation}\label{Distribution_classical_coherent}
\rho_{\boldsymbol{z}}(\boldsymbol{\zeta},\bar{\boldsymbol{\zeta}})=N(\boldsymbol{z})\exp\left(\sum_{k=1}^n\omega_k\left(z_k\bar{\zeta}_k+\bar{z}_k\zeta_k\right)\right)\rho_{can}(\boldsymbol{\zeta},\bar{\boldsymbol{\zeta}})\bigg|_{\beta=1},
\end{equation}
where the normalization factor $N(\boldsymbol{z})$ reads as
\begin{equation*}
N(\boldsymbol{z})=\prod_{k=1}^n\e^{-\omega_k|z_k|^2}.
\end{equation*}

The symplectic tomogram distribution corresponding to $\rho_{\boldsymbol{z}}(\boldsymbol{\zeta},\bar{\boldsymbol{\zeta}})$ is a product
\phantomsection\label{Tomogram_zeta_ps}\begin{equation}\label{Tomogram_zeta}
\mathcal{W}_{\rho_{\boldsymbol{z}}}(\boldsymbol{X},\boldsymbol{\mu},\boldsymbol{\nu},\boldsymbol{z})=\prod_{k=1}^n\mathcal{W}^{(k)}_{\rho_{\boldsymbol{z}}}(X_k,\mu_k,\nu_k,z_k),
\end{equation}
where the tomogram $\mathcal{W}^{(k)}_{\rho_{\boldsymbol{z}}}$ of a single degree of freedom is a Gaussian distribution
\phantomsection\label{Tomogram_zeta_2_ps}\begin{multline}\label{Tomogram_zeta_2}
\mathcal{W}^{(k)}_{\rho_{\boldsymbol{z}}}(X_k,\mu_k,\nu_k,z_k)=\sqrt{\frac{\omega_k}{2\pi(\mu_k^2+\nu_k^2)}}\\
\cdot\exp\left(-\frac{\omega_k\big(X_k-\left\langle X_k(\mu_k,\nu_k,z_k)\right\rangle\big)^2}{2(\mu_k^2+\nu_k^2)}\right)
\end{multline}
of the random variable $X_k$ with mean value
$$
\left\langle X_k\big(\mu_k,\nu_k,z_k)\right\rangle=\sqrt{2}\big(\mu_k\mathfrak{R}(z_k)+\nu_k\mathfrak{I}(z_k)\big)
$$
and variance
$$
\sigma_{X_k}=\sqrt{\frac{\mu_k^2+\nu_k^2}{\omega_k}}\,.
$$

If we compare the formulas \hyperref[Gibbs_Tomogram_ps]{(\ref*{Gibbs_Tomogram})} and \hyperref[Tomogram_zeta_2_ps]{(\ref*{Tomogram_zeta_2})} with the tomograms obtained in \hyperref[section_Toms_q_h_o]{{\color{black}\textbf{chapter 2}}} for the quantum harmonic oscillator \hyperref[NDimqharm_ps]{(\ref*{NDimqharm})}, we see that are similar to the tomogram corresponding to the ground state of  \hyperref[NDimqharm_ps]{(\ref*{NDimqharm})}.

\subsection{A new class of states: Gauss--Laguerre states}

We will introduce now a family of classical states, called Gauss--Laguerre (GL) states, inspired on the Wigner functions of the \hyperref[NDimqharm_ps]{{\color{black}\textbf{excited states}}} of a quantum harmonic oscillator. These functions are only quasi-distributions on phase space, as we have seen in \hyperref[Wigner_state_square]{{\color{black}\textbf{section~\ref*{section_real_QM_ps}}}}, however their square is related to the purity of the corresponding quantum states and are true probability distributions \cite{Do89}. These kind of states appear, for instance, in a physical system in which an electron is moving on a perpendicular plane to a constant magnetic field with quantum azimuthal number $m=0$ (see Landau states in \cite{La30}).

The family of classical states we are considering is defined as:
\phantomsection\label{Laguerre-states_ps}\begin{equation}\label{Laguerre-states}
\rho_{\textrm{GL},\boldsymbol{m}}(\boldsymbol{\xi},\boldsymbol{\eta})=\prod_{k=1}^n\rho^{(k)}_{\textrm{GL},m_k}(\xi_k,\eta_k),
\end{equation}
where $\boldsymbol{m}=(m_1,m_2,\ldots,m_n)$ is a multi-index that is not related with the azimuthal quantum number we mentioned in the previous paragraph, and
\begin{equation}
\rho^{(k)}_{\textrm{GL},m_k}(\xi_k,\eta_k)=\frac{\omega_k}{2\pi}L^2_{m_k}\left(\frac{\omega_k}{2}\left(\xi_k^2+\eta_k^2\right)\right)\e
.
\end{equation}
Here, the function $L_{m_k}$ is the Laguerre polynomial of degree $m_k$. Notice that $\rho^{(k)}_{\textrm{GL},m_k}(\xi_k,\eta_k)$ is a classical state on a bidimensional phase space.

The symplectic Radon Transform of the state factorizes as the state does:
\phantomsection\label{Tomogram_Laguerre_product_ps}\begin{equation}\label{Tomogram_Laguerre_product}
\mathcal{W}_{\textrm{GL},\boldsymbol{m}}(\boldsymbol{X},\boldsymbol{\mu},\boldsymbol{\nu})=\prod_{k=1}^n\mathcal{W}^{(k)}_{\textrm{GL},m_k}(X_k,\mu_k,\nu_k),
\end{equation}
with
\phantomsection\label{Tomogram_Laguerre_k_ps}\begin{multline}\label{Tomogram_Laguerre_k}
\mathcal{W}^{(k)}_{\textrm{GL},m_k}(X_k,\mu_k,\nu_k)=\frac{1}{\sqrt{\pi}\sigma_k}\exp\left(-\frac{X_k^2}{\sigma_k^2}\right)\\
\cdot\sum_{s=0}^{m_k}\frac{1}{2^{2m_k}}\begin{pmatrix}
2(m_k-s)\\
m_k-s
\end{pmatrix}\begin{pmatrix}
2s\\
s
\end{pmatrix}\frac{H^2_{2s}\left(\frac{X_k}{\sigma_k}\right)}{2^{2s}(2s)!}
\end{multline}
and
\begin{equation*}
\sigma_k=\sqrt{\frac{2\left(\mu_k^2+\nu_k^2\right)}{\omega_k}},
\end{equation*}
while $H_{2s}$ is the Hermite polynomial of degree $2s$. The above result can be obtain as follows.

First, we will drop the label $k$ and will write $\mathcal{W}_m(X,\mu,\nu)$ in place of $\mathcal{W}^{(k)}_{\textrm{GL},m_k}(X_k,\mu_k,\nu_k)$ to simplify the notation. Thus,
\begin{equation*}
\hspace{-2.7cm}\mathcal{W}_m(X,\mu,\nu)=\hspace{0.0cm}\frac{\omega}{2\pi}\hspace{-0.3cm}\int\limits_{\hspace{.5cm}\mathbb{R}^{2}}\hspace{-0.2cm}\,L_m^2\left(\frac{\omega}{2}\left(\xi^2+\eta^2\right)\right)\e
\e^{-iK(\mu\xi+\nu\eta)}\diff\xi\diff\eta\diff K.
\end{align}
Now, we put $\sqrt{\mu^2+\nu^2}=r_{\mu\nu}$ and
\begin{alignat*}{3}
\mu&=r_{\mu\nu}\cos\alpha_{\mu\nu},& \hspace{1.2cm}\nu&=r_{\mu\nu}\sin\alpha_{\mu\nu},\\
\xi&=r\sin\theta,& \hspace{1.2cm}\eta&=r\cos\theta,
\end{alignat*}
and we recast the previous formula as:
\phantomsection\label{Tomogram_Rec_Laguerre_ps}\begin{equation}\label{Tomogram_Rec_Laguerre}
\mathcal{W}_m(X,\mu,\nu)=\hspace{.0cm}\frac{1}{2\pi}\hspace{-.2cm}\int\limits_{\hspace{-.04cm}-\infty}^{\hspace{.45cm}\infty}\hspace{-0.15cm}\,\e^{iKX}\widehat{\mathcal{W}}_m(K,\mu,\nu)\diff K,
\end{equation}
where the Fourier Transform $\widehat{\mathcal{W}}_m$ is given by
\begin{equation*}
\widehat{\mathcal{W}}_m(K,\mu,\nu)=\hspace{0cm}\frac{\omega}{2\pi}\hspace{-.2cm}\int\limits_{\hspace{0.15cm}0}^{\hspace{.35cm}2\pi}\hspace{-0.33cm}\int\limits_{\hspace{0.15cm}0}^{\hspace{.35cm}\infty}\hspace{-0.15cm}\,\,L_m^2\left(\frac{\omega r^2}{2}\right)\e
\e^{-iKr_{\mu\nu}r\sin(\theta+\alpha_{\mu\nu})}r\diff r\diff\theta.
\end{equation*}
The integral over the angular variable $\theta_{\mu\nu}=\theta+\alpha_{\mu\nu}$ yields the Bessel function $J_0$, so:
\begin{equation}
\widehat{\mathcal{W}}_m(K,\mu,\nu)=\hspace{-.2cm}\int\limits_{\hspace{0.15cm}0}^{\hspace{.35cm}\infty}\hspace{-0.15cm}\,\,L_m^2\left(\frac{x^2}{2}\right)\e%
J_0\left(\frac{Kr_{\mu\nu}}{\sqrt{\omega}x}\right)x\diff x.
\end{equation}
The above integral can be evaluated and gives \hyperref[ps_Gr07]{{\color{red}[\citen*{Gr07}, {n.\hspace{0,05cm}7.422 2}]}}:
\begin{align}
&\hspace{-0.2cm}\widehat{\mathcal{W}}_m(K,\mu,\nu)=\e%
\hspace{0.1cm}L_{2s}\left(\frac{Kr_{\mu\nu}}{\sqrt{\omega}}\right)^2\!\!,
\end{align}
where the last line has been obtained by a well known addition formula of Laguerre polynomials \hyperref[ps_Gr07]{{\color{red}[\citen*{Gr07}, {n.\hspace{0,05cm}8.976 3}]}}.

We remark that the above equation yields, by multiplication over the restored label $k$, the Fourier Transform $\widehat{\mathcal{W}}_{\textrm{GL},\boldsymbol{m}}(\boldsymbol{K},\boldsymbol{\mu},\boldsymbol{\nu})$ of the tomogram $\mathcal{W}_{\textrm{GL},\boldsymbol{m}}(\boldsymbol{K},\boldsymbol{\mu},\boldsymbol{\nu})$ with $\boldsymbol{K}=(K_1,K_2,\ldots,K_n)$.

Besides, as $\widehat{\mathcal{W}}_{\textrm{GL},\boldsymbol{m}}(0,\boldsymbol{\mu},\boldsymbol{\nu})=1$, we get at once the normalization property of the tomogram $\mathcal{W}_{\textrm{GL},\boldsymbol{m}}(\boldsymbol{X},\boldsymbol{\mu},\boldsymbol{\nu})$.

Finally, we are able to perform the integral~\hyperref[Tomogram_Rec_Laguerre_ps]{(\ref*{Tomogram_Rec_Laguerre})} by means of the integral over $y=Kr_{\mu\nu}/\sqrt{\omega}$, \hyperref[ps_Gr07]{{\color{red}[\citen*{Gr07}, {n.\hspace{0,05cm}7.418 3}]}}:
\begin{multline}
\frac{1}{\pi}\frac{\sqrt{\omega}}{r_{\mu\nu}}\hspace{-.2cm}\int\limits_{\hspace{.15cm}0}^{\hspace{.35cm}\infty}\hspace{-0.15cm}\,\,L_{2s}\left(y^2\right)\e
\cos\left(\frac{\sqrt{\omega}}{r_{\mu\nu}}Xy\right)\diff y\\
                                                                                                                                                                                                                                                                                                  =\frac{\sqrt{\omega}}{\sqrt{2}\pi r_{\mu\nu}}\exp\left(-\frac{\omega}{2r_{\mu\nu}^2}X^2\right)\frac{1}{2^{2s}(2s)!}H_{2s}^2\left(\frac{\sqrt{\omega}}{\sqrt{2}r_{\mu\nu}}X\right).
\end{multline}
Therefore, we have got the predicted expression of $\mathcal{W}_m(X,\mu,\nu)$.\newpage

\section{The tomographic picture of Liouville's equation}\label{section_Tom_Liouville}

Let us discuss in this section the tomographic form of the evolution equation for states, the Liouville's equation~\hyperref[Liouville_eq_ps]{(\ref*{Liouville_eq})}. The evolution equation in the tomographic description was obtained in \cite{Ch07} in relation with a relativistic wave function description of harmonic oscillators. We will describe it here in the realm of our previous discussion. Notice that because of the symplectic reconstruction formula~\hyperref[Reconstruction_symplectis_tomogram_ps]{(\ref*{Reconstruction_symplectis_tomogram})}, we can compute:
\phantomsection\label{diff_state_ps}\begin{multline}\label{diff_state}
\frac{\diff}{\diff t}\rho(\boldsymbol{\xi},\boldsymbol{\eta},t)=\frac{1}{(2\pi)^{2n}}\hspace{-.46cm}\int\limits_{\hspace{.64cm}\mathbb{R}^{3n}}\hspace{-.3cm}\left[\frac{\diff}{\diff t}\mathcal{W}_{sym}(\boldsymbol{X},\boldsymbol{\mu},\boldsymbol{\nu},t)\right]\\
\cdot\e^{-i\sum\limits_{k=1}^n(X_k-\mu_k\xi_k-\nu_k\eta_k)}\diff^n \boldsymbol{X}\diff^n\boldsymbol{\mu}\diff^n\boldsymbol{\nu},
\end{multline}
(is important to highlight that the symplectic tomogram is computed at a given fixed time) and, on the other hand:
\phantomsection\label{poisson_Liouville_ps}
\begin{align*}
&\hspace{-0.6cm}\left\{\vphantom{a^\dagger}\rho,H\right\}=\sum_{k=1}^n\left[\frac{\partial H}{\partial\eta_k}\frac{\partial}{\partial\xi_k}-\frac{\partial H}{\partial\xi_k}\frac{\partial}{\partial\eta_k}\right]\rho=\frac{1}{(2\pi)^{2n}}\sum_{k=1}^n\hspace{-.46cm}\int\limits_{\hspace{.64cm}\mathbb{R}^{3n}}\hspace{-.3cm}\mathcal{W}_{sym}(\boldsymbol{X},\boldsymbol{\mu},\boldsymbol{\nu},t)\\
&\hspace{1cm}\cdot\left[\frac{\partial H}{\partial\eta_k}\frac{\partial}{\partial\xi_k}-\frac{\partial H}{\partial\xi_k}\frac{\partial}{\partial\eta_k}\right]\e^{-i\sum\limits_{k=1}^n(X_k-\mu_k\xi_k-\nu_k\eta_k)}\diff^n \boldsymbol{X}\diff^n\boldsymbol{\mu}\diff^n\boldsymbol{\nu}
\end{align*}

\phantom{a}
\vspace{-0.9cm}
\begin{align}\label{poisson_Liouville}
&\hspace{1cm}=\frac{1}{(2\pi)^{2n}}\sum_{k=1}^n\hspace{-.46cm}\int\limits_{\hspace{.64cm}\mathbb{R}^{3n}}\hspace{-.3cm}\mathcal{W}_{sym}(\boldsymbol{X},\boldsymbol{\mu},\boldsymbol{\nu},t)\left[\frac{\partial H}{\partial\xi_k}\nu_k\frac{\partial}{\partial X_k}-\frac{\partial H}{\partial\eta_k}\mu_k\frac{\partial}{\partial X_k}\right]\nonumber\\
&\hspace{4.7cm}\cdot\e^{-i\sum\limits_{k=1}^n(X_k-\mu_k\xi_k-\nu_k\eta_k)}\diff^n \boldsymbol{X}\diff^n\boldsymbol{\mu}\diff^n\boldsymbol{\nu}.
\end{align}
If we equal the formulas~\hyperref[diff_state_ps]{(\ref*{diff_state})} and~\hyperref[poisson_Liouville_ps]{(\ref*{poisson_Liouville})}, we get:
\phantomsection\label{Tomographic_Liouville_ps}
\begin{align*}
&\hspace{-0.6cm}\frac{\diff}{\diff t}\mathcal{W}_{sym}(\boldsymbol{X},\boldsymbol{\mu},\boldsymbol{\nu},t)=\\
&\hspace{0.1cm}\sum_{k=1}^n\left[\frac{\partial H}{\partial\eta_k}\left(\left\{\xi_j\rightarrow-\left[\frac{\partial}{\partial X_j}\right]^{-1}\hspace{-0.2cm}\frac{\partial}{\partial\mu_j}\right\},\left\{\eta_j\rightarrow-\left[\frac{\partial}{\partial X_j}\right]^{-1}\hspace{-0.2cm}\frac{\partial}{\partial\nu_j}\right\}\right)\mu_k\right.
\end{align*}

\phantom{a}
\vspace{-1.1cm}\begin{multline}\label{Tomographic_Liouville}
-\left.\frac{\partial H}{\partial\xi_k}\left(\left\{\xi_j\rightarrow-\left[\frac{\partial}{\partial X_j}\right]^{-1}\hspace{-0.2cm}\frac{\partial}{\partial\mu_j}\right\},\left\{\eta_j\rightarrow-\left[\frac{\partial}{\partial X_j}\right]^{-1}\hspace{-0.2cm}\frac{\partial}{\partial\nu_j}\right\}\right)\nu_k\right]\\
\cdot\frac{\partial}{\partial X_k}\mathcal{W}_{sym}(\boldsymbol{X},\boldsymbol{\mu},\boldsymbol{\nu},t).
\end{multline}
The operator $\displaystyle{\left[\frac{\partial}{\partial X}\right]^{-1}}$ is defined in terms of a Fourier Transform as
\begin{equation}
\left[\frac{\partial}{\partial X}\right]^{-1}\hspace{-.3cm}\int\limits_{\hspace{-.04cm}-\infty}^{\hspace{.45cm}\infty}\hspace{-0.15cm}F(K)\e^{iKX}\diff k=\hspace{-.2cm}\int\limits_{\hspace{-.04cm}-\infty}^{\hspace{.45cm}\infty}\hspace{-0.15cm}\frac{F(K)}{iK}\e^{iKX}\diff K,
\end{equation}
and the notation $\left\{\vphantom{a^\dagger}*\rightarrow * \right\}$ means that the variable in the left side, which belongs to the phase space, is replaced by the operator at the right side, which belongs to the space of hyperplanes $\Pi_{X_j}(\mu_j,\nu_j)$. This fact happens, in a similar way, in the Fourier framework when one wants to change variables from the time domain to the frequency domain. This replacing law can be deduced in the Fourier framework in this way:
\begin{equation*}
-i\widehat{tf(t)}(k)=\hspace{.0cm}\frac{-i}{\sqrt{2\pi}}\hspace{-.2cm}\int\limits_{\hspace{-.04cm}-\infty}^{\hspace{.45cm}\infty}\hspace{-0.15cm}tf(t)\e^{-ikt}\diff t\\
=\frac{1}{\sqrt{2\pi}}\frac{\diff}{\diff k}\hspace{-.2cm}\int\limits_{\hspace{-.04cm}-\infty}^{\hspace{.45cm}\infty}\hspace{-0.15cm}f(t)\e^{-ikt}\diff t=\frac{\diff}{\diff k}\widehat{f}(k).
\end{equation*}
Therefore, the variable $t$ in time domain becomes the operator $i\displaystyle{\frac{\diff}{\diff k}}$ in frequency domain. If we make the same in the Radon framework, we get:
\begin{align*}
&\hspace{-1.5cm}\mathcal{R}[\xi\rho(\xi,\eta](X,\mu,\nu)=\hspace{-.26cm}\int\limits_{\hspace{.5cm}\mathbb{R}^2}\hspace{-.19cm}\xi\rho(\xi,\eta)\delta(X-\mu\xi-\nu\eta)\diff\xi\diff\eta\\
&\hspace{0.2cm}=\hspace{-.26cm}\int\limits_{\hspace{.5cm}\mathbb{R}^3}\hspace{-.19cm}\xi\rho(\xi,\eta)\e^{ik(X-\mu\xi-\nu\eta)}\diff k\diff\xi\diff\eta=\frac{\partial}{\partial\mu}\hspace{-.26cm}\int\limits_{\hspace{.5cm}\mathbb{R}^3}\hspace{-.19cm}\frac{\rho(\xi,\eta)}{-ik}
\end{align*}

\phantom{a}
\vspace{-1.3cm}\begin{equation*}
\hspace{2.5cm}\cdot\e^{ik(X-\mu\xi-\nu\eta)}\diff k\diff\xi\diff\eta=-\left[\frac{\partial}{\partial X}\right]^{-1}\hspace{-0.2cm}\frac{\partial}{\partial\mu}\mathcal{W}_{sym}(X,\mu,\nu),
\end{equation*}
hence,
\phantomsection\label{xi_to_op_ps}\begin{equation}\label{xi_to_op}
\xi\rightarrow-\left[\frac{\partial}{\partial X}\right]^{-1}\hspace{-0.2cm}\frac{\partial}{\partial\mu}\,.
\end{equation}
Doing the same for $\eta$, we have that
\begin{equation*}
\mathcal{R}[\eta\rho(\xi,\eta](X,\mu,\nu)=-\left[\frac{\partial}{\partial X}\right]^{-1}\hspace{-0.2cm}\frac{\partial}{\partial\nu}\mathcal{W}_{sym}(X,\mu,\nu),
\end{equation*}
then,
\phantomsection\label{eta_to_op_ps}\begin{equation}\label{eta_to_op}
\eta\rightarrow-\left[\frac{\partial}{\partial X}\right]^{-1}\hspace{-0.2cm}\frac{\partial}{\partial\nu}\,.
\end{equation}
Let us also remark that for obtaining the formula~\hyperref[Tomographic_Liouville_ps]{(\ref*{Tomographic_Liouville})}, we have used the property of the derivative of a product:
\begin{align*}
\mathcal{W}_{sym}(\boldsymbol{X},\boldsymbol{\mu},\boldsymbol{\nu},t)\frac{\partial}{\partial X_k}\e^{-i\sum\limits_{k=1}^n(X_k-\mu_k\xi_k-\nu_k\eta_k)}&\nonumber\\
&\hspace{-4.6cm}=\frac{\partial}{\partial X_k}\left(\mathcal{W}_{sym}(\boldsymbol{X},\boldsymbol{\mu},\boldsymbol{\nu},t)\e^{-i\sum\limits_{k=1}^n(X_k-\mu_k\xi_k-\nu_k\eta_k)}\right)\nonumber\\
&\hspace{-4.1cm}-\left(\frac{\partial}{\partial X_k}\mathcal{W}_{sym}(\boldsymbol{X},\boldsymbol{\mu},\boldsymbol{\nu},t)\right)\e^{-i\sum\limits_{k=1}^n(X_k-\mu_k\xi_k-\nu_k\eta_k)},
\end{align*}
and the fact that the tomogram must satisfy
\begin{equation*}
\lim_{X_j\rightarrow\pm\infty}\mathcal{W}_{sym}(\boldsymbol{X},\boldsymbol{\mu},\boldsymbol{\nu},t)=0.
\end{equation*}

Due to the presence of the terms~\hyperref[xi_to_op_ps]{(\ref*{xi_to_op})} and~\hyperref[eta_to_op_ps]{(\ref*{eta_to_op})}, for a generic Hamiltonian $H$ the tomographic evolution equation~\hyperref[Tomographic_Liouville_ps]{(\ref*{Tomographic_Liouville})} is integro-differential. In the particular instance of $H$ given by~\hyperref[Hamilonian_cov_ps]{(\ref*{Hamilonian_cov})}, the tomographic evolution equation takes the form of a differential equation:
\begin{equation}
\frac{\diff}{\diff t}\mathcal{W}_{sym}(\boldsymbol{X},\boldsymbol{\mu},\boldsymbol{\nu},t)=\sum_{k=1}^n\omega_k\left[\nu_k\frac{\partial}{\partial\mu_k}-\mu_k\frac{\partial}{\partial\nu_k}\right]\mathcal{W}_{sym}(\boldsymbol{X},\boldsymbol{\mu},\boldsymbol{\nu},t).
\end{equation}\newpage

\section{Tomography of the Klein--Gordon classical field in a cavity}\label{sec_KG_fields}

\phantomsection\label{resumen_field}
We have just described the tomographic description of states of a classical system with finite degrees of freedom, however this description is not enough to deal with systems involving infinite degrees of freedom. To deal with this problem, we need to introduce the concept of \textit{field}. A field, please excuse the redundancy, is a generalization of the generalized coordinates $q^i$, introduced at the \hyperref[chap_clas]{{\color{black}\textbf{beginning of this chapter}}}, that describes a certain continuous system supported on a given space-time $\mathcal{M}=\mathbb{R}\times\mathcal{V}$, where $\mathcal{V}$ is a manifold of dimension $d$. In the present context, a field will be a real function $\varphi(t,\boldsymbol{x})$ of space and time that represents the configurations of the system, hence they describe the configurations of a system with infinite degrees of freedom (for example, see the section of Classical fields in chapter 2 of \cite{Pe95} and \cite{Ti99} or \hyperref[Ca15_ps]{\RED{[\citen*{Ca15}, sec.\,1.3]}}).

In this new context, in which the space is also a parameter, it is natural to consider systems whose dynamics are defined again by means of an action functional of the form
\begin{equation}
S[\varphi]=\hspace{-0.2cm}\int\limits_{\hspace{0.3cm}\mathcal{M}}\hspace{-0.1cm}\mathscr{L}\big(\varphi(t,\boldsymbol{x}),\partial_\mu\varphi(t,\boldsymbol{x})\big)\diff^d\boldsymbol{x}\diff t.
\end{equation}
The functional $\mathscr{L}$ is called the \textit{Lagrangian density}. 

Applying the principle of least action in the same way as done for a system of finite degrees of freedom~\hyperref[Least_action_ps]{(\ref*{Least_action})}, we get the Euler--Lagrange equation of a classical field:
\phantomsection\label{Euler-Lagrange_field_ps}\begin{equation}\label{Euler-Lagrange_field}
\frac{\delta S}{\delta\varphi(t,\boldsymbol{x})}=\frac{\partial}{\partial x^\mu}\left(\frac{\partial \mathscr{L}}{\partial(\partial_\mu\varphi)}\right)-\frac{\partial \mathscr{L}}{\partial \varphi}=0,
\end{equation}
where $\partial_\mu$ denotes the partial derivative with respect to $x^\mu=(t,\boldsymbol{x})$.\footnote{Notice that we have removed the symbol of sum along $\mu=0,\ldots,d$ in~\hyperref[Euler-Lagrange_field_ps]{(\ref*{Euler-Lagrange_field})} according with \textit{Einstein's convention}.}

In the same way that we have introduced the Lagrangian density, we can introduce the \textit{Hamiltonian density}:
\phantomsection\label{Hamiltinian_density_ps}\begin{equation}\label{Hamiltinian_density}
\mathscr{H}=\frac{\partial\mathscr{L}}{\partial({\partial_t\varphi})}\frac{\partial\varphi}{\partial t}-\mathscr{L},
\end{equation}
where the total Hamiltonian of the system is the integral over the space $\mathcal{V}$ of its density:
\phantomsection\label{Hamiltonian_density_integral_ps}\begin{equation}\label{Hamiltonian_density_integral}
H=\hspace{-0.2cm}\int\limits_{\hspace{0.3cm}\mathcal{V}}\mathscr{H}\textrm{d}^d\boldsymbol{x}.
\end{equation}

Recalling the definition of canonical momentum~\hyperref[canonical_momentum_ps]{(\ref*{canonical_momentum})}, we can define the \textit{canonical momentum field} as:
\phantomsection\label{canonical_momentum_field_ps}\begin{equation}\label{canonical_momentum_field}
\pi(t,\boldsymbol{x})=\frac{\partial \mathscr{L}}{\partial(\partial_t\varphi)}\,,
\end{equation}
therefore, we can write the Euler--Lagrange equations in a similar way to~\hyperref[Hamilton_eq_ps]{(\ref*{Hamilton_eq})} in terms of the Hamiltonian density:
\phantomsection\label{eq_motion_Hamiltonian_field_ps}\begin{equation}\label{eq_motion_Hamiltonian_field}
\frac{\partial\varphi}{\partial t}=\frac{\partial\mathscr{H}}{\partial\pi},\qquad\frac{\partial\pi}{\partial t}=-\frac{\partial\mathscr{H}}{\partial\varphi}+\frac{\partial}{\partial x^\mu}\left(\frac{\partial\mathscr{H}}{\partial(\partial_\mu\varphi)}\right)-\frac{\partial}{\partial t}\left(\frac{\partial\mathscr{H}}{\partial(\partial_t\varphi)}\right),
\end{equation}
or using the concept of variational derivative defined previously~\hyperref[functional_derivative_ps]{(\ref*{functional_derivative})}, these equations can be written in the following way:
\phantomsection\label{eq_motion_Hamiltonian_field_2_ps}\begin{equation}\label{eq_motion_Hamiltonian_field_2}
\frac{\partial\varphi}{\partial t}=\frac{\delta H}{\delta\pi},\qquad\frac{\partial\pi}{\partial t}=-\frac{\delta H}{\delta\varphi},
\end{equation}
which have the canonical structure of equations~\hyperref[Hamilton_eq_ps]{(\ref*{Hamilton_eq})}.

Having shown that an interesting family of states for a finite ensemble of harmonic oscillators is amenable to be described tomographically, we will discuss now a particular instance of a field theory, the Klein--Gordon equation for a real scalar field $\varphi$ in a cavity on $1+d$ Minkowski space-time, and we will describe tomographically a remarkable state of the theory: the canonical state \cite{Ib12}. 

Thus, we will consider Minkowski space-time $\mathcal{M}\coloneq\mathbb{M}=\mathbb{R}^{1+d}$, where $d$ is the dimension of space with the standard Minkowski metric of signature $(-,+,\ldots,+)$ (see \hyperref[section_Minkowsiki]{{\color{black}\textbf{appendix~\ref*{section_Minkowsiki}}}}).

The dynamics of the real scalar field $\varphi(t,\boldsymbol{x})$ is defined now by the Lagrangian density:
\begin{equation}
\mathscr{L}[\varphi]=\frac{1}{2}\partial_\mu\varphi\partial^\mu\varphi-V[\varphi],
\end{equation}
where $V[\varphi]$ is the potential functional of the system. The Euler--Lagrange equation~\hyperref[Euler-Lagrange_field_ps]{(\ref*{Euler-Lagrange_field})} in this case is:
\phantomsection\label{Euler_Lagrange_Classical_field_2_ps}\begin{equation}\label{Euler_Lagrange_Classical_field_2}
\partial_\mu\partial^\mu\varphi=-\frac{\partial V[\varphi]}{\partial\varphi}\,.
\end{equation}
Considering $\displaystyle{V[\varphi]=\frac{1}{2}m^2\varphi^2}$ (the potential field of an harmonic oscillator), we get the \textit{Klein--Gordon} equation:
\phantomsection\label{KG_classical_ps}\begin{equation}\label{KG_classical}
\varphi_{tt}-\Delta\varphi+m^2\varphi=0,
\end{equation}
with $\Delta$ the $d$-dimensional Laplacian in $\mathbb{R}^d$.

As we have extensively discussed before (\hyperref[The canonical ensemble_sec_ps]{{\color{black}\textbf{section \ref*{section_Tomograms_clas_osc}}}}), tomographic methods are described on phase space where conjugated variables and Poisson brackets are available. On such carrier, space dynamical equations are described by a vector field~\hyperref[sym_dynamics_ps]{(\ref*{sym_dynamics})}. Thus, for our Klein--Gordon equation, we have to introduce a larger carrier space where the equations will be of first order in time. 

The transition from second order equations to first order differential equations in time may be done in many ways \cite{Ma85}, here we shall consider one in which the new variables will make the equations of motion more ``symmetric''. We would stress that by using a specific splitting of space-time into the space part and the time part, then the explicit Poincar\'{e} invariance of the theory breaks, but of course our description is still relativistic invariant. 

To proceed, we will consider the Cauchy hypersurface $\mathcal{C}=\{0\}\times\mathbb{R}^d$ and the finite cavity will be defined as $\mathcal{V}\subset\mathcal{C}$. We consider the restriction of the field to the cavity $\mathcal{V}$ using the same notation $\varphi(\boldsymbol{x})\coloneq\varphi(0,\boldsymbol{x})$, $\boldsymbol{x}\in\mathcal{V}$, see {\changeurlcolor{mygreen}\hyperref[Cavity]{Figure \ref*{Cavity}}}, and also consider that the field $\varphi(\boldsymbol{x})$ evolves in time according to $\varphi_t(\boldsymbol{x})\coloneq\varphi(t,\boldsymbol{x})$. Therefore, the Klein--Gordon equation becomes the evolution equation, of second order in time, in the space of fields $\varphi(\boldsymbol{x})$:
\phantomsection\label{KG_classical_2_ps}\begin{equation}\label{KG_classical_2}
\frac{\diff^2\varphi}{\diff t^2}=-(-\Delta+m^2)\varphi.
\end{equation}
\begin{figure}[htbp]
\centering
\includegraphics{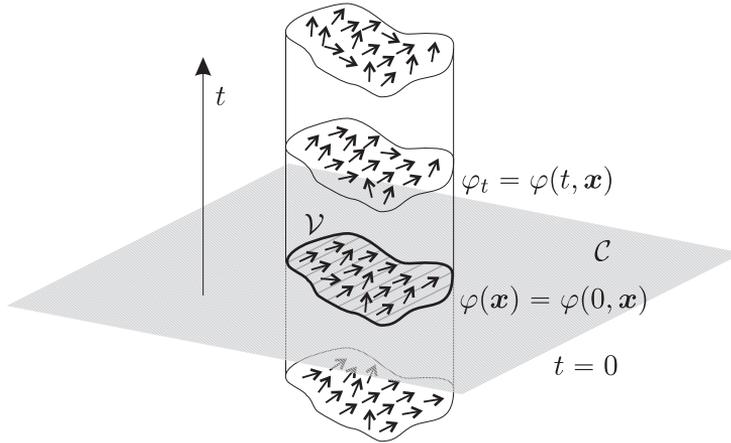}
\caption{\hfil\MYGREEN{Figure 4.5.1}: Evolution of a field restricted to a cavity.\hfil}
\label{Cavity}
\end{figure}

For solutions of Euler--Lagrange equations to be critical points of the action functional $S$, the following boundary term must vanish at the boundary $\partial\mathcal{V}$:
\begin{equation}
\int\limits_{\hspace{.3cm}\partial\mathcal{V}}\hspace{-.1cm}\delta\varphi\hspace{0.03cm}\dot{\varphi}\,\diff^{d-1}\boldsymbol{x}=0,
\end{equation}
where $\dot{\varphi}$ denotes the normal derivative of $\varphi$ at the boundary $\partial\mathcal{V}$.

We can choose boundary conditions such that this integral vanishes and the operator $-\Delta+m^2$ will be strictly positive and self-adjoint on square integrable functions on $\mathcal{V}$ with respect to the Lebesgue measure. Thus, we can define the invertible positive self-adjoint operator $B=\sqrt{-\Delta+m^2}$. We will also assume for simplicity that boundary conditions are chosen in such a way the spectrum of $B$ is nondegenerate, therefore the eigenvalues of $B$ have the form $0<\omega_1<\omega_2<\cdots<\omega_n<\cdots$ with eigenfunctions $\Phi_k(\boldsymbol{x})$:
\phantomsection\label{ps_spectrum_B}
\begin{equation}
B\Phi_k(\boldsymbol{x})=\omega_k\Phi_k(\boldsymbol{x}),\qquad k=1,2,\ldots\,.
\end{equation}

Actually, assuming that the cavity $\mathcal{V}$ is a compact manifold with boundary $\partial\mathcal{V}$, then the space of self-adjoint extensions of the Laplace--Beltrami operator $\Delta$ is parametrized by unitary operators at the boundary $U\in\mathcal{U}\big(L^2(\partial\mathcal{V})\big)$. The structure of such extensions is discussed in \cite{Ib15} (for instance, see also \cite{As05}). We will select among them the class $\mathscr{C}$ of self-adjoint semibounded extensions with lower bound $\mathscr{C}\hspace{-0.02cm}>\hspace{-0.02cm}-m^2$, then the operator $-\Delta+m^2\geq 0$ and $B$ will be well-defined. The domain of $B$ will be contained in the Sobolev space $H^1(\mathcal{V})$ and will be given by functions satisfying the boundary conditions given by Asorey's equation:
\begin{equation*}
\varphi-i\dot{\varphi}=U(\varphi+i\dot{\varphi}).
\end{equation*}  
In addition, if we choose boundary conditions in such a way that the extended operator $\Delta_U$ is elliptic, i.e., $U$ belongs to the \textit{Elliptic Grassmannian space} of elliptic self-adjoint extensions of $\Delta$ \cite{As05}, then because $\mathcal{V}$ is compact, well-known results assert that the spectrum of $\Delta_U$ is discrete, finitely degenerated and its eigenfunctions are smooth functions.

Hence, we may assume, as it was done \hyperref[ps_spectrum_B]{{\color{black}\textbf{above}}}, that the spectrum of $B$ is discrete, non-degenerate  and $\Phi_k(\boldsymbol{x})$ are smooth functions in $L^2(\mathcal{V})$.

Equation \hyperref[KG_classical_2_ps]{(\ref*{KG_classical_2})} may be transformed into a first order evolution differential system by introducing the new fields:
\phantomsection\label{reparemetrization_fields_ps}\begin{equation}\label{reparemetrization_fields}
\xi=B^{1/2}\varphi,\qquad\eta=B^{-1/2}\varphi_t,
\end{equation}
(notice that $B^{1/2}$ is well defined because $B$ is positive and invertible) and the equation of motion~\hyperref[KG_classical_2_ps]{(\ref*{KG_classical_2})} for the field $\varphi$ takes the simple symmetric form:
\phantomsection\label{system_evolution_field_ps}\begin{equation}\label{system_evolution_field}
\frac{\diff}{\diff t}\begin{pmatrix}
\xi\\
\eta
\end{pmatrix}=\begin{pmatrix}
0 & B\\
-B & 0
\end{pmatrix}\begin{pmatrix}
\xi\\
\eta
\end{pmatrix}.
\end{equation}

Then, the equations of motion for the Klein--Gordon field constitute an infinite-dimensional extension of the dynamics of a finite number of independent oscillators, \hyperref[sym_dynamics_ps]{(\ref*{sym_dynamics})}. Using the Fourier expansion of the real fields $\xi$ and $\eta$ (that here we may consider to be in $L^2(\mathcal{V})$\hspace{0.02cm}) with respect to the eigenfunctions $\Phi_k$ of $B$:
\phantomsection\label{xi_eta_classical_fields_cavity_ps}\begin{equation}\label{xi_eta_classical_fields_cavity}
\xi(\boldsymbol{x})=\sum_{k=1}^\infty\xi_k\Phi_k(\boldsymbol{x}),\qquad\eta(\boldsymbol{x})=\sum_{k=1}^\infty\eta_k\Phi_k(\boldsymbol{x}),
\end{equation}
with
\begin{equation*}
\xi_k=\hspace{-.16cm}\int\limits_{\hspace{.3cm}\mathcal{V}}\hspace{-.1cm}\xi(\boldsymbol{x})\Phi_k(\boldsymbol{x})\diff^{d}\boldsymbol{x},\qquad\eta_k=\hspace{-.16cm}\int\limits_{\hspace{.3cm}\mathcal{V}}\hspace{-.1cm}\eta(\boldsymbol{x})\Phi_k(\boldsymbol{x})\diff^{d}\boldsymbol{x},
\end{equation*}
then, the mechanical variables
\begin{equation}
q_k=\sqrt{\omega_k}\xi_k,\qquad p_k=\frac{\eta_k}{\sqrt{\omega_k}}
\end{equation}
can be interpreted as position and momentum for a one-dimensional oscillator of frequency $\omega_k$ and their evolution in time, given by eq.\,\hyperref[dynamics_oscillator_new_ps]{(\ref*{dynamics_oscillator_new})}, as a trajectory in phase space $\Omega=\mathbb{R}^{2\infty}$.

If we compute the Hamiltonian of the system, we have that
\begin{equation*}
\hspace{-3.5cm}H[\varphi]=\frac{1}{2}\hspace{-.16cm}\int\limits_{\hspace{.3cm}\mathcal{V}}\hspace{-.1cm}\left(\varphi_t^2+\nabla\varphi\cdot\nabla\varphi+m^2\varphi^2\right)\diff^d\boldsymbol{x}
\end{equation*}

\phantom{a}
\vspace{-1.5cm}
\begin{equation}
\hspace{3.5cm}=\frac{1}{2}\left(\|\varphi_t\|^2+\|B\varphi\|^2\right)=\frac{1}{2}\sum_{k=1}^\infty\omega_k(\xi_k^2+\eta_k^2).
\end{equation}

In the presence of field fluctuations, we introduce a statistical interpretation to the mechanical degrees of freedom $(q_k,p_k)$ of the field $\varphi(\boldsymbol{x})$. Thus, the classical statistical description of the field, whose physical meaning corresponds to the probability that a certain fluctuation of the field takes place, will be provided by a probability law $\rho$ on the infinite-dimensional phase space $\mathbb{R}^{2\infty}$. Therefore, in the presence of field fluctuations, the state of the field will induce a marginal probability density on each mode $\rho_k(q_k,p_k)$ defined by
\begin{equation*}
\rho_k(q_k,p_k)=\hspace{-.36cm}\int\limits_{\hspace{.5cm}\mathbb{R}^{2\infty}}\hspace{-.19cm}\rho(q_1,q_2,\ldots,q_k,\ldots;p_1,p_2,\ldots,p_k,\ldots)\prod_{l\neq k}^\infty\diff q_l\diff p_l.
\end{equation*}
Such marginal probability could be understood as a probability density for the $k$-th mode of the field $\varphi$ described by the one-dimensional oscillator with Hamiltonian $H_k(\xi_k,\eta_k)$. Similar considerations could be applied to finite-dimensional subspaces of modes of the field, whose statistical and tomographic description would be made as in \hyperref[section_Tomograms_clas_osc]{{\color{black}\textbf{section~\ref*{section_Tomograms_clas_osc}}}}.

The canonical or Gibbs state for the field $\varphi(\boldsymbol{x})$ is given by the probability distribution on the infinite-dimensional phase space of the system as
\phantomsection\label{Gaussian_measure_t_ps}
\begin{equation}\label{Gaussian_measure_t}
\rho_{can}(\xi_1,\xi_2,\ldots;\eta_1,\eta_2,\ldots)=N\e
,
\end{equation}
with $N$ the normalization factor of the state. If we compute the integral to get the normalization factor, we have:
\begin{multline}
\hspace{-.36cm}\int\limits_{\hspace{.5cm}\mathbb{R}^{2\infty}}\hspace{-.19cm}\e%
\prod_{k=1}^\infty\diff\xi_k\diff\eta_k\\
                                                                                                                                                                                                                                                                                                  =\left[\prod_{k=1}^\infty\left(\frac{1}{2\pi}\beta\omega_k\right)\right]^{-1}\hspace{-0.2cm}=\left[\textrm{Det}\left(\frac{\beta}{2\pi}B\right)\right]^{-1}.
\end{multline}

Because of the infinite product, we must regularize the determinant of the operator $B$ by using, for instance, the $\zeta$-function regularization of determinants \hyperref[El94_ps]{\RED{[\citen*{El94}, page 9]}} defined by:
\phantomsection\label{zeta_Riem_ps}
\begin{equation}\label{zeta_Riem}
\textrm{Det}\left(\frac{\beta}{2\pi}B\right)=\exp\left[-\zeta^\prime_{\frac{\beta}{2\pi} B}(0)\right],
\end{equation}
where $\zeta_{\frac{\beta}{2\pi} B}(s)$ denotes the generalized Riemann's $\zeta$-function:
\vspace{-0.2cm}\begin{equation*}
\zeta_{\frac{\beta}{2\pi} B}(s)=\sum_{k=1}^\infty\left(\frac{\beta}{2\pi}\omega_k\right)^{-s}.
\end{equation*}
\phantomsection\label{zeta_proof}
The justification of equation \hyperref[zeta_Riem_ps]{(\ref*{zeta_Riem})} is very simple. Let us define
$$
\widetilde{B}=\frac{\beta}{2\pi}B,
$$
with spectrum $\sigma(\widetilde{B})=\{0<\tilde{\omega}_1<\tilde{\omega}_2<\cdots<\tilde{\omega}_k<\cdots\}$, where
$$
\tilde{\omega}_k=\frac{\beta}{2\pi}\omega_k,
$$
hence, if we differentiate the $\zeta$-function
$$
\zeta_{\widetilde{B}}(s)=\sum_{n=1}^\infty\frac{1}{\tilde{\omega}_k^s},
$$
we get:
$$
\zeta^\prime_{\widetilde{B}}(s)=-\sum_{k=1}^\infty\frac{\log\tilde{\omega}_k}{\tilde{\omega}_k^s}\quad\Longrightarrow\quad\zeta^\prime_{\widetilde{B}}(0)=-\sum_{k=1}^\infty\log\tilde{\omega}_k=-\log\left(\prod_{k=1}^\infty\tilde{\omega}_k\right),
$$
thus, we will get:
\begin{equation*}
\e^{-\zeta^\prime_{\widetilde{B}}(0)}=\exp\left(\log\left(\prod_{k=1}^\infty\tilde{\omega}_k\right)\right)=\prod_{k=1}^\infty\tilde{\omega}_k=\textrm{Det}\,\widetilde{B}.
\end{equation*}

\phantom{a}
\vspace{-0.8cm}\hfill{\hyperref[zeta_proof]{{\color{black}$\blacksquare$}}}

\vspace{0.1cm}Let us remark that if the operator $\widetilde{B}$ is finite-dimensional, defined on a space of dimension $n$, then
\begin{equation}
\textrm{Det}\,\widetilde{B}=\prod_{k=1}^n\tilde{\omega}_k=\det\widetilde{B}.
\end{equation}

Let us compute the value of $\textrm{Det}\,\widetilde{B}$, eq.\,\hyperref[zeta_Riem_ps]{(\ref*{zeta_Riem})}, in an explicit case. Suppose that $\tilde{\omega}_k=k$ with $k=1,2,\ldots$, then
$$
\zeta_{\widetilde{B}}(s)=\sum_{k=1}^\infty\frac{1}{k^s}=\zeta_R(s),
$$
i.e., $\zeta_{\widetilde{B}}(s)$ is the Riemann's $\zeta$-function and then, it is well-known \hyperref[An99_ps]{\RED{[\citen*{An99},}} \hyperref[An99_ps]{\RED{page 16]}} that
$$
\zeta_R(0)=-\frac{1}{2},\qquad\zeta^\prime_{R}(0)=-\frac{1}{2}\log 2\pi.
$$ 
Hence,
\begin{equation}
\textrm{Det}\left(\frac{\beta}{2\pi}B\right)=\exp\left(\frac{1}{2}\log 2\pi\right)=\sqrt{2\pi}.
\end{equation}

Thus, the canonical state for the real scalar Klein--Gordon field is defined as the Gaussian measure \hyperref[Gaussian_measure_t_ps]{(\ref*{Gaussian_measure_t})} with variance
\begin{equation*}
\sigma=\frac{1}{\sqrt{2\pi}}\left[\textrm{Det}\left(\frac{\beta}{2\pi}B\right)\right]^{-1}
\end{equation*}
on $\mathbb{R}^{2\infty}$. The canonical ensemble for the Klein--Gordon field at finite temperature $T$ will be written in the usual form:
\phantomsection\label{canonical_ensemble_field_ps}\begin{equation}\label{canonical_ensemble_field}
\diff\mu_{can}[\varphi]=N\e^{-\beta H[\varphi]}\mathcal{D}\varphi,
\end{equation}
with the symbol $\mathcal{D}\varphi$ denoting the ``infinite dimensional'' measure:
\begin{equation*}
\mathcal{D}\varphi=\prod_{k=1}^\infty\diff q_k\diff p_k.
\end{equation*}
In what follows we will use this notation to remove infinite products. 

Moreover, if $F[\varphi]$ denotes an observable over the field $\varphi$ (like energy, momentum, etc.), then the expected value of $F$ on the canonical distribution will be given by:
\begin{equation}
\langle F\rangle_{can}=\frac{\displaystyle{\hspace{-.16cm}\int\limits_{\hspace{.5cm}\mathbb{R}^{2\infty}}\hspace{-.19cm}F[\varphi]\e^{-\beta H[\varphi]}}\mathcal{D\varphi}}{\displaystyle{\hspace{-.16cm}\int\limits_{\hspace{.5cm}\mathbb{R}^{2\infty}}\hspace{-.19cm}\e^{-\beta H[\varphi]}\mathcal{D}\varphi}}\,.
\end{equation}

The tomographic description of the states of the Klein--Gordon field will be performed, as in the case of an ensemble of harmonic oscillators, by choosing the spaces
$$\mathcal{M}=\mathbb{R}^{2\infty}\quad\mbox{and}\quad\displaystyle{\mathcal{N}=\prod_{k=1}^\infty}\mathcal{N}_k,$$ with $\mathcal{N}_k$ the space of straight lines on the phase space of the one-dimensional oscillator $(\xi_k,\eta_k)$. Then, as in eq.\,\hyperref[symplectic_cov_ps]{(\ref*{symplectic_cov})}, we will define:
\phantomsection\label{Tomogram_KG_field_ps}\begin{multline}\label{Tomogram_KG_field}
\mathcal{W}_{\rho_{can}}[X,\mu,\nu]=\hspace{-.4cm}\int\limits_{\hspace{.5cm}\mathbb{R}^{2\infty}}\hspace{-.19cm}\rho_{can}[\xi,\eta]\prod_{k=1}^\infty\delta(X_k-\mu_k\xi_k-\nu_k\eta_k)\diff\xi_k\diff\eta_k\\
=\hspace{-.4cm}\int\limits_{\hspace{.5cm}\mathbb{R}^{2\infty}}\hspace{-.19cm}\e^{-\beta H[\xi,\eta]}\delta\left[X(\boldsymbol{x})-\mu(\boldsymbol{x})\xi(\boldsymbol{x})-\nu(\boldsymbol{x})\eta(\boldsymbol{x})\right]\mathcal{D}\xi\mathcal{D}\eta.
\end{multline}
Here, the Dirac functional distribution must be understood as an infinite product\footnote{Notice that we are omitting here a normalization factor that will be absorbed by the regularization used in the definition of the integration over the spaces of fields $\xi$ and $\eta$.}:
\vspace{-0.2cm}\begin{multline}
\delta\left[X(\boldsymbol{x})-\mu(\boldsymbol{x})\xi(\boldsymbol{x})-\nu(\boldsymbol{x}\eta(\boldsymbol{x}))\right]=\prod_{k=1}^\infty\delta(X_k-\mu_k\xi_k-\nu_k\eta_k)\\
=\hspace{-.36cm}\int\limits_{\hspace{.5cm}\mathbb{R}^{\infty}}\hspace{-.19cm}\exp\left[i\hspace{-.16cm}\int\limits_{\hspace{.3cm}\mathcal{V}}\hspace{-.19cm}K(\boldsymbol{x})\big(X(\boldsymbol{x})-\mu(\boldsymbol{x})\xi(\boldsymbol{x})-\nu(\boldsymbol{x})\eta(\boldsymbol{x})\big)\diff^d \boldsymbol{x}\right]\mathcal{D}K,
\end{multline}
where $X(\boldsymbol{x})$, $\mu(\boldsymbol{x})$ and $\nu(\boldsymbol{x})$ are fields whose expansions over the modes $\omega_k$ of the field $\varphi(\boldsymbol{x})$ are given by:

\vspace{-0.4cm}\phantomsection\label{X_mu_nu_fields_ps}\begin{align}\label{X_mu_nu_fields}
&\hspace{1.7cm} X(\boldsymbol{x})=\sum_{k=1}^\infty X_k\Phi_k(\boldsymbol{x}),\nonumber\\ 
&\hspace{-0.3cm}\mu(\boldsymbol{x})=\sum_{k=1}^\infty \mu_k\Phi_k(\boldsymbol{x}),\qquad\nu(\boldsymbol{x})=\sum_{k=1}^\infty \nu_k\Phi_k(\boldsymbol{x}).
\end{align}
Notice that
$$
\|X\|_2=\Big\|\sum_{k=1}^\infty X_k\Phi_k\,\Big\|_2\leq\sum_{k=1}^\infty|X_k|\|\Phi_k\|_2=\sum_{k=1}^\infty|X_k|,
$$ 
therefore, the field $X(\boldsymbol{x})$ is in $L^2(\mathcal{V})$ if the $l^1$-norm of the vector $\boldsymbol{X}=(X_1,X_2,\ldots)$ is finite. Moreover, we may consider $X(\boldsymbol{x})$ to be a distribution as follows. Let $\phi$ be a test function, then let us define
\begin{equation}
X[\phi]=\sum_{k=1}^\infty X_k\langle\Phi_k,\phi\rangle,
\end{equation}
and we see that
\begin{multline*}
\|X[\phi]\|\leq\sum_{k=1}^\infty|X_k||\langle\Phi_k,\phi\rangle|\leq\sum_{k=1}^\infty|X_k|\|\Phi_k\|_2\|\phi\|_2\\
=\left(\sum_{k=1}^\infty|X_k|\right)\|\phi\|_2=\|\boldsymbol{X}\|_{l^1}\|\phi\|_2,
\end{multline*}
then, it makes sense to consider that the fields $X(\boldsymbol{x}),\,\mu(\boldsymbol{x})$ and $\nu(\boldsymbol{x})$ are distributions on $\mathcal{V}$.

Notice also that the time dependence of the various fields is encoded in the coefficients of the corresponding expansions. Hence, from eq.\,\hyperref[Gibbs_Tomogram_ps]{(\ref*{Gibbs_Tomogram})} putting $n=\infty$ and using the homogeneity property~\hyperref[homogeneity_Tomogram_1_ps]{(\ref*{homogeneity_Tomogram_1})}, we have that the tomogram of the canonical state of the Klein--Gordon field is:
\begin{multline*}
\mathcal{W}_{\rho_{can}}[X,\mu,\nu]=\prod_{k=1}^\infty\sqrt{\frac{\beta\omega_k}{2\pi(\mu_k^2+\nu_k^2)}}\exp\left(-\frac{\beta\omega_kX_k^2}{2(\mu_k^2+\nu_k^2)}\right)\\
=\left[\textrm{Det}\left(\frac{\beta}{2\pi}B\right)\right]^{1/2}\Bigg[\prod_{k=1}^\infty\sqrt{\frac{1}{\mu_k^2+\nu_k^2}}\Bigg]\e
,
\end{multline*}
with
\begin{equation*}
\widetilde{X}_k=\frac{X_k}{\sqrt{\mu_k^2+\nu_k^2}}\,.
\end{equation*}
And if we define the self-adjoint operator $A[\mu,\nu]$ depending on the fields $\mu(\boldsymbol{x})$ and $\nu(\boldsymbol{x})$ as
\begin{equation}
A[\mu,\nu]\Phi_k(\boldsymbol{x})=\sqrt{\mu_k^2+\nu_k^2}\,\Phi_k(\boldsymbol{x}), 
\end{equation}
hence, we finally have:
\phantomsection\label{Tom_with_det_ps}
\begin{equation}\label{Tom_with_det}
\mathcal{W}_{\rho_{can}}[X,\mu,\nu]=\left[\textrm{Det}\left(\frac{\beta}{2\pi}B\right)\right]^{1/2}\hspace{-0.1cm}\frac{1}{\textrm{Det}\big(A[\mu,\nu]\big)}\e
,
\end{equation}
and the reconstruction formula of the state can be performed by means of eq.\,\hyperref[Reconstruction_symplectis_tomogram_ps]{(\ref*{Reconstruction_symplectis_tomogram})}:
\begin{multline}
\rho_{can}[\xi,\eta]=\hspace{-0.46cm}\int\limits_{\hspace{.64cm}\mathbb{R}^{3\infty}}\hspace{-.3cm}\mathcal{W}_{\rho_{can}}[X,\mu,\nu]\\
\cdot\exp\left[-i\hspace{-.16cm}\int\limits_{\hspace{.3cm}\mathcal{V}}\hspace{-.19cm}\big(X(\boldsymbol{x})-\mu(\boldsymbol{x})\xi(\boldsymbol{x})-\nu(\boldsymbol{x})\eta(\boldsymbol{x})\big)\diff^d\boldsymbol{x}\right]\mathcal{D}X\mathcal{D}\mu\mathcal{D}\nu.
\end{multline}

\section{Tomographic picture of continuous modes}\label{section_Tomographic_continuous_modes}

If we consider the scalar field in a finite volume cavity or in the full Minkowski space-time for instance, many or all of the modes of the system will become continuous. For simplicity, we will assume that we are discussing the field again in the $1+d$ Minkowski space-time $\mathbb{M}=\mathbb{R}^{1+d}$ and the continuous modes of the fields $\varphi(\boldsymbol{x})$, $\xi(\boldsymbol{x})$ and $\eta(\boldsymbol{x})$ are described by the wave vector $\boldsymbol{k}$, that is:
\begin{equation}
\xi(\boldsymbol{x})=\hspace{-.16cm}\int\limits_{\hspace{.3cm}\mathcal{V}}\hspace{0cm}(\xi_k\e^{-i\boldsymbol{k}\cdot\boldsymbol{x}}+\xi_{-k}\e^{i\boldsymbol{k}\cdot\boldsymbol{x}})\diff^d\boldsymbol{k},\qquad \mbox{etc}.
\end{equation}
Now, a state of the field $\varphi$ will be represented by a probability measure $\rho[\xi,\eta]$, again non-negative and normalized. Notice that the fields $\xi(\boldsymbol{x})$ and $\eta(\boldsymbol{x})$ are, in general, any parametrization of the phase space of fields which leaves invariant the canonical measure, i.e., $\mathcal{D}\varphi=\mathcal{D}\xi\mathcal{D}\eta$ and satisfy the canonical commutation relations:
\begin{equation}
\left\{\vphantom{a^\dagger}\xi(\boldsymbol{x}),\eta(\boldsymbol{y})\right\}=\delta^d(\boldsymbol{x}-\boldsymbol{y}),\quad\left\{\vphantom{a^\dagger}\xi(\boldsymbol{x}),\xi(\boldsymbol{y})\right\}=\left\{\vphantom{a^\dagger}\eta(\boldsymbol{x}),\eta(\boldsymbol{y})\right\}=0.
\end{equation}
An example of such states will be given by the canonical ensemble, that is, the Gaussian measure whose covariance is given by the operator $B$ as in~\hyperref[canonical_ensemble_field_ps]{(\ref*{canonical_ensemble_field})}:
\begin{equation}
\diff\mu_{can}[\varphi]=\e^{-\beta H[\varphi]}\mathcal{D}\varphi=\e^{-\beta H[\xi,\eta]}\mathcal{D}\xi\mathcal{D}\eta,
\end{equation}
with the normalization constant $N$ absorbed in the definition of the measure.

We will consider, as analogue of Gibbs states, states that are absolutely continuous with respect to the canonical state, i.e., states of the form:
\phantomsection\label{state_continuous_modes_ps}\begin{equation}\label{state_continuous_modes}
\rho_f[\varphi]\mathcal{D}\varphi=f[\xi,\eta]\diff\mu_{can},
\end{equation}
with
\begin{equation}
f[\xi,\eta]\geq 0,\qquad\hspace{-.16cm}\int\limits_{\hspace{.5cm}\mathbb{R}^{2\infty}}\hspace{-.22cm}f[\xi,\eta]\e^{-\beta H[\xi,\eta]}\mathcal{D}\xi\mathcal{D}\eta=1.
\end{equation}

Even though, at a formal level, we may introduce as in~\hyperref[Tomogram_KG_field_ps]{(\ref*{Tomogram_KG_field})} a tomographic probability density for a state of a field of the  form~\hyperref[state_continuous_modes_ps]{(\ref*{state_continuous_modes})} as a functional of three auxiliary tomographic fields $X(\boldsymbol{x})$, $\mu(\boldsymbol{x})$ and $\nu(\boldsymbol{x})$ and apply, at the functional level, the usual Radon Transform. The expansions~\hyperref[X_mu_nu_fields_ps]{(\ref*{X_mu_nu_fields})} will be replaced by the Fourier Transform:
\begin{equation}
X(\boldsymbol{x})=\frac{1}{(2\pi)^{d/2}}\hspace{-.16cm}\int\limits_{\hspace{.3cm}\mathcal{V}}\hspace{0cm}(X_k\e^{-i\boldsymbol{k}\cdot\boldsymbol{x}}+X_{-k}\e^{i\boldsymbol{k}\cdot\boldsymbol{x}})\diff^d\boldsymbol{k},\qquad\mbox{etc.}
\end{equation}
Therefore,
\phantomsection\label{Tomogram_continuous_modes_2_ps}\begin{equation*}
\mathcal{W}_f[X,\mu,\nu]=\hspace{-.39cm}\int\limits_{\hspace{.5cm}\mathbb{R}^{2\infty}}\hspace{-.22cm}f[\xi,\eta]\delta[X(\boldsymbol{x})-\mu(\boldsymbol{x})\xi(\boldsymbol{x})-\nu(\boldsymbol{x})\eta(\boldsymbol{x})]\e^{-\beta H[\xi,\eta]}\mathcal{D}\xi\mathcal{D}\eta,
\end{equation*}
\phantom{a}
\vspace{-0.8cm}\begin{equation}\label{Tomogram_continuous_modes_2}
\phantom{a}
\end{equation}
and the Inverse Radon Transform is given again by:
\phantomsection\label{Inverse_Radon_continuous_modes_2_ps}\begin{multline}\label{Inverse_Radon_continuous_modes_2}
\rho_f[\xi,\eta]=\hspace{-.46cm}\int\limits_{\hspace{.64cm}\mathbb{R}^{3\infty}}\hspace{-.3cm}\mathcal{W}_{f}[X,\mu,\nu]\\
\cdot\exp\left[-i\hspace{-.16cm}\int\limits_{\hspace{.3cm}\mathcal{V}}\hspace{-.19cm}\big(X(\boldsymbol{x})-\mu(\boldsymbol{x})\xi(\boldsymbol{x})-\nu(\boldsymbol{x})\eta(\boldsymbol{x})\big)\diff^d\boldsymbol{x}\right]\mathcal{D}X\mathcal{D}\mu\mathcal{D}\nu.
\end{multline}
The tomographic probability functional~\hyperref[Tomogram_continuous_modes_2_ps]{(\ref*{Tomogram_continuous_modes_2})} has, by construction, the well-known properties of non-negativity and normalization:
\begin{equation}
\mathcal{W}_f[X,\mu,\nu]\geq 0,\qquad\hspace{-.16cm}\int\limits_{\hspace{.5cm}\mathbb{R}^{2\infty}}\hspace{-.21cm}\mathcal{W}_f[X,\mu,\nu]\mathcal{D}X=1.
\end{equation}
These formulas hold true for any value of the auxiliary fields $X(\boldsymbol{x})$, $\mu(\boldsymbol{x})$ and $\nu(\boldsymbol{x})$.

In the current case, the manifold $\mathcal{N}$ used to construct the Generalized Positive Transform, is described by the tomographic fields $X(\boldsymbol{x})$, $\mu(\boldsymbol{x})$ and $\nu(\boldsymbol{x})$, which would be a continuum version of the finite-mode version of the straight lines
\begin{equation}
X_k-\mu_k\xi_k-\nu_k\eta_k=0.
\end{equation}

We will end this discussion by emphasizing again the homogeneity property of the tomographic description of the scalar field we have just presented, homegeneity that is described by the condition:
\begin{equation}
\left[X(\boldsymbol{x})\frac{\delta}{\delta X(\boldsymbol{x})}+\mu(\boldsymbol{x})\frac{\delta}{\delta \mu(\boldsymbol{x})}+\nu(\boldsymbol{x})\frac{\delta}{\delta \nu(\boldsymbol{x})}+1\right]\mathcal{W}_f[X(\boldsymbol{x}),\mu(\boldsymbol{x}),\nu(\boldsymbol{x})]=0.
\end{equation}

\markboth{The tomographic picture of classical systems}{4.7.\,\hspace{0.07cm}The\hspace{0.03cm} tomographic\hspace{0.03cm} picture\hspace{0.03cm} of\hspace{0.03cm} the\hspace{0.03cm} evolution\hspace{0.03cm} equation\hspace{0.03cm} for\hspace{0.03cm} classical\newline{\color{white}.......}\hspace{0.02cm}fields}
\section{The tomographic picture of the evolution equation for classical fields}\label{tom_evo_arx}
\markboth{The tomographic picture of classical systems}{4.7.\,\hspace{0.07cm}The\hspace{0.03cm} tomographic\hspace{0.03cm} picture\hspace{0.03cm} of\hspace{0.03cm} the\hspace{0.03cm} evolution\hspace{0.03cm} equation\hspace{0.03cm} for\hspace{0.03cm} classical\newline{\color{white}.......}\hspace{0.02cm}fields}

In the \hyperref[section_Tomographic_continuous_modes]{{\color{black}\textbf{previous section}}}, we have seen that the state of the classical scalar field $\varphi$ can be described either by a probability density functional $\rho_f[\xi,\eta]$ on the field phase space or by the tomographic probability density functional $\mathcal{W}_f[X,\mu,\nu]$. Both probability density functionals are connected by the invertible functional Radon Transform eqs.\,\hyperref[Tomogram_continuous_modes_2_ps]{(\ref*{Tomogram_continuous_modes_2})} and \hyperref[Inverse_Radon_continuous_modes_2_ps]{(\ref*{Inverse_Radon_continuous_modes_2})} and in view of this, they both contain equivalent information about the random field states. The dynamical evolution of the states of the field $\varphi(t,\boldsymbol{x})$ can be determined by the Euler--Lagrange equation~\hyperref[Euler_Lagrange_Classical_field_2_ps]{(\ref*{Euler_Lagrange_Classical_field_2})}.

If the Hamiltonian providing the evolution of the field is given by $H[\varphi]=H[\xi,\eta]$, in a similar way as Liouville's equation of a system of finite degrees of freedom~\hyperref[Liouville_eq_ps]{(\ref*{Liouville_eq})} is a consequence of the Hamiltonian equations of motion~\hyperref[Hamilton_eq_ps]{(\ref*{Hamilton_eq})}, then from the similar ones of a Hamiltonian field~\hyperref[eq_motion_Hamiltonian_field_2_ps]{(\ref*{eq_motion_Hamiltonian_field_2})}, we can obtain the Liouville functional differential equation
\begin{equation}
\frac{\partial}{\partial t}\rho_f=\{\vphantom{a^\dagger}\rho_f,H\},
\end{equation}
where here, the functional Poisson bracket $\{\vphantom{a^\dagger}\cdot,\cdot\}$ is given by:
\begin{equation}
\left\{\vphantom{a^\dagger}F,G\right\}=\hspace{-.16cm}\int\limits_{\hspace{.3cm}\mathcal{V}}\hspace{-.1cm}\left(\frac{\delta F}{\delta\xi(\boldsymbol{x})}\frac{\delta G}{\delta\eta(\boldsymbol{x})}-\frac{\delta F}{\delta\eta(\boldsymbol{x})}\frac{\delta G}{\delta\xi(\boldsymbol{x})}\right)\diff^d\boldsymbol{x},
\end{equation}
for any functionals $F$ and $G$.

Therefore, working as in \hyperref[section_Tom_Liouville]{{\color{black}\textbf{section~\ref*{section_Tom_Liouville}}}}, we get the tomographic evolution equation for fields:
\phantomsection\label{Tomographic_Liouville_fields_ps}\begin{align}\label{Tomographic_Liouville_fields}
\frac{\diff}{\diff t}\mathcal{W}_f[X,\mu,\nu]=&\hspace{-.16cm}\int\limits_{\hspace{.3cm}\mathcal{V}}\hspace{-.09cm}\left[\frac{\delta H}{\delta\eta(\boldsymbol{x})}\left(\left\{\xi(\boldsymbol{x})\rightarrow-\left[\frac{\delta}{\delta X(\boldsymbol{x})}\right]^{-1}\hspace{-0.2cm}\frac{\delta}{\delta\mu(\boldsymbol{x})}\right\},\right.\right.\nonumber\\
&\hspace{1cm}\left.\left.\left\{\eta(\boldsymbol{x})\rightarrow-\left[\frac{\delta}{\delta X(\boldsymbol{x})}\right]^{-1}\hspace{-0.2cm}\frac{\delta}{\delta\nu(\boldsymbol{x})}\right\}\right)\mu(\boldsymbol{x})\right.\nonumber\\
&\hspace{0cm}-\left.\frac{\delta H}{\delta\xi(\boldsymbol{x})}\left(\left\{\xi(\boldsymbol{x})\rightarrow-\left[\frac{\delta}{\delta X(\boldsymbol{x})}\right]^{-1}\hspace{-0.2cm}\frac{\delta}{\delta\mu(\boldsymbol{x})}\right\},\right.\right.\nonumber\\
&\hspace{1cm}\left.\left.\left\{\eta(\boldsymbol{x})\rightarrow-\left[\frac{\delta}{\delta X(\boldsymbol{x})}\right]^{-1}\hspace{-0.2cm}\frac{\delta}{\delta\nu(\boldsymbol{x})}\right\}\right)\nu(\boldsymbol{x})\right]\nonumber\\
&\hspace{4cm}\cdot\frac{\delta}{\delta X(\boldsymbol{x})}\mathcal{W}_f[X,\mu,\nu]\diff^d\boldsymbol{x}.
\end{align}
And for the case in which the field is a collection of non-interacting oscillators described by the potential energy
\begin{equation}
V[\varphi]=\frac{1}{2}m^2\varphi^2
\end{equation}
parametrized with the fields \hyperref[reparemetrization_fields_ps]{(\ref*{reparemetrization_fields})}, therefore the equation \hyperref[Tomographic_Liouville_fields_ps]{(\ref*{Tomographic_Liouville_fields})} becomes:
\begin{equation}
\frac{\diff}{\diff t}\mathcal{W}_f[X,\mu,\nu]=\hspace{-.16cm}\int\limits_{\hspace{.3cm}\mathcal{V}}\hspace{-.09cm}B\left[\nu(\boldsymbol{x})\frac{\delta}{\delta\mu(\boldsymbol{x})}-\mu(\boldsymbol{x})\frac{\delta}{\delta\nu(\boldsymbol{x})}\right]\mathcal{W}_f[X,\mu,\nu]\diff^d\boldsymbol{x}.
\end{equation}

%% file: Tesischap5.tex
\chapter{Tomography in Quantum Field Theory}\label{chap_QFT}

\section{From Quantum Mechanics to Quantum Field Theory}

In \hyperref[chap_birth]{{\color{black}\textbf{chapter \ref*{chap_birth}}}}, it was shown how Quantum Tomography is in a sense a natural continuation of ideas born in classical telecommunications. For that, it was shown that after a canonical quantization of the E.M. field, some of these ideas could be extended to describe the states of photons.

The quantization of the E.M. field used in \hyperref[section_Quantization_elec]{{\color{black}\textbf{section \ref*{section_Quantization_elec}}}} is perhaps the simplest way to proceed when dealing with a classical field and was traditionally called \textit{canonical second quantization}. There is not a natural transition from Quantum Mechanics to a Quantum Theory of Fields, thus we will not pretend here to do a full development of a tomographic description of arbitrary quantum fields. However, what we will do will be, using as inspiration the canonical quantization of the E.M. field, to work, systematically, the example of a free quantum scalar field in a cavity and provide a tomographic description of some of its quantum states. Notice that such quantum field can be described as an infinite ensemble of harmonic oscillators and then, using again the techniques developed in \hyperref[chap_tom_qu]{{\color{black}\textbf{chapter \ref*{chap_tom_qu}}}}, we will be able to obtain nice tomographic pictures for it.

Quantum fields can be described axiomatically in various (equivalent) ways. We will use here the Wightman--Streater axioms \cite{St64} which are closer to the formalism developed so far and we will use an adaptation of \textit{Group Quantum Tomography} based on Poincar\'e group, similar to the one developed in \hyperref[section_particular group]{{\color{black}\textbf{section \ref*{section_particular group}}}}, to construct a tomographic description of them.

It is a fundamental axiom of the theory that the space of quantum states supports a unitary representation of Poincar\'e group. Then, we will use such data as a main ingredient in the construction. In this way, the constructed tomographic theory is explicitly Poincar\'e invariant, something that is not obvious at all in the canonical picture.

We will end this chapter by proving a reconstruction theorem that shows that the tomographic picture is equivalent to the Wightman--Streater axiomatic picture.

\section{The holomorhpic quantization of the scalar field in a cavity}\label{resumen_quantization}

Similar to the classical setting discussed in \hyperref[sec_KG_fields]{{\color{black}\textbf{section~\ref*{sec_KG_fields}}}}, let us consider Minkowski space-time $\mathbb{M}=\mathbb{R}^{1+d}$ with metric of signature $(-,+\ldots,+)$. Let us consider again the Cauchy hypersurface $\mathcal{C}=\{0\}\times\mathbb{R}^d$ and the compact smooth cavity $\mathcal{V}\subset\mathcal{C}$.

We will consider a scalar field $\varphi:\mathbb{R}\times\mathcal{V}\rightarrow\mathbb{R}$ where the dynamics is given by Klein--Gordon equation \hyperref[KG_classical_ps]{(\ref*{KG_classical})}. Such field can be used to describe a number of physical systems, but for the moment, we will be considering just its mathematical aspects.

The canonical quantization of the classical field $\varphi$ and its momentum $\pi$ is obtained by defining the canonical commutators for the quantum fields $\boldsymbol{\varphi}$ and $\boldsymbol{\pi}$ with the Dirac's correspondence principle:
\begin{equation}
\{\vphantom{a^\dagger}\cdot,\cdot\}\longrightarrow-i[\vphantom{a^\dagger}\cdot,\cdot].
\end{equation}
Hence, the equal time canonical commutators are:
\begin{equation}
\left[\vphantom{a^\dagger}\boldsymbol{\varphi}(\boldsymbol{x}),\boldsymbol{\pi}(\boldsymbol{x}')\right]=i\delta(\boldsymbol{x}-\boldsymbol{x}'),\quad\left[\vphantom{a^\dagger}\boldsymbol{\varphi}(\boldsymbol{x}),\boldsymbol{\pi}(\boldsymbol{x}')\right]=\left[\vphantom{a^\dagger}\boldsymbol{\pi}(\boldsymbol{x}),\boldsymbol{\pi}(\boldsymbol{x}')\right]=0.
\end{equation}
The fields $\xi(\boldsymbol{x})$ and $\eta(\boldsymbol{x})$ defined in~\hyperref[xi_eta_classical_fields_cavity_ps]{(\ref*{xi_eta_classical_fields_cavity})} become now quantum field operators:
\begin{equation}
\boldsymbol{\xi}(\boldsymbol{x})=\sum_{k=1}^\infty\boldsymbol{\xi}_k\Phi_k(\boldsymbol{x}),\qquad\boldsymbol{\eta}(\boldsymbol{x})=\sum_{k=1}^\infty\boldsymbol{\eta}_k\Phi_k(\boldsymbol{x}),
\end{equation}
where the operators $\boldsymbol{\xi}_k$ and $\boldsymbol{\eta}_k$ satisfy the commutation relations:
\begin{equation}
\left[\vphantom{a^\dagger}\boldsymbol{\xi}_k,\boldsymbol{\eta}_l\right]=i\delta_{kl},\qquad\left[\vphantom{a^\dagger}\boldsymbol{\xi}_k,\boldsymbol{\xi}_l\right]=\left[\vphantom{a^\dagger}\boldsymbol{\eta}_k,\boldsymbol{\eta}_l\right]=0,\qquad k,l=1,2,\ldots\,.
\end{equation}

As it was done in the study of the harmonic oscillator in \hyperref[section_Toms_q_h_o]{{\color{black}\textbf{subsec-}}} \hyperref[section_Toms_q_h_o]{{\color{black}\textbf{tion~\ref*{section_Toms_q_h_o}}}}, it would be convenient to deal with the quantum scalar field by using the corresponding extension of the holomorphic quantization (or Bargmann--Segal quantization scheme) discussed in \hyperref[holom_chap_2]{{\color{black}\textbf{subsection~\ref*{holom_chap_2}}}}.

For that, we will proceed first to construct the Fock Hilbert space of the system. Consider the family of creation and annihilation operators $a_k^\dagger$, $a_k$ by means of:
\begin{equation}
a_k=\frac{1}{\sqrt{2}}(\boldsymbol{\xi}_k+i\boldsymbol{\eta}_k),\qquad a_k^\dagger=\frac{1}{\sqrt{2}}(\boldsymbol{\xi}_k-i\boldsymbol{\eta}_k).
\end{equation}
Clearly, we have the canonical commutation relations:
\phantomsection\label{properties_1_a_ops_ps}\begin{equation}\label{properties_1_a_ops}
\left[\vphantom{a^\dagger}\right.\hspace{-0.1cm}a_k,a_l^\dagger\hspace{-0.1cm}\left.\vphantom{a^\dagger}\right]=\delta_{kl},\qquad\left[\vphantom{a^\dagger}a_k,a_l\right]=\left[\vphantom{a^\dagger}\right.\hspace{-0.1cm}a_k^\dagger,a_l^\dagger\hspace{-0.1cm}\left.\vphantom{a^\dagger}\right]=0,\qquad\forall k,l.
\end{equation}

The Fock space of the theory is the Hilbert space $\mathcal{F}_\infty$ generated by the set of vectors
\begin{equation}
\left\{\vphantom{a^\dagger}\right.\hspace{-0.1cm}a_{k_1}^\dagger\cdots a_{k_N}^\dagger|0\rangle\hspace{-0.1cm}\left.\vphantom{a^\dagger}\right\},
\end{equation}
where $|0\rangle$ denotes the \textit{ground state} or \textit{vacuum} of the theory that satisfies:
\phantomsection\label{property_2_a_ops_ps}\begin{equation}\label{property_2_a_ops}
a_{k}|0\rangle=0,\qquad\forall k.
\end{equation}

The multipartite state corresponding to the modes $k_1,\ldots,k_N$ will be denoted as $|1_{k_1},\ldots,1_{k_N}\rangle$ and:
\begin{equation}
|1_{k_1},\ldots,1_{k_N}\rangle=a_{k_1}^\dagger\cdots a_{k_N}^\dagger|0\rangle,\qquad k_1,\ldots, k_N=1,2,\ldots\,.
\end{equation}
The inner product is defined as:
\begin{equation}
\langle 1_{k_1},\ldots,1_{k_N}|1_{l_1},\ldots,1_{l_M}\rangle=\delta_{NM}\cdot\left\{\begin{matrix}
1&\mbox{if}\quad\exists\sigma\in S_N\quad\mbox{s.t.}\quad k_i=l_{\sigma(i)},\\
0&\mbox{otherwise.}
\end{matrix}\right.
\end{equation}
Thus, the particle state corresponding to the mode $\Phi_1$ will be denoted as:
\begin{equation}
|1\rangle=a_1^\dagger|0\rangle,
\end{equation}
and because of the bosonic nature of the field, a state of $N$ particles will be:
\begin{multline}
|n_{k_1},n_{k_2},\ldots,n_{k_N}\rangle=\frac{1}{\sqrt{n_{k_1}\hspace{-0.06cm}!\,n_{k_2}\hspace{-0.04cm}!\,\cdots\, n_{k_N}\hspace{-0.05cm}!\,}}\left(\vphantom{a^\dagger}\right.\hspace{-0.1cm}a_{k_1}^\dagger\hspace{-0.1cm}\left.\vphantom{a^\dagger}\right)^{n_{k_1}}\hspace{-0.13cm}\left(\vphantom{a^\dagger}\right.\hspace{-0.1cm}a_{k_2}^\dagger\hspace{-0.1cm}\left.\vphantom{a^\dagger}\right)^{n_{k_2}}\\
\cdots\left(\vphantom{a^\dagger}\right.\hspace{-0.1cm}a_{k_N}^\dagger\hspace{-0.1cm}\left.\vphantom{a^\dagger}\right)^{n_{k_N}}|0\rangle.
\end{multline}
Therefore, the Fock space $\mathcal{F}_\infty$ can be written as follows:
\begin{equation}
\mathcal{F}_\infty=\overline{\textrm{span}}\left\{\vphantom{a^\dagger}\right.\hspace{-0.1cm|}n_{k_1},\ldots,n_{k_N}\rangle\,|\,\forall N\in\mathbb{N},\,k_1,\ldots,k_N\geq 1,\, n_{k_1},\ldots,n_{k_N}\geq 0\hspace{-0.1cm}\left.\vphantom{a^\dagger}\right\}.
\end{equation}

The canonical state may be defined as the ``quantization'' of the classical one:
\begin{equation}
\rho(\boldsymbol{\xi},\boldsymbol{\eta})=\e^{-\beta H(\boldsymbol{\xi},\boldsymbol{\eta})},
\end{equation}
then, we can define the \textit{quantum canonical tomogram} for the canonical state of the theory in analogy with~\hyperref[Tomogram_KG_field_ps]{(\ref*{Tomogram_KG_field})}:
\begin{multline}
\mathcal{W}_{q,can}[X,\mu,\nu]=\Tr\!\left(\e^{-\beta H(\boldsymbol{\xi},\boldsymbol{\eta})}\right.\\
\left.\cdot\delta\big(X(\boldsymbol{x})\mathds{1}-\mu(\boldsymbol{x})\boldsymbol{\xi}(\boldsymbol{x})-\nu(\boldsymbol{x})\boldsymbol{\eta}(\boldsymbol{x})\big)\vphantom{\e^{-\beta H(\boldsymbol{\xi},\boldsymbol{\eta})}}\right).
\end{multline}
In general, if $\boldsymbol{\rho}$ denotes a state of the quantum field $\boldsymbol{\varphi}$, that is, a density operator on the Fock space of $\boldsymbol{\varphi}$, we will define its quantum canonical tomogram as:
\begin{equation}
\mathcal{W}_{q,\rho}[X,\mu,\nu]=\Tr\!\left(\boldsymbol{\rho}\delta\big(X(\boldsymbol{x})\mathds{1}-\mu(\boldsymbol{x})\boldsymbol{\xi}(\boldsymbol{x})-\nu(\boldsymbol{x})\boldsymbol{\eta}(\boldsymbol{x})\big)\right)\!,
\end{equation}
in close analogy with the definition of quantum tomogram given in \hyperref[tom_cart]{{\color{black}\textbf{chap-}}} \hyperref[tom_cart]{{\color{black}\textbf{ter~\ref*{chap_tom_qu}}}}.

It is easy to show that defining the complex fields
\begin{equation*}
\hspace{-1.35cm}w=\frac{1}{\sqrt{2}}(\mu(\boldsymbol{x})+i\nu(\boldsymbol{x}))=\frac{1}{\sqrt{2}}\sum_{k=1}^\infty(\mu_k+i\nu_k)\Phi_k(\boldsymbol{x})
\end{equation*}

\phantom{a}
\vspace{-0.7cm}
\noindent and
\begin{equation}
\overline{w}=\frac{1}{\sqrt{2}}(\mu(\boldsymbol{x})-i\nu(\boldsymbol{x}))=\frac{1}{\sqrt{2}}\sum_{k=1}^\infty(\mu_k-i\nu_k)\Phi_k(\boldsymbol{x}),
\end{equation}
with components
$$
w_k=\mu_k+i\nu_k,\qquad \overline{w}_k=\mu_k-i\nu_k,
$$
the quantum canonical tomogram of the ground state $\boldsymbol{\rho}_0=|0\rangle\langle 0|$ of the free quantum field $\boldsymbol{\varphi}$ is given by:
\phantomsection\label{tom_0_quantum_scalar_ps}
\begin{equation}\label{tom_0_quantum_scalar}
\mathcal{W}_{q,0}[X,w,\overline{w}]=\Tr\!\left(\boldsymbol{\rho}_0\delta\big(X(\boldsymbol{x})\mathds{1}-\overline{w}(\boldsymbol{x})a(\boldsymbol{x})-w(\boldsymbol{x})a^\dagger(\boldsymbol{x})\big)\right)\!,
\end{equation}
where
\begin{equation*}
w(\boldsymbol{x})a^\dagger(\boldsymbol{x})=\sum_{k=1}^\infty w_ka_k^\dagger=\frac{1}{\sqrt{2}}\sum_{k=1}^\infty(\mu_k+i\nu_k)\Phi_k(\boldsymbol{x})a_k^\dagger
\end{equation*}
and
\begin{equation*}
\overline{w}(\boldsymbol{x})a(\boldsymbol{x})=\sum_{k=1}^\infty\overline{w}_ka_k=\frac{1}{\sqrt{2}}\sum_{k=1}^\infty(\mu_k-i\nu_k)\Phi_k(\boldsymbol{x})a_k.
\end{equation*}
Hence, we will obtain:
\begin{multline}
\mathcal{W}_{q,0}[X,w,\overline{w}]=\Tr\!\left[\hspace{-.26cm}\int\limits_{\hspace{.5cm}\mathbb{R}^\infty}\hspace{-0.2cm}\boldsymbol{\rho}_0\prod_{r=1}^\infty\e^{ik_r(X_r\mathds{1}-\overline{w}_ra_r-w_ra_r^\dagger)}\frac{\diff k_r}{2\pi}\right]\\
=\prod_{r=1}^\infty\hspace{-0cm}\int\limits_{\hspace{-.04cm}-\infty}^{\hspace{.45cm}\infty}\hspace{-0.15cm}\e^{ik_rX_r}\langle0|D(z_r)|0\rangle\frac{\diff k_r}{2\pi},
\end{multline}
where $z_r=-ik_rw_r$ and $D(z_r)$ denotes the displacement operator:
\begin{equation}
D(z_r)=\e^{z_ra_r^\dagger-\bar{z}_ra}_r.
\end{equation}
Notice that
\begin{equation}
D(z_r)|0\rangle=|z_r\rangle,
\end{equation}
where $|z_r\rangle$ is the coherent state defined in~\hyperref[coherent_state_ps]{(\ref*{coherent_state})}. Therefore, from \hyperref[prod_coherent_ps]{(\ref*{prod_coherent})} we have that
\begin{equation}
\langle 0|z_r\rangle=\e
\,.
\end{align}
Notice that the divergent factor appearing in the r.h.s of this formula can be analyzed as in \hyperref[sec_KG_fields]{{\color{black}\textbf{section \ref*{sec_KG_fields}}}} using the $\zeta$-function regularization of determinants \hyperref[zeta_Riem_ps]{(\ref*{zeta_Riem})}. Thus, if we consider the self-adjoint operator $\widetilde{A}[\mu,\nu]$ with eigenvalues
\begin{equation}
|w_k|=\sqrt{\mu_k^2+\nu_k^2}\,,
\end{equation}
then, if $|\pi w_k|>k$, we have that the formula \hyperref[Tom_q_0_chap_5_ps]{(\ref*{Tom_q_0_chap_5})} is finally written as:
\begin{equation}
\mathcal{W}_{q,0}[X,w,\overline{w}]=\frac{1}{\textrm{Det}\big(\widetilde{A}[\mu,\nu]\big)}\e^{-\|\widetilde{X}\|^2},
\end{equation}
where recall that the definition of the coefficients of $\widetilde{X}$ is
\begin{equation*}
\widetilde{X}_k=\frac{X_k}{\sqrt{\mu_k^2+\nu_k^2}}\,.
\end{equation*}
\newpage

\section{Wightman--Streater axioms of a Quantum Field Theory}\label{section_WS_axioms}

The Wightman--Streater axioms \hyperref[St64_ps]{\RED{[\citen*{St64}, ch.\,3]}} for a Quantum Field Theory provide an axiomatic setting to describe a quantum scalar field $\boldsymbol{\varphi}$ in Minkowski space-time $\mathbb{M}=\mathbb{R}^{1+3}$. The main ingredients of the theory are a complex separable Hilbert space $\mathcal{F}_\infty$, a distribution in Minkowski space-time with values in bounded operators in $\mathcal{F}_\infty$:
\begin{equation}
\boldsymbol{\varphi}:\mathcal{D}(\mathbb{M})\rightarrow\mathcal{B}(\mathcal{F}_\infty),\qquad f\rightsquigarrow\boldsymbol{\varphi}(f),\quad\forall f\in\mathcal{D}(\mathbb{M}),
\end{equation}
and a unitary representation $U$ of the proper orthochronous Poincar\'e group $\mathcal{P}_+^\uparrow=\mathbb{R}^{4}\times\mathcal{L}_0$ (see \hyperref[section_Minkowsiki]{{\color{black}\textbf{appendix A}}}) in $\mathcal{F}_\infty$:
\begin{equation}
U:\mathcal{P}_+^\uparrow\rightarrow\mathcal{U}(\mathcal{F}_\infty),\qquad\left\{\Lambda,a\right\}\rightsquigarrow U\left(\Lambda,a\right),\quad\forall\left\{\Lambda,a\right\}\in\mathcal{P}_+^\uparrow.
\end{equation}

The states of the theory are described by unit rays in the Hilbert space $\mathcal{F}_\infty$. The relativistic transformation law of the states is given by the continuous unitary representation $U(\Lambda,a)$ of the proper orthochronous Poincar\'e group $\mathcal{P}_+^\uparrow$. Since $U(\mathds{1},a)$ is unitary, it can be written as 
$$
U(\mathds{1},a)=\e^{ia^\mu \textbf{P}_\mu},
$$
where $\textbf{P}_\mu$ are the unbounded self-adjoint operators representing the energy ($\mu=0$) and the momentum ($\mu=1,2,3$) of the theory. These operators satisfy $\textbf{P}^\mu\textbf{P}_\mu=m^2$. The spectrum of the energy-momentum operators lies inside the future component of the hyperboloid $H^{+}_m$:
\begin{equation*}
\sigma(\textbf{P}_\mu)\subset H^{+}_m=\left\{p_\mu\in\mathbb{R}^4\left|\:-p_0^2+p^2= m^2\right.,\:p_0\geq0\right\}.
\end{equation*} 
All of this is subjected to the following axioms:

\begin{enumerate}

\phantomsection\label{vacuum_prop_p}\item[\MYBROWN{1.}] Existence of the vacuum:

This first axiom requires the existence of a unique state $\left|0\right\rangle$ in the Hilbert space $\mathcal{F}_\infty$ such that
\begin{equation}
U\left(\Lambda,a\right)\left|0\right\rangle=\left|0\right\rangle.
\end{equation}

\item[\MYBROWN{2.}] Completeness: 

The set of vectors in the Fock Hilbert space $\mathcal{F}_\infty$,
\begin{equation}
\left\{\vphantom{a^\dagger}\boldsymbol{\varphi}(f_n)\cdots\boldsymbol{\varphi}(f_1)\left|0\right\rangle\right\},
\end{equation}
for all $n\in\mathbb{N}$ and any set of functions $f_i\in\mathcal{D}(\mathbb{M})$, $i=1,\ldots,n$, is dense in $\mathcal{F}_\infty$.

\phantomsection\label{Covariance_axiom_WS_ps}
\item[\MYBROWN{3.}] Covariance:

The scalar field $\boldsymbol{\varphi}$ must transform, under the action of the Poincar\'e group $\mathcal{P}_+^\uparrow$, in the natural covariant way:
\begin{equation}
U\left(\Lambda,a\right)\boldsymbol{\varphi}(f)U\left(\Lambda,a\right)^\dagger= \boldsymbol{\varphi}\left(\left(\Lambda,a\right)^*f\right),
\end{equation}
where $(\Lambda,a)^*f$ denotes $(\Lambda,a)^*f(x)=f\big((\Lambda,a)^{-1}\cdot x\big)$. Notice that in the rest of this thesis, $\boldsymbol{x}$ will denote vectors in the spatial part of $\mathbb{R}^3$ and $x$ will denote events, that is, $x\in\mathbb{M}$.

\phantomsection\label{microscopic_prop_p}\item[\MYBROWN{4.}] Microscopic causality:

If the support of the test functions $f$ and $g$ are space-like separated, i.e., if $f(x)g(y)=0$ for all pairs of points $x$ and $y$ in $\mathbb{M}$ such that $x-y$ is space-like (see \hyperref[spatial_vector_appendix]{\MYBROWN{2}} in \hyperref[section_Minkowsiki]{{\color{black}\textbf{appendix A}}}) then:
\begin{equation}
\left[\boldsymbol{\varphi}(f),\boldsymbol{\varphi}(g)\right]=0.
\end{equation}

\item[\MYBROWN{5.}] Asymptotic completeness:

In a collision of particles, we will require that Hilbert spaces before and after the collision are equal:
\begin{equation}
\mathcal{F}_\infty^{\,in}=\mathcal{F}_\infty=\mathcal{F}_\infty^{\,out}.
\end{equation}
There are several approaches to this notion, but the one which is closer to the other axioms is due to Haag and Ruelle. Ruelle has shown that axioms \hyperref[vacuum_prop_p]{\MYBROWN{1}}\MYBROWN{--}\hspace{0.03cm}\hyperref[microscopic_prop_p]{\MYBROWN{4}} imply the existence of collision states, that is, incoming and outgoing states of one, two, or more particles provided that the one-particle states may be created by a polynomial in the fields. Then, one can formulate this axiom in terms of these collision states. More information may be found in \RED{[}\citen{Ha59}\RED{,}\hspace{0.05cm}\citen{Ru62}\RED{]}.

\end{enumerate}

\phantom{a}

\markboth{Tomography in Quantum Field Theory}{5.4. Smeared\hspace{0.02cm} covariant\hspace{0.01cm} characters\hspace{0.02cm} and\hspace{0.02cm} tomograms\hspace{0.02cm} of\hspace{0.02cm} a\hspace{0.02cm} quantum\hspace{0.01cm} real\hspace{0.02cm} scalar\newline{\color{white}.....}\hspace{0.07cm} field}
\vspace{-1cm}\section{Smeared covariant characters and tomograms of a quantum real scalar field}\label{section_characters_QFT}
\markboth{Tomography in Quantum Field Theory}{5.4. Smeared\hspace{0.02cm} covariant\hspace{0.01cm} characters\hspace{0.02cm} and\hspace{0.02cm} tomograms\hspace{0.02cm} of\hspace{0.02cm} a\hspace{0.02cm} quantum\hspace{0.01cm} real\hspace{0.02cm} scalar\newline{\color{white}.....}\hspace{0.07cm} field}\label{section_characters_QFT}

In analogy with the smeared characters of a quantum state discussed at length in \hyperref[section_particular group]{{\color{black}\textbf{section \ref*{section_particular group}}}}, eq.\,\hyperref[smeared_character_chap_2_ps]{(\ref*{smeared_character_chap_2})}, we can extend this notion to quantum states of fields using the unitary representation $U$ of the Poincar\'e group $\mathcal{P}^{\uparrow}_{+}$ provided by the theory. Given a family of test functions $f_1,\ldots,f_n$, we will define the \textit{smeared covariant character} of the state $\boldsymbol{\rho}$ corresponding to the vector $\boldsymbol{\varphi}(f_n)\cdots\boldsymbol{\varphi}(f_1)|0\rangle$ in $\mathcal{F}_\infty$ as follows:
\begin{equation}
\chi_g(f_1,\ldots f_n)=\Tr\!\big(\boldsymbol{\rho}U(g)\big),\qquad g=(\Lambda,a)\in\mathcal{P}_+^\uparrow.
\end{equation}
Notice that in this case (where the scalar field $\boldsymbol{\varphi}$ is real):
\phantomsection\label{smeared_character_Wightman_ps}\begin{equation}\label{smeared_character_Wightman}
\chi_g(f_1,\ldots,f_n)=\left\langle0\right|\boldsymbol{\varphi}(f_1)\cdots\boldsymbol{\varphi}(f_n) U(g)\boldsymbol{\varphi}(f_n)\cdots\boldsymbol{\varphi}(f_1)\left|0\right\rangle.
\end{equation}
In this formula, we should divide by the factor $\left\langle0\right|\boldsymbol{\varphi}(f_1)\cdots\boldsymbol{\varphi}(f_n)\boldsymbol{\varphi}(f_n)\cdots$ $\boldsymbol{\varphi}(f_1)\left|0\right\rangle$ to normalize the state $\boldsymbol{\rho}$ but for simplicity in the writing, we will assume in what follows that the state is normalized. 

Applying axiom \hyperref[Covariance_axiom_WS_ps]{\MYBROWN{3}}, we get after a simple computation that
\phantomsection\label{phipurestate_ps}\begin{equation}\label{phipurestate}
\chi_g(f_1,\ldots,f_n)=\left\langle0\right|\boldsymbol{\varphi}(f_1)\cdots\boldsymbol{\varphi}(f_n)\boldsymbol{\varphi}(g^*f_n)\cdots\boldsymbol{\varphi}(g^*f_1)\left|0\right\rangle,
\end{equation}
where $g^*f(x)=f\big(g^{-1}\cdot x\big)$.

The r.h.s. of the equation~\hyperref[phipurestate_ps]{(\ref*{phipurestate})} defines a distribution in $\displaystyle{\mathbb{M}\times\cdots\times\mathbb{M}^{\hspace{-1.23cm}2n}}$\hspace{0.95cm}\phantom{,} that can be written as:
\phantomsection\label{Wightman_funct_nn_ps}\begin{multline}\label{Wightman_funct_nn}
\left\langle0\right|\boldsymbol{\varphi}(f_1)\cdots\boldsymbol{\varphi}(f_n)\boldsymbol{\varphi}(g^*f_n)\cdots\boldsymbol{\varphi}(g^*f_1)\left|0\right\rangle\\
=W_{nn}(f_1,\ldots,f_n,g^*f_n,\ldots,g^*f_1),
\end{multline}
where $W_{nn}$ are Wightman functions \hyperref[St64_ps]{\RED{[\citen*{St64}, page 106]}}. We can write this distribution as a linear functional:
\begin{multline}
W_{nn}(f_1,\ldots,f_n,g^*f_n,\ldots,g^*f_1)=\hspace{-.4cm}\int\limits_{\hspace{.5cm}\mathbb{M}^{2n}}\hspace{-0.3cm} \overline{f_1(x_1)}\cdots\overline{f_n(x_n)}W_{nn}\big(x_1,\ldots,x_n,\\
g^{-1}\cdot x_n',\ldots,g^{-1}\cdot x_1'\big)f_n(x_n')\cdots f_1(x_1')\,\diff x_1\cdots \diff x_n\,\diff x_n'\cdots \diff x_1',
\end{multline}
and, in what follows, we will just use the kernels
\phantomsection\label{Wightman_funct_nn_kernel_ps}
\begin{multline}\label{Wightman_funct_nn_kernel}
W_{nn}\big(x_1,\ldots,x_n,g^{-1}\cdot x_n',\ldots,g^{-1}\cdot x_1'\big)\\
=\left\langle0\right|\boldsymbol{\varphi}(x_1)\cdots\boldsymbol{\varphi}(x_n)\boldsymbol{\varphi}(g^{-1}\cdot x_n')\cdots\boldsymbol{\varphi}(g^{-1}\cdot x_1')\left|0\right\rangle
\end{multline}
instead of the full functions. Now, we will use the notation $x=(x_1,\ldots,x_n)$ to indicate the collection of $n$ points in Minkowski space-time, $x_i\in\mathbb{M}$, $i=1,\ldots,n$, i.e., $x\in\mathbb{M}^n$. Then, $g\cdot x=(g\cdot x_1,\ldots, g\cdot {x_n})$ is the diagonal action of $\mathcal{P}^\uparrow_{+}$ on $x$.

By analogy, we can write the kernel of the smeared characters as follows:
\begin{multline}
\chi_g(x,x')\coloneq\chi_g(x_1,\ldots,x_n,x_n',\ldots,x_1')\\
=W_{nn}\big(x_1,\ldots,x_n,g^{-1}\cdot x_n',\ldots,g^{-1}\cdot x_1'\big).
\end{multline}

In the following enumeration, we will write the properties that satisfy the kernels of the smeared characters induced by the axioms presented in the \hyperref[section_WS_axioms]{{\color{black}\textbf{previous section}}} and the properties of Wightman functions.
\begin{enumerate}

\phantomsection
\label{ps_Covariance}
\item[\MYBROWN{(a)}] Covariance:
\phantomsection\label{Covariance_sec_5_4_ps}
\begin{equation}\label{Covariance_sec_5_4}
\chi_g(hx,hx')=\chi_{h^{-1}gh}(x,x'),\qquad\forall g,h\in\mathcal{P}_+^\uparrow.
\end{equation}
This property comes directly from a direct application of the covariance axiom \hyperref[Covariance_axiom_WS_ps]{\MYBROWN{3}}. Notice that the Wightman functions~\hyperref[Wightman_funct_nn_kernel_ps]{(\ref*{Wightman_funct_nn_kernel})} are invariant  under the action of the group, this is:
\phantomsection\label{Invariance_Wightamann_Group_action_ps}\begin{equation}\label{Invariance_Wightamann_Group_action}
W_{nn}(x,x')=g\cdot W_{nn}(x,x')=W_{nn}(g\cdot x,g\cdot x'),
\end{equation}
hence, from this fact, \hyperref[Covariance_sec_5_4_ps]{(\ref*{Covariance_sec_5_4})} is easily obtained.\newpage

\phantomsection\label{Hermiticity_con_ps}
\item[\MYBROWN{(b)}] Hermiticity:
\begin{equation}
\overline{\chi_g(x,x')}=\chi_{g^{-1}}(x',x).
\end{equation}
This property is also obtained from the invariance of the Wightman functions under the action of the group~\hyperref[Invariance_Wightamann_Group_action_ps]{(\ref*{Invariance_Wightamann_Group_action})} and the unitarity of the representation $U$.

\phantomsection\label{Positivity_cond_Wight_ps}
\item[\MYBROWN{(c)}] Positivity:
\begin{equation}
\int\limits_{\hspace{.5cm}\mathbb{M}^{2n}}\hspace{-0.0cm}\sum_{i,j=1}^N\overline{f_i(x)}\chi_{g_i^{-1}g_j}(x,x')f_j(x')\diff^nx\,\diff^nx'\geq 0,
\end{equation}
for all $N$ in $\mathbb{N}$, $f_i$ in $\mathcal{D}(\mathbb{M})$ and $g_i$ in $\mathcal{P}_+^\uparrow$, $i=1,\ldots,N$.

\phantomsection\label{positivity_Wight_proof_ps}
The proof of this property is a direct consequence of the positivity (see also \hyperref[tom_positive]{\MYBROWN{\textbf{Thm.\,\ref*{tom_positive}}}}) of the smeared characters $\chi_g(f_1,\ldots,f_n)$:
\begin{align*}
0\leq&\sum_{i,j=1}^N\bar{\xi}_i\xi_j\chi_{g_i^{-1}g_j}(f_1,\ldots,f_n)=\hspace{-0.1cm}\sum_{i,j=1}^N\bar{\xi}_i\xi_jW_{nn}\big(f_1,\ldots,f_n,(g_i^{-1}g_j)^*f_n,\\
&\hspace{0cm}\ldots,(g_i^{-1}g_j)^*f_1\big)=\hspace{-0.1cm}\sum_{i,j=1}^N\bar{\xi}_i\xi_j\hspace{-.4cm}\int\limits_{\hspace{.5cm}\mathbb{M}^{2n}}\hspace{-0.3cm} \overline{f_1(x_1)}\cdots\overline{f_n(x_n)}\\
&\hspace{2.5cm}\cdot W_{nn}\big(x,(g_i^{-1}g_j)^{-1}\cdot x'\big)f_n(x_n')\cdots f_1(x_1')\diff^n x\,\diff^n x'\\
&\hspace{0.2cm}=\hspace{-0.3cm}\int\limits_{\hspace{.5cm}\mathbb{M}^{2n}}\hspace{-0.0cm}\sum_{i,j=1}^N\overline{\big(\xi_if_1(x_1)\cdots f_1(x_n)\big)}W_{nn}\big(x,(g_i^{-1}g_j)^{-1}\cdot x'\big)
\end{align*}

\phantom{a}
\vspace{-1.5cm}\begin{multline*}
\hspace{0.25cm}\cdot\big(\xi_if_n(x_n')\cdots f_1(x_1')\big)\diff^n x\,\diff^n x'\\
=\hspace{-0.3cm}\int\limits_{\hspace{.5cm}\mathbb{M}^{2n}}\hspace{-0.0cm}\sum_{i,j=1}^N\overline{h_i(x)}\chi_{g_i^{-1}g_j}(x,x')h_j(x')\diff^n x\,\diff^n x'.
\end{multline*}

\vspace{-0.7cm}{\hfill\hyperref[positivity_Wight_proof_ps]{{\color{black}$\blacksquare$}}}\newpage

\item[\MYBROWN{(d)}] Local symmetry:
\begin{multline*}
\chi_g(x_1,\ldots,x_j,x_{j+1},\ldots,x_n,x_n',\ldots x_1')\\
=\chi_g(x_1,\ldots,x_{j+1},x_j,\ldots,x_n,x_n',\ldots x_1'),
\end{multline*}
provided that $\|x_j-x_{j+1}\|>0$. This property follows immediately from microscopic causality (axiom \hyperref[microscopic_prop_p]{\MYBROWN{4}}).

\phantomsection
\label{ps_Clustering}
\item[\MYBROWN{(e)}] Clustering:
\begin{equation}
\lim_{a\rightarrow\pm\infty}\chi_{(\mathds{1},a)}(x,x')=W_n(x)\overline{W_n(x')},
\end{equation}
where $W_n(x)$ is the Wightman function:
\begin{equation}
W_n(x)=\left\langle0\right|\boldsymbol{\varphi}(x_1)\cdots\boldsymbol{\varphi}(x_n)|0\rangle.
\end{equation}

This property is obtained from the clustering property of Wightman functions, that is, Wightman functions factorize when the points $x$ and $x'$ are asymptotically apart (see \hyperref[St64_ps]{\RED{[\citen*{St64}, page 111]}} for mathematical details). We will provide a physical interpretation of this property. 

\phantomsection\label{Clustering_proof_ps}
Let us see that
\begin{multline}
\chi_{(\mathds{1},a)}(x,x')=W_{nn}(x,x'-a)\\
=\langle0|\boldsymbol{\varphi}(x_1)\cdots\boldsymbol{\varphi}(x_n)\boldsymbol{\varphi}(x_n'-a)\cdots\boldsymbol{\varphi}(x_1'-a)\left|0\right\rangle.
\end{multline}
The smeared character $\chi_{(\mathds{1},a)}(x,x')$ is the expected value of the state $\boldsymbol{\rho}$ when the second $n$ arguments are translated by the 4-vector $a$. However, if the events in $x$ and $x'-a$ are separated enough to not be causally related, what happens in $x$ will be independent to what happens in $x'-a$, then the expected value should be the product of the expected values in both events: 
\begin{multline}
\langle0|\boldsymbol{\varphi}(x_1)\cdots\boldsymbol{\varphi}(x_n)\boldsymbol{\varphi}(x_n'-a)\cdots\boldsymbol{\varphi}(x_1'-a)\left|0\right\rangle=\\
\langle0|\boldsymbol{\varphi}(x_1)\cdots\boldsymbol{\varphi}(x_n)\left|0\right\rangle\langle0|\boldsymbol{\varphi}(x_n'-a)\cdots\boldsymbol{\varphi}(x_1'-a)\left|0\right\rangle,
\end{multline}\newpage
\noindent which is the property we have enunciated. 

\vspace{-0.2cm}{\hfill\hyperref[Clustering_proof_ps]{{\color{black}$\blacksquare$}}}

The Wightman function $W_n(x)$ for odd $n$ is equal to zero. It is a fact that can be directly determined from the explicit form of the solution of the scalar free field Klein--Gordon equation~\hyperref[KG_classical_2_ps]{(\ref*{KG_classical_2})} with $\boldsymbol{\varphi}$ an operator,
\phantomsection\label{field_operator_Minkowsky_ps}\begin{equation}\label{field_operator_Minkowsky}
\boldsymbol{\varphi}(t,\boldsymbol{x})=\hspace{-0.3cm}\int\limits_{\hspace{.5cm}\mathbb{R}^{3}}\hspace{-0.1cm}\frac{1}{\sqrt{2E_p}}\left(a_p\e^{-ip^\mu x_\mu}+a_p^\dagger \e^{ip^\mu x_\mu}\right)\frac{\diff^3p}{(2\pi)^3},
\end{equation}
where $E_p=p^0=\sqrt{\boldsymbol{p}^2+m^2}$\,, because of the commutation relations of the annihilation and creation operators~\hyperref[properties_1_a_ops_ps]{(\ref*{properties_1_a_ops})} and because the action of the annihilation operator on the vacuum vanishes~\hyperref[property_2_a_ops_ps]{(\ref*{property_2_a_ops})}.

Then, the clustering property for odd $n$ gives:
\begin{equation}
\lim_{a\rightarrow\pm\infty}\chi_{(\mathds{1},a)}(x,x')=0,
\end{equation}
and for even $n$, we may write it in terms of smeared characters:
\begin{multline}
\lim_{a\rightarrow\pm\infty}\chi_{(\mathds{1},a)}(x,x')=\chi_{(\mathds{1},0)}(x_1,\ldots,x_{n/2},x_{n/2+1},\ldots x_{n})\\
\cdot\overline{\chi_{(\mathds{1},0)}(x'_1,\ldots,x'_{n/2},x'_{n/2+1},\ldots x'_{n})}.
\end{multline}

\end{enumerate}

Finally, let us finish this section by writing the tomogram of a pure state of the quantum scalar field corresponding to an element $\xi$ of the Lie algebra of the Poincar\'e group $\mathcal{P}_+^\uparrow$. The tomogram is obtained by applying \hyperref[tom_cart]{\MYBROWN{\textbf{Thm.\,\ref*{tom_cart}}}}:
\phantomsection\label{tomogram_Poincare_ps}\begin{multline}\label{tomogram_Poincare}
\mathcal{W}_{q,f_1,\ldots,f_n}(X;\xi)=\Tr\!\big(\boldsymbol{\rho}\delta(X\mathds{1}-\langle\boldsymbol{\Theta},\xi\rangle)\big)\\
=\left\langle0\right|\boldsymbol{\varphi}(f_1)\cdots\boldsymbol{\varphi}(f_n)|\delta(X\mathds{1}-\langle\boldsymbol{\Theta},\xi\rangle)|\boldsymbol{\varphi}(f_n)\cdots\boldsymbol{\varphi}(f_1)\left|0\right\rangle.
\end{multline}
Recall that the Lie algebra of the Poincar\'e group $\mathcal{P}^\uparrow_{+}$ is the ten-dimensional Lie algebra generated by $\textbf{P}_\mu$ (corresponding to the translation part of $\mathbb{R}^{1+3}$) and $\textbf{M}_{\mu\nu}$ (corresponding to the Lorentz group $SO(1,3)$), where $\mu,\nu=0,\ldots,3$, and with commutation relations:
\begin{align}
&\hspace{0.8cm}\left[\vphantom{a^\dagger}\right.\hspace{-0.1cm}\textbf{P}_\mu,\textbf{P}_\nu\hspace{-0.1cm}\left.\vphantom{a^\dagger}\right]=0,\qquad\left[\vphantom{a^\dagger}\right.\hspace{-0.1cm}\textbf{M}_{\mu\nu},\textbf{P}_\alpha\hspace{-0.1cm}\left.\vphantom{a^\dagger}\right]=i\big(g_{\alpha\mu}\textbf{P}_{\nu}-g_{\alpha\nu}\textbf{P}_{\mu}\big),\nonumber\\
&\left[\vphantom{a^\dagger}\right.\hspace{-0.1cm}\textbf{M}_{\mu\nu},\textbf{M}_{\alpha\beta}\hspace{-0.1cm}\left.\vphantom{a^\dagger}\right]=i\big(g_{\mu\alpha}\textbf{M}_{\nu\beta}-g_{\mu\beta}\textbf{M}_{\nu\alpha}-g_{\nu\alpha}\textbf{M}_{\mu\beta}+g_{\nu\beta}\textbf{M}_{\mu\alpha}\big),
\end{align}
where $\alpha,\beta=0,\ldots,3$, $g_{\mu\nu}$ is the metric tensor, and $\textbf{M}_{\mu\nu}$  is the infinitesimal generator of the proper orthochronous Lorentz group $\mathcal{L}_0$\footnote{See \hyperref[section_Minkowsiki]{{\color{black}\textbf{appendix~\ref*{section_Minkowsiki}}}} or \hyperref[Ca15_ps]{\RED{[\citen*{Ca15}, section 4.6]}} for more details.}:
\begin{equation}
U(\Lambda,a)=\e^{ia_\mu\textbf{P}^\mu}\e
,
\end{equation}
where $a_\mu$ is the four-vector denoting the infinitesimal translations of the origin and $\omega_{\mu\nu}$ is a rank-2 antisymmetric tensor defined by the matrix elements of the infinitesimal Lorentz transformation $\Lambda$:
\begin{equation}
\Lambda_{\mu\nu}=\delta_{\mu\nu}+\omega_{\mu\nu}.
\end{equation}
Therefore, if $\langle\boldsymbol{\Theta},\xi\rangle=\boldsymbol{\xi}$ belongs to the Poincar\'e algebra $\mathcal{P}^\uparrow_{+}$, we get:
\begin{equation}
\boldsymbol{\xi}=a_\mu\textbf{P}^\mu-\frac{1}{2}\omega_{\mu\nu}\textbf{M}^{\mu\nu}
\end{equation}
and the tomogram becomes:
\phantomsection\label{tomogram_Poincare_exp_ps}
\begin{equation}\label{tomogram_Poincare_exp}
\mathcal{W}_{q,f_1,\ldots,f_n}(X;\omega_{\mu\nu},a_\mu)=\Tr\!\big(\boldsymbol{\rho}\delta\big(X\mathds{1}-a_\mu\textbf{P}^\mu+\frac{1}{2}\omega_{\mu\nu}\textbf{M}^{\mu\nu}\big)\big).
\end{equation}
Also recall that again the smeared characters $\chi_{(\omega_{\mu\nu},a_\mu)}(f_1,\ldots,f_n)$ and the tomograms $\mathcal{W}_{q,f_1,\ldots,f_n}(X;\omega_{\mu\nu},a_\mu)$ are related by a Fourier Transform:
\begin{equation}
\mathcal{W}_{q,f_1,\ldots,f_n}(X;\omega_{\mu\nu},a_\mu)=\frac{1}{2\pi}\hspace{-.25cm}\int\limits_{\hspace{-.04cm}-\infty}^{\hspace{.45cm}\infty}\hspace{-0.15cm}\e^{-ikX}\chi_{(k\omega_{\mu\nu},ka_\mu)}(f_1,\ldots,f_n)\diff k.
\end{equation} 
These formulas also hold for the corresponding kernels $\mathcal{W}_{q,x,x'}(X;\omega_{\mu\nu},a_\mu)$ and $\chi_{(\omega_{\mu\nu},a_\mu)}(x,x').$\newpage

\section{A reconstruction theorem for states in Quantum Field Theory}\label{resumen_rec_field}

Here, we will present a variation of the reconstruction theorem of Wight-man--Streater stated in terms of smeared characters and we will give a sketch of the proof. After that, to finish the reconstruction process for states, we will see in which cases we can find an orthogonality condition that allows us to obtain the state $\boldsymbol{\rho}$ through an expansion in the representation elements $U(g)$, $g$ in $\mathcal{P}_+^\uparrow$.
\\

\noindent\MYBROWN{\textbf{Theorem\hspace{0.05cm} 5.5.1.}} \emph{Let $\left\{\vphantom{a^\dagger}\chi_g(x_1,\ldots,x_n,x_n',\ldots,x_1')\right\}$, $n=1,2,\ldots$ be a family of distributions with $g=(\Lambda,a)\in\mathcal{P}^\uparrow_{+}$ for any $x_1,x_2,\ldots,x_n$ and $x_1',x_2',\ldots,x_n'$ in Minkowski space-time $\mathbb{M}$. Suppose that these distributions satisfy the properties} \hyperref[ps_Covariance]{{\color{mybrown}{\textrm{(a)}}}}\hspace{0.02cm}\MYBROWN{--}\hspace{0.03cm}\hyperref[ps_Clustering]{{\color{mybrown}{\textrm{(e)}}}}\hspace{-0.02cm}\emph{ stated before for all finite sequences $f_1(x_1),f_2(x_2),\ldots$ of test functions. Then, there exist a separable Hilbert space $\mathcal{F}_\infty$, a continuous unitary representation $U$ of $\mathcal{P}_+^\uparrow$ in that Hilbert space, a unique state $|0\rangle$ invariant under $U(\Lambda,a)$, and a scalar field $\boldsymbol{\varphi}$ such that:
$$
\left\langle0\right|\boldsymbol{\varphi}(x_1)\cdots\boldsymbol{\varphi}(x_n)\boldsymbol{\varphi}(g^{-1}\cdot x_n')\cdots\boldsymbol{\varphi}(g^{-1}\cdot x_1')\left|0\right\rangle=\chi_g(x_1,\ldots,x_n,x_n',\ldots,x_1').
$$}
\setcounter{theorem}{1}

\vspace{-0.6cm}
\phantomsection\label{proof_rec_wight_ps}
\noindent\MYBROWN{\textbf{Proof}}: The proof of this theorem is inspired directly in the original theorem stated in \hyperref[St64_ps]{\RED{[\citen*{St64}, page 117]}}. We will show only part of the proof, mainly the reconstruction of the Hilbert space $\mathcal{F}_\infty$, the vacuum state $|0\rangle$, the representation $U$ and the scalar field $\boldsymbol{\varphi}$.

Let us begin with a vector space $H$ formed by sequences $(h_0,h_1,\ldots)$ of test functions where $h_0$ is any constant function and $h_n\in\mathbb{M}^n$, $n=1,2,\ldots$\,:
\begin{align}
h_1(x_1)&=f_1(x_1),\nonumber\\
h_2(x_1,x_2)&=f_1(x_1)f_2(x_2),\nonumber\\
&\hspace{0.2cm}\vdots\nonumber\\
h_n(x_1,x_2,\ldots,x_n)&=f_1(x_1)f_2(x_2)\cdots f_n(x_n).
\end{align}
Addition and multiplication by scalars, are defined in the usual way:
\begin{align}
(h_0,h_1,\ldots)+(k_0,k_1,\ldots)&=(h_0+k_0,h_1+k_1,\ldots),\nonumber\\
\alpha(h_0,h_1,\ldots)&=(\alpha h_0,\alpha h_1,\ldots).
\end{align}

To obtain a Hilbert space, we need to define an inner product $\langle\cdot,\cdot\rangle$. Because of the positivity property \hyperref[Positivity_cond_Wight_ps]{{\color{mybrown}{{(c)}}}}, it is natural to define:
\phantomsection\label{inner_rec_the_ps}\begin{multline}\label{inner_rec_the}
\langle h,k\rangle=\sum_{i,j=0}^\infty\hspace{-0.35cm}\int\limits_{\hspace{.5cm}\mathbb{M}^{2n}}\hspace{-0.2cm}\overline{h_i(x_1,\ldots,x_i)}\chi_{g_i^{-1}g_j}(x_1,\ldots,x_i,x_j',\ldots,x_1')\\
\cdot k_j(x_1',\ldots,x_j')\diff x_1\cdots\diff x_i\diff x_j'\cdots\diff x_1'.
\end{multline}
Notice that this inner product satisfies:
\begin{equation}
\langle h,k\rangle=\overline{\langle k,h\rangle},
\end{equation}
thanks to the Hermiticity condition \hyperref[Hermiticity_con_ps]{{\color{mybrown}{\textrm{(b)}}}}. And also, from the positivity condition \hyperref[Positivity_cond_Wight_ps]{{\color{mybrown}{\textrm{(c)}}}}, $\|h\|^2=\langle h,h\rangle\geq0$.

Let us define now the linear transformation $U(\Lambda,a)$ on the vector space given by:
\begin{equation}
U(\Lambda,a)(h_0,h_1,h_2.\ldots)=(h_0,(\Lambda,a)^*h_1,(\Lambda,a)^*h_2,\ldots),
\end{equation}
where
\begin{equation}
(\Lambda,a)^*h_n(x_1,\ldots,x_n)=h_n\big(\Lambda^{-1}(x_1-a),\ldots,\Lambda^{-1}(x_n-a)\big).
\end{equation}
If we denote the vector $(1,0,0,\ldots)$ by $|0\rangle$, we have:
\begin{equation}
U(\Lambda,a)|0\rangle=|0\rangle.
\end{equation}
Notice also that the operator $U(\Lambda,a)$ leaves invariant the inner product defined before in~\hyperref[inner_rec_the_ps]{(\ref*{inner_rec_the})} by virtue of the covariance condition \hyperref[ps_Covariance]{{\color{mybrown}{\textrm{(a)}}}}, and it is a representation of the Poincar\'e group $\mathcal{P}_+^\uparrow$ because it verifies:
\phantomsection\label{transformation_law_poincare_2_ps}\begin{equation}\label{transformation_law_poincare_2}
U(\Lambda_1,a_1)U(\Lambda_2,a_2)=U(\Lambda_1\Lambda_2,a_1+\Lambda_1a_2).
\end{equation}

Let us introduce now the linear operator $\boldsymbol{\varphi}(f)$ defined for each test function $f$:
\begin{equation}
\boldsymbol{\varphi}(f)(h_0,h_1,h_2,\ldots)=(0,fh_0,f\otimes h_1,f\otimes h_2,\ldots),
\end{equation}
where
\begin{equation}
(f\otimes h_n)(x_1,\ldots,x_{n+1})=f(x_1)h_n(x_2,\ldots,x_{n+1})
\end{equation}
is a test function too. If we apply the operator $U(\Lambda,a)$ to the operator $\boldsymbol{\varphi}(f)$, we get:
\begin{align}
U(\Lambda,a)\boldsymbol{\varphi}(f)(h_0,h_1,\ldots)&=U(\Lambda,a)(0,fh_0,f\otimes h_1,f\otimes h_2,\ldots)\nonumber\\
&=(0,(\Lambda,a)^*fh_0,(\Lambda,a)^*f\otimes h_1,\ldots)\nonumber\\
&=\boldsymbol{\varphi}\big((\Lambda,a)^*f\big)(h_0,(\Lambda,a)^*h_1,\ldots)\nonumber\\
&=\boldsymbol{\varphi}\big((\Lambda,a)^*f\big)U(\Lambda,a)(h_0,h_1,\ldots),
\end{align}
hence, we have that
\begin{equation*}
U(\Lambda,a)\boldsymbol{\varphi}(f)U(\Lambda,a)^\dagger=\boldsymbol{\varphi}\big((\Lambda,a)^*f\big),
\end{equation*}
i.e., the linear operator $\boldsymbol{\varphi}(f)$ satisfy the covariant transformation law (axiom \hyperref[Covariance_axiom_WS_ps]{\MYBROWN{3}}), then $\boldsymbol{\varphi}$ is a quantum scalar field.

At this point, we have found a representation $U$ of the Poincar\'e group $\mathcal{P}_+^\uparrow$, a vector $|0\rangle$ that is invariant under the transformation of the group, a scalar field $\boldsymbol{\varphi}$ and a vector space $H$ that can be created by the recurrent action of the field over the state $|0\rangle$. Therefore, to conclude the proof we should show that the Hilbert space $\mathcal{F}_\infty$ is the completion of the quotient of that vector space $H$ with the space of distributions different from zero with norm 0. This process can be done in a similar way as the \hyperref[ps_GNS]{{\color{black}\textbf{GNS}}} \hyperref[ps_GNS]{{\color{black}\textbf{construction}}} described in \hyperref[section_C-algebras]{{\color{black}\textbf{section~\ref*{section_C-algebras}}}}, for that, we will not repeat it. 

Also, it remains to prove that the state $|0\rangle$, invariant under the group, is unique, however we will not show that in this text.

\vspace{-0cm}{\hfill\hyperref[proof_rec_wight_ps]{{\color{black}$\blacksquare$}}}

To finish this section, we will find a biorthogonal condition for the unitary representation of the Poincar\'e group to obtain the formula that allows to reconstruct the state $\boldsymbol{\rho}$. Because the states prepared in the laboratory are static in time, we will consider the subgroup of the Poincar\'e group with no translation in time, then let us compute the following trace:
\phantomsection\label{trace_1_poinc_ps}
\begin{equation}\label{trace_1_poinc}
\Tr\!\big(U(\Lambda,a)U(\Lambda',a')^\dagger\big),
\end{equation}
where $a=(0,\boldsymbol{a})$. 

First of all, let us make a few remarks about the standard situation when $d=3$ ($\mathbb{M}=\mathbb{R}^{1+3}$), however the generalization to any $d$ can be made in a natural way. We will consider the Lorentz invariant resolution of the identity and the inner product of Lorentz invariant momentum vectors:
\begin{equation}
\mathds{1}=\frac{1}{(2\pi)^3}\hspace{-0.28cm}\int\limits_{\hspace{.5cm}\mathbb{R}^3}\hspace{-0.17cm}|\boldsymbol{p}\rangle\langle\boldsymbol{p}|\frac{\diff^3\boldsymbol{p}}{2E_p},\qquad \langle\boldsymbol{p}|\boldsymbol{p}'\rangle=2E_p(2\pi)^3\delta(\boldsymbol{p}-\boldsymbol{p}'),
\end{equation}
with $E_p=\sqrt{\boldsymbol{p}^2+m^2}$\,. Now, we can compute the trace~\hyperref[trace_1_poinc_ps]{(\ref*{trace_1_poinc})}:
\begin{equation*}
\Tr\big(U(\Lambda,a)U(\Lambda',a')^\dagger\big)=\frac{1}{(2\pi)^3}\hspace{-0.28cm}\int\limits_{\hspace{.5cm}\mathbb{R}^3}\hspace{-0.17cm}\langle\boldsymbol{p}|U(\Lambda,a)U(\Lambda',a')^\dagger|\boldsymbol{p}\rangle\frac{\diff^3\boldsymbol{p}}{2E_p}.
\end{equation*}
Hence, using the transformation law of the Poincar\'e group~\hyperref[transformation_law_poincare_2_ps]{(\ref*{transformation_law_poincare_2})}, and splitting the representation $U$ in the part corresponding to the translation subgroup and the part corresponding to the Lorentz group:
\begin{equation}
U(\Lambda,a)=\e^{ia^\mu \textbf{P}_\mu}U(\Lambda),
\end{equation}
we have:
\begin{align*}
\hspace{0cm}\frac{1}{(2\pi)^3}\hspace{-0.28cm}\int\limits_{\hspace{.5cm}\mathbb{R}^3}\hspace{-0.17cm}\langle\boldsymbol{p}|&U(\Lambda,a)U(\Lambda',a')^\dagger|\boldsymbol{p}\rangle\frac{\diff^3\boldsymbol{p}}{2E_p}\\
&\hspace{0cm}=\frac{1}{(2\pi)^3}\hspace{-0.28cm}\int\limits_{\hspace{.5cm}\mathbb{R}^3}\hspace{-0.17cm}\langle\boldsymbol{p}|U(\Lambda\Lambda'^{-1},a-\Lambda\Lambda'^{-1}a')|\boldsymbol{p}\rangle\frac{\diff^3\boldsymbol{p}}{2E_p}\\
&=\frac{1}{(2\pi)^3}\hspace{-0.28cm}\int\limits_{\hspace{.5cm}\mathbb{R}^3}\hspace{-0.17cm}\langle\boldsymbol{p}|\e^{i( a^\mu-(\Lambda\Lambda'^{-1})_\nu^\mu a'^\nu)\textbf{P}_\mu} U(\Lambda\Lambda'^{-1})|\boldsymbol{p}\rangle\frac{\diff^3\boldsymbol{p}}{2E_p}
\end{align*}
\begin{align}
&=\frac{1}{(2\pi)^3}\hspace{-0.28cm}\int\limits_{\hspace{.5cm}\mathbb{R}^3}\hspace{-0.17cm}\langle\boldsymbol{p}|\e^{i( a^\mu-(\Lambda\Lambda'^{-1})_\nu^\mu a'^\nu)\textbf{P}_\mu}|\Lambda'\Lambda^{-1}\boldsymbol{p}\rangle\frac{\diff^3\boldsymbol{p}}{2E_p}\nonumber\\
&=\frac{1}{(2\pi)^3}\hspace{-0.28cm}\int\limits_{\hspace{.5cm}\mathbb{R}^3}\hspace{-0.17cm}\e^{i( a^\mu-(\Lambda\Lambda'^{-1})_\nu^\mu a'^\nu)(\Lambda'\Lambda^{-1})_\mu^{\tau}p_\tau}\langle\boldsymbol{p}|\Lambda'\Lambda^{-1}\boldsymbol{p}\rangle\frac{\diff^3\boldsymbol{p}}{2E_p}\nonumber\\
&=\hspace{-0.28cm}\int\limits_{\hspace{.5cm}\mathbb{R}^3}\hspace{-0.17cm}\e^{i( a^\mu-(\Lambda\Lambda'^{-1})_\nu^\mu a'^\nu)(\Lambda'\Lambda^{-1})_\mu^{\tau}p_\tau}\delta\big((\mathds{1}-\Lambda'\Lambda^{-1})\boldsymbol{p}\big)\diff^3\boldsymbol{p}.
\end{align}
Notice that the delta function $\delta\big((\mathds{1}-\Lambda'\Lambda^{-1})\boldsymbol{p}\big)$ is different from zero only if $\Lambda=\Lambda'$, hence we have formally:
\begin{equation}
\delta\big(\Lambda'\Lambda^{-1}\big)=\left\{\begin{matrix}
\infty&\Lambda=\Lambda',\\
0&\Lambda\neq\Lambda'.
\end{matrix}\right.
\end{equation}
Thus,
$$
\delta\big(\Lambda'\Lambda^{-1}\big)\hspace{-0.28cm}\int\limits_{\hspace{.5cm}\mathbb{R}^3}\hspace{-0.17cm}\e^{i( a^\mu-(\Lambda\Lambda'^{-1})_\nu^\mu a'^\nu)(\Lambda'\Lambda^{-1})_\mu^{\tau}p_\tau}\diff^3\boldsymbol{p}=\delta\big(\Lambda'\Lambda^{-1}\big)\hspace{-0.28cm}\int\limits_{\hspace{.5cm}\mathbb{R}^3}\hspace{-0.17cm}\e^{i( a^\mu-a'^\mu)p_\mu}\diff^3\boldsymbol{p},
$$
and because we are supposing that there is not temporal translation, we get:
\begin{multline}
\delta\big(\Lambda'\Lambda^{-1}\big)\hspace{-0.28cm}\int\limits_{\hspace{.5cm}\mathbb{R}^3}\hspace{-0.17cm}\e^{i( a^\mu-a'^\mu)p_\mu}\diff^3\boldsymbol{p}=\delta\big(\Lambda'\Lambda^{-1}\big)\hspace{-0.28cm}\int\limits_{\hspace{.5cm}\mathbb{R}^3}\hspace{-0.17cm}\e^{i( \boldsymbol{a}-\boldsymbol{a}'^)\boldsymbol{p}}\diff^3\boldsymbol{p}\\
=(2\pi)^3\delta(\Lambda'\Lambda^{-1})\delta(\boldsymbol{a}-\boldsymbol{a}').
\end{multline} 
Then
\phantomsection\label{trace_poincare_ps}\begin{equation}\label{trace_poincare}
\Tr\!\big(U_0(\Lambda,\boldsymbol{a})U_0(\Lambda',\boldsymbol{a})^\dagger\big)=(2\pi)^3\delta\big(\Lambda'\Lambda^{-1}\big)\delta(\boldsymbol{a}-\boldsymbol{a}'),
\end{equation}
where $U_0(\Lambda,\boldsymbol{a})$ is the representation of the Poincar\'e group with no temporal displacement:
\begin{equation}
U_0(\Lambda,\boldsymbol{a})=\e^{i\boldsymbol{a}\cdot\textbf{P}}\e
.
\end{equation}
Finally, we can write a formula similar to~\hyperref[reconstruction_rho_HW_ps]{(\ref*{reconstruction_rho_HW})} for the subgroup of the Poincar\'e group with no temporal displacement:
\begin{multline}
\boldsymbol{\rho}=\frac{1}{(2\pi)^3}\hspace{-0.1cm}\int\limits_{\hspace{-.26cm}-\infty}^{\hspace{.15cm}\infty}\hspace{-.1cm}\int\limits_{\hspace{-.08cm}-\infty}^{\hspace{.15cm}\infty}\hspace{0.02cm}\chi_{(\Lambda,\boldsymbol{a})}(f_1,\ldots,f_n)U_0(\Lambda,\boldsymbol{a})^\dagger\diff^6\Lambda\diff^3\boldsymbol{a}\\
=\frac{1}{(2\pi)^3}\hspace{-0.3cm}\int\limits_{\hspace{.5cm}\mathbb{R}^3}\hspace{-0.17cm}\e^{iX\mathds{1}}U_0(\Lambda,\boldsymbol{a})\mathcal{W}_{q,f_1,\ldots,f_n}(X;\Lambda,\boldsymbol{a})\diff X\diff^ 6\Lambda\diff^3\boldsymbol{a},
\end{multline}
where $\mathcal{W}_{q,f_1,\ldots,f_n}(X;\Lambda,\boldsymbol{a})$ is the tomogram given by~\hyperref[tomogram_Poincare_exp_ps]{(\ref*{tomogram_Poincare_exp})} with $a=(0,\boldsymbol{a})$.

\section{Canonical tomograms of a real scalar field}\label{resumen_bosonic}

Let us consider the commutator of the annihilation and creation operators given by:
\phantomsection\label{commutator_creation_annihilation_bosonic_field_2_ps}\begin{equation}\label{commutator_creation_annihilation_bosonic_field_2}
\left[\vphantom{a^\dagger}\right.\hspace{-0.1cm}a_{\boldsymbol{p}},a_{\boldsymbol{p}'}^\dagger\hspace{-0.1cm\left.\vphantom{a^\dagger}}\right]=(2\pi)^3\delta^{(3)}(\boldsymbol{p}-\boldsymbol{p}'),\qquad\left[\vphantom{a^\dagger}\right.\hspace{-0.1cm}a_{\boldsymbol{p}},a_{\boldsymbol{p}'}\hspace{-0.1cm\left.\vphantom{a^\dagger}}\right]=\left[\vphantom{a^\dagger}\right.\hspace{-0.1cm}a_{\boldsymbol{p}}^\dagger,a_{\boldsymbol{p}'}^\dagger\hspace{-0.1cm\left.\vphantom{a^\dagger}}\right]=0,
\end{equation}
and let be the vector $\boldsymbol{\varphi}(f)|0\rangle$ and its corresponding pure state $\boldsymbol{\rho}$:
\begin{equation}
\boldsymbol{\rho}=|\boldsymbol{\varphi}(f)|0\rangle\langle 0|\boldsymbol{\varphi}(f)|=\hspace{-.4cm}\int\limits_{\hspace{.5cm}\mathbb{M}^{2}}\hspace{-0.3cm} \overline{f(x)}|\boldsymbol{\varphi}(x)|0\rangle\langle 0|\boldsymbol{\varphi}(y)|f(y)\diff x\diff y.
\end{equation}
The canonical tomogram of the state $\boldsymbol{\rho}$ of a scalar field is defined as (recall eq.\,\hyperref[tom_0_quantum_scalar_ps]{(\ref*{tom_0_quantum_scalar})} \hyperref[resumen_quantization]{{\color{black} in \textbf{section \ref*{resumen_quantization}}}}):
\begin{multline}
\mathcal{W}_{q,f}(X,w,\overline{w})=\Tr\big(\boldsymbol{\rho}\delta\big(X\mathds{1}-\overline{w}a_{\boldsymbol{p}}-wa^\dagger_{\boldsymbol{p}}\big)\big)\\
=\hspace{-.4cm}\int\limits_{\hspace{.5cm}\mathbb{M}^{2}}\hspace{-0.3cm} \overline{f(x)}\left\langle0\right|\boldsymbol{\varphi}(x)|\delta(X\mathds{1}-\overline{w}a_{\boldsymbol{p}}-wa^\dagger_{\boldsymbol{p}})|\boldsymbol{\varphi}(y)|\left|0\right\rangle f(y)\diff x\diff y,
\end{multline}
hence, let us compute the kernel of the tomogram:
\begin{equation}
\mathcal{W}_{q,x,y}(X,w,\overline{w})=\left\langle0\right|\boldsymbol{\varphi}(x)|\delta(X\mathds{1}-\overline{w}a_{\boldsymbol{p}}-wa^\dagger_{\boldsymbol{p}})|\boldsymbol{\varphi}(y)\left|0\right\rangle.
\end{equation}

In a similar way to the computation of an ensemble of harmonic oscillators of \hyperref[ps_tom_q_h_o]{{\color{black}\textbf{subsection~\ref*{section_Toms_q_h_o}}}} and using the BCH formula~\hyperref[BCH_ps]{(\ref*{BCH})}, we get:

\phantom{a}
\vspace{-1cm}\phantomsection\label{tomogram_bosonic_field_1_ps}\begin{multline}\label{tomogram_bosonic_field_1}
\mathcal{W}_{q,x,y}(X,w,\overline{w})=\frac{1}{2\pi}\hspace{-0.26cm}\int\limits_{\hspace{-.04cm}-\infty}^{\hspace{.45cm}\infty}\hspace{-0.15cm}\e^{ikX}\e
\left\langle0\right|\boldsymbol{\varphi}(x)|\\
                                                                                                                                                                                                                                                                                                  \cdot\e^{-ikwa^\dagger_{\boldsymbol{p}}}\e^{-ik\overline{w}a_{\boldsymbol{p}}}|\boldsymbol{\varphi}(y)\left|0\right\rangle\diff k,
\end{multline}
therefore, because of the action of the annihilation and creation operators on the vacuum:
\phantomsection\label{creation_annihilation_vaccum_ps}\begin{equation}\label{creation_annihilation_vaccum}
a_{\boldsymbol{p}}|0\rangle=0,\qquad a^\dagger_{\boldsymbol{p}}|0\rangle=\frac{1}{\sqrt{2E_{\boldsymbol{p}}}}|\boldsymbol{p}\rangle,
\end{equation}
if we use the definition of the scalar field~\hyperref[field_operator_Minkowsky_ps]{(\ref*{field_operator_Minkowsky})}, the equation \hyperref[tomogram_bosonic_field_1_ps]{(\ref*{tomogram_bosonic_field_1})} becomes:
\phantomsection\label{tomogram_bosonic_field_3_ps}
\begin{multline}\label{tomogram_bosonic_field_3}    
\hspace{0cm}\mathcal{W}_{q,x,y}(X,w,\overline{w})=\frac{1}{2\pi}\hspace{-0.26cm}\int\limits_{\hspace{-.04cm}-\infty}^{\hspace{.45cm}\infty}\hspace{-0.15cm}\e^{ikX}\e
\Bigg(\frac{1}{(2\pi)^6}\hspace{-0.28cm}\int\limits_{\hspace{.5cm}\mathbb{R}^6}\hspace{-0.17cm}\e^{-ip'\cdot x}\e^{ip''\cdot y}\\
\hspace{0cm}\cdot\langle 0|a_{\boldsymbol{p}'}\e^{-ikwa^\dagger_{\boldsymbol{p}}}\e^{-ik\overline{w}a_{\boldsymbol{p}}}a^\dagger_{\boldsymbol{p''}}|0\rangle\frac{\diff^3\boldsymbol{p}'}{\sqrt{2E_{\boldsymbol{p}'}}}\frac{\diff^3\boldsymbol{p}''}{\sqrt{2E_{\boldsymbol{p}''}}}\Bigg)\diff k.                                                                            
\end{multline}
Notice that to simplify the notation, we have written
$$
p\cdot x\coloneq p^\mu x_\mu.
$$

From the definition of the commutator of the creation and annihilation operators~\hyperref[commutator_creation_annihilation_bosonic_field_2_ps]{(\ref*{commutator_creation_annihilation_bosonic_field_2})}, we get that
\begin{equation*}
\left[\vphantom{a^\dagger}\right.\hspace{-0.1cm}a^n_{\boldsymbol{p}},a^\dagger_{\boldsymbol{p}'}\hspace{-0.1cm\left.\vphantom{a^\dagger}}\right]=na^{n-1}_p(2\pi)^3\delta^{(3)}(\boldsymbol{p}-\boldsymbol{p}'),
\end{equation*}
hence,
\begin{align*}
\langle 0|a_{\boldsymbol{p}'}\e^{-ikwa^\dagger_{\boldsymbol{p}}}&=\langle 0|\big(a_{\boldsymbol{p}'}-ikw(2\pi)^3\delta^{(3)}(\boldsymbol{p}-\boldsymbol{p}')\big),\\
\e^{-ik\overline{w}a_{\boldsymbol{p}}}a^\dagger_{\boldsymbol{p}'}|0\rangle&=\big(a^\dagger_{\boldsymbol{p}'}-ik\overline{w}(2\pi)^3\delta^{(3)}(\boldsymbol{p}-\boldsymbol{p}')\big)|0\rangle.
\end{align*}
Then, substituting this result in~\hyperref[tomogram_bosonic_field_3_ps]{(\ref*{tomogram_bosonic_field_3})}, we get:
\begin{align*}
&\hspace{-0.4cm}\mathcal{W}_{q,x,y}(X,w,\overline{w})=\frac{1}{2\pi}\hspace{-0.26cm}\int\limits_{\hspace{-.04cm}-\infty}^{\hspace{.45cm}\infty}\hspace{-0.15cm}\e^{ikX}\e
\Bigg(\frac{1}{(2\pi)^6}\hspace{-0.28cm}\int\limits_{\hspace{.5cm}\mathbb{R}^6}\hspace{-0.17cm}\e^{-ip'\cdot x}\e^{ip''\cdot y}\nonumber\\
&\hspace{1.5cm}\cdot\langle 0|\big(a_{\boldsymbol{p}'}-ikw(2\pi)^3\delta^{(3)}(\boldsymbol{p}-\boldsymbol{p}')\big)\nonumber\\                                                                                                                                                                                                                                                                           &\hspace{1.5cm}\cdot\big(a^\dagger_{\boldsymbol{p}''}-ik\overline{w}(2\pi)^3\delta^{(3)}(\boldsymbol{p}-\boldsymbol{p}'')\big)|0\rangle\frac{\diff^3\boldsymbol{p}'}{\sqrt{2E_{\boldsymbol{p}'}}}\frac{\diff^3\boldsymbol{p}''}{\sqrt{2E_{\boldsymbol{p}''}}}\Bigg)\diff k.
\end{align*}
Using the formula \hyperref[commutator_creation_annihilation_bosonic_field_2_ps]{(\ref*{commutator_creation_annihilation_bosonic_field_2})}, we obtain:
\begin{multline}
\langle 0|\big(a_{\boldsymbol{p}'}-ikw(2\pi)^3\delta^{(3)}(\boldsymbol{p}-\boldsymbol{p}')\big)\big(a^\dagger_{\boldsymbol{p}''}-ik\overline{w}(2\pi)^3\delta^{(3)}(\boldsymbol{p}-\boldsymbol{p}'')\big)|0\rangle\\
=(2\pi)^3\delta^{(3)}(\boldsymbol{p}''-\boldsymbol{p}')-(2\pi)^6k^2|w|^2\delta^{(3)}(\boldsymbol{p}-\boldsymbol{p}')\delta^{(3)}(\boldsymbol{p}-\boldsymbol{p}''),
\end{multline}
and finally, using the representation for the propagator for the Klein--Gordon field:
\begin{equation}
D(x-y)\coloneq\langle 0|\boldsymbol{\varphi}(x)\boldsymbol{\varphi}(y)|0\rangle=\frac{1}{(2\pi)^3}\hspace{-0.2cm}\int\limits_{\hspace{-.04cm}-\infty}^{\hspace{.45cm}\infty}\hspace{-0.1cm}\e^{-ip\cdot(x-y)}\frac{\diff^3\boldsymbol{p}}{2E_{\boldsymbol{p}}},
\end{equation}
and after integrating over the variable $k$, we get the following expression for the kernel of the tomogram:
\begin{multline}
\mathcal{W}_{q,x,y}(X,w,\overline{w})=\frac{1}{\sqrt{2\pi}|w|}\Bigg(D(x-y)\\
+\frac{X}{|w|^2}\frac{\e^{-ip\cdot(x-y)}}{2E_{\textbf{p}}}\left(X-|w|^2\right)\Bigg)\e
.
\end{multline}

%% file: Appendix.tex
\appendix

\chapter{The Minkowski space-time and the Poincar\'e group}\label{section_Minkowsiki}
\markboth{The Minkowski space-time and the Poincar\'e group}{Appendix}

Newtonian Mechanics proposes a concept in which the time is absolute, however it is well-known that closer to the speed of light this concept is no longer available. That framework is known as \textit{special relativity} and it is based in two fundamental postulates: the speed of light $c$ in vaccum is a universal constant and that the laws of physics are the same in all inertial frames (systems moving at constant velocity).

When dealing with fields that propagate at speed of light, we are in the special relativity framework and there, instead of working in an Euclidean space, we have to work in a Minkowski space-time (see for instance \hyperref[Ca15_ps]{\RED{[\citen*{Ca15},}} \hyperref[Ca15_ps]{\RED{sec.\,4.6]}}).

Minkowski space-time is a four dimensional manifold $\mathbb{M}=\mathbb {R}^{1+3}$ with the pseudo-Riemannian metric:
\phantomsection\label{pseudo_metric_ps}\begin{equation}\label{pseudo_metric}
g_{\mu\nu}=\begin{pmatrix}
-1& & &\\
& 1 & &\\
& & 1 &\\
& & & 1
\end{pmatrix}.\tag{A.1}
\end{equation}

Events in this space-time are points in Minkowski space-time $\mathbb{M}$, $x^\mu=(ct,x,y,z)$, that are usually called in the literature ``contravariant vectors''. The elements of the cotangent bundle are usually called ``covariant vectors'' and they are usually written with the index below, $x_\mu=(-ct,x,y,z)$, to emphasize that the indexes of contravariant and covariant vectors cancel out when an element of the cotangent bundle acts on an element of the tangent bundle (here we are identifying vectors in the tangent bundle with points in $\mathbb{M}$):
\begin{equation}
x_\mu=g_{\mu\nu}x^\nu.\tag{A.2}
\end{equation}

The distance between two events $A, B\in\mathbb{M}$ is usually called \textit{interval} and is defined as the product of the vector $x^\mu$, that joins the two events, with itself with respect to the metric~\hyperref[pseudo_metric_ps]{(\ref*{pseudo_metric})}:
\phantomsection\label{interval_ps}\begin{equation}\label{interval}
s^2_{AB}=\|x\|^2\coloneq x_\mu x^\mu.\tag{A.3}
\end{equation}
Vectors can be classified as:

\begin{enumerate}

\item[\MYBROWN{1.}] Temporal: vectors $x\in\mathbb{M}$ such that $\|x\|<0$.

\phantomsection\label{spatial_vector_appendix}
\item[\MYBROWN{2.}] Spatial: vectors $x\in\mathbb{M}$ such that $\|x\|>0$.

\item[\MYBROWN{3.}] Light, isotropic or null: vectors $x\in\mathbb{M}$ such that $\|x\|=0$ if $x\neq 0$.

\item[\MYBROWN{4.}] $x=0$.

\end{enumerate}

\begin{figure}[h]
\centering
\includegraphics{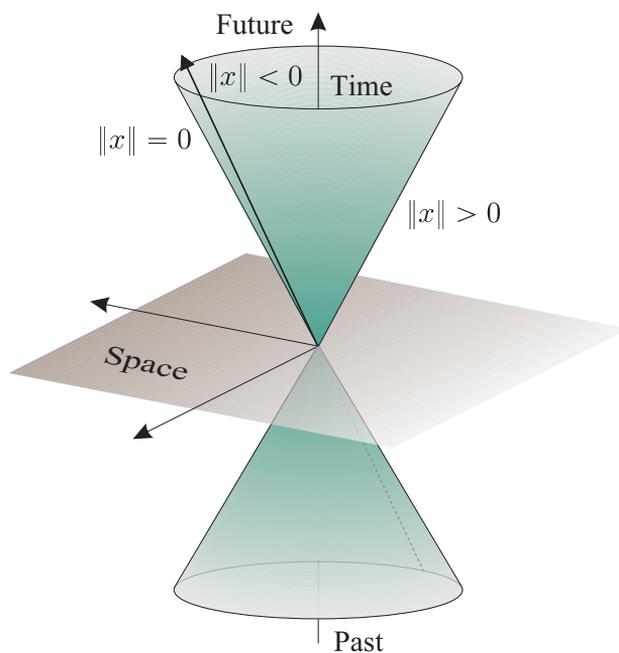}
\caption{\hfil\MYGREEN{Figure A.1}: Light cone.\hfil}
\label{Light_cone}
\end{figure}

Two events are said to be temporal, spatial or light related if the associated vector is temporal, spatial or light. Light events are events, as its name tells, that only can be linked by particles traveling at speed of light, temporal events are events that are related in a causal-effect way, and spatial events are events that can not be related in a causal-effect way, that is, that they would be only linked by particles traveling ``faster'' than light. These facts are usually shown in the so called light cone diagram, {\changeurlcolor{mygreen}\hyperref[Light_cone]{Figure~A.1}}.

The Poincar\'{e} group $\mathcal{P}$ is the group of Minkowski isometries, i.e., is the group of affine transformations in $\mathbb{M}$ that leaves invariant the interval between two events. This group of isometries can be written explicitly as the set of transformations
\phantomsection\label{Poincare_group_ps}\begin{equation}\label{Poincare_group}
x'_\mu=a^\mu+\Lambda^\mu_{\;\;\nu} ºx^\nu,\tag{A.4}
\end{equation}
where $\Lambda^\mu_{\;\;\nu}$ denotes the transformations leaving the origin invariant and is called the Lorentz group $\mathcal{L}$, and the four-vector $a^\mu$, $\mu=0,\ldots,3$, that denotes the group of translations of the origin.

The Lorentz transformation $\Lambda^\mu_{\;\;\nu}$ satisfies:
\phantomsection\label{Lorentz_1_ps}\begin{equation}\label{Lorentz_1}
g_{\mu\nu}\Lambda^\mu_{\;\;\alpha}\Lambda^\nu_{\;\;\beta}=g_{\alpha\beta},\tag{A.5}
\end{equation}
and its explicit form for a Lorentz boost in $z$-direction is the following:
\phantomsection\label{Lorentz_2_ps}\begin{equation}\label{Lorentz_2}
\begin{pmatrix}
x'^0\\
x'^1\\
x'^2\\
x'^3
\end{pmatrix}=\begin{pmatrix}
\sinh\beta& 0& 0&\cosh\beta\\
0&1&0&0\\
0&0&1&0\\
\cosh\beta&0&0&\sinh\beta
\end{pmatrix}\begin{pmatrix}
x^0\\
x^1\\
x^2\\
x^3
\end{pmatrix},\tag{A.6}
\end{equation}
where the physical meaning of $\beta$ is given by:
\phantomsection\label{beta_ps}\begin{equation*}
\tanh\beta=\frac{v}{c}\,,
\end{equation*}
where $v$ is the relative velocity between the two systems $x_\mu$ and $x'_\mu$.

The condition \hyperref[Lorentz_1_ps]{(\ref*{Lorentz_1})} can be written in matrix form as
\begin{equation}
\Lambda^TG\Lambda=G,\tag{A.7}
\end{equation}
hence, it is immediate to see that
\begin{equation}
|\!\det\Lambda|=1.\tag{A.8}
\end{equation}

Transformations such that $\det\Lambda=+1$ define a unimodular subgroup in $\mathcal{L}$ denoted by $\mathcal{L}_+$. 

Because Lorentz transformations preserve the metric, then in particular $|\Lambda_0^0|\geq 1$, and as a topological space, $\mathcal{L}$ is not connected, however it has four connected components characterized as follows:
\begin{alignat}{4}
\mathcal{L}_0=&\,\mathcal{L}^{\uparrow}_+& &\hspace{0.9cm}\mbox{Proper orthochronous,} &\hspace{0.3cm} \det\Lambda&=1, &\hspace{0.4cm}\Lambda_0^0&\geq 1,\nonumber\\
&\,\mathcal{L}^{\uparrow}_{-}& &\hspace{0.7cm}\mbox{Improper orthochronous,}&\hspace{0.3cm} \det\Lambda&=-1, &\hspace{0.4cm}\Lambda_0^0&\geq 1,\nonumber\\
&\,\mathcal{L}^{\downarrow}_{+}& &\hspace{0.5cm}\mbox{Proper antiorthochronous,}&\hspace{0.3cm} \det\Lambda&=1, &\hspace{0.4cm}\Lambda_0^0&\leq-1,\nonumber\\
&\,\mathcal{L}^{\downarrow}_{-}& &\hspace{0.3cm}\mbox{Improper antiorthochronous,}&\hspace{0.3cm} \det\Lambda&=-1, &\hspace{0.4cm}\Lambda_0^0&\leq-1.\tag{A.9}
\end{alignat}
Only the proper orthochronous component $\mathcal{L}_0$, that contains the identity, is a subgroup.

If we consider an infinitesimal Lorentz transformation 
\begin{equation}
\Lambda_{\mu\nu}=\delta_{\mu\nu}+\omega_{\mu\nu},\tag{A.10}
\end{equation}
if we approximate the Lorentz condition \hyperref[Lorentz_1_ps]{(\ref*{Lorentz_1})} to first order in $\omega$, we get:
\begin{equation}
g_{\alpha\beta}=g_{\mu\nu}(\delta^\mu_\alpha+\omega^\mu_\alpha)(\delta^\nu_\beta+\omega^\nu_\beta)=g_{\alpha\beta}+\omega_{\alpha\beta}+\omega_{\beta\alpha}+O(\omega^2),\tag{A.11}
\end{equation}
but since the metric $g_{\mu\nu}$ is symmetric with respect to the change of its two indexes, we have:
\begin{equation}
\omega_{\alpha\beta}+\omega_{\beta\alpha}=0,\tag{A.12}
\end{equation}
hence, $\omega_{\mu\nu}$ is a rank-2 antisymmetric tensor with $6$ independent components that determines the Lie algebra of the Lorentz group. Notice that then, the Lorentz group has dimension $6$.

The Poincar\'e group is the semidirect product of the Lorentz group and the group of translations $\mathcal{P}=\mathcal{L}\times\mathbb{R}^4$. We will denote by $\mathcal{P}_{+}^\uparrow$, or simply $\mathcal{P}_0$, its connected component that corresponds to $\mathcal{L}_0\times\mathbb{R}^4$.

Accordingly, the Lie algebra of Poincar\'e group is (as a vector space) the direct sum of the Lie algebra of $\mathcal{L}$ and the Lie algebra of $\mathbb{R}^4$ (identified with $\mathbb{R}^4$ itself). Then, we conclude that the Lie algebra of $\mathcal{P}$ (also $\mathcal{P}_0$) is determined by the generators $\textbf{P}_\mu$ of $\mathbb{R}^4$ and six generators $\textbf{M}_{\mu\nu}$ of the Lie algebra of the Lorentz group satisfiying the commutation relations:
\begin{align}
&\hspace{0.8cm}\left[\vphantom{a^\dagger}\right.\hspace{-0.1cm}\textbf{P}_\mu,\textbf{P}_\nu\hspace{-0.1cm}\left.\vphantom{a^\dagger}\right]=0,\qquad\left[\vphantom{a^\dagger}\right.\hspace{-0.1cm}\textbf{M}_{\mu\nu},\textbf{P}_\alpha\hspace{-0.1cm}\left.\vphantom{a^\dagger}\right]=i\big(g_{\alpha\mu}\textbf{P}_{\nu}-g_{\alpha\nu}\textbf{P}_{\mu}\big),\nonumber\\
&\left[\vphantom{a^\dagger}\right.\hspace{-0.1cm}\textbf{M}_{\mu\nu},\textbf{M}_{\alpha\beta}\hspace{-0.1cm}\left.\vphantom{a^\dagger}\right]=i\big(g_{\mu\alpha}\textbf{M}_{\nu\beta}-g_{\mu\beta}\textbf{M}_{\nu\alpha}-g_{\nu\alpha}\textbf{M}_{\mu\beta}+g_{\nu\beta}\textbf{M}_{\mu\alpha}\big),\tag{A.13}
\end{align}
hence, a generic element $\boldsymbol{\xi}$ of the Poincar\'e algebra can be written as follows:
\begin{equation}
\boldsymbol{\xi}=a_\mu\textbf{P}^\mu-\frac{1}{2}\omega_{\mu\nu}\textbf{M}^{\mu\nu}.\tag{A.14}
\end{equation}

%% file: Conclusions.tex
\refstepcounter{chapter}
\chapter*{The end of a long journey and the beginning of another: conclusions and further work}\label{long_journey_arx}
\addcontentsline{toc}{chapter}{The end of a long journey and the beginning of another: conclusions and further work}
\markboth{The end of a long journey and the beginning of another}{Conclusions}
\refstepcounter{section}
\section*{Conclusions} 
\addcontentsline{toc}{section}{Conclusions}

In this work, we have discussed a still not so well developed way of describing the state of a quantum system different from the usual Schr\"{o}dinger and Heisenberg pictures, and known as the Tomographic picture. In general, observables in Quantum Mechanics are treated as self-adjoint operators on a Hilbert space $\mathcal{H}$. Here, we have considered a different scenario of Quantum Mechanics in which the algebra of observables is a $C^*$--algebra $\mathcal{A}$. We have taken advantage of tools in that algebra, mainly the \hyperref[ps_GNS]{{\color{black}\textbf{GNS}}} \hyperref[ps_GNS]{{\color{black}\textbf{construction}}} \RED{[}\citen{Ge43}\RED{,}\hspace{0.05cm}\citen{Se47}\RED{]} and \hyperref[prop_positive_proof_ps]{{\color{black}\textbf{Naimark's theorem}}} \cite{Na64}, to reach the goal of getting a reconstruction formula of the state $\rho$ of a quantum system by means of a family of observables in the algebra $\mathcal{A}$.

At the beginning of \hyperref[chap_tom_qu]{{\color{black}\textbf{chapter \ref*{chap_tom_qu}}}} in sections \hyperref[section_Sampling_C]{{\color{black}\textbf{\ref*{section_Sampling_C}}}} and \hyperref[ps_GPT]{{\color{black}\textbf{\ref*{ps_GPT}}}}, we have shown that the tomographic description of Quantum Mechanics may be achieved by using two main ingredients: a \hyperref[section_Sampling_C]{{\color{black}\textbf{\textit{Generalized Sampling Theory}}}} and a \hyperref[ps_GPT]{{\color{black}\textbf{\textit{Generalized Positive Transform}}}}. 

The first one consists on recovering the state of the system by sampling it with a family of observables, called a \hyperref[section_Sampling_C]{{\color{black}\textbf{\textit{tomographic set}}}}. The \hyperref[resumen_sampling_fun]{{\color{black}\textbf{\textit{sampling}}}} \hyperref[resumen_sampling_fun]{{\color{black}\textbf{\textit{function}}}} $F_\rho$ introduced there is nothing but the \hyperref[mean_value_conclusion]{{\color{black}\textbf{expected value}}} of the elements of that family of observables on the state $\rho$. One of the problems there is that the \hyperref[resumen_sampling_fun]{{\color{black}\textbf{sampling function}}}, in general, is not a quantity that can be measured directly in the laboratory because is not a probability distribution, for that reason, a second ingredient is needed.

The \hyperref[ps_GPT]{{\color{black}\textbf{Generalized Positive Transform}}} consists basically on a map that transforms \hyperref[ps_conclusions_s_map]{{\color{black}\textbf{sampling functions}}} in probability distributions. Such distributions are the so called \textit{tomograms} in this picture \hyperref[ps_tomogram_conclusion]{(\ref*{tomogram_conclusion})}. This second tool is clearly inspired on the \textit{Classical Radon Transform} \cite{Ra17}.

Once we have identified a procedure to get our purpose of reconstructing a state by means of the \hyperref[ps_tomogram_conclusion]{{\color{black}\textbf{tomograms}}} that can be measured in the laboratory, we have to deal with the problem of how to recover the state of a concrete system and which family of observables should be used. One answer of these questions can be given using Group Theory, \hyperref[section_particular group]{{\color{black}\textbf{sec.\,\ref*{section_particular group}}}}. 

We have seen that for any finite and compact Lie group, we can implement this \hyperref[section_particular group]{{\color{black}\textbf{tomographic picture}}} and \hyperref[adapted_state_a_ps]{{\color{black}\textbf{reconstruct}}} the state of any system, although we have seen too that there are other important groups that allow to reconstruct the state of a system, as for example the \hyperref[HW_conclusions]{{\color{black}\textbf{Heisenberg--Weyl}}} \hyperref[HW_conclusions]{{\color{black}\textbf{group}}}, which is neither finite nor compact, that appears in many problems in Quantum Mechanics, \hyperref[section_recosntruction_states_groups]{{\color{black}\textbf{sec.\,\ref*{section_recosntruction_states_groups}}}}. We have also shown an \hyperref[Q_homodyne]{{\color{black}\textbf{experimental}}} \hyperref[Q_homodyne]{{\color{black}\textbf{setting}}} to get the desired tomograms in such case in \hyperref[section_hom_det_clas]{{\color{black}\textbf{section \ref*{section_hom_det_clas}}}}.

Let us also point out that a tomographic picture beyond standard Quantum Mechanics giving a tomographic description of classical systems of infinite degrees of freedom have been started, \hyperref[chap_clas]{{\color{black}\textbf{chapter \ref*{chap_clas}}}} (in particular, the free scalar field in a cavity, \hyperref[sec_KG_fields]{{\color{black}\textbf{section \ref*{sec_KG_fields}}}}). We have also given a reconstruction theorem, \hyperref[resumen_rec_field]{{\color{black}\textbf{sec.\,\ref*{resumen_rec_field}}}}, in the case of a \hyperref[section_WS_axioms]{{\color{black}\textbf{quantum scalar field}}} described using the Wightman--Streater \hyperref[section_WS_axioms]{{\color{black}\textbf{axiomatic description}}}.

To finish this summary, it is important to say that thanks to this tomographic description of Quantum Mechanics and inspired by methods and ideas developed in this context, we have been able to solve a problem that, in principle, is not related to this, which is the decomposition of reducible representations of groups into their irreducible components. To achieve that it has been developed the \hyperref[section_smily_alg]{{\color{black}\textbf{SMILY}}} algorithm presented in \hyperref[chap_smily]{{\color{black}\textbf{chapter \ref*{chap_smily}}}}.

This numerical algorithm solves the problem of computing the irreducible components of any finite dimensional unitary representation of a compact Lie group, with respect to a closed subgroup, without any a priori knowledge of its \hyperref[resumen_rep]{{\color{black}\textbf{irreducible representations}}}. We have realized that there is a family of states associated to the unitary representation we want to decompose, called \hyperref[adapted_state_a_ps]{{\color{black}\textbf{\textit{adapted states}}}}, which can be decomposed in states in those irreducible representations. The result we got is that when one transforms generic \hyperref[adapted_state_a_ps]{{\color{black}\textbf{adapted states}}} with a unitary matrix that diagonalizes another one, it emerges a \hyperref[rho_2structure_ps]{{\color{black}\textbf{block structure}}} that is shared with the rest of the \hyperref[adapted_state_a_ps]{{\color{black}\textbf{adapted states}}}. That makes easy to find a \hyperref[CGmatrix_ps]{{\color{black}\textbf{unitary transfor-}}} \hyperref[CGmatrix_ps]{{\color{black}\textbf{mation}}} which transforms all the adapted states in block diagonal matrices such that on each block, we get the states \hyperref[structure_rho_ps]{{\color{black}\textbf{spanned}}} by the corresponding irreducible representation. Thus, that \hyperref[CGmatrix_ps]{{\color{black}\textbf{unitary transformation}}}, usually called the \hyperref[resumen_CGP]{{\color{black}\textbf{Clebsh--Gordan matrix}}}, reduces in \hyperref[resumen_diag]{{\color{black}\textbf{block diagonal}}} matrices all the elements of the initial representation and each block gives one of the irreducible representations decomposing it.

\phantomsection\label{resumen_future}
\refstepcounter{section}
\section*{Further work} 
\addcontentsline{toc}{section}{Further work}
\markboth{The end of a long journey and the beginning of another}{Further work}

We have seen that the tomographic picture we have been explaining may have a good perspectives of future because is a theory that depends directly on the technological capacities in Quantum Optics. For instance, one immediate application of this theory is in the domain of detection of radiation in Quantum Information technologies or for medical purposes.

In \hyperref[chap_birth]{{\color{black}\textbf{chapter \ref*{chap_birth}}}}, an effort has been done to offer a historical perspective of the birth of this theory and we have seen that Quantum Tomography is, in the sense discussed there, a natural prolongation of classical telecommunications. Thus, one path to follow in the future is to adapt techniques from classical telecommunications to quantum optical devices. 

We have shown in this work the implementation of \hyperref[Q_homodyne]{{\color{black}\textbf{homodyne}}} and \hyperref[Q_heterodyne]{{\color{black}\textbf{heterodyne}}} detectors to Quantum Optics by means of suitable configurations of \hyperref[Bs]{{\color{black}\textbf{beam-splitters}}} and \hyperref[res_foto]{{\color{black}\textbf{photodetectors}}}, hence the idea is to extend this configurations to other techniques of detection. 

In the latest section of \hyperref[chap_smily]{{\color{black}\textbf{chapter \ref*{chap_smily}}}} (\hyperref[conclusion_create_states]{{\color{black}\textbf{section \ref*{conclusion_create_states}}}}), we have presented a way to obtain adapted states with $SU(2)$ symmetry. However, we would like to proceed further and see if, in a similar way as we remove frequencies of electric signals, we can try to get only a desired part of a mixed state by removing part of it.    

Another further work is the study of the reconstruction formula of states from the point of view of Sampling Theory. It would be relevant to estimate the errors made in the reconstruction of states when we approximate the integrals that appear in the reconstruction formulas by finite sums.

Future work related with \hyperref[chap_QFT]{{\color{black}\textbf{chapter \ref*{chap_QFT}}}} is, first, to generalize the canonical tomograms obtained in \hyperref[resumen_bosonic]{{\color{black}\textbf{section \ref*{resumen_bosonic}}}} for other fields. Also, it would be interesting to see in which cases it is possible to compute explicitly the covariant tomograms and also to extend the theory beyond the scalar free field developing a tomographic description of perturbation theory.

Also, there are a lot of interesting questions related with the \hyperref[section_smily_alg]{{\color{black}\textbf{SMILY}}} algorithm presented in \hyperref[chap_smily]{{\color{black}\textbf{chapter \ref*{chap_smily}}}}. It is important to highlight that because we have used only arguments of Quantum Mechanics to create the algorithm and we only just need unitary transformations to obtain the CG matrix of a representation of a group, the \hyperref[section_smily_alg]{{\color{black}\textbf{SMILY}}} algorithm has a natural extension to be implemented as a quantum algorithm in a quantum computer. Thus, we could try to prepare the \hyperref[adapted_state_a_ps]{{\color{black}\textbf{adapted states}}} and implement the unitary transformations we need with quantum gates to run it into a quantum computer.

Finally, let us mention another problem that can be addressed using this algorithm, the characterization of quantum entanglement, that is, finding out whether a state is entangled or separable. This is one of the most important open problems in Quantum Information Theory. The idea in which we are thinking is trying to see when we would be able to decompose a state as a linear combination of adapted states of several subgroups of the product group $G\times\cdots\times G$.\newpage  

\refstepcounter{section}
\section*{Publications}
\addcontentsline{toc}{section}{Publications}
\markboth{The end of a long journey and the beginning of another}{Publications}

\begin{itemize}

\item Chapter 2: A. Ibort, A. L\'opez--Yela. Quantum Radon Transform. (In preparation).

\item Chapter 3: A. Ibort, A. L\'opez--Yela and J. Moro. The SMILY algorithm to compute the reduction of unitary representations and their Clebsh--Gordan coefficients. (In preparation).

\item Chapter 4: A. Ibort, A. L\'{o}pez--Yela, V.I. Man'ko, G. Marmo, A. Simoni, E.C.G. Sudarshan and F. Ventriglia. On the tomographic description of classical fields. Phys. Let. A. \textbf{376}, 1417--1425 (2012).

\item Chapter 5: A. Ibort, A. L\'opez--Yela. On the tomographic description of a quantum real scalar field. (In preparation).

\end{itemize} 